THE NATIONAL ACADEMY OF SCIENCES OF UKRAINE

INSTITUTE OF APPLIED PHYSICS

R.I. KHOLODOV
O.P. NOVAK
M.M. DIACHENKO

# RESONANT AND POLARIZATION EFFECTS IN THE PROCESSES OF QUANTUM ELECTRODYNAMICS IN A STRONG MAGNETIC FIELD

Monograph

Sumy - 2021

The monograph considers resonance and polarization effects in quantum electrodynamics processes that take place in a strong external magnetic field. A method for analyzing spin-polarization effects has been developed. The factorization of process cross sections in resonant conditions and the representation of these cross sections in the form of Breit-Wigner are considered. The possibility of testing these effects in modern international projects to test quantum electrodynamics in strong fields is shown.

For researchers, teachers, graduate students and students of physical and physical-technical specialties.

У монографії розглянуто резонансні і поляризаційні ефекти в процесах квантової електродинаміки, що перебігають у сильному зовнішньому магнітному полі. Розроблено метод аналізу спін-поляризаційних ефектів. Розглянуто факторизацію перерізів процесів у резонансних умовах і представлення цих перерізів у формі Брейта-Вігнера. Показана можливість перевірки даних ефектів у сучасних міжнародних проектах з перевірки квантової електродинаміки у сильних полях.

Для наукових робітників, викладачів, аспірантів і студентів фізичних та фізико-технічних спеціальностей.

# FOREWORD

**Quantum ElectroDynamics** (QED) is a theory of electromagnetic interaction that studies processes involving photons, electrons and positrons (leptons), it is the most developed part of **Quantum Field Theory** (QFT) and tested experimentally with high accuracy, in particular by measuring the anomalous magnetic moment of an electron (muon) and the Lamb shift of atomic levels. QED processes with external electromagnetic fields, in particular, the magnetic field, play an important role in the following phenomena: the formation of synchrotron radiation, the use of which has increased significantly in scientific and applied research; radiation of pulsars associated with strongly magnetized neutron stars; collision of heavy ions. The question of experimental verification of quantum electrodynamics in strong electromagnetic fields comparable to the critical Schwinger value ($m^2/e$=4.41·$10^{17}$ *V/m*) is still open. The FAIR mega-construction project (Facility for Antiproton and Ion Research), which is currently under construction, includes in its research program the verification of quantum electrodynamics in extremely strong electromagnetic fields. A characteristic feature of QED processes in a strong electromagnetic field is a strong influence of particles polarization on the process, a presence of resonant peculiarities.

Quantum electrodynamics in an external magnetic field as independent direction of theoretical research begins in the middle of the last century, which is due to the need to create a theory of synchrotron radiation. Synchrotron radiation is widely used today. Sources of X-ray synchrotron radiation are used in radioscopy with elemental analysis, in micromechanics, in microelectronics, in X-ray diffraction analysis of biomolecules. Synchrotron radiation is used to obtain polarized beams of electrons (positrons), for radiation cooling of electrons and positrons, in particular, to create B-factories. Therefore, finding new properties of the synchrotron radiation is, of course, an urgent task of theoretical physics, despite the in-depth study of this issue.

Another object of researches using methods of QED in a magnetic field is a magnetosphere of neutron stars, where the magnetic field reaches a value $10^{12}$-$10^{13}$ *Gs* and above. To date, more than 2,500 pulsars, about two hundred X-rays and gamma pulsars of our galaxy have been detected. The magnetosphere of a neutron star is a unique laboratory to study QED processes in a strong near-critical magnetic field. Both the first-order processes (synchrotron radiation, $e^+e^-$ pair production by a photon (annihilation)) and second-order processes (a photon scattering by an electron, two-photon synchrotron radiation, two-photon production (annihilation) of $e^+e^-$ pair, vacuum birefringence, etc.) of quantized field perturbation theory play a key role in formation of the electron-positron plasma of the magnetosphere, the presence of which explains synchrotron radiation of X-ray pulsars. Cyclotron lines, as well as annihilation lines in the emission spectra of pulsars, comptonization processes, cascades and electromagnetic showers are quite fully studied. However, the subject of studying X-ray pulsars, which includes both astronomical observations and theoretical calculations, remains relevant to this day, as there is no single theory of X-ray pulsar radiation. In particular, the questions of a spin population of the $e^+e^-$ plasma and its effect on synchrotron radiation of X-ray pulsar, as well as on influence of a field of cyclotron photons on the process of resonant formation of the $e^+e^-$ plasma have not been resolved. The process of photon propagation in a strong magnetic field, when there is a cascade production of the $e^+e^-$ pair with subsequent annihilation, has not been studied.

The SPARC collaboration (Stored Particles Atomic Physics Research Collaboration) is one of the key collaborations of the FAIR megaproject under construction. It plans to study the QED phenomena in extremely strong electromagnetic fields. In particular, studies involving photons, electrons and atoms in the presence of strong fast-changing electromagnetic fields, structural studies of heavy ions, as well as studies of the dynamics of collisions of heavy ions in strong fields, including the process of $e^+e^-$ pair production, will be conducted.

It should be noted that when heavy ions collide with target parameters that are larger than the size of the nucleus $\rho \geq R_{nuc}$, quantum-electrodynamic processes in the region between the nuclei in the strong magnetic field of the nuclei are possible. Between the nuclei, the magnetic fields of the moving nuclei add up, and the electric ones compensate. For impact parameters of order $10^{-10}cm$, heavy nuclei with a charge $Z=90$, that collide and move with speed $\sim c/10$, create a magnetic field of order $10^{12}Gs$. In this area, it is quite possible that QED processes with the participation of a strong magnetic field can take place, which makes their study relevant to the SPARC research of the FAIR project.

The first experiments where first- and second-order quantum-electrodynamic processes in the external field of an intense laser wave were studied were the 1996-1997 SLAC experiments. In these experiments, a beam of electrons with energies $\sim 50 GeV B$ was directed toward the laser beam with intensity $10^{18} W/cm^2$. As a result, gamma quanta with energy $>2mc^2$ ($m$ is electron mas) were detected, as well as positrons. There is no complete theory of positron production in these experiments. It should be noted that the external electromagnetic field of any configuration in the ultrarelativistic motion of an electron in its own frame of reference looks like a constant crossed electromagnetic field. In this regard, it is important to solve the problem of the generation of $e^+e^-$ pairs by an ultrarelativistic electron in a magnetic field with subsequent application to SLAC experiments.

The above QED processes that occur in the magnetosphere of X-ray pulsars, processes in the SLAC experiment, processes that are planned to be investigated in the tasks of testing QED in strong electromagnetic fields in the FAIR project are a single class of QED processes in a magnetic field. It is important to construct a unified approach to the analysis of these processes, taking into account the polarization of particles and resonant phenomena.

The paper is devoted to theoretical research of elementary processes of quantum electrodynamics in a strong magnetic field with polarized particles and photons. Spin-polarization and resonance effects in the processes under study are analyzed in detail in the approximation of the lowest Landau levels in a subcritical magnetic field.

The authors are grateful to V.Yu. Storizhko, N.F. Shulga, V.P. Gusynin, A.Yu. Korchyn, V.V. Skalozub, N.P. Merenkov, A.V. Lysenko as well as employees of the theoretical department of the Institute of Applied Physics of the National Academy of

# CONTENTS

# CHAPTER 1
# PROCESSES OF QUANTUM ELECTRODYNAMICS
# IN A STRONG MAGNETIC FIELD

## 1.1. Introduction

A review of papers on the study of quantum electrodynamics processes in a strong magnetic field is carried out. Papers on synchrotron radiation and photon e⁺e⁻ pair production, on QED processes on pulsars, on resonant effects of QED processes of the second order, as well as papers related to the international FAIR project and SLAC experiments are analyzed.

## 1.2. Synchrotron radiation. Electron-positron pair production by a photon

<u>Synchrotron radiation.</u> The emergence of science which is called "quantum electrodynamics in an external magnetic field", apparently, is associated with the discovery of **Synchrotron Radiation** (SR). In 1946, Blewett measured an electron radiation loss in the induction accelerator [1]. A year later, Floyd Haber (Pollock's assistant) experimentally discovered a glow on the synchrotron [2]. This glow is called synchrotron radiation. SR is the electromagnetic radiation of a relativistic charged particle when moving in a uniform constant magnetic field. It is also called magnetobremsstrahlung radiation. In the case of nonrelativistic motion, the radiation is called cyclotron radiation. A few years later, a complete relativistic theory of this phenomenon was written by Sokolov, Ternov (1953), and others [3-6]. The radiation intensity is inversely proportional to the fourth power of mass of the moving particle:

$$I_{SR} = \frac{2e^4 H^2 E^2}{3m^4 c^7} \sim \frac{1}{m^4}. \qquad (1.1)$$

As a result, the synchrotron radiation is significant for lightest charged particles (electrons and positrons). For example, a proton emits $10^{13}$ times less energy than an electron moving with the same energy. SR for protons is essential if their energy $E > 10\ TeV$.

Synchrotron radiation is widely used today: sources of X-ray SR (radioscopy with elemental analysis, micromechanics, microelectronics, X-ray diffraction analysis of biomolecules); radiation cooling of electrons and positrons, in particular, to create B-factories; obtaining polarized beams of electrons (positrons). It should be noted the negative side of SR, it is a significant obstacle to acceleration of high-energy particles.

The process of electron emission of a photon in a magnetic field in the framework of quantum electrodynamics, which corresponds to the Feynman diagram in Fig.1.1 was calculated by M.Demeur (1953) [7] and Klepikov (1954) [8].

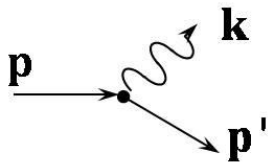

Fig.1.1. Feynman diagram of photon emission by an electron

In Fig.1.1. the initial electron with a 4-momentum **p** emits the photon with a momentum **k** and goes to the final state with a momentum **p'**.
Solid lines are wave functions of the electrons in the initial and the final states in an external magnetic field. Obtained expressions of radiation intensity correspond to a quasiclassical motion of electrons at high energy levels in ultrarelativistic approximation.

The **Lowest Landau Level** approach (LLL approach) is an opposite of the ultra relativistic case. To implement the LLL approximation requires a strong magnetic field, which is comparable in magnitude with the critical Schwinger value $H_0 = m^2/e = 4.4 \cdot 10^{13}\ Gs$.

In the textbook [9], such an approximation in supercritical fields is called the ultraquantum approximation to emphasize that in this case both a process of photon emission and a motion of an electron are purely quantum processes. The process of photon emission by an electron in the LLL approximation was considered in papers [10-12]. In the case of magnetic fields ~$10^{12}Gs$, which are characteristic of the magnetosphere of neutron stars, the SR process was studied in [13-16], and in [17] an absorption of radiation was considered.

<u>Polarization effects in SR.</u> In the SR process, polarization phenomena are significant. It is important to consider both the polarization of radiation and the polarization of particles [18]. The SR polarization was calculated in [19] and tested experimentally in [20]. The spin of electrons was taken into account in [21-24]. Analysis of evolution of the spin of electron (positron) in the SR process was done in [6,25,26]. The phenomenon of radiation self-polarization of particles moving in storage rings has been discovered by Sokolov A.O. and Ternov I.M.. As a result of the SR process in the storage rings 92% of the electrons have a spin oriented against the direction of magnetic field.

**One-photon electron-positron Pair Production (OPP)**. OPP is the process of electron-positron pair production by a single photon. This process is a cross-channel to the process of photon emission by an electron. For the first time in an ultrarelativistic approximation it was considered in [8]. In a strong magnetic field of pulsars, the process was studied in [13,27-32]. In [27], the OPP theory was developed and a photon attenuation coefficient was found taking into account an additional electric field directed along the magnetic field to model the field configuration in the region of the pulsar poles. In [28,29], the OPP process was studied near the threshold $\omega \approx 2m$ and it was shown that the attenuation coefficient as a function of the photon frequency has a sawtooth dependence. In [30], the OPP process was studied taking into account the polarization of both the photon and the electron (positron). In [31], the probability of this process in a strong magnetic field that changes over time was calculated, taking into account an additional constant gravitational field.

<u>Modified propagation function and vacuum polarization.</u> After the appearance of Schwinger's paper on gauge invariance and vacuum polarization [33], a series of papers was published, where the propagation function of a charged particle in an external magnetic field was found on the basis of Schwinger's own time method [34-38]. In [34, 35], the problem was performed for a charged particle with spin 1/2 and 0, respectively. Real and imaginary parts of the found mass operator give a radiation correction to the particle energy and a total probability of the SR, respectively. In [36,37] a polarization operator was calculated and the photon absorption coefficient due to the OPP process was found using the imaginary part of the polarization operator.

Operator method. In [39-47], an operator method was used to study quantum effects in the motion of a charged high-energy particle in an external magnetic field. For ultrarelativistic energies, the motion of a particle is quasiclassical. The noncommutativeness of dynamic variables associated with the quantization of particle motion has an order of magnitude of the ratio of the cyclotron frequency to the energy of the particle $\omega_H/\varepsilon$, which is a small value. The advantage of the method is its application for external fields of any configuration and for charged particles with any spin value. The obvious disadvantage is its unsuitability for non-relativistic particles in a strong magnetic field, as, for example, in the case of electron-positron gas particles at the lowest Landau levels in a magnetosphere of neutron stars. In [39, 40] the processes of SR and OPP, as well as the process of annihilation of $e^+e^-$ pair in one photon were considered. In [41,43,44] a mass operator was found, and in [42] lepton loops with $n$-photon lines were calculated, in particular, in the case $n = 2$ a polarization operator was found. Relatively recently, a paper was published [46], where the operator method considered the OPP process in a strong magnetic field. It was shown that both in the case of weak fields $H \ll H_0$ and in the case of superstrong fields $H \geq H_0$ "the results of quasiclassical calculations are very close to the average probabilities of the exact theory in a wide range of photon energies". It is also worth mentioning [48, 49], where the intrinsic energy of an electron in a strong magnetic field was calculated and the effect of spin on the total probability of SR was studied. In the monograph [50] a wide range of QED issues in a strong external electromagnetic field was considered, all QED processes of the first order were analyzed in detail.

Despite the fact that the general theory of synchrotron radiation was built more than half a century ago, and still in this process, in the process described by the simplest Feynman diagram in Fig. 1, there are unsolved problems. In particular, the spin-polarization effects, ie the effect of spin of particles on the polarization of a finite photon, when the particles are at the lowest Landau levels in both nonrelativistic and ultrarelativistic cases in strong magnetic fields, have not been studied.

The polarization of SR in the ultraquantum ultrarelativistic approximation has not been studied. Also relevant is the problem of studying the effect of polarization of an initial photon on spin directions of electron and positron in the process of single-photon production of an electron-positron pair in these cases.

### 1.3. Physics in a strong magnetic field of neutron stars

<u>The discovery of pulsars.</u> The next burst of scientific activity in the direction of QED in a magnetic field after the discovery of SR occurred in the 70s of last century, after in 1971 the first orbital X-ray observatory Uhuru discovered powerful radiation with a well-defined periodicity, the so-called X-ray pulsar [51]. Shortly before that, the first pulsar (radio pulsar) was discovered on the radio telescope of the Mullard Radio Astronomical Observatory of Cambridge University [52]. X-ray pulsar emission comes from rotating neutron stars, which are located in binary systems in which the accretion process takes place (see Fig. 1.2). Dust and charged particles flow from an ordinary star to a compact, neutron star, causing radiation. All this happens in a strong magnetic field created by a neutron star. The magnetic field of magnetosphere of neutron stars reaches a value $10^{12}$-$10^{13}$ *Gs*. Also a distinctive feature of pulsar is the high speed of rotation of a neutron star. High speed of rotation, as well as a strong magnetic field, are the consequences of the laws of conservation of momentum and conservation of magnetic flux during compression of supernova nucleus during its explosion to the size of a neutron star.

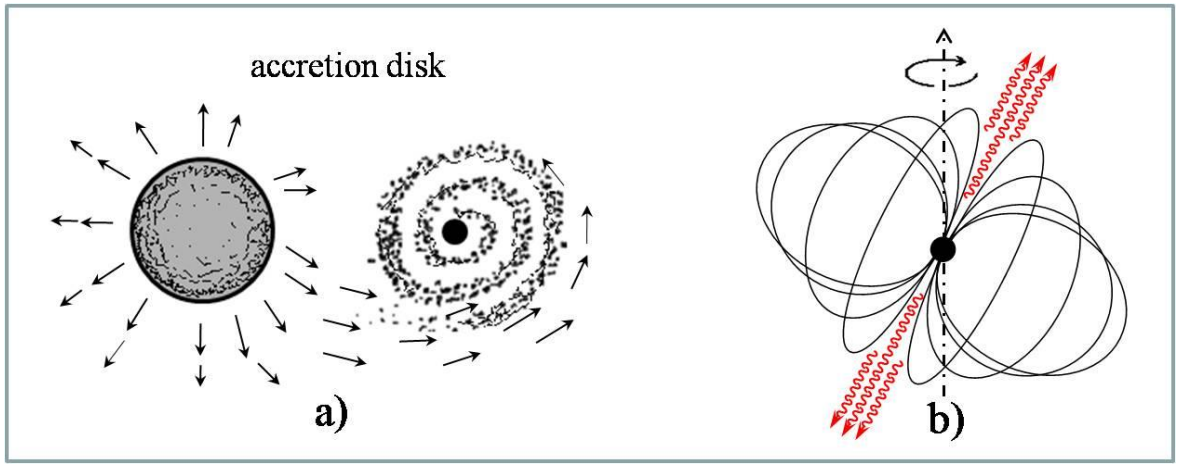

Fig.1.2. X-ray pulsar in a dual system, a) accretion disk, b) pulsar radiation (pencil type)

So far, more than 2,500 pulsars, about two hundred X-rays and gamma pulsars of our galaxy have been discovered, and the first information about pulsars of other galaxies appears. Catalogs of pulsars have been created, for example [53-54].

Thus, the magnetosphere of a neutron star is a unique laboratory for the course of QED processes in strong magnetic fields of the order of the critical Schwinger field. <u>Cyclotron lines. Comptination.</u> In 1973 Gnedin Yu.N. and Sunyaev RA pointed to the possibility of direct measurement of magnitude of a magnetic field of pulsar magnetosphere. They predicted gyrolines (cyclotron lines) in the emission (absorption) spectrum of an electron-positron gas of neutron star's magnetosphere [55], which were discovered by Trümper and colleagues in 1978. [56] (see Fig. 1.3).

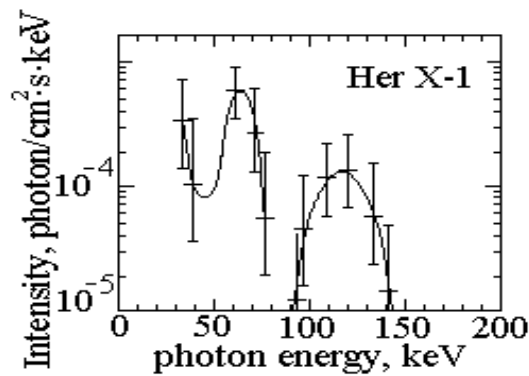

Fig.1.3. Cyclotron X-ray pulsar Hercules X-1 [54]

From this moment began the study of QED processes in the magnetic field associated with cyclotron lines of X-ray pulsars and processes more complex than SR and OPP, with higher-order processes of perturbation theory. In particular, the issue of comptonization, ie the change in the photon frequency due to a series of successive Compton scattering by electrons and positrons, has been studied [57-62]. Comptonization in thermal plasma can lead to characteristic power-law spectra of X-rays emission. It is also responsible for changing the intensity of radiation of a relic background of radio range of hot electrons of interstellar and intergalactic gas (the Sunyaev-Zeldovich effect).

In [59], the polarization of an optical and X-ray compact thermal source in a magnetic field was calculated. In [60], the calculation of the typical spectrum of radiation from plasma cloud by mechanism of comptonization showed the effect of comptonization on the profile of spectral lines of iron. In [61], the creation of cyclotron lines in the X-ray spectrum of the Hercules X-1 pulsar was evaluated taking into account the effect of photons passing through hot plasma. In [62], the spectrum and shape of cyclotron lines of the X-ray pulsar atmosphere were calculated taking into account the combination of the effects of comptonization and anisotropy.

Orbital observatories systematically collect data on cyclotron lines, which are registered in most X-ray pulsars. In [63-66] the data are presented from the international orbital X-ray station RXTE (Rossi X-ray Timing Explorer), and in [67-73] the data on cyclotron lines were obtained at the station BeppoSAX (Satellite per Astronomia a raggi X).

<u>Annihilation lines.</u> Annihilation lines were detected in the spectrum of pulsars, in addition to cyclotron lines and spectral lines of atoms and ions in the gamma range. These lines are associated with the process of annihilation of an electron-positron pair in one photon ($1\gamma$ process) and in two photons ($2\gamma$ process) in the pulsar magnetosphere. Papers [74-81] are devoted to the study of these processes. In [74-75] it was assumed that the particles are initially in the ground energy states $l = 0$. It was shown that $1\gamma$ process dominates over $2\gamma$ process for $10^{13}$ Gs magnetic field. In [76], the excited energy levels of the electron and positron l> 0 were taken into account in this process. Asymptotics for the reaction rate in the case of large quantum numbers of the $e^+e^-$ pair were also obtained. In [77-78], the process of $e^+e^-$ pair annihilation into one photon was considered taking into account spins of particles and arbitrary polarization of radiation. It was shown that the ground spin states (the electron spin is directed against the field and the positron spin is directed along the field) make the main contribution to the probability of the process. In [79], the polarization of radiation in the process of $e^+e^-$ pair annihilation into two photons ($2\gamma$ process) was studied. In magnetic field $H \sim 0.1H_0$, the radiation is linearly polarized, the degree of polarization reaches several tens of percent and depends on the frequency, direction and magnitude of the field. The $2\gamma$ process in supercritical fields in an average relativistic mode was considered in [80-81]. Note the paper [82], which studied the process of $e^+e^-$ pair annihilation with the formation of neutrinos and antineutrinos.

The disadvantages of these papers include the lack of analysis of resonances in the $2\gamma$ process.

<u>Cascades and electromagnetic showers.</u> If in a magnetosphere of a neutron star in the pole region the electron is accelerated to energies $\sim 10^{12} eV$ due to the presence of an electric field parallel to the magnetic one, then the motion of this electron along curved magnetic field lines will lead to the emission of a hard photon (curvature radiation). This photon, in turn, generates a high-energy electron-positron pair in a magnetic field. The result is a cascade of processes: curvature radiation → electron-positron pair production. Many theoretical studies are devoted to the analysis of the cascade process scenario. In particular, the papers [83-92] are devoted to calculations of cascades. In most of these works, numerical simulations were performed. In [83], a radio emission of a system, where an electron positron cascade self-supporting process that was realized in parallel magnetic $H$ and electric $E$ fields in the case $E<<H$, was analyzed. In [84-87], the calculation of up to two dozen steps of a cascade was performed by a Monte Carlo method and the effect of cascades on the spectra of $e^+e^-$ pairs and gamma rays was analyzed. In [88], the theory of cascade mechanisms of electromagnetic showers was constructed, which was based on kinetic equations for the distribution functions of particles and photons. Modeling of electromagnetic showers in the case $(\varepsilon/m)\cdot(H/H_0) \gg 1$ (hard ultrarelativism) was carried out in [89]. Simulation of cascades in fast-rotating pulsars was performed in [90]. The cascade included the curvature radiation by a primary electron, a conversion of photons emitted by the primary and secondary particles into e⁺e⁻ pairs, a quantization of the SR, and inverse Compton scattering by secondary pairs. In [91], a self-consistent kinetic modeling of e⁺e⁻ cascades was performed taking into account the shielding of the electric field by particles. In [92], modeling of e⁺e⁻ cascades taking into account the polarization of photons, as well as modeling of resonant Compton scattering in strongly magnetized neutron stars were performed.

Electron-positron plasma of the magnetosphere of pulsars. As a result of repeated cascades together with the processes of pair annihilation and Compton scattering, e⁺e⁻ gas (e⁺e⁻ plasma) of the pulsar magnetosphere is formed. The papers [93-97] are devoted to modeling the process of e⁺e⁻ plasma formation. One of the features of radiation from pulsars is coherent radiation in radio domains. In [96] it was shown that the source of coherence is instability of the plasma arising in the magnetosphere. In [97], the simulation of e⁺e⁻ plasma in supercritical fields of magnetars $H \sim 10^{14}$-$10^{15} Gs$ was performed. Note also the papers [98,99], where the thermodynamic properties of e⁺e⁻ gas in a magnetic field are studied.

Quantum-electrodynamic processes near neutron stars in strong magnetic fields were considered in reviews [100-103] and monographs [104-105].

The parameters of the pulsar radiation, which are measured during observations (pulsar brightness curves, its period and pulse profile, energy spectrum and its variability at different time scales, polarization, etc.) certainly depend on the elementary QED processes occurring in the electron-positron plasma in the neutron star magnetosphere. The subject of studying X-ray pulsars, which includes both astronomical observations and theoretical calculations, remains relevant to this day, as there is no single theory of X-ray pulsar radiation. In particular, the determination of a spin population of a magnetized e⁺e⁻ gas and its effect on the SR of an X-ray pulsar is an unsolved problem in the framework of quantum electrodynamics.

## 1.4. Second-order QED processes near resonances

Second-order quantum electrodynamic processes, such as **Compton Scattering** (CS), that is the scattering of a photon by an electron or positron, **Double Synchrotron Radiation** (DSR), that is the emission of two photons by an electron, **Two photons $e^+e^-$ Pair Production** (TPP), that is the production of electron-positron pair by two photons, **One-photon $e^+e^-$ Pair Production with a photon Emission** (OPPE), pass through the intermediate electron (positron) state. These processes are cross-channels in relation to each other. An external electromagnetic field allows to bring the intermediate state to the mass surface, which corresponds to resonant conditions of the processes [106-112]. The cross-section of the QED process in resonant conditions can exceed the cross-section in non-resonant cases by several orders of magnitude, which is of great physical interest. The resonant divergence is eliminated by the Breit-Wigner rule [113]. Papers [114-116] is devoted to elimination of resonant divergences in the case of a stable intermediate state.

Compton scattering. The change in the frequency of a photon during its scattering by a free electron is known to have been discovered by Arthur Compton in 1923 (Compton effect). The complete quantum relativistic theory of the CS process was built by O. Klein and Y. Nishina in 1929 and independently by I. Tamm in 1930. An external magnetic field modifies the CS process. As mentioned above, initially the study of the influence of a field on the CS process was associated with the processes of comptonization that take place in the magnetosphere of neutron stars.

The CS process in a strong magnetic field was studied in [109], [117-128]. In [117], the total cross section of the CS was found by the Schwinger method when a photon propagates along a field. In [118], the differential cross section of the CS was obtained if initial and final electrons are in the ground energy states $l = l' = 0$. The CS process was considered in [119] with the study of cyclotron resonances when the initial electron is in the ground energy state $l = 0$. The general case in the LLL approximation was considered in [120] taking into account the process of two-photon Compton scattering. In particular, it was shown that the probability of the latter in cyclotron resonance is comparable to nonresonant single-photon scattering. In [121], the scattering of soft photons by relativistic electrons was considered when electrons move along the direction of a strong magnetic field. In [122], a comparison of two relativistic processes was performed taking into account particle spins: photon absorption by an electron and CS. It was shown that at the point of cyclotron resonance these cross sections are equal. In [124], the CS process in resonant and nonresonant modes was considered, taking into account the polarization of radiation when a relativistic electron moves along a supercritical field. In [125-126], resonant and interresonant Compton scattering were considered, taking into account the electron spin. Under resonant conditions, the differential cross section was represented in the form of Breit-Wigner. In the magnetic field $H \sim 10^{12} Gs$, the resonant cross-section is several orders of magnitude larger than the interresonant cross-section, the interresonant cross-section is of the order of the Thomson cross-section, and the resonance width is tens of electron volts. In a relatively recently published papers [127], differential cross section of this process was obtained in the case of both subcritical $H < H_0$ and supercritical $H > H_0$ magnetic fields, taking into account the electron spin, when the initial photon propagates along the field. In [128], the scattering of a photon by an electron was considered through the study of an intermediate process of photon emission by an electron in a magnetic field and an electromagnetic wave directed along the field (Redmond configuration).

The disadvantages of this papers include the lack of analysis of the effect of polarization of the initial photon, both on polarization of the radiation and on the spin states of the final electron for different spin states of the initial electron under resonant conditions.

An actual problem is to find the mechanism of polarization of electron beams (positrons) by a linearly polarized electromagnetic wave based on the process of resonant CS in a magnetic field.

Two photons electron-positron pair production. The TPP process in a magnetic field was studied in [129-133]. In [129], a probability amplitude of the fourth-order process (the scattering of a photon by a photon) was found, and the total probability of TPP was found by using the optical theorem in the case when both initial photons are directed along the field. In [130-132], expressions were obtained for the probability of the OPP process in the Redmond configuration field with subsequent selection of one initial photon from the electromagnetic wave field. In [132], the total probability of the OPP process was obtained through optical theorem from the polarization operator and it was shown that a magnetic field causes oscillations in the cross section that have an amplitude that significantly exceeds corrections obtained by perturbation theory. To the end consistently as a process of the second order, the process of TPP was considered in [133]. The case when the energy of each of photons does not exceed 2m was considered to remove the OPP process. The resonant behavior of the OPP cross section and the dependence of the cross section on polarization of photons were analyzed. Comparison of the processes of electron-positron pair production by one and two photons (OPP and TPP) was performed in [134-138] with respect to X-ray pulsars and soft gamma-ray bursts. In particular, in [134] it was shown that the OPP process dominates for most pulsars, when the magnetic field is of the order of $H \sim 10^{12} Gs$, the photon density of a magnetosphere does not exceed $n_\gamma < 10^{25} cm^{-3}$, the e$^+$e$^-$ plasma temperature exceeds $kT > m$. The disadvantage of these papers is the use of formulas for the probability of TPP without an external magnetic field. That is, the estimates did not take into account the resonant course of the TPP process and therefore should be revised. In [139], the process of electron-positron pair production by a photon during its propagation in a thermal bath in a superstrong magnetic field $H >> H_0$ was studied.

It should be emphasized that there are two types of divergences (resonances) in the TPP process in the external magnetic field.
The first divergence is related to the the case when an intermediate electron (positron) is on the mass shell. In this case, the second-order process splits into two independent first-order processes. The second type of divergences is associated with discrete motion of an electron (positron) in the plane transverse to direction of a magnetic field. It takes place at the reaction threshold when particles are produced with zero longitudinal momenta. The OPP process contains a similar divergence. The second type of resonances was studied in [133]. The resonant course of the TPP process in the first scenario was not studied.

The unsolved problems in study of the TPP process are as follows:
- Investigation of the resonant TPP process when an intermediate particle comes to its mass shell.
- Analysis of the effect of initial photon polarization on the degree of polarization of final particle beams in the resonant TPP process.
- Comparison of the processes of OPP and TPP in the conditions of a $e^+e^-$ plasma of a magnetosphere of X-ray pulsars taking into account the resonant behaviour of the process of two-photon production of $e^+e^-$ pair in a magnetic field. Searching for parameters when the last process can compete with first-order process (production of $e^+e^-$ pair by one photon).

Double synchrotron radiation. Relatively few works were devoted to the process of emission of two photons by an electron (the DSR process), when the electron moves in a magnetic field [140-142]. In [140], the process of radiation of one photon by an electron in the Redmond configuration field with the emission of one photon from an electromagnetic wave was considered. In [141], the DSR process was studied in the second Born approximation in the quasiclassical case with ultrarelativistic particles. Finally, second-order calculations of the perturbation theory were performed in [142] for the process of emission of two photons by an electron in a strong magnetic field. The calculations were applied to the fields of magnetars $H \sim H_0$ and dependences of the process probability on both the electron spin direction and the photons polarization were found. The conditions for the occurrence of resonances were analysed and numerical estimates of probabilities were performed.

The disadvantages of these works include the lack of analysis of comparisons of SR processes and resonant DSR process in the LLL approximation (in a strong magnetic field). The conditions of probability factorization in resonant cases and the influence of a spin-flip process in resonant DSR on radiation polarization have not been studied.

One-photon production of e⁺e⁻ pair with photon emission (OPPE). This process in a strong magnetic field should be characterized by the same basic properties as its cross-channels CS, TPP, DSR, namely: the presence of resonances, the strong dependence of the process rate on the spin direction of the particles and the polarization of photons, the achievement of the process rate in resonance to the magnitude of the first-order processes. Therefore, the study of the OPPE process is an urgent task. However, there is no mention of this process in the literature.

### 1.5. QED test in the FAIR project

An international convention was signed in 2010 to establish a new scientific mega-project - FAIR accelerator complex (Facility for Antiproton and Ion Research) with a total budget of over one billion euros which is being built in Germany (near Darmstadt) on the basis of the GSI Helmholtz Centre for Heavy Ion Research. The FAIR project includes the study of QED phenomena in extremely strong electromagnetic fields, in addition to the program of studying hadronic matter, the study of fundamental symmetries and interactions, as well as the study of the dynamics of multiparticle systems taking into account collective effects [143,144]. This direction is performed by SPARC collaboration (Stored Particles Atomic Physics Research Collaboration). In particular, studies involving photons, electrons and atoms in the presence of strong rapidly changing electromagnetic fields, as well as structural studies of heavy ions will be performed. It is planned to performed experiments on interaction of hydrogen and helium ions with an intense pulsed laser field of the PHELIX laser, when an electron in the inner orbit of a heavy ion experiences a field of the order of the critical one $E_0=m^2/e$. Verification of quantum electrodynamics in strong fields within the FAIR project involves consideration of the following tasks: Lamb shift, fine structure of heavy ion levels, anomalous magnetic moment of bound electrons, ions and electrons in intense field of laser. Adjustment of the laser to transitions between electron energy levels is possible in experiments with counter beams (intense laser beam and beam of hydrogen-like heavy ions) due to Doppler shift. The study of the dynamics of collisions in strong fields in the FAIR program includes the following tasks: radiation capture of an electron in ion-atomic collisions, formation of quasimolecules in ion-ion collisions, production of $e^+e^-$ pairs in the collision of heavy ions.

Supercritical charge of nucleus. The solution of the Dirac equation has a singularity for the bound states of an electron in the field of a point nucleus with charge $Z$ [145]. According to Sommerfeld's formula for the fine structure of atom, the energy of the lower electronic level $1S_{1/2}$ has the form [146]:

$$\varepsilon_0 = m\sqrt{1-(Ze^2)^2} \ . \tag{1.2}$$

The energy (1.2) approaches zero at $Z = Z_c$, where $Z_c = 137$. When $Z > Z_c$, it becomes an imaginary value, which indicates incorrectness of the problem without taking into account the boundary conditions for the wave function at zero. Note that the imaginary additive to energy is known to describe a nonstationary process, decay, ie the state with $Z > Z_c$ must be unstable. A more realistic model with a finite nucleus size allows to go beyond $Z > 137$ [147-148]. In [147], the notion of the critical charge of the nucleus $Z_c$ (charge at which the energy of the $1S_{1/2}$ level of electron reaches the value of energy of the lower continuum) was introduced and this value was estimated $Z_c = 200$. Later in [149-151] a more accurate analysis of the Dirac equation with $Z \sim Z_c$ was performed, whence follows $Z_c \approx 170$. In these papers it was noted that in the case $Z > Z_c$ for a bare nucleus and with a vacancy on the $K$-shell there is a spontaneous process of capture of electrons from the negative continuum (Dirac basement) to the $K$-shell and the formation of a hole in the continuum, which manifests itself as spontaneous positron production. The effect of spontaneous quasi-static positron production at $Z > 170$ can be observed in the collision of two nuclei with a total charge of $Z_1 + Z_2 > Z_c$, for example, in the collision of two bare uranium nuclei [149]. The calculation of cross section of the process of spontaneous positrons production in the heavy ions collision by the above mechanism was carried out in [152-153].

Darmstadt peaks. In the late 1970s, the GSI Helmholtz Centre for Heavy Ion Research launched a program for the UNILAC accelerator to study the processes of heavy ions collisions accompanied by the spontaneous production of positrons. As part of this program, two groups EPOS [154-157] and ORANGE [158-161] carried out experiments on the collision of fast with energy near the Coulomb barrier (~6 *MeV/nucl*) heavy ($Z \sim 90$) ions with the formation for a short time $\sim 10^{-21}$s of a heavy composite nucleus with a supercritical charge. During experiments in both groups, the spectra of positrons formed as a result of nuclear collisions were measured. The main mechanisms of positron зкщвгсешщт in ion collisions are as follows: 1) internal pair conversion when removing the excitation of the nuclei with a radiation time of $\sim 10^{-15}$s (conversion positrons), 2) extraction of electrons from the negative continuum under the action of rapidly changing in time and space deep ($\sim 20$ *MeV*) Coulomb potential with a radiation time of $\sim 10^{-21}$s (dynamic positrons) and, finally, 3) the process of spontaneous production by supercritical charge with a radiation time of $\sim 10^{-19}$s (spontaneous positrons). The first two processes are the background to the sought process of spontaneous positrons production. The experimentally measured background was in good agreement with theoretical calculations. The time of spontaneous transition of an electron from a negative continuum is two orders of magnitude greater than the time of ion collision, ie the time of existence of a supercritical charge. And this was probably the main difficulty in trying to detect spontaneous positrons. However, in the positron spectrum of both groups, abnormal narrow peaks (several tens of kiloelectronvolts wide) were detected in the region of kinetic energies of positrons of 200–400 *keV*. To study electron-positron matches, EPOS and ORANGE spectrometers were further modified with the addition of electron spectrometers [162-164]. In the spectra of electron-positron pairs, narrow lines corresponding to the lines in the positron spectra, but with a smaller width, were also detected (see Fig. 1.4). The location of the lines did not depend on a magnitude of total charge of the composite nucleus, while the cross section was proportional $\sim (Z_1+Z_2)^{20}$.

In [165] an overview of the experiments of both groups is given. Experiments of the APEX group with improved electron and positron spectrometers did not reveal any abnormal features [166] and later experiments in this direction were stopped. A fairly complete review of both the Darmshadst experiments and related issues was conducted in [167].

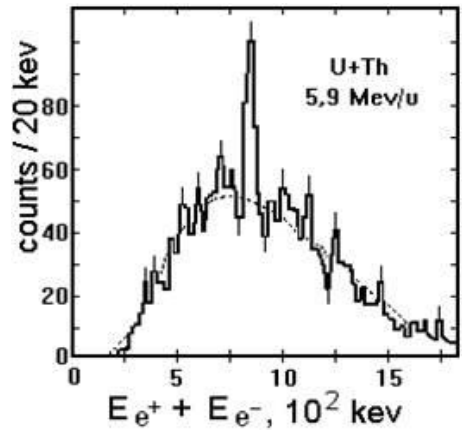

Fig. 1.4. Anomalous peak in the energy spectrum of e$^+$e$^-$ pair production, which was observed in Darmstadt experiments on the collision of heavy ions [164]

Strong magnetic field of colliding heavy nuclei. It should be noted that quantum-electrodynamic processes can take place in the region between nuclei in a strong magnetic field of nuclei when heavy ions collide with impact parameters that have an order of magnitude (or more) of the nucleus size $\rho \geq R_{nuc}$ (ultraperipheral collisions). Between the nuclei, the magnetic fields of moving nuclei are added, and the electric ones are compensated (see Fig. 1.5.). For impact parameters of the order of $10^{-10}$ cm, colliding heavy nuclei with a charge $Z = 90$, which move at a speed of ~ c / 10, create a magnetic field of the order of $10^{12} Gs$. In this area, the course of QED processes involving a strong magnetic field is quite possible. In the process of production of an electron-positron pair in a collision of nuclei, if the pair is produced between the nuclei, it will be at the Landau levels. The distance between adjacent Landau levels for the field $H = 5 \cdot 10^{12} Gs$ is equal to 50 keV, which can be determined experimentally. In the energy spectrum of e$^+$e$^-$ pairs generated by the collision of nuclei, narrow lines should be observed, located quasi-equidistantly, which correspond to the Landau levels.

Cyclotron lines in the X-ray radiation range accompanying electron transitions to neighboring Landau levels should also be observed.

We assume that the series of quasi-equidistant peaks observed in the Darmstadt experiments could exactly correspond to the indicated Landau levels [168]. Note also the work [169], which analyzed the effect of a strong magnetic field of moving nuclei on the generation of positrons in the collision of heavy ions, as well as the possible participation of the magnetic field in the formation of narrow peaks. In this paper the stationary two-center Dirac equation, which included the magnetic fields of moving charges, was solved and the method of adiabatic phase-correlation diagrams was used. The electron-positron pair production rates obtained in two-center calculations, including the magnetic field, were close to rates calculated in the monopoly approximation without a magnetic field and did not contain a resonant structure.

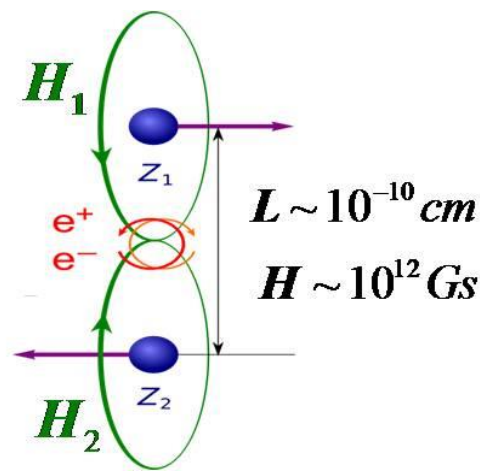

Fig.1.5. When heavy ions collide in the region between the nuclei, the $e^+e^-$ pair is produced in a strong magnetic field of moving nuclei

Thus, it is important to study the QED with magnetic field in relation to the SPARC of the FAIR project, including a program to revive the old Darmstadt experiments. This paper studies the process of generating $e^+e^-$ pair at low Landau levels by two photons in a strong magnetic field.

## 1.6. Photon propagation in a magnetic field

Photon dispersion. The process of photon propagation in a region with an external classical electromagnetic field (in our case, a magnetic field) is similar to the propagation of an electromagnetic wave in an anisotropic optically active medium. Therefore, the study of this process is reduced to the analysis of an optical properties of the active medium using a renormalized polarization tensor (photon polarization operator). The polarization operator was mentioned above when considering first-order processes, where it was used to obtain the full probability of the OPP process.

Works [170-176] are devoted to the analysis of analytical properties of the polarization tensor. In [170], the solution of dispersion equation for eigenmodes was found using the polarization tensor. The solutions near a threshold of $e^+e^-$ pair production (cyclotron resonance) were analyzed in detail. The monograph in FIAN proceedings [171] is devoted to the study of a polarization of vacuum and quantum relativistic gas in a constant external electromagnetic field. The singular behavior of the polarization operator due to the OPP process in a magnetic field was studied in more detail. In [172], the process of photon propagation in a supercritical field $H \gg H_0$ was considered. The behavior of a photon in an arbitrary superposition of constant magnetic and electric fields was studied in [173]. The polarization operator in the 2+1 dimensional QED with nonzero fermion density in a constant uniform magnetic field was calculated in [174]. In [175], the polarization operator of $e^+e^-$ gas in a magnetic field was found using the Matsubar technique for Green's temperature functions. Radiation corrections to the polarization operator were calculated in [176].

**Vacuum Birefringence** (VB). Birefringence is known to be an effect of splitting a beam of light into two components (ordinary and extraordinary) in anisotropic media (calcite crystals). If we choose the conditions under which the directions of the ordinary and extraordinary rays coincide, then there is an effect of changing the polarization. In addition to crystals, the birefringence is observed in isotropic media placed in an external electromagnetic field. So in the electric field there is a Kerr effect. In the magnetic field, a Cotton-Mouton effect and a Faraday effect are analogs. In the field of a laser wave, this is a Kerr optical effect. Since the external magnetic field, which has a value of an order of the critical field $H_0$, polarizes the vacuum and the vacuum as a result exhibits properties of an anisotropic medium, it is possible vacuum birefringence of light. This effect was first predicted and theoretically described in [177]. The effect is based on a fourth-order QED process, namely the process of light scattering on light. A cross section of this process was found in [178,179].

Further development of methods using the nonlinear Heisenberg-Euler Lagrangian and the study of the VB effect was carried out in [180-188]. The polarization of vacuum and VB in an external electromagnetic field of arbitrary configuration were considered in [180]. In [182], the process of propagation of a linearly polarized laser beam in an external transverse magnetic field was studied. The response of QED vacuum and its instability in asymptotically large magnetic fields $H \gg H_0$ was studied in [183]. Papers [184-185] are devoted to the VB effect in a strong magnetic field, where properties of a photon polarization tensor and a complex refractive index are numerically obtained and analyzed in detail. In [186], polarization effects caused by photon-photon interaction in laser experiments were considered. The laser beam propagated in an area with a magnetic field or it collided with another laser beam. In [187], experiments (astronomical observations) were performed at the Very Large Telescope (Chile), where experimental confirmation of the VB effect in optical-polarimetric measurements of an isolated neutron star was obtained for the first time. Note also the monograph [188], where the process of particles production from vacuum, as well as the process of polarization and rearrangement of vacuum by an external field was considered on the basis of Bogolyubov's transformations and the monograph [189], where the theory of perturbations was developed and methods for finding the Green's function have been developed.

**Photon Splitting** (PS). The photon splitting process is a third-order QED process in the single-loop approximation and cannot occur without an external field according to Farry's theorem [190, 191]. The difference from the polarization operator is the presence of an additional photon in the final state. A strong external magnetic field makes such a process possible and interesting in astrophysics as a mechanism for generating linearly polarized gamma rays. Papers [192-197] are devoted to the study of this process. In [192], a refractive index of photon propagation and an absorption coefficient , as well as the rules for selecting polarization for photon splitting were determined. In [193], the probability per unit time of photon decay into two photons was found. The process of merging two photons into one photon is the reverse process to the photon splitting. It was considered in [194]. In [195], numerical results for a probability of photon splitting as a function of photon energy below the energy threshold of the $e^+e^-$ pair for different values of a magnetic field of neutron stars were obtained, and a recalculation of previously obtained results was performed. In [196], S-matrix approach, which was used in [195] to consider the PS process in a magnetized vacuum, was critically discussed and the problem of amplitude convergence in the case of weak magnetic fields was solved. In [197], the main physical aspects of two processes with the same initial conditions (PS process and OPP process) were presented. Their manifestation in a magnetosphere of neutron stars was discussed.

It should be emphasized that when a photon propagates in a magnetic field under resonant conditions corresponding to production of a real electron-positron pair, the reverse process to annihilate the pair into one photon is possible. Thus, a cascade of processes of production and annihilation of the $e^+e^-$ pair is formed, which has not been studied before. Relevant issues are:

- study of the process of photon propagation in a strong magnetic field, when there is a **Cascade of processes of the $e^+e^-$ Pair Production and subsequent the Pair Annihilation** (CPPPA);
- calculation of the change in photon polarization in the CPPPA process in resonant and interresonant regions.

### 1.7. QED processes in a laser wave field. SLAC experiments

<u>QED processes in a plane electromagnetic wave field.</u> The procedure for calculating the processes of quantum electrodynamics in a plane electromagnetic wave field is methodologically the same as in the external magnetic field, so it is advisable to mention this topic without claiming the completeness of the review. Indeed, in both cases the Farry picture diagram technique is used [191], when the external classical field is precisely taken into account through the solution of the Dirac equation for the electron (positron). And the interaction of particles with quantum photons is taken into account by perturbation theory. Also note, as will be shown below, under certain conditions, the results of calculations in the laser wave field and in the constant magnetic field coincide.

The general solution of the Dirac equation for an electron in a plane electromagnetic wave was found by Volkov in the 1930s [198]. After the advent of lasers, Volkov functions began to be used extensively to study the elementary processes of QED in the laser field.

One of the main parameters of the problem is a relativistically invariant parameter of multiphotonity, which has the form:

$$\eta = \frac{|e|F}{m\omega},\qquad(1.3)$$

where $F$, $\omega$ are strength (intensity) and frequency of the external electromagnetic field. This parameter has the physical meaning of a work of the field at the wavelength in units of rest energy of electron. In the case $\eta \ll 1$, the QED process involves a small number of virtual photons of the external field, in the extreme case, one photon. If $\eta \sim 1$ or more, multiphoton processes are essential. In order to reach a value $\eta \sim 1$ in the optical range, the field strength must be of the order of magnitude $F \sim 10^{10}\ V/cm$. This corresponds to intensity of laser $\sim 10^{18} W/cm^2$. Modern high-power femtosecond lasers are already reaching the required intensity to observe multiphoton nonlinear QED processes. Value $2\cdot 10^{22} W/cm^2$ was obtained in [199] as a record value of intensity for 2008 in optical range $\eta \sim 100$ at a laser with a power of 300 TW. The LFEX (Laser for Fast Ignition Experiments) optical laser with a record power of 2 PW was built in Japan (Osaka University). Two European projects ELI (Extreme Light Infrastructure) [200] and XCELS (Exawatt Center for Extreme Light

Studies] [201] are under construction, where it is planned to achieve exavatt power and field strength $10^{24} W/cm^2$ and above.

Consideration of first-order QED processes in a laser field (spontaneous photon radiation by an electron, e⁺e⁻ pair production by a photon) is presented, for example, in [44], [202-205]. Recently, such processes have been studied in the field of ultrashort laser pulse with $\eta \geq 1$ [206] and in the ultra-strong laser field with $\eta >> 1$ [207]. Scattering of an electron at Coulomb center in a pulsed field of a laser wave was considered in [208]. Papers [209,210] were devoted to the cross-channel to electron scattering, namely the e⁺e⁻ pair production by a Coulomb field in a laser field in tunnel mode. Resonances in the process of photon scattering by an electron in the field of a plane monochromatic wave were studied in [108], and in the field of a pulsed laser wave in [211-214]. The process of bremsstrahlung of an electron at Coulomb center in the field of a pulsed laser was studied in [215-217]. The papers [111,112], [218,219] are devoted to lepton-lepton scattering. In [220], [221] the process of spontaneous emission of two photons by an electron in an intense laser field was considered. In [222], the total cross section of high-energy electron-positron pair production by a photon in the combined field of an atom and an intense laser was calculated. A review of studies of the processes of relativistic quantum dynamics, quantum electrodynamics, nuclear physics, and elementary particles in extremely intense laser fields was performed in [223]. In the monographs [224] and [225], the resonant and coherent effects of QED in a light field (field of a plane monochromatic wave) and in a strong pulsed laser field, respectively, were studied.

QED processes in the Redmond configuration field. The Dirac equation for an electron in an external classical field, which is a superposition of the constant magnetic field and the field of a plane electromagnetic wave directed along the magnetic field (Redmond configuration), has an exact solution [226]. The Green's function of an electron in the Redmond configuration field was found in [227]. This allowed to perform a number of works on the study of QED processes in this field [109], [128], [130-132], [140], [204], [228-234]. The process of spontaneous emission of a photon by an electron was considered in [204], [228-230]. In [229], the problem was solved when an electron was at the lowest Landau levels. The process of e⁺e⁻ pair production by one photon was studied in [231].

The obtained results of researches of QEDs of first-order processes can be used as auxiliary for studying QEDs of higher-order processes in a purely magnetic field. To do this, you need to select a finite number of photons from the field of the plane electromagnetic wave. The problem of photon scattering by an electron in a magnetic field was performed in this way in [128], [140], [232-233]. In [233], the scattering cross sections found were further used to obtain a cross section of bremsstrahlung of an ultrarelativistic electron on a nucleus using the equivalent photon method. The problem of e⁺e⁻ pair production by two photons in a magnetic field was solved in [130,131]. In [234], the problem of e⁺e⁻ pair production by a single photon on a nucleus in a magnetic field was considered. The study of cyclotron resonances and the process of scattering in a magnetic field was carried out in [109], analyzing the mass operator of electron in the Redmond field and using the optical theorem. By a similar method, investigating the polarization operator in the Redmond configuration field, the probabilities of electron pair production by two photons were found in [132].

It should be noted that papers with direct calculations of the second-order QED processes in the Redmond field with the analysis of resonant and polarization properties in the literature are absent. To perform them, as a first step, it is necessary to find the expression for the Green's electron function in the Redmond configuration field in a form convenient for calculations of resonant processes.

<u>SLAC experiments.</u> The first experiments to study first- and second-order quantum electrodynamic processes in the external field of an intense laser wave were the SLAC experiments of 1996-1997 [235-237].

In these experiments, a beam of electrons with energies of ~ 50 *GeV* was directed toward a laser beam with an intensity of $10^{18}$ *W/cm*$^2$ (see Fig.1.6). As a result, hard photons with energies > $2mc^2$ were detected [235], as well as positrons [236]. High-energy electrons moving in an external field, in this case a laser wave field, inevitably spontaneously emit photons, in the spectrum of which there is a rigid component. In turn, a hard photon with an energy $> 2mc^2$, moving in the same external laser field, is able to generate an electron-positron pair, which was experimentally observed.

The authors of SLAC experiments identify two possible channels for the electron-positron pair production. The first channel is the Breit Wheeler process, [191] the process of $e^+e^-$ pair production by two photons. One photon is a rigid photon formed by inverse Compton scattering of wave photons by an electron. The role of the second photon is performed by *n* photons of the laser wave. The second channel is the **trident process** (TP), where the initial electron generates an electron-positron pair through an intermediate virtual photon.

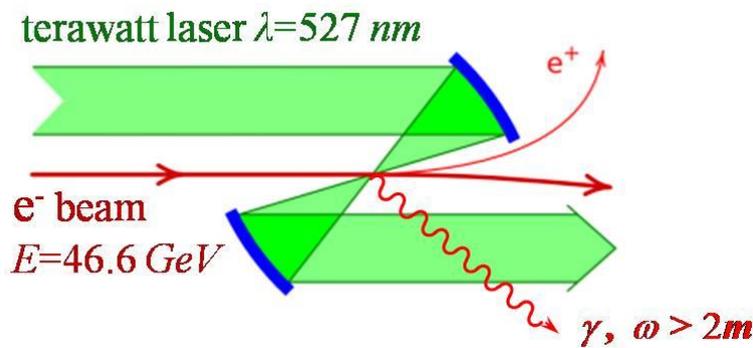

Fig. 1.6. Scheme of collision of the electron beam with the laser beam in the SLAC experiment on the production of $e^+e^-$ pairs [236]

Calculations of the multiphoton trident process, which occurs when an intense laser beam interacts with an electron beam, in the Vazzecker-Williams approximation [191] were performed in [238]. A complete theory of the multiphoton trident process of $e^+e^-$ pair production in a strong laser field in the second Born approximation was presented in [239]. The numerically obtained probability per unit time of $e^+e^-$ pair production $R = 4 \cdot 10^{-4} c^{-1}$ was four times higher than the experimental value [236]. In [240], a review of QED processes in the intense field of a laser beam was presented, where nonlinear collective effects in photon-photon and photon-plasma interactions were analyzed.

The trident process of e⁺e⁻ pair generation in a strong external magnetic field in the Vazzecker-Williams approximation of equivalent photons with ultrarelativistic particles was studied in [13]. The excitation of Landau levels of an electron due to collisions in a strong magnetic field was considered in [241]. Note that in the papers of A.I.Nikishov, V.I.Ritus [202,203] it was shown that the expression for probability of spontaneous emission of a photon by an electron in a laser wave field coincides with the expression for probability of SR process (photon emission by an electron in a magnetic field) in the case of ultrarelativistic electron motion. The fact is that an external electromagnetic field of any configuration in ultrarelativistic motion of an electron in its own frame of reference looks like a constant crossed electromagnetic field.

In this regard, it is important to solve the following problems:
- To find the probability of e⁺e⁻ pair production by an ultrarelativistic electron in a magnetic field.
- To estimate the yield of positrons in the SLAC experiment [236] using the Nikishov and Ritus theorem (on the equivalence of external fields for ultrarelativistic processes) [202,203].

**1.8. The record values of magnetic field strength and the processes in superstrong magnetic fields**

<u>The record magnetic fields.</u> Strong constant magnetic fields, up to $4.5 \cdot 10^5 Gs$, are achieved in laboratory conditions in the Bitter electromagnets installed inside superconducting magnets. Pulsed magnetic systems with the magnetic field of $8.9 \cdot 10^5 Gs$ were developed at the National High Magnetic Field Laboratory (Los Alamos) [242]. In the middle of the last century A.D. Sakharov proposed an idea of magnetic cumulation and the basic design of explosive magnetic generators [243,244]. As a result of rapid deformation by an explosion of conductive circuits, the seed initial magnetic field of ~ 100 *kGs* was compressed to a value of 25 *MGs*. The advent of powerful pulsed petawatt lasers has led to the fixation of a new record value of the magnetic field in experiment. A laser beam with an intensity of $10^{21}$ *W/cm²* using pulse compression technology to intervals below picoseconds when interacting with laser materials forms a clot of dense plasma ~ $10^{21}$ *cm*$^{-3}$ and above (laser plasma), which creates a dynamic electric field of 100 MV/micron with corresponding magnetic field ~ 1 *GGs* [245-247]. The method of generating a strong magnetic field using magnetic coils with a laser driver is quite promising. The magnetic field ~ 10 *MGs* was achieved at the Gekko XII laser facility (Osaka University's Institute for Laser Engineering, Japan) with a capacitor-coil target [248]. This approach has been adopted in a number of laboratories with different targets and the magnetic field strength in such experiments ranges from 10 *kGs* to 10 *MGs*. The differences are determined by the geometry of the target and the parameters of the incident laser. A similar experiment was conducted to study the generation of a magnetic field by a laser when irradiating a snail-type target at the PHELIX (Petawatt High Energy Laser for Heavy Ion Research, GSI, Germany) at wavelength of 1056 *nm*, the duration of 0.5 *ps* and the intensity of 2·$10^{19}$ *W/cm²* [249]. The maximum value of the magnetic field in this experiment was 8 *MGs*. The work [250] is devoted to a review of the results of recent experimental studies of the magnetic field generation. It should be noted that for the first time the magnetic field of wake waves with a laser driver at wavelength of 800 *nm*, the duration of 85 *fs* and the intensity of 3·$10^{18}$ *W/cm²* was experimentally studied at the JETI (JEna TItanium laser, Helmholtz Institute Jena, Germany). The Faraday rotation was used to measure the values of the magnetic field. According to the results obtained at this facility, the magnetic field was about 10 *MGs* [251]. Similar experiments were also performed in LOA (Laboratoire d'Optique Appliquee, France) [252] and at the European XFEL (European X-ray Free-Electron Lasers, Germany) [253]. Characteristic magnetic fields of magnetosphere of neutron stars, as noted earlier, are fields of ~ $10^{12}$*Gs*. The record surface magnetic fields of ~ $10^{15}$*Gs* are characteristic fields for magnetars, which include anomalous X-ray pulsars and soft gamma-ray repeaters [254-258]. The maximum magnetic field in which standard quantum electrodynamics operates, ~ $10^{42}$*Gs*, is associated with the effect of positron collapse. The magnetic field significantly increases the Coulomb attraction between electron and positron. This happens until the electron and positron fall on each other, which corresponds to the collapse [259, 260].

Processes in supercritical magnetic fields. In the previously mentioned papers [80], [81], [97], [172], [183] the processes of QED (annihilation of e$^+$e$^-$ pair, photon propagation) were considered in magnetic fields above the critical Schwinger value $H_0$. The calculation of ionization energy of atoms in such fields was performed in [261]. Photon emission and relativistic shift of energy levels for a hydrogen-like atom in a supercritical magnetic field were considered in [262]. The process of photon capture by a strong magnetic field and suppression of e$^+$e$^-$ pair production process was studied in [263, 264]. The properties of relativistic positronium in a superstrong magnetic field were considered in [265]. The problem of matter and radiation in very strong magnetic fields was devoted to review works [266, 267], as well as a monograph [268].

The papers [269-274] were devoted to the problem of dynamic breaking of symmetries (symmetry of flavors in dimension 2 + 1, chiral symmetry in QED) and generation of mass of fermions in a constant magnetic field. The role of the magnetic field in (2 + 1) dimensional models is similar to the role of the Fermi surface in the Bardin-Cooper-Schriffer theory of superconductivity. This phenomenon is based on the spatial reduction D → D-2 in the dynamics of fermion pairing in a magnetic field, since the motion of a charged particle in a plane perpendicular to the field is limited. It should be noted that the nonperturbative QED dynamics is significant here. In the Farry picture, when the quantized interaction is taken into account perturbatively, a dynamic violation of symmetries in QED processes is not observed. Analysis of the static potential between a particle and an antiparticle in a strong magnetic field in the LLL approximation was performed in [275]. It was been shown that Coulomb's standard law is modified due to the appearance of vacuum polarization in a magnetic field. The problem of spontaneous magnetization of vacuum of non-Abelian calibration fields at high temperatures was considered in [276]. In 2018-2020, a series of works [277-279] was published, in which QED processes in supercritical magnetic fields were studied. In particular, the process of an electron-positron pair annihilation into two photons was studied in detail. Based on the obtained cross sections, the radiation of magnetars was simulated by the Monte Carlo method and it was shown that the radiation spectrum has a long low-frequency tail and it coincides with the spectrum of magnetars.

### 1.10. Conclusions to the Chapter 1

The above QED processes, namely that run in the magnetosphere of X-ray pulsars, processes in the SLAC experiment [235,236], processes that are planned to be studied in the field of quantum electrodynamics in strong electromagnetic fields in the FAIR project are a single class of QED processes of Furry picture. They are described by Feynman diagrams of the first and the second orders, where solid lines are wave and Green's functions of an electron in a homogeneous external magnetic field, and wavy lines are wave functions of a photon (see Fig. 1.7). The figure shows diagrams of the following processes: 1) the synchrotron radiation (the SR process), 2) the electron-positron pair production by one photon (the OPP process), 3) the Compton scattering (the CS process), 4) the double synchrotron radiation (the DSR process), 5) the electron-positron pair production by two photons (the TPP process), 6) the electron-positron pair production by one photon with photon emission (the OPPE process) 7) the photon propagation in a magnetic field 8) the trident process.

These processes will be studied in this paper. Note that in addition to those shown in Fig. 1.7. in principle, such processes are also possible, the Feynman diagrams of which are presented in Fig. 1.8. These are 1) scattering of an electron on a nucleus, 2) bremsstrahlung of an electron on a nucleus, 3) scattering of an electron on an electron. The processes shown in Fig. 1.8 are characterized by a cross section and require to introduce of value of a flow of initial electrons. The addition of a strong magnetic field, which quantizes the transverse motion of particles, significantly changes the concept of the flow of initial electrons.

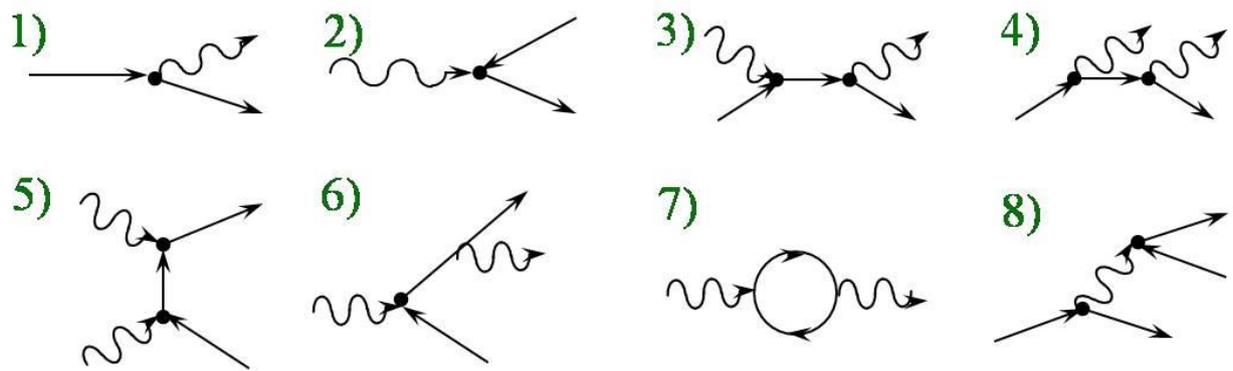

Figure 1.7. Feynman diagrams of the first and the second order quantum electrodynamics processes in an external magnetic field

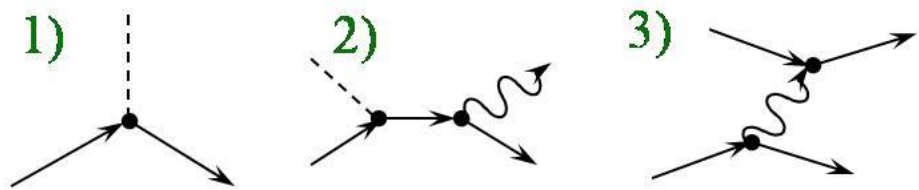

Figure 1.8. Feynman diagrams of processes: 1) scattering of an electron on a nucleus, 2) bremsstrahlung of an electron on a nucleus, 3) scattering of an electron on an electron

# CHAPTER 2
# SPIN AND POLARIZATION EFFECTS
# IN PROCESSES OF SYNCHROTRON RADIATION (SR)
# AND ONE PHOTON e⁺e⁻ PAIR PRODUCTION (OPP)

## 2.1. Introduction

Research methods of quantum electrodynamics elementary processes are the standard rules of QED for finding probabilities of processes. To perform calculations of QED processes in an external magnetic field, a diagram technique within the framework of Farry picture is used, when the interaction of charged particles with a classical magnetic field is taken into account accurately, and interaction with photons treated within the perturbation theory. The external magnetic field measured in the units of the critical (Schwinger) field is a small parameter of the problem in the ultraquantum approximation, which makes it possible to obtain simple analytical expressions for the probabilities of QED processes.

Relativistic unit system is used hereafter, $\hbar = c = 1$.

In the study of QED processes, the direction of electron (positron) spins and photon polarization are taken into account. In a magnetic field, the motion of an electron is characterized by a specific value of the spin projection on the direction of the field, +1/2, or -1/2. The probability of a process involving an electron with a spin projection plus 1/2 (minus 1/2) will be denoted as $W^+$ ($W^-$). The polarization of photons is described by the Stokes parameters $\xi_1$, $\xi_2$, $\xi_3$, which determine the degree of polarization $P$ by the relation

$$P = \sqrt{\xi_1^2 + \xi_2^2 + \xi_3^2} \ . \tag{2.1}$$

In processes with an external magnetic field, important cases of photon polarization are: (i) normal linear polarization, when the photon polarization plane is perpendicular to the plane of the wave vector and the direction of the magnetic field, in this case $\xi_3=-1$; (ii) anomalous linear polarization, when the field plane the photon coincides with the plane of the wave vector and the direction of the magnetic field, with $\xi_3=+1$.

***The spin-polarization effect*** is the effect of coupling of the polarization of the initial photons to the spins of the final particles, and vice versa, coupling of the spins of the initial particles to the polarization of the final photons.

Figure 2.1 shows schematic view of processes with polarized particles considered within quantum scattering theory. Here, *a* and *b* are the particles in the initial and final states respectively; $P_a$ and $P_b$ are the parameters describing polarization of the particles *a* and *b* respectively; $f_c$ is an analyzer filter that selects particles with given polarization $P_c$.

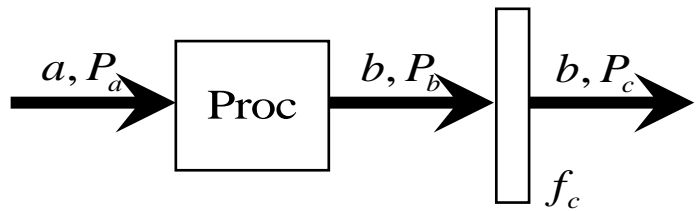

Fig. 2.1. Schematic view of a QED process with polarized particles

Thus, as a result of scattering of the particles of set *a*, they pass into the set of particles *b*. After application of the filter $f_c$, the final particles are characterized by the parameters $P_c$. In this formulation of the problem, the scattering theory gives an expression for the probability of transition from the initial state with fixed polarization parameters $P_a$ to the final state with fixed polarization parameters $P_c$. It is not correct to obtain the probabilities of transition to the final state *b*, $P_b$ without an additional filter, because the final polarization parameters are not set, but are the result of the process. However, the parameters $P_b$ can be found by determining the values of $P_c$ at which the degree of polarization is maximum,

$$P_b = \max\{P_c\}. \qquad (2.2)$$

As an example, consider a process with a photon in the final set of particles and let us determine the polarization and the degree of polarization of this photon. In general, the probability resulting from the quantum theory can be written in the form

$$W_\xi = A(1 + b_1\xi_1 + b_2\xi_2 + b_3\xi_3) = A(1 + \vec{b}\vec{\xi}), \qquad (2.3)$$

where $\xi_i$ are the Stokes parameters defined by the polarization vector of the wave function of the final photon (see sec. 2.2). In fact, these are the polarization parameters of the filter $f_c$ (see fig. 2.1). Further, $A$ is a factor that does not contain $\xi_i$, $b_i$ — are the coefficients at the parameters $\xi_i$. The explicit form of $A$ and $b_i$ is determined by the process. The degree of polarization is

$$P = (W_\xi - W_{-\xi}) / (W_\xi + W_{-\xi}) = \vec{b}\vec{\xi}. \qquad (2.4)$$

This expression is maximum when the vectors $\vec{b}$ are $\vec{\xi}$ collinear, and the filter $f_c$ transmits purely polarized states only, e.g. $|\vec{\xi}| = 1$. The obtained quantity is the sought polarization degree of the final photon,

$$P_{max} = |\vec{b}| = \sqrt{b_1^2 + b_2^2 + b_3^2}, \qquad (2.5)$$

It follows from the definition (2.1) that coefficients $b_i$ are the sought Stokes parameters of a final photon.

<u>Wave functions and parameters of particle polarization.</u> Let us set the wave functions of an electron and a photon with defined polarization parameters. For the electron wave function we will use the following expression [125],

$$\Psi^- = \frac{A_l}{\sqrt{S}} e^{-i(\varepsilon_l t - p_y y - p_z z)}[i\sqrt{m_l - \mu m}U_l(\zeta) + \mu\sqrt{m_l + \mu m}U_{l-1}(\zeta)\gamma^1]u_l^-,$$

$$(2.6)$$

where $\varepsilon_l = (m^2 + 2lhm^2 + p_z^2)^{1/2}$ is the electron energy in magnetic field; $p_y, p_z$ are the projections of the generalized electron momentum to the axes $x$ and $y$, respectively; $S$ is normalizing area in the $xy$ plane; $A_l = (h^{1/2}m / 4m_l\varepsilon_l)^{1/2}$ is the normalizing constant;

$$m_l = m\sqrt{1+2lh};  \qquad (2.7)$$

$h = H/H_0 = eH/m^2$ is the magnetic field strength in the units of the critical one; $\mu$ is the sign of the spin projection on the field direction; $l$ is the Landau level number;

$$U_l(\zeta) = \frac{1}{\sqrt{\pi^{1/2}2^l l!}} \exp(-\zeta^2/2) H_l(\zeta) \qquad (2.8)$$

is the Hermite function and $H_l(\zeta)$ is the Hermite polynomial;

$$\zeta = \sqrt{hm^2}(x + \frac{p_y}{hm^2}) \qquad (2.9)$$

is the offset dimensionless coordinate along the quantized direction $x$;

$\gamma^1$ is the Dirac matrix in the standard representation; $u_l^-$ is a constant bispinor of the form

$$u_l^- = \frac{1}{\sqrt{\varepsilon_l - \mu m_l}} \begin{pmatrix} 0 \\ \mu m_l - \varepsilon_l \\ 0 \\ p_z \end{pmatrix}. \qquad (2.10)$$

Let the constant magnetic field $H$ be directed along $z$ axis. In this case, the electron wave function of the form (2.6) corresponds the following gauge of the external electromagnetic potential (Landau gauge),

$$A_{ext}^0 = 0, \ \vec{A}_{ext} = (0, xH, 0). \qquad (2.11)$$

We use usual expression for the wave function of a photon [191],

$$A^i = \sqrt{\frac{2\pi}{\omega V}} e^i \exp(-i\mathbf{kx}), \qquad (2.12)$$

where $V$ is the normalization volume; $\omega$, $\mathbf{k}$ are the frequency and 4-momentum of a photon; $e^i$ is the polarization vector.

In the case when the photon propagates along $z$ axis, the photon polarization vector $e^i$ has the form

$$e^i = (0, \vec{e}), \quad \vec{e} = (\cos\alpha, \sin\alpha \cdot e^{i\beta}, 0), \tag{2.13}$$

where $\alpha$ and $\beta$ are the photon polarization parameters. Vector $e^i{}_{\text{ort}}$ corresponding to the orthogonal polarization can be obtained from (2.13) by the substitution

$$\alpha \to \alpha + \pi/2. \tag{2.14}$$

Obviously, the following relation is true,

$$e^i e^*_{i,\text{ort}} = 0. \tag{2.15}$$

In the general case of elliptic polarization, the rotation direction of the polarization plane is given by the sign of

$$direct = \text{sgn}(\text{tg}(\alpha)\sin(\beta)), \tag{2.16}$$

and the angle between the semi-major axis of the ellipse and $x$-axis is given by the expression

$$\text{tg}(2 \cdot angle) = \text{tg}(2\alpha)\cos(\beta). \tag{2.17}$$

For example, the polarization vector is $\vec{e} = (1,0,0)$ when $\alpha=0$ and $\beta=0$, that corresponds to the linear polarization along $x$ axis. Similarly, if $\alpha=\pi/4$ and $\beta=\pi/2$, the polarization vector is $\vec{e} = (1,i,0)/\sqrt{2}$, that corresponds to right circular polarization. Photon polarization matrix is defined via $x$ and $y$ components of the polarization vector $e^\mu$ according to

$$\rho_{ik} = e_i e_k^* = \begin{pmatrix} \cos^2\alpha & \sin 2\alpha \cdot e^{-i\beta}/2 \\ \sin 2\alpha \cdot e^{i\beta}/2 & \sin^2\alpha \end{pmatrix}. \tag{2.18}$$

Note that summation over two orthogonal polarizations yields the unitary matrix,

$$\rho_{ik} + (\rho_{ik})_{\text{ort}} = \delta_{ik}. \tag{2.19}$$

The most known set of the polarization parameters are the Stokes parameters $\vec{\xi} = (\xi_1, \xi_2, \xi_3)$, that are defined as

$$\vec{\xi} = \text{Sp}\rho\vec{\sigma}, \tag{2.20}$$

where $\vec{\sigma} = (\sigma_1, \sigma_2, \sigma_3)$ are the Pauli matrices. Inserting into this definition the explicit expression of the density matrix (2.18), we obtain the relationship between the Stokes parameters and the parameters $\alpha$ and $\beta$:

$$\xi_1 = \sin 2\alpha \cdot \cos \beta, \; \xi_2 = \sin 2\alpha \cdot \sin \beta, \; \xi_3 = \cos 2\alpha. \qquad (2.21)$$

The transition to the orthogonal polarization vector according to rule (2.14) results in the Stokes parameters with the opposite sign:

$$\xi_i \to -\xi_i, \; i = 1, 2, 3. \qquad (2.22)$$

Note that the polarization degree (2.1), determined by these parameters, is equal to one corresponding to the pure state:

$$P = \sqrt{\xi_1^2 + \xi_2^2 + \xi_3^2} = 1. \qquad (2.23)$$

In the case of an arbitrarily directed photon, with a wave vector

$$\vec{k} = k(\sin\theta\cos\varphi, \sin\theta\sin\varphi, \cos\theta), \qquad (2.24)$$

the spatial components of the photon polarization vector can be obtained by rotating the coordinate system, if the z axis is oriented along the wave vector,

$$\vec{e} = \begin{pmatrix} \cos\theta\cos\varphi & -\sin\varphi & \sin\theta\cos\varphi \\ \cos\theta\sin\varphi & \cos\varphi & \sin\theta\sin\varphi \\ -\sin\theta & 0 & \cos\theta \end{pmatrix} \begin{pmatrix} \cos\alpha \\ \sin\alpha \cdot e^{i\beta} \\ 0 \end{pmatrix},$$

or

$$\vec{e} = \begin{pmatrix} \cos\theta\cos\varphi\cos\alpha - \sin\varphi\sin\alpha \cdot e^{i\beta} \\ \cos\theta\sin\varphi\cos\alpha + \cos\varphi\sin\alpha \cdot e^{i\beta} \\ -\sin\theta\cos\alpha \end{pmatrix}. \qquad (2.25)$$

## 2.2. Spin-polarization effects in synchrotron radiation

Probability amplitude of SR. To construct the probability amplitude of the process of photon radiation by an electron $A_{if}$ we use the standard QED rules according to the Feynman diagram shown in Fig.2.2,

$$A_{if} = -ie\int d^4x \bar{\Psi}'(\xi)\gamma^i A_i^* \Psi(\zeta),\qquad(2.26)$$

where $\gamma_\mu$ are the Dirac matrices; $\bar{\Psi}'(\xi)$ are the Dirac conjugated wave function of the final electron of the form

$$\bar{\Psi}' = \frac{A_{l'}}{\sqrt{S}} e^{i(\varepsilon'_{l'}t - p'_y y - p'_z z)} \bar{u}^-_{l'}[-i\sqrt{m_{l'} - \mu'm}\, U_{l'}(\xi) + \mu'\sqrt{m_{l'} + \mu'm}\, U_{l'-1}(\xi)\gamma^1].\quad(2.27)$$

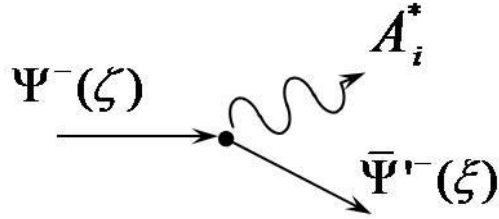

Fig.2.2. Feynman diagram of the SR process

Here, $\mu'$ is the sign of the spin projection of the final electron on the magnetic field, $m_{l'} = m(1 + 2l'h)^{1/2}$,

$$\xi = \sqrt{hm^2}(x + p'_y / hm^2).\qquad(2.28)$$

Integrals in Eq. (2.26) over the variables $t$, $z$, $y$ yields three Dirac delta functions that corresponds to three conservation laws,

$$\varepsilon_l = \varepsilon'_{l'} + \omega,\ \ p_z = p'_z + k_z,\ \ p_y = p'_y + k_y.\qquad(2.29)$$

As follows from Eqs. (2.9) and (2.28), $y$ components of the generalized momentum of the initial and final electrons have the meaning of $x$ coordinates of the rotation center of the classical particle orbits. The later equation in (2.29) determines the relationship of these coordinates with $y$ component of the photon momentum. The former two equations in (2.29) are the conservation laws for the energy and the longitudinal momentum particles. From these relations we can obtain the expression for the photon frequency in the form

$$\omega = \frac{1}{1-u^2}(\varepsilon_l - p_z u - \sqrt{(\varepsilon_l - p_z u)^2 - 2hm^2(l-l')(1-u^2)}), \quad (2.30)$$

where

$$u = \cos\theta \quad (2.31)$$

is the cosine of the polar escape angle of the photon.

In the ultraquantum approximation, when the electron occupies the lowest Landau levels as well as in the case of a small angle $\theta$, the frequency $\omega$ can be reduced to the form

$$\omega = \frac{hm^2(l-l')}{\varepsilon_l - p_z u}, \quad \text{якщо} \quad \theta \ll 1 \text{ або } h(l-l') \ll 1. \quad (2.32)$$

It should be noted that the frequency vanishes if one inserts the condition $l = l'$ into Eq. (2.30). This means that the SR process is possible only when the electron transits to a lower Landau level.

Integrating the amplitude (2.26) in the $x$ variable result in appearance of the special functions

$$I(l',l;\vec{k},p_y) \equiv I(l',l),$$

that occur in the description of all QED processes in an external magnetic field. They are defined as follows:

$$I(l',l) = \sqrt{hm^2} \int_{-\infty}^{+\infty} dx e^{-ik_x x} U_{l'}(\xi) U_l(\zeta). \quad (2.33)$$

The variables $\xi$ and $\zeta$ are related to each other as

$$\xi = \zeta - k_y / \sqrt{hm^2}. \quad (2.34)$$

The complex function $I(l',l)$ can be expressed via its absolute value and phase as

$$I(l',l) = J(l',l)e^{i\Phi}, \quad (2.35)$$

where the phase $\Phi$ is

$$\Phi = \frac{k_x(2p_y - k_y)}{2hm^2} + (l - l')(\varphi - \frac{\pi}{2}). \qquad (2.36)$$

Here, $\varphi$ - is the azimuth angle of the final photon and $J(l',l)$ is a real valued function of the variable $\eta$, defined as

$$J(l',l) = e^{-\frac{\eta}{2}} \eta^{\frac{|l-l'|}{2}} \begin{cases} \sqrt{\dfrac{l!}{l'!} \dfrac{1}{(l-l')!}} F(-l', l-l'+1, \eta), & l > l' \\ (-)^{l+l'} \sqrt{\dfrac{l'!}{l!} \dfrac{1}{(l'-l)!}} F(-l, l'-l+1, \eta), & l' > l \end{cases}, \qquad (2.37)$$

where $F$ is the confluent hypergeometric function, and

$$\eta = \frac{k_x^2 + k_y^2}{2hm^2} = \frac{\omega^2(1-u^2)}{2hm^2}. \qquad (2.38)$$

Given the wave functions (2.6), (2.12) and (2.27), the amplitude of the SR process can be reduced to the general form

$$A_{if} = \frac{-ie 2\pi^3 \sqrt{2\pi} \delta^3(p-p'-k)}{S\sqrt{V}\sqrt{\omega \varepsilon_l \varepsilon'_{l'} m_l m_{l'}}} e^{i\Phi} \sum_{a=1}^{4} Q_a, \qquad (2.39)$$

where $Q_1 = -J(l',l) M_m M_m' D e_z$, $Q_2 = -J(l'-1, l-1) \mu M_p \mu' M_p' D e_z$,
$Q_3 = -J(l', l-1) \mu M_p M_m' C H_m^*$, $Q_4 = -J(l'-1, l) M_m \mu' M_p' C H_p^*$.

In the above expressions we denote

$$M_m = \sqrt{m_l - \mu m}, \quad M_p = \sqrt{m_l + \mu m}, \qquad (2.40)$$

$$C = -E_m E_m' + \operatorname{sgn}(p_z)\operatorname{sgn}(p_z') E_p E_p', \quad D = \operatorname{sgn}(p_z') E_m E_p' + \operatorname{sgn}(p_z) E_p E_m', \qquad (2.41)$$

$$E_m = \sqrt{\varepsilon_l - \mu m_l}, \quad E_p = \sqrt{\varepsilon_l + \mu m_l}, \qquad (2.42)$$

$$e_z = -\sin\theta\cos\alpha, \quad H_m = \cos\theta\cos\alpha - i\sin\alpha \cdot e^{i\beta}, \quad H_p = \cos\theta\cos\alpha + i\sin\alpha \cdot e^{i\beta}. \qquad (2.43)$$

Here, primed quantities $M_m$, $M_p$, $E_m$, $E_p$ corresponds to the final electron. An asterisk near $H_m$ and $H_p$ denotes complex conjugation. Note that the expression for the amplitude (2.39) is a scalar complex value, i.e. calculations of all spinor expressions have already been performed in the amplitude.

Probability of the SR process. The SR probability equals to the squared amplitude multiplied by the number of final states. In general case, the expression of the process rate for SR with arbitrary spins and photon polarization and fixed values of the Landau levels takes the form

$$\frac{dW^{\mu\mu'}_{ll';\xi_2,\xi_3}}{du} = \frac{\alpha\omega}{16 m_l m'_{l'} \varepsilon_l (\varepsilon'_{l'} + \omega u^2 - p_z u)} \left|\sum Q\right|^2. \qquad (2.44)$$

Below we analyze these expressions in two cases: the ultraquantum, or LLL (Lowest Landau Level), approximation [9], and the ultrarelativistic approximation.

SR process in the ultraquantum approximation. The probabilities of QED processes significantly depend on the spin projection of particles, so it is necessary to analyze individual probabilities with fixed values of $\mu$ and $\mu'$. Transition to another inertial frame of reference moving along the direction of magnetic field does not change the field itself, thus, without loosing generality we can set the longitudinal momentum of the initial electron $p_z$ to zero.

In the ultraquantum approximation the quantity $lh$ is a small parameter. Hence, the hypergeometric function $F$ in Eq. (2.37) equals to unity, and the square of the transverse frequency of the photon has the form

$$\eta = \frac{1}{2}(l-l')^2 h(1-u^2). \qquad (2.45)$$

We write here the expression for the process rate in four cases of particle spin projectiosn,

$$\frac{dW^{--}}{du} = \alpha m h^2 \frac{l(l-l')}{4l'} R^2_{ll'}[1+u^2 - (1-u^2)\xi_3 + 2u\xi_2], \qquad (2.46)$$

$$\frac{dW^{++}}{du} = \alpha m h^2 \frac{(l-l')}{4} R^2_{ll'}[1+u^2 - (1-u^2)\xi_3 + 2u\xi_2], \qquad (2.47)$$

$$\frac{dW^{+-}}{du} = \alpha m h^3 \frac{(l-l')^3}{8l'} R^2_{ll'}[1+u^2 + (1-u^2)\xi_3 + 2u\xi_2], \qquad (2.48)$$

$$\frac{dW^{-+}}{du} = \frac{\alpha m}{32} h^5 l(l-l')^3 R_{ll'}^2 \left[ (1 - \frac{l-l'}{l-l'+1}(1-u^2))^2 (1+\xi_3) + \right.$$

$$\left. + (1 + \frac{l-l'}{l-l'+1}(1-u^2))^2 u^2 (1-\xi_3) + 2(1 - \frac{(l-l')^2}{(l-l'+1)^2}(1-u^2)^2) u \xi_2 \right]. \quad (2.49)$$

In above expressions, the factor $R^2$ is

$$R_{ll'}^2 = e^{-\eta} \eta^{l-l'-1} \frac{(l-1)!}{(l'-1)!} \frac{1}{(l-l'-1)!^2}. \quad (2.50)$$

Differential radiation intensity is defined as the product of the differential rate and the frequency:

$$\frac{dI}{du} = \omega \frac{dW}{du}. \quad (2.51)$$

The form of the angular dependence significantly depends on the value of the photon polarization and the particle spin projections.

Figure 2.3 shows the angular distribution of radiation intensity in the case of linear photon polarization and electron transition from the level $l=2$ to $l'=1$. Magnetic field strength is $h=0.1$ ($H=4 \cdot 10^{12}$G). Note that the intensity is shown in the units of $I_0$,

$$I_0 = \alpha h^3 m^2 / 4 \sim 10^{21} eB / рад \cdot c. \quad (2.52)$$

In the case of no-spin-flip processes, the intensity has maxima in the direction perpendicular to the magnetic field, in accordance with the previous results [3,4]. In spin-flip processes, the radiation has maximum at angles $\pm 45^0$ with respect to the field direction.

After summation over the particle polarizations in Eqs. (2.46) - (2.49), we get the previously found expressions [19,20]. Note that to obtain expression (2.49) for the probability of the electron transition from the ground state to the state with inversed spin $W^{-+}$, development in the parameter $lh$ to the second order is required. In particular, the approximate expression for the hypergeometric function is

$$F(a,c,\eta) = 1 + \frac{a}{c}\eta. \quad (2.53)$$

In this case, the photon frequency is

$$\omega = (l-l')hm - \frac{1}{2}h^2 m(l-l')(l+l'+(l-l')u^2). \qquad (2.54)$$

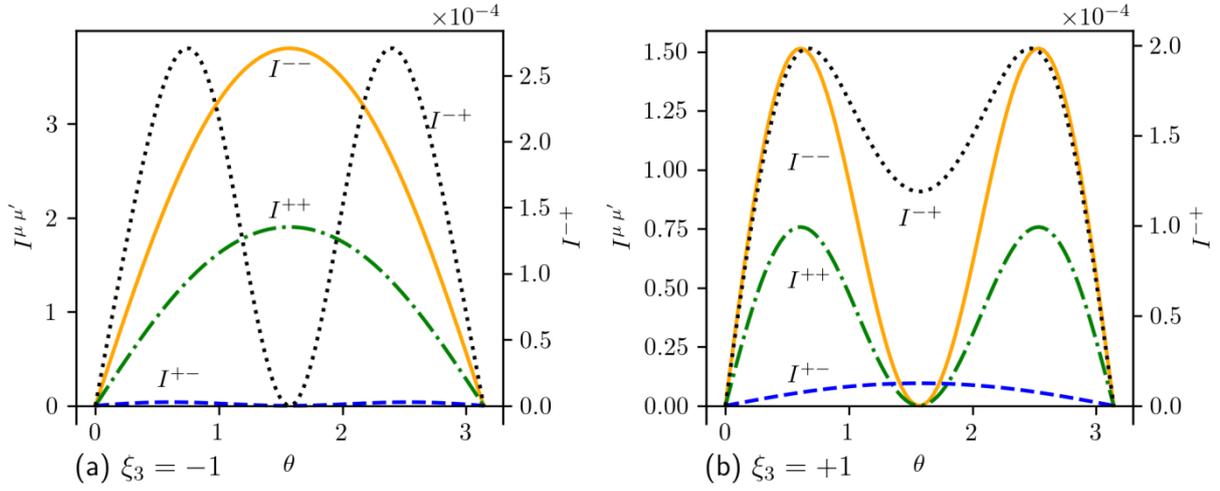

Fig.2.3. Angular dependence of the SR intensity for the electron transition $l=2 \to l'=1$ and the field magnitude $h=0.1$. a) normal photon polarization ($\xi_3=-1$); b) anomalous photon polarization ($\xi_3=+1$)

Let's analyze the obtained relations (2.46) - (2.49). The transition to the closest Landau level is the most probable one in the SR process. Transition to subsequent levels increases power of small parameter $h$. Note that the Stokes polarization parameter $\xi_1$ is absent in these expressions. This result is obviously predictable, because the $\xi_1$ parameter describes the linear polarization at angle $\pm 45^0$ relative to the plane of the vectors $\{\vec{k}, \vec{H}\}$, and these are the equivalent orientations. The most probable are the processes without electron spin inversion. Their probabilities $W^{--}$, $W^{++}$ have identical dependence on the photon polarization.

To find the polarization of the radiation, it is necessary to find the maximum of the polarization degree $P_\xi$, as pointed out in the beginning of Section 1. We write separately the multiplier with the photon Stokes parameters entering the expressions (2.46)-(2.47)

$$U_\xi = 1 + u^2 - (1-u^2)\xi_3 + 2u\xi_2. \qquad (2.55)$$

The degree of polarization is

$$P_\xi = \frac{U_\xi - U_{-\xi}}{U_\xi + U_{-\xi}} = \frac{2u}{1+u^2}\xi_2 - \frac{1-u^2}{1+u^2}\xi_3. \qquad (2.56)$$

The radiation Stokes parameters are the coefficients at $\xi_1, \xi_2, \xi_3$ in Eq. (2.56)

$$\xi_{1SR} = 0, \quad \xi_{2SR} = \frac{2u}{1+u^2}, \quad \xi_{3SR} = -\frac{1-u^2}{1+u^2}. \qquad (2.57)$$

The degree of polarization $P_\xi$ with the polarization parameters (2.57) is equal to one. The polarization of synchrotron radiation at the lowest Landau levels determined by Eqs. (2.57) coincides with the previously obtained results [3,4]. For example, radiation along the direction of the magnetic field ($u = 1$) has the right circular polarization ($\xi_2=1$, $\xi_3=0$) and the perpendicular to the field radiation ($u = 0$) has normal polarization ($\xi_2=0$, $\xi_3=-1$). Inserting the Stokes parameters (2.57) into the expressions for the rates $W^{--}$, $W^{++}$ (2.46), (2.47), we get the same expression as in the case of summation over the photon polarization,

$$W_{\xi=\xi_{SR}} = \sum_\xi W_\xi. \qquad (2.58)$$

Thus, the polarization in the SR process without inversion of the electron spin is the same as in classical electrodynamics. On the other hand, if the electron changes the spin projection (a spin-flip process), then we need to analyze Eq (2.48) to obtain the polarization of radiation,

$$P_\xi^{+-} = \frac{W_\xi^{+-} - W_{-\xi}^{+-}}{W_\xi^{+-} + W_{-\xi}^{+-}} = \frac{2u}{1+u^2}\xi_2 + \frac{1-u^2}{1+u^2}\xi_3, \qquad (2.59)$$

and, consequently,

$$\xi_{2SR}^{+-} = \frac{2u}{1+u^2}, \quad \xi_{3SR}^{+-} = \frac{1-u^2}{1+u^2}. \qquad (2.60)$$

In this case, the perpendicular to the field component of linear polarization has the opposite sign compared to the no-spin-flip process with photon polarization parameters given by (2.57), i.e. spin inversion changes the linear polarization of the radiation to the opposite one.

Let us consider resulting radiation polarization for both main and the spin-flip processes in the case of electron transition to the nearest Landau level. Now the polarization degree looks like

$$P_\xi = \frac{(W^{--}+W^{++}+W^{+-})_\xi - (W^{--}+W^{++}+W^{+-})_{-\xi}}{(W^{--}+W^{++}+W^{+-})_\xi + (W^{--}+W^{++}+W^{+-})_{-\xi}} \qquad (2.61)$$

which results in

$$P_\xi = \frac{2u}{1+u^2}\xi_2 - \frac{1-u^2}{1+u^2}(1-\frac{2h}{2l+1})\xi_3. \qquad (2.62)$$

The maximum polarization degree corresponds to the Stokes parameters values

$$\xi_{2SR} = \frac{2u}{1+u^2}, \quad \xi_{3SR} = -\frac{1-u^2}{1+u^2}(1-\frac{2h}{2l+1}) \qquad (2.63)$$

and equals to

$$P_{\xi\max} = 1 - \frac{4h}{2l+1}\left(\frac{1-u^2}{1+u^2}\right)^2 < 1. \qquad (2.64)$$

Thus, after taking into account the spin-flip process, the linear polarization of the radiation reduces by a value proportional to the small parameter $h$. The radiation is partially polarized. The difference from the pure state is most noticeable at the lowest Landau levels.

<u>Radiation in the case of a nonzero longitudinal electron momentum.</u> Figure 2.4 shows the dependence of the final photon frequency on the longitudinal component of the electron momentum and the photon escape angle (2.30).

If $u=1$ (emission along the field) the frequency increases monotonically with increasing $p_z$, while for a fixed $u<1$ the frequency has a maximum as a function of $p_z$, as shown in Fig.2.4. The energy, the longitudinal momentum of the electron and the photon frequency at the maximum point are

$$\varepsilon_m = \frac{m_l}{\sqrt{1-u^2}}, \quad p_m = \frac{m_l u}{\sqrt{1-u^2}}, \quad \omega_m = \frac{m_l - m_{l'}}{\sqrt{1-u^2}}. \qquad (2.65)$$

In this case, the differential rates of the SR process with fixed spin projections are

$$\left.\frac{dW^{--}}{du}\right|_{\omega=\omega_m} = \frac{\alpha}{4}h\omega_m R_{ll'}^2 \frac{l}{l'}(1-\xi_3),  \qquad (2.66)$$

$$\left.\frac{dW^{++}}{du}\right|_{\omega=\omega_m} = \frac{\alpha}{4}h\omega_m R_{ll'}^2 (1-\xi_3),  \qquad (2.67)$$

$$\left.\frac{dW^{+-}}{du}\right|_{\omega=\omega_m} = \frac{\alpha}{8}h^2\omega_m R_{ll'}^2 \frac{(l-l')^2}{l'}(1+\xi_3),  \qquad (2.68)$$

$$\left.\frac{dW^{-+}}{du}\right|_{\omega=\omega_m} = \frac{\alpha}{32}h^4\omega_m R_{ll'}^2 \frac{l(l-l')^2}{(l-l'+1)^2}(1+\xi_3).  \qquad (2.69)$$

Thus, we conclude that at the point of maximum photon frequency for any photon escape angle except emission along the field, the radiation polarization is normal linear for no-spin-flip processes and anomalous linear for spin-flip processes. Note that expressions (2.66) - (2.69) coincide with (2.46) - (2.49) if we set $u=0$ in the latter.

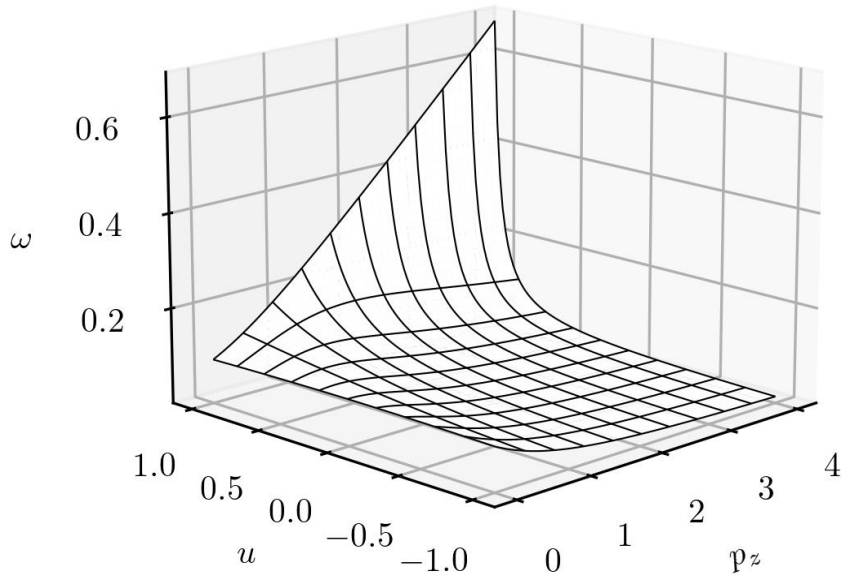

Fig.2.4. Dependence of the photon frequency on the longitudinal momentum of the electron and the photon escape angle

Consider the case of an ultrarelativistic electron moving in the direction close to the direction of the magnetic field. Consequently, the electron occupies the lowest Landau levels and has a large longitudinal momentum $p_z \gg m$. Then, the photon frequency is

$$\omega = \frac{2hm^2(l-l')p_z}{m^2 + (p_z\theta)^2}, \qquad (2.70)$$

where $\theta$ is polar escape angle of the photon. Hence, the critical angle defined as

$$\theta_c = \frac{m}{p_z} \qquad (2.71)$$

distinguishes two cases. If $1 \gg \theta \gg \theta_c$ (relatively large angle), then the radiation frequency is less than synchrotron frequency and the process rate is negligibly small. On the other hand, in the case $\theta \ll \theta_c$ the radiation frequency $\omega = 2(l-l')hp_z$ is much greater than synchrotron frequency. If $\theta = 0$ (forward radiation), the $\eta$ parameter (2.38) vanishes, and the probability amplitude (2.39) is not zero for the nearest Landau levels only. In this case, the radiation rates for the fixed spin projections are

$$\left.\frac{dW^{--}}{du}\right|_{u=1} = \alpha h l \omega (1+\xi_2), \quad \left.\frac{dW^{++}}{du}\right|_{u=1} = \alpha h (l-1) \omega (1+\xi_2), \qquad (2.72)$$

$$\left.\frac{dW^{+-}}{du}\right|_{u=1} = \alpha h^2 \omega (1+\xi_2)/2, \quad \left.\frac{dW^{-+}}{du}\right|_{u=1} = \alpha h^4 l(l-1)\omega(1+\xi_2)/8. \quad (2.73)$$

The radiation is fully polarized with circular polarization for all spin states of the electron.

The SR process in the ultrarelativistic approximation. Now consider the case $l \gg 1$, $l' \gg 1$ and $p_z = 0$. In this case, the radiation is mostly emitted perpendicular to the direction of the field, so it is convenient to introduce a new angle of photon radiation,

$$\theta = \frac{\pi}{2} - \psi. \qquad (2.74)$$

In the general expression for the process amplitude (2.39) we can express the special functions $J'(l,l')$ with different values of $l$ and $l'$ in terms of $J(l,l')$ and its derivative $J'(l,l')$ according to the recurrent relations

$$\sqrt{ll'}J(l'-1,l-1)=0.5(l+l'-\eta)J-\eta J', \qquad (2.75)$$

$$\sqrt{l\eta}J(l',l-1)=0.5(l-l'+\eta)J+\eta J', \qquad (2.76)$$

$$\sqrt{l'\eta}J(l'-1,l)=0.5(l-l'-\eta)J-\eta J'. \qquad (2.77)$$

In the linear approximation, the photon frequency is expressed by the square of a small angle $\psi$ as

$$\omega = m\sqrt{2h}(\sqrt{l}-\sqrt{l'})(1-\frac{1}{4h\sqrt{ll'}}-\frac{1}{2\sqrt{l'}}(\sqrt{l}-\sqrt{l'})\psi^2). \qquad (2.78)$$

Further, the argument of special functions $J(l, l')$ takes on the form

$$\eta = (\sqrt{l}-\sqrt{l'})^2(1-\frac{1}{2h\sqrt{ll'}}(1+2hl\psi^2)). \qquad (2.79)$$

In the ultrarelativistic approximation, the function $J(l,l')$ oscillates rapidly. However, its argument $\eta$ has large magnitude. It would be impossible to separate different Landau levels experimentally. Therefore, a fixed value of electron energy (fixed $l$) implies averaging over the interval of Landau levels $\Delta l \gg 1$. In this case, the rapidly oscillating function $J(l,l')$ and its derivative can be replaced by smoothed MacDonald functions according to the prescription given in Ref. [8]:

$$J(l',l;\eta) = \frac{\sqrt{(\sqrt{l}-\sqrt{l'})^2-\eta}}{\pi\sqrt{3}(\sqrt{l}-\sqrt{l'})}K_{1/3}\left(\frac{2}{3}\frac{\sqrt[4]{ll'}((\sqrt{l}-\sqrt{l'})^2-\eta)^{3/2}}{(\sqrt{l}-\sqrt{l'})^2}\right), \qquad (2.80)$$

$$\frac{\partial J(l',l;\eta)}{\partial \eta} = \frac{\sqrt[4]{ll'}((\sqrt{l}-\sqrt{l'})^2-\eta)}{\pi\sqrt{3}(\sqrt{l}-\sqrt{l'})^3}K_{2/3}\left(\frac{2}{3}\frac{\sqrt[4]{ll'}((\sqrt{l}-\sqrt{l'})^2-\eta)^{3/2}}{(\sqrt{l}-\sqrt{l'})^2}\right). \qquad (2.81)$$

Given the relation (2.79), these functions equals to

$$J = \frac{F}{\pi\sqrt{6h}\sqrt[4]{ll'}}K_{1/3}(\kappa), \quad J' = \frac{F}{\pi\sqrt{6h}\sqrt[4]{ll'}(\sqrt{l}-\sqrt{l'})}K_{2/3}(\kappa), \qquad (2.82)$$

where

$$F = \sqrt{1+2lh\psi^2} = \sqrt{1+\Psi^2} = \sqrt{1+\psi^2/\psi_c^2}, \qquad (2.83)$$

$\psi_c = 1/\sqrt{2lh}$ - is a typical emission angle, and the argument of the MacDonald functions is

$$\kappa = \frac{2}{3} \frac{(\sqrt{l} - \sqrt{l'})}{\sqrt{ll'}\sqrt{2h}^3} F^3. \qquad (2.84)$$

Since inthe individual Landau levels are not separable in the ultrarelativistic case, the Landau level numbers are inconvenient parameters for the description of the process. Instead, we will use the photon frequency and the energy of the initial electron. In the argument of the MacDonald function, the small parameter $\psi$ already enters $F$, so with given accuracy

$$\frac{(\sqrt{l} - \sqrt{l'})}{\sqrt{l'}} = \frac{\varepsilon_l - \varepsilon'_{l'}}{\varepsilon'_{l'}} = \frac{\omega}{\varepsilon_l - \omega}.$$

Hence, the argument $\kappa$ can be transformed to the form

$$\kappa = \frac{\omega}{(\varepsilon_l - \omega)z} F^3, \qquad z = \frac{3h\varepsilon_l}{m}. \qquad (2.85)$$

Here, $z$ is the dimensionless energy of the initial electron. In addition, we define dimensionless photon frequency $y$ as

$$y = \frac{\omega}{\omega_c}, \qquad \omega_c = \frac{z\varepsilon_l}{2+z}. \qquad (2.86)$$

Its maximum value of $y_m = 1 + 2/z$ corresponds the radiation frequency equal to $\varepsilon_l$. If $z \ll 1$ we have $\omega_c = z\varepsilon_l/2$, i.e. $z/2$ is the fraction of the energy that is converted to the radiation. The argument of the MacDonald function can be expressed in terms of the quantities $y$ and $z$ in the form

$$\kappa = yF^3/(2+z-yz), \qquad (2.87)$$

and the parameter $\eta$ equals to

$$\eta = y^2 z^2 (1-\psi^2)/(18h^3(2+z)^2). \qquad (2.88)$$

With the above definitions, the initial and the final Landau level numbers of the electron are

$$l' = \frac{l}{(2+z)^2}((2+z-yz)^2 - y^2z^2\psi^2), \quad l = \frac{z^2}{18h^3}. \qquad (2.89)$$

Finally, in the ultrarelativistic case the differential intensity of the SR can be written in as

$$\frac{dI^{\mu\mu'}}{dyd\Psi} = \frac{9I_0 y^2 F^2}{8\pi^2(2+z)^3(2+z-yz)^2} D^{\mu\mu'}, \qquad (2.90)$$

$$I_0 = \alpha h^2 \varepsilon^2. \qquad (2.91)$$

In Eq. (2.90), the factors $D^{\mu\mu'}$ corresponding to the no-spin-flip processes ($\mu=\mu'$) look like

$$D^{\mu\mu} = [(a\Psi^2 + b)K_{1/3}^2 + aF^2 K_{2/3}^2 - 2\mu c F K_{1/3} K_{2/3}] + 2\xi_2 \Psi[aFK_{1/3}K_{2/3} - \mu c K_{1/3}^2] +$$
$$+\xi_3[(a\Psi^2 - b)K_{1/3}^2 - aF^2 K_{2/3}^2 + 2\mu c F K_{1/3} K_{2/3}], \qquad (2.92)$$

where $a = (4+2z-yz)^2$, $b = y^2 z^2$, $c = \sqrt{ab}$.

For the spin-flip processes ($\mu=-\mu'$), the factors $D^{\mu,-\mu}$ have the form

$$D^{\mu,-\mu} = y^2 z^2 \{[F^2(K_{1/3}^2 + K_{2/3}^2) + 2\mu F K_{1/3} K_{2/3}] + 2\xi_2 \Psi[FK_{1/3}K_{2/3} + \mu K_{1/3}^2] +$$
$$+\xi_3[(1-\Psi^2)K_{1/3}^2 + F^2 K_{2/3}^2 + 2\mu F K_{1/3} K_{2/3}]\}. \qquad (2.93)$$

The dependence of the differential intensity of SR on the photon frequency and its escape angle is shown in Fig. 2.5. Intensity is measured in units $I_0 = \alpha m^2 = 3 \cdot 10^{24}$ $eV/rad \cdot s$. The parameter $z$ is set to $z=3$, that corresponds to the equality $h\varepsilon = m$. In the ultrarelativistic case ($\varepsilon \gg 1$) this means that the magnetic field is much smaller than the critical one ($h \ll 1$). Figure 2.5 *a)* shows the intensity in the case of normal linear polarization ($\xi_3=-1, \xi_2=0$), and Fig. 2.5 *b)* shows the case of anomalous polarization ($\xi_3=1, \xi_2=0$). It is clear from Eqs. (2.92), (2.93) that both of these cases have the maximum probability.

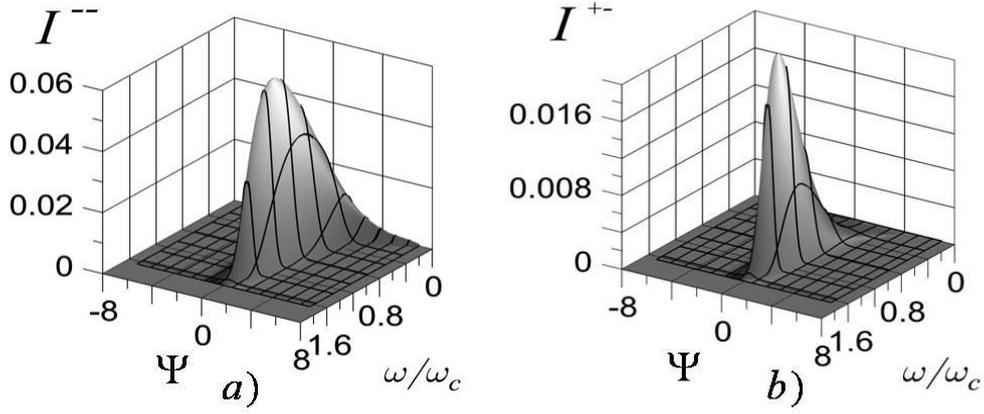

Fig.2.5. Differential intensity of the SR process as a function of the photon frequency and its escape angle in the ultrarelativistic case. (a) the process without spin inversion ($\mu=\mu'=-1$), (b) spin-flip process ($\mu=-\mu'=1$)

The intensity is maximum when $y = 0.8$ that corresponds to the frequency $\omega_{max}=0.5\varepsilon$ (half the energy of the electron). Note that the known expression for the radiation frequency at the point of maximum intensity [4] gives incorrect value of $1.5\varepsilon$,

$$\omega_{max} = \frac{z\varepsilon}{2} = \frac{3h\varepsilon^2}{2m} = 1.5\varepsilon$$

The above expression have been obtained in the approximation $z \ll 1$ and is not applicable in the considered case.

Now let us consider the remaining cases of anomalous photon polarization $\xi_3=1$ without electron spin inversion and normal photon polarization $\xi_3=-1$ with spin inversion. In this cases, the factors $D^{\mu\mu'}$ entering the intensity (2.90) can be transformed to

$$D^{++}_{\xi_3=1} = D^{--}_{\xi_3=1} = 2a\Psi^2 K^2_{1/3}, \quad D^{+-}_{\xi_3=-1} = D^{-+}_{\xi_3=-1} = 2b\Psi^2 K^2_{1/3}. \quad (2.94)$$

Figure 2.6. shows the dependence of the radiation intensity on the escape angle corresponding to Eqs. (2.94) with $z=3$ and $y=1$. It is clear from the figure that the considered processes differ significantly from the main channel shown in Fig.2.5 by the absence of radiation in the direction perpendicular to the magnetic field ($\Psi=0$).

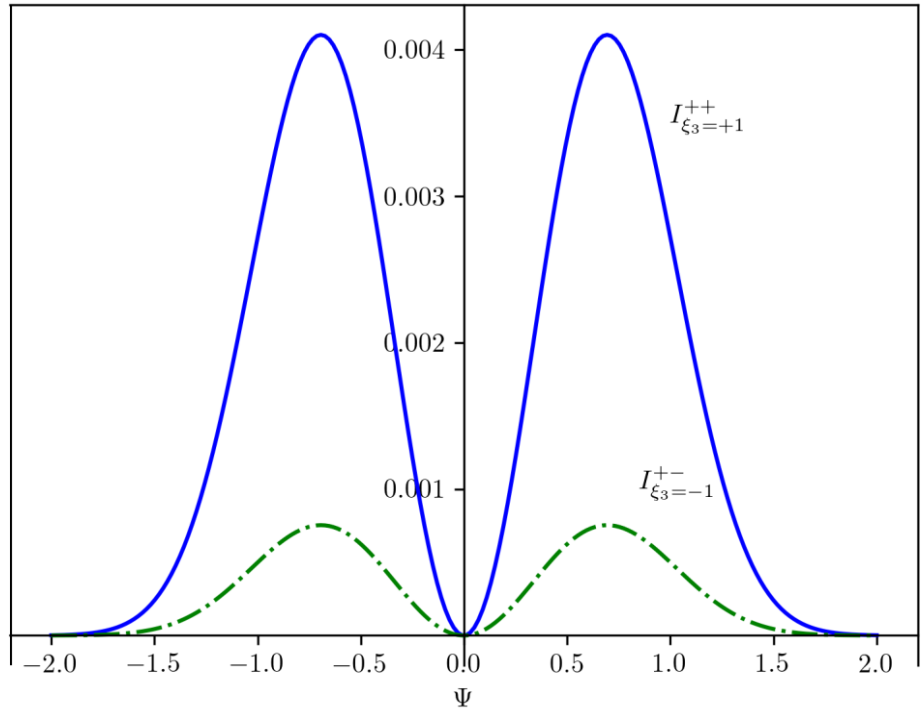

Fig.2.6. The dependence of SR intensity on the photon escape angle for cases of anomalous photon polarization without electron spin inversion and normal photon polarization with spin inversion

Thus, we reproduced a result of [19] obtained in classical approximation: an ultrarelativistic electron rotating on a circular orbit emits photons of normal polarization ($\sigma$ polarization) in a narrow cone with the maximum intensity at the angle $\Psi=0$, and photons of anomalous polarization ($\pi$ polarization) in the same narrow cone, but with total absence of radiation in the direction $\Psi=0$. Note that a similar effect was observed in the ultraquantum case (see Fig.2.3).

In Figure 2.7, the theoretical predictions about angular dependence of $\sigma$ and $\pi$ polarized radiation are compared to experimental data for emission of 250 *MeV* electrons in the visible part of the spectrum [20]. Such value of electron energy and magnetic field strength of several thousands Gauss correspond to a very small value of the parameter $z \sim 10^{-6}$. In this case, the spin-flip processes are small, but at $\Psi=0$ they contribute to the $\pi$ component of linear polarization. Let us compare the radiation intensities without spin inversion $I^{\mu\mu}$ and with a flip of spin, $I^{\mu,-\mu}$, at $\Psi=0$. In this case, the factors (2.92) and (2.93) are, respectively,

$$D^{\mu\mu} = (bK_{1/3}^2 + aK_{2/3}^2 - 2\mu c K_{1/3}K_{2/3})(1-\xi_3),$$

$$D^{\mu,-\mu} = b(K_{1/3}^2 + K_{2/3}^2 + 2\mu K_{1/3}K_{2/3})(1+\xi_3).$$

Note that in this case always $I^{--} > I^{++}$ and $I^{+-} > I^{-+}$. In this expressions, the coefficients near the parameter $\xi_3$ equal to -1 and +1, that indicates that SR with $\sigma$ polarization takes place in the process without spin inversion and SR with $\pi$ polarization results from the spin-flip process. In the case $z<<1$ the following condition is true,

$$\frac{I^{\mu,-\mu}|_{\xi_3=+1}}{I^{\mu\mu}|_{\xi_3=-1}} = I_\pi / I_\sigma \approx \frac{(yz)^2}{8}.$$

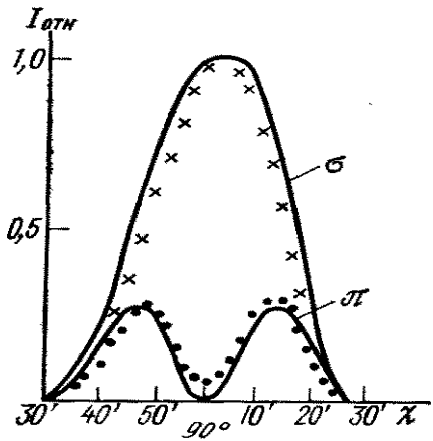

Fig. 2.7. Comparison of theoretical predictions and experimental data for angular dependence of $\sigma$ and $\pi$ polarized SR [20]

In the visible part of spectrum ($y=0.01$), the obtained ratio is negligibly small ($\sim 10^{-17}$) if $z\sim 10^{-6}$. It can be concluded that the experimental presence of nonzero $\pi$ component of the linear polarization at $\Psi=0$ in Fig.2.7 is connected with the finite size of the electron beam. It should be stressed out, that $\pi$ polarization of the SR associated with the spin-flip process can be observed experimentally in the frequency range $\omega\sim\varepsilon$ ($y\sim 1/z$), or by increasing $z$.

Now we proceed to general analysis of polarization of synchrotron radiation for arbitrary direction and frequency and with arbitrary electron spin projections. We start with a process without electron spin inversion $(\mu=\mu')$, the intensity of which is proportional to expressions (2.92). The degree of polarization of the emitted photon is defined as the maximum value of the expression

$$P_{\xi_{SR}}^{\mu\mu} = \max \frac{D_{\xi}^{\mu\mu} - D_{-\xi}^{\mu\mu}}{D_{\xi}^{\mu\mu} + D_{-\xi}^{\mu\mu}}. \qquad (2.95)$$

This allow us to find the Stokes parameters of the emitted photon,

$$\xi_{2SR} = \frac{2\Psi[aFK_{1/3}K_{2/3} - \mu c K_{1/3}^2]}{(a\Psi^2 + b)K_{1/3}^2 + aF^2 K_{2/3}^2 - 2\mu cFK_{1/3}K_{2/3}}, \qquad (2.96)$$

$$\xi_{3SR} = \frac{(a\Psi^2 - b)K_{1/3}^2 - aF^2 K_{2/3}^2 + 2\mu cFK_{1/3}K_{2/3}}{(a\Psi^2 + b)K_{1/3}^2 + aF^2 K_{2/3}^2 - 2\mu cFK_{1/3}K_{2/3}}. \qquad (2.97)$$

Expectedly, the Stokes parameter $\xi_{1SR}$ vanishes. It can be easily checked that the polarization degree determined by the above expressions equals to unity,

$$P_{\xi_{SR}}^{--} = P_{\xi_{SR}}^{++} = 1. \qquad (2.98)$$

Consequently, radiation is fully polarized if electron spin does not change its direction.

Figure 2.8 shows the dependence of the Stokes parameters of the final photon on the escape angle; a) initial energy is 1) $z=0.03$, 2) $z=3$, 3) $z=300$ while $y=1$; b) photon frequency is 1) $y=0.01$, 2) $y=0.1$, 3) $y=1.0$ and the electron spin projection is $\mu=-1$.

It is clear from Fig.2.8 that the polarization changes from normal linear at $\Psi=0$ (emission perpendicular to the field) to right circular at $\Psi=+2$ (periphery of the narrow radiation cone).

Note that in the nonrelativistic ultraquantum approximation, the radiation perpendicular to the field is linearly polarized and radiation along the field is circularly polarized. The ultrarelativistic motion of the electron preserves this picture qualitatively, but "compresses" it into a narrow cone of radiation. The typical radiation angle $\psi_c = m/\varepsilon$ ($\Psi=1$) is determined only by the energy of the initial electron.

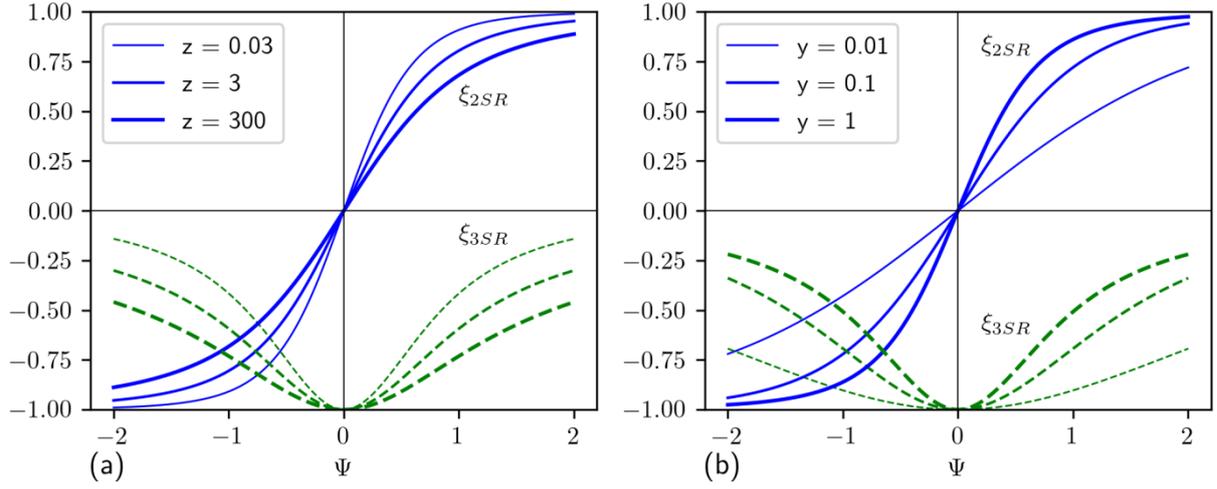

Fig. 2.8. The dependence of the Stokes parameters on the escape angle. a) 1) $z=0.03$, 2) $z=3$, 3) $z=300$; b) 1) $y=0.01$, 2) $y=0.1$, 3) $y=1.0$

Note that for estimation purposes we can set

$$K_{2/3} \approx 1.2 K_{1/3} \qquad (2.99)$$

when $y=1$ and $|\Psi|<1$.

In the case of moderate relativistic energy $z<<1$, we set approximately $a = 16$ and $b = c = 0$. Then, the Stokes parameters takes the simple form

$$\xi_{2SR} = \frac{2\Psi F K_{1/3} K_{2/3}}{\Psi^2 K_{1/3}^2 + F^2 K_{2/3}^2}, \quad \xi_{3SR} = \frac{\Psi^2 K_{1/3}^2 - F^2 K_{2/3}^2}{\Psi^2 K_{1/3}^2 + F^2 K_{2/3}^2} \qquad (2.100)$$

and depends on the emission angle only,

$$\xi_{2SR} \approx \frac{2.4 \Psi F}{1.4 + 2.4 \Psi^2}, \quad \xi_{3SR} \approx -\frac{1.4 + 0.4 \Psi^2}{1.4 + 2.4 \Psi^2}. \qquad (2.101)$$

This well known result has been obtained theoretically and verified experimentally for unpolarized electron beams [19,20]. The reason of such agreement lays in the fact that radiation polarization does not depend on electron spin states in the considered approximation, and main contribution comes from the no-spin-flip processes.

In the case of hard ultrarelativistic energy $z >> 1$, the quantities $a, b, c$ approximately equal to $a = b = c = z^2$ and the Stokes parameters are

$$\xi_{2SR} = \frac{2\Psi(1.2F - \mu)}{2.4F(F - \mu)}, \quad \xi_{3SR} = -\frac{2.4(1 - \mu F) + 0.4\Psi^2}{2.4F(F - \mu)}. \tag{2.102}$$

The obtained approximate equations are in a good agreement with the general expressions (2.96) and (2.97) if the electron occupies the ground spin state ($\mu$=-1). In contrast, the agreement is rather poor when $\mu$=+1.

Figure 2.9 shows the dependence of the Stokes parameters on the emission angle for the case of inverse electron spin population ($\mu$=+1), with y=1. Apparently, the emission angle interval with normal linear polarization of SR contracts into a small vicinity of zero angle ($\Psi$=0) when z reaches hard ultrarelativistic values (z>>1). In the rest of the narrow cone the radiation has also the anomalous component of linear polarization in addition to circular one.

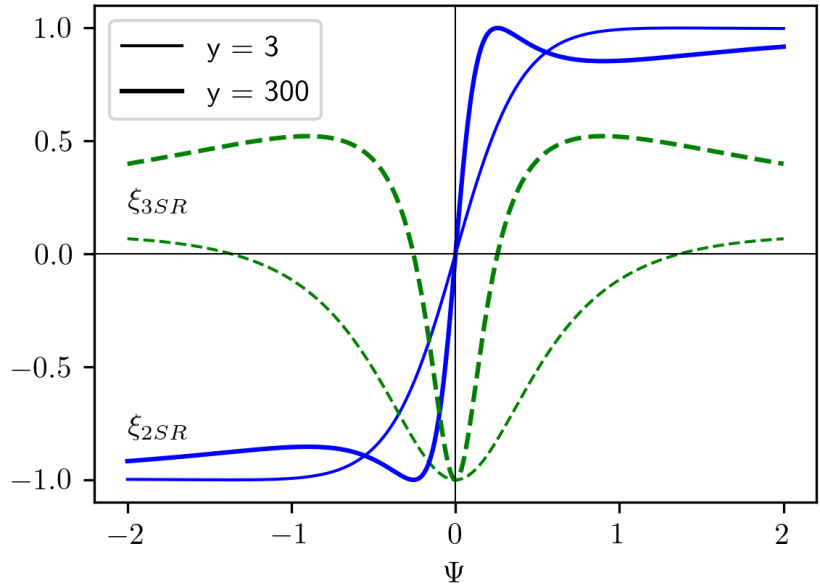

Fig.2.9. Dependence of the Stokes parameters on the emission angle in the case of inverted electron spin. 1) z=3, 2) z=300

If electrons do not change the spin projection and are allowed to be in both spin up ($\mu$=+1) and spin down ($\mu$=-1) states, the polarization degree of SR has the form

$$P_{\xi_{SR}} = \max \frac{D_\xi^{--} + D_\xi^{++} - D_{-\xi}^{--} - D_{-\xi}^{++}}{D_\xi^{--} + D_\xi^{++} + D_{-\xi}^{--} + D_{-\xi}^{++}}. \tag{2.103}$$

The corresponding Stokes parameters of the emitted photon can be found by inserting $\mu=0$ in Eqs (2.96) and (2.97),

$$\xi_{2SR} = \frac{2\Psi aFK_{1/3}K_{2/3}}{(a\Psi^2+b)K_{1/3}^2 + aF^2K_{2/3}^2}, \quad \xi_{3SR} = \frac{(a\Psi^2-b)K_{1/3}^2 - aF^2K_{2/3}^2}{(a\Psi^2+b)K_{1/3}^2 + aF^2K_{2/3}^2}. \quad (2.104)$$

The polarization degree of the final photon determined by the Stokes parameters (2.104),

$$P_{\xi_{SR}} = \sqrt{\xi_{2SR}^2 + \xi_{3SR}^2},$$

is shown in Figure 2.10 as a function of the emission angle for $y=1$. As can be seen, the polarization degree reaches the minimum value of $P=0.88$ when $z\gg 1$ and approximately $\Psi=1$.

In the moderate relativistic energy range, $z\ll 1$, the approximate Stokes parameters are given by Eqs. (2.100). In this case, $P=1$. In the case of hard ultrarelativistic energy, $z\gg 1$, the Stokes parameters are approximated by

$$\xi_{2SR} = \frac{\Psi}{F}, \quad \xi_{3SR} = -\frac{1}{F^2}, \quad (2.105)$$

and the polarization degree is

$$P = \sqrt{\frac{\Psi^4 + \Psi^2 + 1}{\Psi^2 + 1}}. \quad (2.106)$$

Now consider the spin-flip process ($\mu'=-\mu$). In this case, the general expressions for the Stokes parameters follows from Eq. (2.93),

$$\xi_{2SR} = \frac{2\Psi[FK_{1/3}K_{2/3} + \mu K_{1/3}^2]}{F^2(K_{1/3}^2 + K_{2/3}^2) + 2\mu FK_{1/3}K_{2/3}}, \quad (2.107)$$

$$\xi_{3SR} = \frac{(1-\Psi^2)K_{1/3}^2 + F^2K_{2/3}^2 + 2\mu FK_{1/3}K_{2/3}}{F^2(K_{1/3}^2 + K_{2/3}^2) + 2\mu FK_{1/3}K_{2/3}}. \quad (2.108)$$

The resulting polarization degree equals to unity. In the centre of the narrow radiation cone ($\Psi=0$), the parameter $\xi_{3SR}$ has the opposite sign compared to the case without spin inversion ($\mu'=\mu$). Note that in the case of $y=1$, the obtained expressions practically do not depend on the electron energy (or $z$ parameter).

In the case of $\mu=+1$, $\mu'=-1$ and taking into account the relation (2.99), the Stokes parameters take a simple form in the region of maximum radiation ($y=1$),

$$\xi_{2SR} = \frac{\Psi}{F}, \quad \xi_{3SR} = \frac{1}{F}. \tag{2.109}$$

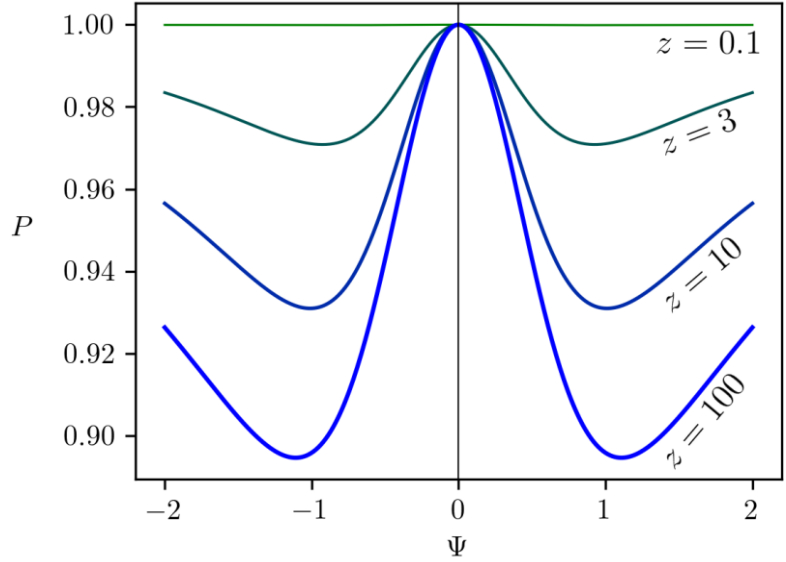

Fig.2.10. Dependence of the polarization degree of SR on the emission angle. An electron occupy both the ground and inverse spin states

In the opposite case of $\mu=-1$ and $\mu'=+1$, the dependence of the Stokes parameters on the emission angle is similar to that shown in Fig. 2.9 (2) but with the opposite sign of $\xi_{3SR}$.

If electrons are allowed to occupy both the ground ($\mu=-1$) and inverted ($\mu=+1$) spin states, the SR polarization in the spin-flip process is

$$\xi_{2SR} = \frac{2\Psi K_{1/3} K_{2/3}}{F(K_{1/3}^2 + K_{2/3}^2)} \approx \frac{\Psi}{F}, \tag{2.110}$$

$$\xi_{3SR} = \frac{(1-\Psi^2)K_{1/3}^2 + F^2 K_{2/3}^2}{F^2(K_{1/3}^2 + K_{2/3}^2)} \approx \frac{1}{F^2}. \tag{2.111}$$

The radiation is partially polarized and the degree of polarization is determined by expression (2.106), similarly to the no-spin-flip process with hard ultrarelativistic electron energy.

Consider the SR process in which the electron has a fixed spin projection in the initial state, and take sum of contributions from all final spin states. In other words, we aim to find the total polarization of all radiation from the initially polarized electron. In this case, the degree of polarization is defined as

$$P^{\mu}_{\xi_{SR}} = \max \frac{D^{\mu\mu}_{\xi} + D^{\mu,-\mu}_{\xi} - D^{\mu\mu}_{-\xi} - D^{\mu,-\mu}_{-\xi}}{D^{\mu\mu}_{\xi} + D^{\mu,-\mu}_{\xi} + D^{\mu\mu}_{-\xi} + D^{\mu,-\mu}_{-\xi}}. \quad (2.112)$$

The Stokes parameters are

$$\xi_{2SR} = \frac{2\Psi[(a+b)FK_{1/3}K_{2/3} + \mu(b-c)K^2_{1/3}]}{K^2_{1/3}(a\Psi^2 + b + bF^2) + K^2_{2/3}(a+b)F^2 + 2\mu(b-c)FK_{1/3}K_{2/3}}, \quad (2.113)$$

$$\xi_{3SR} = \frac{(a-b)\Psi^2 K^2_{1/3} - (a-b)F^2 K^2_{2/3} + 2\mu(c+b)FK_{1/3}K_{2/3}}{K^2_{1/3}(a\Psi^2 + b + bF^2) + K^2_{2/3}(a+b)F^2 + 2\mu(b-c)FK_{1/3}K_{2/3}}. \quad (2.114)$$

In the limit of moderate relativistic electron energy ($z \ll 1$), polarization of radiation at characteristic frequency $y = 1$ does not depend on spin projection and is given by Eq. (2.100). The radiation is fully polarized in this case, $P=1$.

If $z>1$, the radiation is partially polarized and the degree of polarization strongly correlates with the radiation frequency $y$ electron spin $\mu$. Figure 2.11 shows the degree of polarization as a function of electron energy $z$ at $\Psi=0$, $y=0.8$ (a), and as a function of electron energy $z$ and photon frequency $y$ at $\Psi=0$ (b). As can be seen from Fig.2.11 (a), if electrons are polarized opposite to the field direction ($\mu=-1$), the SR polarization degree decreases monotonically with increasing electron energy $z$ and reaches the minimum value of $P_m=0.4$. In contrast, if electrons are polarized along the field ($\mu=+1$), $z$ dependence of $P$ is essentially non-monotonic, and the polarization degree vanishes twice at the points $z_1=5.06$ and $z_2=877$ while approaching the same value $P_m=0.4$ in the limit case $z\rightarrow\infty$.

It is obvious from Fig. 2.11 (b) that completely unpolarized radiation $P = 0$ corresponds to some "canyon" on the graph of $f = P(y,z)$. On $y,z$ plane, this line has a maximum and intersects with the line $y=y_0$ at two points, which explains the presence of two minima in Fig. 2.11 (a). Note that the value of $P_m$ depends on $y$ and vanishes if $y\to 1$.

Figure 2.12 shows the dependence of the Stokes parameters of the final photon on the emission angle for various values of the electron energy and the frequency of $y=0.8$. Panels (a) and (b) depict graphs for the cases $\mu=+1$ and $\mu=-1$ respectively. We can see that in the case $\mu=-1$, $\mu=+1$ in the considered energy range from $z=1$ to $z=30$ the parameter $\xi_{2SR}$ weakly depends on the emission angle, as well as $\xi_{3SR}$ in the case $\mu=-1$. At the same time, in the case of $\mu=+1$, linear polarization $\xi_{3SR}$ strongly depends on the angle and changes its sign.

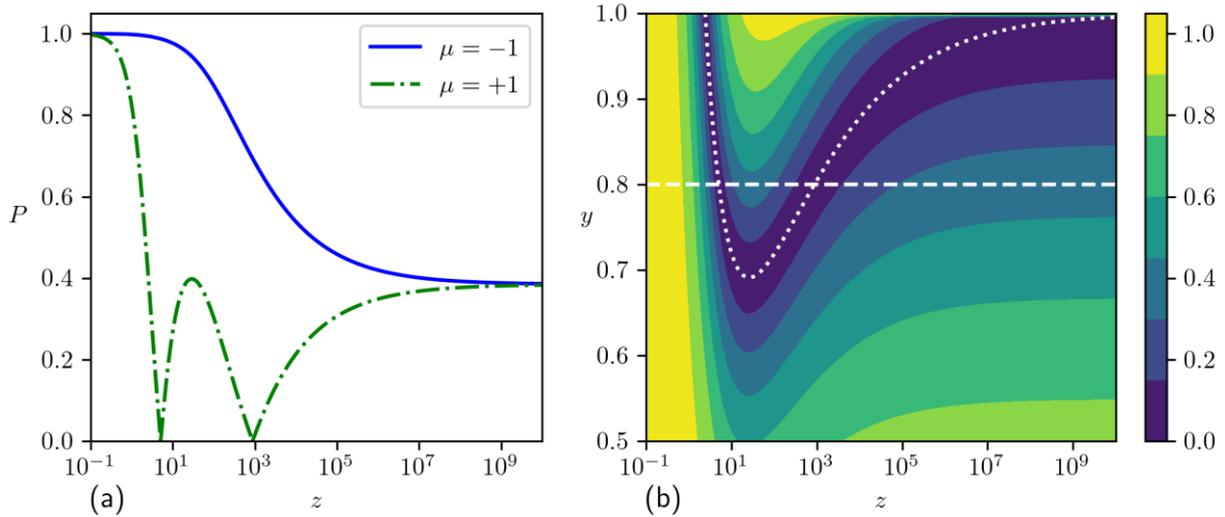

Fig. 2.11. The degree of polarization: (a) as a function of electron energy $z$ at $\Psi=0$, $y=0.8$; (b) as a function of electron energy $z$ and photon frequency $y$ at $\Psi=0$

Let us consider the question of the polarization of SR from an unpolarized electron. To find the Stokes parameters, we set $\mu=0$ in expressions (2.112) and (2.113):

$$\xi_{2SR} = \frac{2\Psi(a+b)FK_{1/3}K_{2/3}}{K_{1/3}^2(a\Psi^2+b+bF^2)+K_{2/3}^2(a+b)F^2}, \qquad (2.115)$$

$$\xi_{3SR} = \frac{(a-b)(\Psi^2 K_{1/3}^2 - F^2 K_{2/3}^2)}{K_{1/3}^2(a\Psi^2+b+bF^2)+K_{2/3}^2(a+b)F^2}. \qquad (2.116)$$

In the limit of moderate relativistic energy $z \ll 1$, the polarization degree equals to unity, and the Stokes parameters are approximated by Eq. (2.100), in agreement with known result [19]. In the hard relativistic limit $z \ll 1$, the radiation is polarized partially. Figure 2.13 shows the dependence of the polarization degree (a) on electron energy $z$ and photon frequency $y$ at $\Psi=0$ and (b) on electron energy $z$ and emission angle $\Psi$ at $y=0.8$.

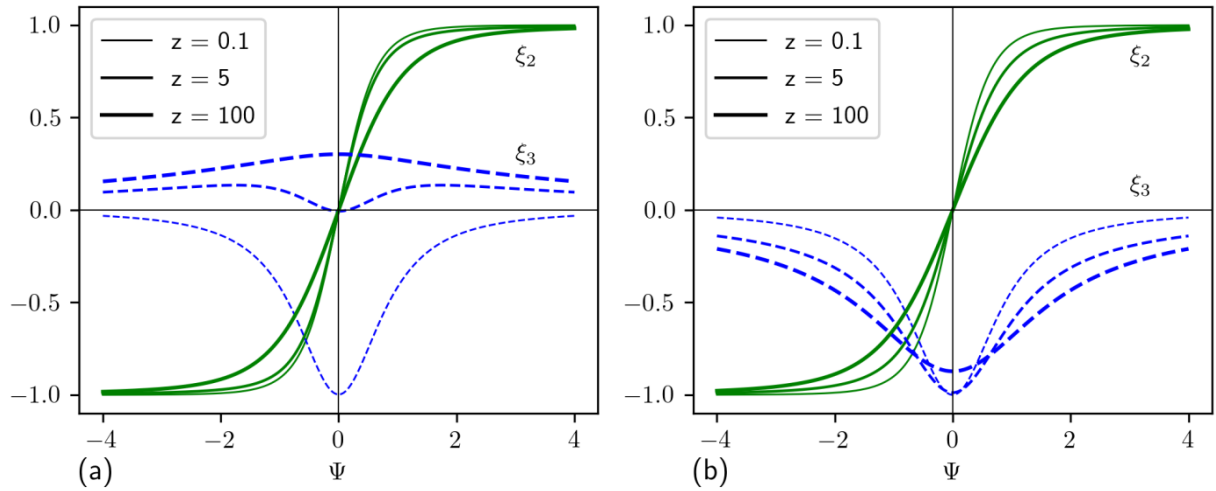

Fig. 2.12 The Stokes parameters as a function of the emission angle, (a) $\mu=+1$ and (b) $\mu=-1$. The frequency is $y=0.8$

The polarization degree reaches its minimum for radiation in the orbit plane ($\Psi=0$) near the intensity maximum ($y = 1$) and approximately equals to

$$P = \frac{11.2(z+2)}{4.8z^2 + 11.2(z+2)} \approx \frac{2.3}{z}, \qquad (2.117)$$

hence, the radiation is completely unpolarized in the limit case of large electron energy, $z \to \infty$.

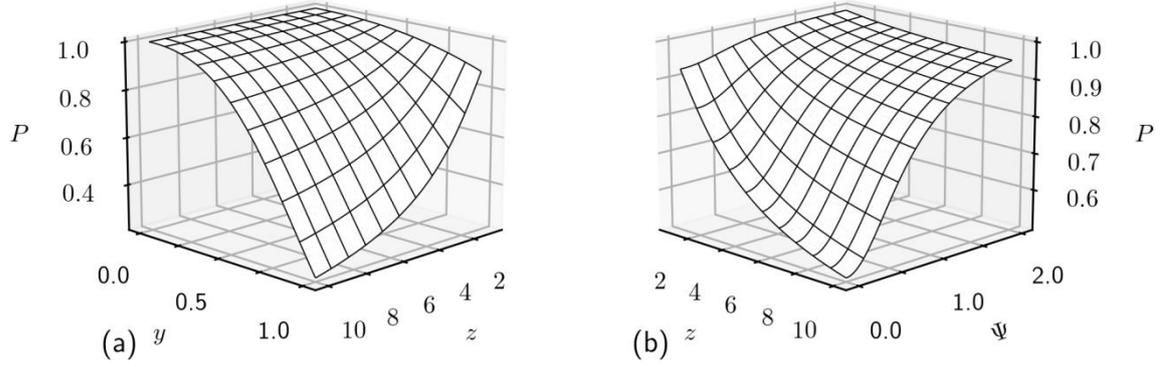

Fig. 2.13. The dependence of the polarization degree (a) on electron energy $z$ and photon frequency $y$ at $\Psi=0$ and (b) on electron energy $z$ and emission angle $\Psi$ at $y=0.8$

Note that in laboratory conditions the typical value of electron energy is of order of several GeV and magnetic strength is of order of $H\sim 10^4$G. This corresponds to $z=10^{-5}$, i.e. energy lays in the moderate relativistic range. However, magnetic fields up to hundreds of Tesla and electron energies up to several TeV are achievable, which results in a value of $z \sim 1$ and higher. Hence, observation of the considered interplay between electron spin and SR polarization in highly ultrarelativistic domain becomes experimentally feasible.

## 2.3. Spin and polarization effects in the process of one photon $e^+e^-$ pair production

The probability amplitude of one photon $e^+e^-$ pair production (OPP). We construct the probability amplitude of the considered process according to the Feynman diagram in Fig.2.14.

$$A_{if} = -ie\int d^4x \overline{\Psi}^-(\zeta)\gamma^i A_i \Psi^+(\xi), \qquad (2.118)$$

where $\bar{\Psi}^-(\zeta)$ is the wave function of a final electron (we replace prime with "-" sign in Eq. (2.27)), $\zeta = \sqrt{hm^2}(x + p_y^-/hm^2)$ and $\Psi^+(\xi)$ is the wave function of a positron of the form

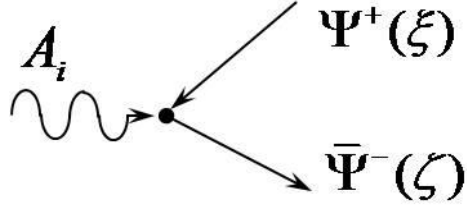

Fig.2.14. Feynman diagram of the OPP process

$$\Psi^+(\xi) = \frac{A_{l^+}}{\sqrt{S}} e^{i(\varepsilon_{l^+}^+ t - p_y^+ y - p_z^+ z)} [i\sqrt{m^+ + \mu^+ m}\, U_{l^+}(\xi) -$$

$$-\mu^+\sqrt{m^+ - \mu^+ m}\, U_{l^+-1}(\xi)\gamma^1]u_{l^+}^+, \qquad (2.119)$$

where $\xi = \sqrt{hm^2}(x - p_y^+/hm^2)$, $m^\pm = m(1 + 2l^\pm h)^{1/2}$.

The process kinematics is determined by the conservation laws of energy and longitudinal momentum,

$$\omega = \varepsilon_{l^-}^- + \varepsilon_{l^+}^+, \quad k_z = \omega u = p_z^- + p_z^+, \qquad (2.120)$$

where $u = \cos\theta$ is cosine of a polar angle of the incident photon. To find the threshold energy and momentum of the final particles we search for a maximum of the function

$$f(p_z^-) = \omega - \sqrt{(m^-)^2 + (p_z^-)^2} - \sqrt{(m^+)^2 + (\omega u - p_z^-)^2}. \qquad (2.121)$$

Figure 2.15 shows the dependence of the function $f(p_z^-)$ on the electron momentum, (a) the case of $u = 0$ and (b) the case of $u = 1$, the magnetic field is $h=0.1$ in both cases.

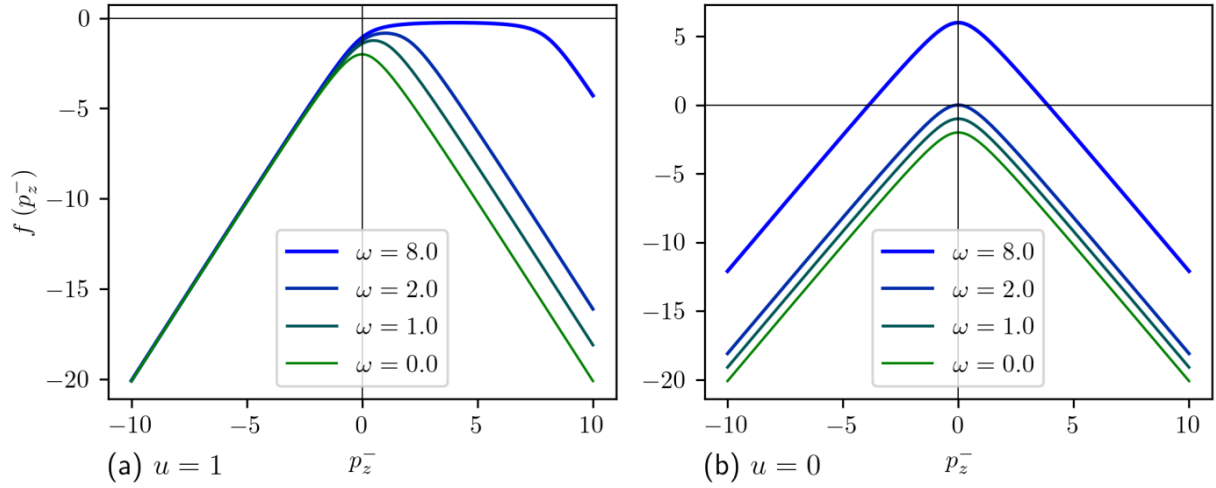

Fig.2.15. Plot of the function $f(p^-_z)$ for various values of frequency; a) $u=1$, b) $u=0$; magnetic field is $h = 0.1$ in both cases.

According to the conservation laws (2.120), the process is possible when $f(p^-_z)=0$, and the process threshold is reached when the maximum value of $f(p_m)$ equals to zero. Equating the derivative of $f(p^-_z)$ to zero yields

$$f(p_m) = \frac{\omega(p_m - u\varepsilon_m)}{p_m}. \qquad (2.122)$$

Further, Eq. (2.122) equals to zero at the limiting threshold values of the photon frequency and energies and longitudinal momenta of the electron and the positron,

$$\omega_m = (m^- + m^+)/\sqrt{1-u^2}, \quad \varepsilon_m^\pm = m^\pm/\sqrt{1-u^2}, \quad p_m^\pm = u\varepsilon_m^\pm. \qquad (2.123)$$

As follows from these relations, if the photon is directed perpendicular to the field ($u = 0$), then the threshold values of longitudinal momenta vanish and $\omega_m = m^+ + m^-$, i.e. the particles are "at rest" at the Landau levels. If the photon is directed along the field ($u = 1$), then expression (2.122) is always negative and never vanishes, hence OPP process is impossible in this case. It is useful to express the particle energies and momenta in terms of the photon frequency. For a fixed frequency, the threshold polar angle is given by the ratio

$$u_m = \pm\sqrt{\omega^2 - (m^+ + m^-)^2}/\omega. \qquad (2.124)$$

Let the Landau level numbers of the particles be fixed and the photon frequency satisfies the condition $u_m<1$. Then, the process is impossible if the photon angle meets the condition $u>u_m$. The threshold energies and momenta are

$$\varepsilon_m^{\pm} = \frac{m^{\pm}}{m^+ + m^-}\omega, \quad p_m^- = \pm\frac{m^-}{m^+ + m^-}\sqrt{\omega^2 - (m^+ + m^-)^2}, \quad p_m^+ = \pm\frac{m^+}{m^+ + m^-}\sqrt{\omega^2 - (m^+ + m^-)^2}. \quad (2.125)$$

Taking into account the relation $k_m = \omega u_m$ we obtain

$$\omega^2 = (m^+ + m^-)^2 + k_m^2.$$

Hence, at the process threshold, the relation between the photon energy and its longitudinal momentum is of the same form as that of an electron in a magnetic field, provided that we ascribe transversal energy of $m_\omega = m^+ + m^-$ to the photon.

To clarify the physical meaning of the obtained expressions (2.123), it is convenient to pass to a new (primed) reference frame moving along the magnetic field, such that the longitudinal photon momentum vanishes, $(p^-_m)'=0$. The Lorentz transformations of the energy and momentum have the form

$$(\varepsilon_m^-)' = \gamma(\varepsilon_m^- - Vp_m^-), \quad (p_m^-)' = \gamma(p_m^- - V\varepsilon_m^-), \quad \gamma = 1/\sqrt{1-V^2}. \quad (2.126)$$

The velocity and gamma factor are defined by equating $(p^-_m)'$ to zero,

$$V = u, \quad \gamma = \omega_m / (m^+ + m^-). \quad (2.127)$$

As a result, in the new reference frame the energies and momenta are

$$(\varepsilon_m^{\pm})' = m^{\pm}, \quad (p_m^{\pm})' = 0. \quad (2.128)$$

From the conservation laws (2.120) we obtain

$$(\omega_m)' = m^+ + m^-, \quad u' = 0. \quad (2.129)$$

Thus, if the photon frequency equals to the limit value, then the longitudinal momenta vanish and the photon frequency equals to the sum of the effective masses of the electron and the positron in the reference frame where the photon propagates perpendicular to the field.

When the photon frequency exceeds the threshold value, $\omega > \omega_m$, the conservation laws (2.120) result in the following expressions for energy and momenta of an electron and a positron,

$$\varepsilon^-_{1,2} = \frac{a^- \pm b^- u}{2\omega(1-u^2)}, \quad p^-_{1,2} = \frac{a^- u \pm b^-}{2\omega(1-u^2)}, \quad \varepsilon^+_{1,2} = \frac{a^+ \mp b^+ u}{2\omega(1-u^2)}, \quad p^+_{1,2} = \frac{a^+ u \mp b^+}{2\omega(1-u^2)}, \quad (2.130)$$

where $a^\pm = \omega^2(1-u^2) \pm (m^+)^2 \mp (m^-)^2$, $b^\pm = \sqrt{(a^\pm)^2 - 4(m^\pm)^2 \omega^2(1-u^2)}$, and the condition $b^+ = b^-$ is true.

Note that two values of the particles energy and momentum are possible for each value of the polar angle of the initial photon. In the particular case of pair production to equal Landau levels the expressions (2.130) take the form

$$\varepsilon^-_{1,2} = \frac{\omega}{2}[1 \pm u\sqrt{1-(\frac{\omega_m}{\omega})^2}], \quad p^-_{1,2} = \frac{\omega}{2}[1 \pm u\sqrt{1-(\frac{\omega_m}{\omega})^2}], \quad \varepsilon^+_{1,2} = \varepsilon^-_{2,1}, \quad p^+_{1,2} = p^-_{2,1}, \quad (2.131)$$

Suppose the photon propagates perpendicular to the field, $u = 0$. Then Eq. (2.139) transforms to

$$\varepsilon^- = \frac{1}{2\omega}(\omega^2 + (m^-)^2 - (m^+)^2), \quad p^- = \frac{\pm 1}{2\omega}\sqrt{\omega^2 - (m^+ + m^-)^2}\sqrt{\omega^2 - (m^+ - m^-)^2}, \quad (2.132)$$

$$\varepsilon^+ = \frac{1}{2\omega}(\omega^2 + (m^+)^2 - (m^-)^2), \quad p^+ = -p^-. \quad (2.133)$$

Let the photon frequency be a sum of the limit value and a small term,

$$\omega = \omega_m + \delta\omega, \quad \omega_m = m^+ + m^-, \quad \delta\omega \ll \omega_m, \quad (2.134)$$

then the energies and momenta take the simple form

$$\varepsilon^- = m^- + \frac{m^+ \delta\omega}{\omega_m}, \quad \varepsilon^+ = m^+ + \frac{m^- \delta\omega}{\omega_m}, \quad p^- = -p^+ = \pm\sqrt{\frac{2m^+ m^- \delta\omega}{\omega_m}}. \quad (2.135)$$

If the particles occupy the lowest Landau levels we need to analyze three separate cases of the additional term magnitude. In the first case, let

$$\omega = \omega_m + \delta\omega, \quad \delta\omega = \alpha_1 h^2 m, \quad \alpha_1 \sim 1, \quad (2.136)$$

$$\varepsilon^- = m(1 + l^- h), \quad \varepsilon^+ = m(1 + l^+ h), \quad p^- = -p^+ = \pm\sqrt{\alpha_1} hm, \quad (2.137)$$

i.e. particles momenta are comparable with the cyclotron frequency and they can be neglected in the expressions for particle energies. In the second case,

$$\omega = \omega_m + \delta\omega, \quad \delta\omega = \alpha_1 hm, \quad \alpha_1 \sim 1, \qquad (2.138)$$

$$\varepsilon^- = m(1+l^-h+\alpha_1 h/2), \; \varepsilon^+ = m(1+l^+h+\alpha_1 h/2), \; p^- = -p^+ = \pm\sqrt{\alpha_1 hm}. \; (2.139)$$

Hence, the corresponding term in the particle energies is comparable with the distance between Landau levels. In the third case,

$$\omega = \omega_m + \delta\omega, \quad \delta\omega = \alpha_1 m \sim \omega_m, \qquad (2.140)$$

$$\varepsilon^- = \varepsilon^+ = m(1+\alpha_1/2), \quad p^- = -p^+ = \pm m\sqrt{4\alpha_1 + \alpha_1^2}/2. \qquad (2.141)$$

As was mentioned before, we can let the photon be directed perpendicular to the field ($u = 0$) without loss of generality. Below we analyze OPP with assumption that this condition is true. Performing integration in (2.118) with wave functions (2.12), (2.27), (2.119) we obtain the general probability amplitude of OPP in the form

$$A_{if} = \frac{M_{if}\delta^3(k-p^+-p^-)}{S\sqrt{V}}, \quad M_{if} = \frac{-ie2\pi^3\sqrt{2\pi}e^{i\Phi}}{\sqrt{\omega\varepsilon^+\varepsilon^- m^+ m^-}}\sum_{a=1}^{4}Q_a, \quad (2.142)$$

where

$$Q_1 = J(l^+, l^-)M_m^- M_p^+ D\sin\theta\cos\alpha,$$

$$Q_2 = -J(l^+ - 1, l^- - 1)\mu^- M_p^- \mu^+ M_m^+ D\sin\theta\cos\alpha,$$

$$Q_3 = -J(l^+, l^- - 1)\mu^- M_p^- M_p^+ CH_m,$$

$$Q_4 = J(l^+ - 1, l^-)M_m^- \mu^+ M_m^+ CH_p.$$

Here, we denote

$$M_m^\pm = \sqrt{m^\pm - \mu^\pm m}, \; M_p^\pm = \sqrt{m^\pm + \mu^\pm m}, \qquad (2.143)$$

$$C = -E_m^- E_m^+ + \text{sgn}(p^-)\text{sgn}(p^+)E_p^- E_p^+, \; D = \text{sgn}(p^+)E_m^- E_p^+ + \text{sgn}(p^-)E_p^- E_m^+, \; (2.144)$$

$$E_m^\pm = \sqrt{\varepsilon^\pm - \mu^\pm m^\pm}, \; E_p^\pm = \sqrt{\varepsilon^\pm + \mu^\pm m^\pm}. \qquad (2.145)$$

The quantities $H_m$ and $H_p$ are defined by Eq. (2.43). The phase $\Phi$ is

$$\Phi = \frac{-k_x(2p_y^- - k_y)}{2hm^2} + (l^+ - l^-)(\varphi - \frac{\pi}{2}). \qquad (2.146)$$

The special function $J(l^+,l^-)$ is defined by the expression (2.37) and depends on the quantity $\eta$ given by (2.38) with $u=0$.

Rate of the OPP process. Let us now consider the normalizing constants $S$, $V$ in the amplitude (2.142). We write the amplitude in the form

$$A_{if} = \frac{M_{if}}{S\sqrt{V}} \delta(\omega - \varepsilon^+ - \varepsilon^-)\delta(k_z - p_z^+ - p_z^-)\delta(k_y - p_y^+ - p_y^-). \qquad (2.147)$$

The differential rate is the product of squared absolute value of (2.147) and the number of final states,

$$dN = dN^+ dN^- = \frac{dp_y^+ dp_z^+ S}{(2\pi)^2} \cdot \frac{dp_y^- dp_z^- S}{(2\pi)^2}. \qquad (2.148)$$

After transformations, we have

$$dW = \frac{STp_y^-}{V} \cdot \frac{|M_{if}|^2}{(2\pi)^7} \delta(\omega - \varepsilon^+ - \varepsilon^-) dp_z^-. \qquad (2.149)$$

To obtain the above formula we used

$$\left(\delta(\omega - \varepsilon^+ - \varepsilon^-)\right)^2 = \frac{T}{2\pi} \delta(\omega - \varepsilon^+ - \varepsilon^-),$$

$$\left(\delta(k_z - p_z^+ - p_z^-)\delta(k_y - p_y^+ - p_y^-)\right)^2 = \frac{S}{(2\pi)^2} \delta(k_z - p_z^+ - p_z^-)\delta(k_y - p_y^+ - p_y^-),$$

and calculated integrals in $dp_y^+$ and $dp_z^+$. As was mentioned before, in the Landau gauge the "quantized" $x$ coordinate enters the electron wave function in the form of the quantity (2.9),

$$\zeta = (hm^2)^{1/2}(x-x_0),$$

where the parameter $x_0 = -p_y/hm^2$ defines the centre of an electron orbit in the classical approximation. Identifying this quantity as the normalizing length $L_p$, we can write

$$p_y^- / L_x = Sp_y^- / V = hm^2. \qquad (2.150)$$

The resulting differential rate is

$$dW^{\mu^-\mu^+} = hm^2 \cdot \frac{|M_{if}|^2}{(2\pi)^7} \delta(\omega - \varepsilon^+ - \varepsilon^-) dp_z^-. \qquad (2.151)$$

The resulting expression contains a single Dirac delta function and a single differential. In the LLL approximation the individual Landau levels can be resoluted, so it is sensible to perform the integration in $p_z$ using the delta function. In this case, the OPP process rates $W^{\mu^-\mu^+}(\xi_i)$ are the functions of Landau level numbers of the particles, their spin projections and the Stokes parameters of the photon.

On the other hand, in the ultrarelativistic approximation we need to average the rate over the interval of levels. The number of final states is defined by Eq. (2.148) multiplied by $dl^+ \cdot dl^-$, so the delta function can be used to perform the integration in $dl^+$. In this case the OPP process is described by the differential rate of production into the interval of Landau level numbers and into the interval of $z$-component of the momentum of the electron, $dW^{\mu^-\mu^+}/dl^- dp_z$. It is convenient to pass from $dl^- dp_z^-$ to $d\varepsilon d\Psi$, where

$$\varepsilon = h\varepsilon^-/m, \quad \Psi = p_z^-/\varepsilon^-. \qquad (2.152)$$

<u>The OPP process in the LLL approximation.</u> To find the integrals in Eq. (2.151) we use the known property of the Dirac delta function,

$$\delta(\omega - \varepsilon^- - \varepsilon^+) = \sum_{i=1}^{2} \frac{\varepsilon^+ \varepsilon^-}{|\varepsilon^+ p_z^- - \varepsilon^- p_z^+|} \delta(p_z^- - p_i^-), \qquad (2.153)$$

where $p_i^-$ are two possible values of the longitudinal momentum of the electron (2.130).

Note that the denominator of (2.153) goes to zero if $p_z^- = p_z^+ = 0$, i.e. at the threshold of $e^+e^-$ pair production by a photon propagating perpendicular to the field. The appearance of singularities in the process rate is explained by the emission of soft photons, which always accompany QED processes but which is neglected in our consideration. This phenomenon is similar to the so called "infrared catastrophe" considered in the theory of electron scattering by a Coulomb centre [190] (see Fig. 2.16). The infrared divergence appears because the perturbation theory cannot describe soft photon emission. The bremsstrahlung cross-section is proportional to the inverse frequency of the emitted photon, $\sigma \sim 1/\omega'$, and it goes to infinity when $\omega' \to 0$. Similarly, if we take into account emission of an additional photon, the $e^+e^-$ pair production rate has the same dependency on the final photon frequency, $W \sim 1/\omega'$. Note that in this case the divergence at the process threshold ($p_z^- = p_z^+ = 0$) disappears.

In the LLL approximation, the special function $J(l^+, l^-)$ entering the process amplitude (2.142) reduces to

$$J(l^+, l^-) \equiv J_0 = (-1)^{l^+} \frac{e^{-\eta/2}}{\sqrt{l^+! l^-!}} \eta^{(l^+ + l^-)/2}, \qquad (2.154)$$

where $\eta = \omega^2/2hm^2 = 2/h$ in zero-order approximation.

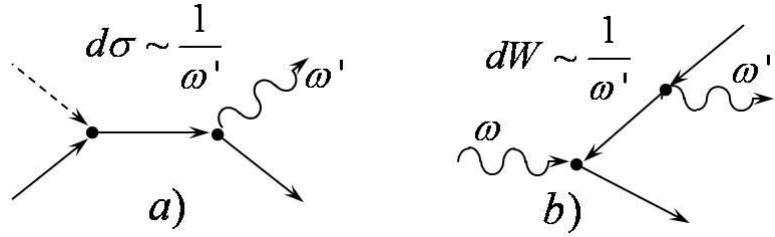

Fig.2.16. Feynman diagrams of processes: a) bremsstrahlung in electron scattering by the Coulomb centre, b) single-photon $e^+e^-$ pair production with photon emission.

Similarly, the other special function in (2.142) are

$$J(l^+ - 1, l^- - 1) = -\frac{\sqrt{l^+ l^-}}{\eta} J_0, \ J(l^+, l^- - 1) = \sqrt{\frac{l^-}{\eta}} J_0, \ J(l^+ - 1, l^-) = -\sqrt{\frac{l^+}{\eta}} J_0. \qquad (2.155)$$

Let us now write the OPP rates for fixed spin projections of the electron and the positron. The analysis of three kinematical cases (2.136) - (2.141) shows that the rates $W^{--}$, $W^{++}$ and $W^{-+}$ have the same dependence on the longitudinal electron momentum $p^-$. They are, respectively,

$$W^{-+} = \frac{\alpha m^4 h}{2\omega\varepsilon^- |p^-|} J_0^2 (1 + \xi_3), \qquad (2.156)$$

$$W^{--} = \frac{\alpha m^4 h^2 l^+}{4\omega\varepsilon^- |p^-|} J_0^2 (1 - \xi_3), \qquad (2.157)$$

$$W^{++} = \frac{\alpha m^4 h^2 l^-}{4\omega\varepsilon^- |p^-|} J_0^2 (1-\xi_3). \qquad (2.158)$$

In contrast, the rate of pair production to the inverted spin states ($\mu^-=+1, \mu^+=-1$) substantially depends on the difference $\delta\omega=\omega-\omega_m$. In the case of $\delta\omega=\alpha_1 mh^2$, particle energies and momenta are given by Eq. (2.137), and the corresponding rate is

$$W^{+-} = \frac{\alpha m^3 h^5 l^- l^+}{32\omega\varepsilon^- |p^-|} J_0^2 (1 + \frac{16(p^-)^2}{m^2 h^2} + \xi_3(1 - \frac{16(p^-)^2}{m^2 h^2})). \qquad (2.159)$$

In the case of $\delta\omega=\alpha_1 mh$, particle energies and momenta are given by Eq. (2.139), and the rate is

$$W^{+-} = \alpha m h^3 |p^-| l^- l^+ J_0^2 (1-\xi_3)/2\omega. \qquad (2.160)$$

In the case of $\delta\omega=\alpha_1 m$, particle energies and momenta are given by Eq. (2.141), and the rate has the form

$$W^{+-} = \frac{\alpha m^4 h^3 |p^-| l^- l^+}{4\omega(\varepsilon^-)^3} J_0^2 (2 + \frac{(p^-)^2}{2(\varepsilon^-)^2} - \xi_3(2 - \frac{(p^-)^2}{2(\varepsilon^-)^2})). \qquad (2.161)$$

Note that the most probable process channel is pair production to the ground spin state with $\mu^-=-1$ and $\mu^+=+1$. In this case, the corresponding rate contains the least power of the small parameter $h$. In all process channels, the rate depends on the polarization parameter $\xi_3$ only. The absence of $\xi_1$, $\xi_2$ is obvious from the consideration of the problem symmetries, because in the considered frame of reference the photon propagates perpendicular to magnetic field. In this reference frame, the cases of linear polarization at an angle of $\pm 45^0$ and right and left circular polarization are equivalent.

The total OPP process rate summed over the final particles states and averaged over the photon polarization equals to

$$<W> = \alpha m^4 h J_0^2 / 2\omega\varepsilon^- |p^-|, \qquad (2.162)$$

that coincides with results of [46]. The dependence of the total rate of the OPP process on the square of the photon frequency is shown in Fig.2.17 for the case of $h = 0.1$. The dotted line shows the data taken from Ref. [46]. As the Landau level numbers and the photon frequency increase, the curve determined by formula (2.162) deviates from the results of Ref. [46], which is due to the violation of the applicability of the LLL approximation.

As mentioned above, the most probable channel is the pair production to the ground spin states. However, the rate $W^{-+}$ (2.156) contains a factor $(1+\xi_3)$ and vanishes for normal linear polarization of the photon, $\xi_3 = -1$. At the same time, the rates $W^{--}$, $W^{++}$ are proportional to the factor $(1-\xi_3)$ and they are the main channels in the case of $\xi_3 = -1$.

In order to correctly compare the rates $W^{-+}$, $W^{--}$, $W^{++}$ in the case of $\xi_3 = -1$, let us write the expression for $W^{-+}$ with account of additional power in $h$,

$$W^{-+} = \frac{\alpha m^4 h}{2\omega \varepsilon^- g |p^-|} J_0^2 (1+\xi_3)(1+\frac{h}{2}(3(l^+ + l^-) - \frac{2l^+ l^-}{g^2})), \quad (2.163)$$

where $g = 1+(p^-/m)^2$. Note that the refined expression (2.163) has the same dependence on the photon polarization as (2.156). The analysis shows that this is also true for expressions that take into account a higher degree of $h$.

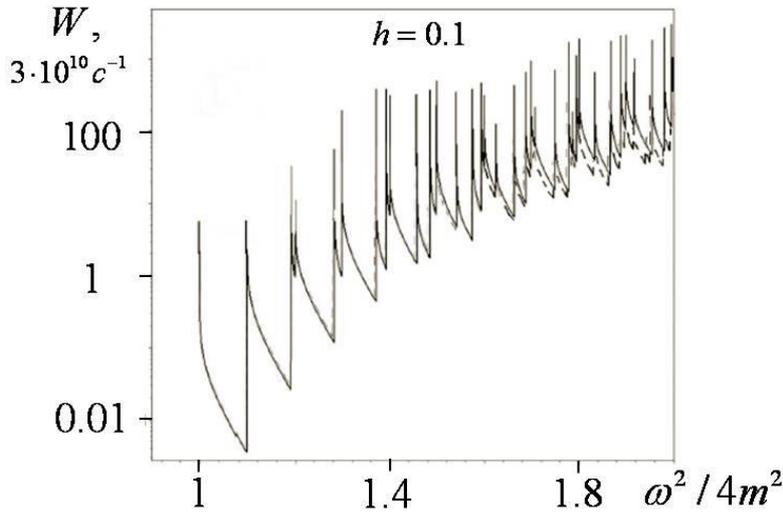

Fig.2.17. Dependence of the total rate of OPP process on the square of the photon frequency.

Thus, $e^+e^-$ pair are produced in the ground spin state except the small vicinity of photon polarization $\delta\xi$ near the value $\xi_3=-1$. Therefore, the created particle beams are practically purely polarized. Let us consider a relation between the magnitude of this polarization interval $\delta\xi$ and mutual orientation of the external magnetic field $\vec{H}$ and the electric field of the photon $\vec{E}_{ph}$ (see Fig. 2.18). We assume that the angle between these vectors is close to $\pi/2$, and the angle $\vartheta$ is correspondingly small. The Stokes parameter can be defined in the terms of the projections of the photon electric field on $x$ and $z$ axes as follows (see Fig. 2.18):

$$\xi_3 = \frac{E_1^2 - E_2^2}{E_1^2 + E_2^2}. \qquad (2.164)$$

For small values of $\vartheta$ we can write $E_1 = E_{ph}\vartheta$ and $E_2 = E_{ph}$. Then,

$$\xi_3 = -1 + \delta\xi = -1 + 2\vartheta^2. \qquad (2.165)$$

Consequently, the sought interval $\delta\xi$ equals to $2\vartheta^2$ and it should be of the order of magnitude of the small parameter $h$ so that the rate of the main channel $W^{-+}$ be of the same order of magnitude with $W^{--}$ and $W^{++}$,

$$\delta\xi_c = 2\vartheta_c^2 = \alpha_c h, \quad \alpha_c \sim 1. \qquad (2.166)$$

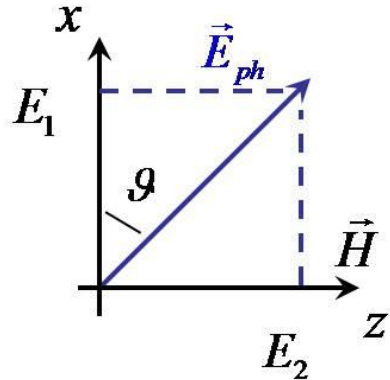

Fig.2.18. Mutual arrangement of $\vec{E}_{ph}, \vec{H}$ vectors

Let us now find the degree of polarization of the final particles. The degree of polarization of electrons (positrons) is the degree of orientation of theirs spin and it is defined as

$$P_{e^-} = \frac{(W^{++} + W^{+-}) - (W^{-+} + W^{--})}{(W^{++} + W^{+-}) + (W^{-+} + W^{--})}, \qquad (2.167)$$

$$P_{e^+} = \frac{(W^{++} + W^{-+}) - (W^{+-} + W^{--})}{(W^{++} + W^{-+}) + (W^{+-} + W^{--})}. \qquad (2.168)$$

If the angle between $\vec{E}_{ph}$ and $\vec{H}$ (see Fig. 2.18) is greater than $\vartheta_c$, then $(1+\xi_3)>\delta\xi_c$. Taking into account Eqs. (2.157), (2.158) and (2.163), the polarization degree (2.167), (2.168) to the first order in $h$ are

$$P_{e^-} = -1 + hl^- \frac{1-\xi_3}{1+\xi_3}, \quad P_{e^+} = +1 - hl^+ \frac{1-\xi_3}{1+\xi_3}. \qquad (2.169)$$

For an anomalous photon polarization ($\xi_3=1$), as well as at the zero Landau level (obviously), both electrons and positrons are created purely polarized to the ground spin state, where electron spins are directed against the field and positron spins are directed along the field. In other cases except the narrow cone $\vartheta < \vartheta_c$, the pure polarization is slightly disturbed by an order of magnitude $h$. The magnitude of depolarization of electrons (positrons) is proportional to the Landau level number of the electrons (positrons).

If the angle between $\vec{E}_{ph}$ and $\vec{H}$ is much less than the critical $\vartheta_c$, then the polarization degree with account of (2.165) is

$$P_{e^-} = \frac{l^- - l^+ - 2\vartheta/h}{l^- + l^+ + 2\vartheta/h} = \frac{l^- - l^+}{l^- + l^+} - \vartheta \frac{4l^-}{h(l^- + l^+)^2}, \qquad (2.170)$$

$$P_{e^+} = \frac{l^- - l^+ + 2\vartheta/h}{l^- + l^+ - 2\vartheta/h} = \frac{l^- - l^+}{l^- + l^+} + \vartheta \frac{4l^+}{h(l^- + l^+)^2}. \qquad (2.171)$$

It follows from expressions (2.170) and (2.171) that the polarization degree depends only on the Landau level numbers to within a small angle $\vartheta$. For $l^-<l^+$ electrons are created mostly to the ground spin state, opposite to the case of $l^->l^+$ when they are created in the inverse state. The same is true for positrons, i.e. a particle with a higher energy level is born to the inverse spin state, and a particle with a lower level is born to the ground state. If particles are produced to the same level, the polarization is absent within the accuracy of $\vartheta$. When the angle $\vartheta$ increases, it results in decrease of $P_{e^-}$ and increase $P_{e+}$, which means approaching the ground spin state for both electrons and positrons.

Suppose the photon frequency (2.134) is fixed, and the limit value equals to

$$\omega_m = m^- + m^+ = 2m + hm(l^- + l^+). \qquad (2.172)$$

A photon of this frequency is able to produce $e^+e^-$ pairs to Landau levels with the following conditions,

$$\begin{cases} 1)\ l^+ = l^-: & l^+ = l,\ l^- = l, \\ 2)\ l^+ > l^-: & l^+ = l + \Delta l,\ l^- = l - \Delta l, \\ 3)\ l^+ < l^-: & l^+ = l - \Delta l,\ l^- = l + \Delta l. \end{cases} \qquad (2.173)$$

Let us compare the rates $W^{-+}$, $W^{--}$ and $W^{++}$ in these cases. Landau level numbers enter the rate of the main channel $W^{-+}$ (2.156) only in the quantity $J_0^2$ (2.154). The rate can be written as

$$W^{-+} = A \frac{1}{(l^-)!(l^+)!}. \qquad (2.174)$$

Denoting the quantity (2.174) for the case of equal level numbers as $W_0$, we obtain the following expression of the rate for the cases (2) and (3) in (2.173),

$$W^{-+} = \frac{l(l-1)...(l-\Delta l+1)}{(l+1)...(l+\Delta l)} W_0 < W_0. \qquad (2.175)$$

Thus, the most probable is symmetrical pair production with equal level number of the electron and the positron.

To do similar analysis of the rates $W^{--}$ and $W^{++}$ we write them in the form

$$W^{--} = B \frac{l^+}{(l^-)!(l^+)!},\quad W^{++} = B \frac{l^-}{(l^-)!(l^+)!}. \qquad (2.176)$$

In the case of equal level numbers, let us denote the quantities (2.176) as $W_1$. If the condition (2) of (2.173) is true, the rates reduce to

$$W^{--} = \frac{(l-1)...(l-\Delta l+1)}{(l+1)...(l+\Delta l-1)} W_1,\quad W^{++} = \frac{(l-1)...(l-\Delta l)}{(l+1)...(l+\Delta l)} W_1. \qquad (2.177)$$

Particularly, if $\Delta l=1$ the rates equal to

$$W^{--} = W_1, \quad W^{++} = \frac{(l-1)}{(l+1)} W_1. \tag{2.178}$$

Thus, the efficiency of pair production is the same in the case of equal levels and in the case of the electron being produced to the next lower level in the ground spin state. Other configurations have lesser process rates. In the case (3) of (2.173), the expressions for the rates $W^{--}$ and $W^{++}$ in Eq. (2.177) swap places.

The OPP process in the ultrarelativistic approximation. The differential rate is given by Eq. (2.151) multiplied by $dl^+ \cdot dl^-$. The integration in $dl^+$ can be done using the Dirac delta-function and the relation

$$dl^+ = \frac{\varepsilon^+}{hm^2} d\varepsilon^+. \tag{2.179}$$

We change variables from $l^-, p^-$ to $\varepsilon, \Psi$ (2.152), with

$$dl^- dp^- = \frac{1}{mh^2} d\varepsilon d\Psi. \tag{2.180}$$

The resulting differential process rate takes the form

$$dW^{\mu^-\mu^+} = \frac{|M_{if}|^2 m^2 (\Omega - \varepsilon)\varepsilon}{(2\pi)^7 h^5} d\varepsilon d\Psi, \tag{2.181}$$

where $\Omega = h\omega/m$ is the dimensionless photon frequency.

In the amplitude $M_{if}$ (2.142), the asymptotic expressions of the special functions $J(l^+,l^-)$, $J'(l^+,l^-)$ are expressed via the Macdonald functions similarly to Eqs. (2.80) and (2.81),

$$J(l^+,l^-;\eta) = \frac{\sqrt{\eta - (\sqrt{l^+} + \sqrt{l^-})^2}}{\pi\sqrt{3}(\sqrt{l^+} + \sqrt{l^-})} K_{1/3}\left( \frac{2}{3} \frac{\sqrt[4]{l^+l^-}(\eta - (\sqrt{l^+} + \sqrt{l^-})^2)^{3/2}}{(\sqrt{l^+} + \sqrt{l^-})^2} \right), \tag{2.182}$$

$$\frac{\partial J(l^+,l^-;\eta)}{\partial \eta} = \frac{\sqrt[4]{l^+l^-}(\eta - (\sqrt{l^+} + \sqrt{l^-})^2)}{\pi\sqrt{3}(\sqrt{l^+} + \sqrt{l^-})^3} K_{2/3}\left( \frac{2}{3} \frac{\sqrt[4]{l^+l^-}(\eta - (\sqrt{l^+} + \sqrt{l^-})^2)^{3/2}}{(\sqrt{l^+} + \sqrt{l^-})^2} \right). \tag{2.183}$$

After transformation to the variables $\varepsilon$ and $\Psi$, the argument of Macdonald functions takes the form

$$\kappa = \frac{\Omega}{3\varepsilon(\Omega-\varepsilon)} F^3, \quad F^2 = 1+\Psi^2. \qquad (2.184)$$

The differential rate of the pair production to the interval $d\varepsilon d\Psi$ with fixed spins by a photon with defined polarization looks like

$$\frac{dW^{\mu^-\mu^+}}{d\varepsilon d\Psi} = \frac{\alpha m h F^2}{24\pi^2 \varepsilon^2 \Omega(\Omega-\varepsilon)^2} D^{\mu^-\mu^+}, \qquad (2.185)$$

If the particles have the opposite spin projections ($\mu^+=-\mu^-$), the factors $D^{\mu^-\mu^+}$ in Eq. (2.185) take the form

$$D^{\mu,-\mu} = \Omega^2\{[F^2(K_{1/3}^2 + K_{2/3}^2) + 2\mu F K_{1/3} K_{2/3}] - 2\xi_2 \Psi[F K_{1/3} K_{2/3} + \mu K_{1/3}^2] +$$
$$+\xi_3[(1-\Psi^2)K_{1/3}^2 + F^2 K_{2/3}^2 + 2\mu F K_{1/3} K_{2/3}]\}. \qquad (2.186)$$

The minus sign of $\mu$, $\mu=-1$, corresponds to the ground spin state, and $\mu=+1$ corresponds to the inverse state. In case of equal spin projections, $\mu^+=\mu^-$, the factors $D^{\mu^-\mu^+}$ are

$$D^{\mu\mu} = [(a\Psi^2+b)K_{1/3}^2 + aF^2 K_{2/3}^2 - 2\mu c F K_{1/3} K_{2/3}] - 2\xi_2\Psi[aF K_{1/3} K_{2/3} - \mu c K_{1/3}^2] +$$
$$+\xi_3[(a\Psi^2-b)K_{1/3}^2 - aF^2 K_{2/3}^2 + 2\mu c F K_{1/3} K_{2/3}], \qquad (2.187)$$

where $a=(2\varepsilon-\Omega)^2$, $b=\Omega^2$, $c=(2\varepsilon-\Omega)\Omega$. Note that obtained Eqs. (2.186) and (2.187) coincide with similar expressions (2.93) and (2.92) for SR process with the opposite sign of the $\xi_2$ parameter. Figure 2.19 shows the dependence of the OPP rate on the dimensionless electron energy $\varepsilon$ and its longitudinal momentum $\Psi$ in the cases: (a) $\xi_3=-1$, (b) $\xi_3=+1$; the photon frequency is set to $\Omega=1$.

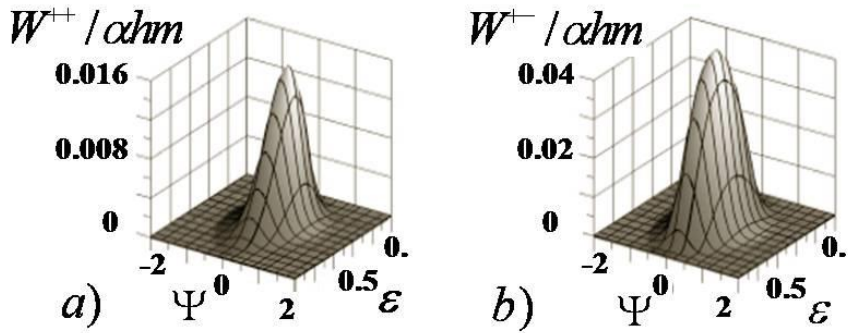

Fig.2.19. The dependence of the OPP rate on the electron energy and its escape angle: a) ξ3 = -1, b) ξ3 = + 1; the photon frequency is Ω=1.

The main channels a (a) pair production to the state with equal spin projections by a photon with normal polarization, and (b) pair production to the inverse spin state by a photon with anomalous polarization. In both cases, the process rate has maximum in the perpendicular to the field plane ($\Psi =0$) with equal energies of the particles ($\varepsilon=0.5$).

After inserting $\Psi =0$ into the factors $D^{\mu^-\mu^+}$ they takes the form

$$D^{\mu,-\mu} = b[K_{1/3}^2 + K_{2/3}^2 + 2\mu K_{1/3} K_{2/3}](1+\xi_3), \qquad (2.188)$$

$$D^{\mu,\mu} = [bK_{1/3}^2 + aK_{2/3}^2 - 2\mu c K_{1/3} K_{2/3}](1-\xi_3). \qquad (2.189)$$

The obtained expressions have the same dependence on the Stokes parameter $\xi_3$ of the initial photon as in the LLL case (2.156) – (2.158). As before, a photon with normal polarization ($\xi_3 =-1$) creates an e$^+$e$^-$ pair with the same spin projections, and a photon with anomalous polarization ($\xi_3 =-1$) produces a pair with opposite spin projections. If an electron and a positron have equal energies ($a=c=0$) the condition $D^{--}=D^{++}$ is true. The same conclusion is true in the LLL approximation. If particles are created with different energies, then the inverted spin state of the less energetic particle is more probable. For example, if $\varepsilon^->\varepsilon^+$, then $c>0$ and $D^{++}< D^{--}$. It follows from Eq. (2.188) that always $D^{+-} > D^{-+}$, hence the particles tend to be produced in the inverse spin state. Note that this results are opposite to those obtained in the LLL approximation.

Let us now find the polarization degree in the case of $\Psi =0$. Inserting Eqs (2.188) and (2.189) into Eqs. (1.167) and (2.168) we obtain

$$P_{e^{\mp}} = \frac{2[-c(1-\xi_3)\pm b(1+\xi_3)]K_{1/3}K_{2/3}}{(bK_{1/3}^2 + aK_{2/3}^2)(1-\xi_3)+b(K_{1/3}^2 + K_{2/3}^2)(1+\xi_3)}, \qquad (2.190)$$

where the upper sign corresponds to the electron and the lower sign corresponds to the positron.

If the photon has normal polarization ($\xi_3 =-1$), the polarization degree of the electron (positron) is

$$P_{e^-} = P_{e^+} = \frac{-2cK_{1/3}K_{2/3}}{(bK_{1/3}^2 + aK_{2/3}^2)} \approx \frac{-2c}{b+a}. \qquad (2.191)$$

If particle energies are equal then $c=0$ and their polarization vanishes. If the energy difference goes its maximum value, $\varepsilon^-=\omega$, $\varepsilon^+=0$, then $a=b=c$ and polarization degrees of electrons and positrons are equal, $P_{e^-} = P_{e^+} = -1$. And vice versa, if $\varepsilon^-=0$, $\varepsilon^+=\omega$ then $P_{e^-} = P_{e^+} = 1$.

In the case of anomalous photon polarization ($\xi_3 = +1$), the electron (positron) polarization degree is

$$P_{e^\mp} = \frac{\pm 2K_{1/3}K_{2/3}}{(K_{1/3}^2 + K_{2/3}^2)} \approx \pm 1. \qquad (2.192)$$

Thus, regardless of the particle energies, they are created in the inverted spin state with 100% polarization degree.

If the photon is polarized at the angle of $\pm 45°$ relative to magnetic field ($\xi_1 = \pm 1$, $\xi_3 = 0$), the electron (positron) polarization degree is

$$P_{e^\mp} = \frac{2(-c \pm b)K_{1/3}K_{2/3}}{2bK_{1/3}^2 + (a+b)K_{2/3}^2} \approx \frac{2(-c \pm b)}{3b+a}. \qquad (2.193)$$

When the particles have equal energies it reduces to $P_{e^\mp} = \pm 2/3$. In the case of maximum energy difference, $\varepsilon^-=\omega$ and $\varepsilon^+=0$, we obtain $P_{e^-} = 0$ and $P_{e^+} = -1$. In the opposite case of $\varepsilon^-=0$ and $\varepsilon^+=\omega$, the polarization degrees are $P_{e^-} = 1$ and $P_{e^+} = 0$.

Thus, by selecting various linear polarization of the incident photon, it is possible to obtain particle beams with controlled polarization ranging from the purely polarized to the fully depolarized state. Photon polarization can be changed by rotating the photon beam around its axis by a fixed angle. According to Fig.2.18, zero angle $\vartheta = 0$ corresponds to normal photon polarization, $\xi_3 = -1$. Increasing this angle by $+45°$ gives the case of $\xi_1 = +1$, and, finally, further increase by $+45°$ results in anomalous linear polarization, $\xi_3 = +1$.

The polarization degree of electrons with arbitrary escape angle $\Psi$ is given by Eqs. (2.186) and (2.187). For circular photon ($\xi_1 = 0$, $\xi_3 = 0$) polarization it equals to

$$P_{e^-} = \frac{2(b-c)K_{1/3}(FK_{2/3} - \xi_2 \Psi K_{1/3})}{K_{1/3}^2(2b + (a+b)\Psi^2) + K_{2/3}^2 F^2(a+b) - 2\xi_2 \Psi(a+b)FK_{1/3}K_{2/3}}. \quad (2.194)$$

For positron polarization degree the $b$ quantity in the nominator of (2.194) should be replaced according to $b \to -b$. It is clear from (2.194) that electron polarization does not change after change of right circular polarization to the left one and simultaneous altering the sign of $\Psi$.

Figure 2.20 shows the dependence of the electron polarization degree on the escape angle $\Psi$ in the OPP by a circularly polarized photon ($\xi_2=1$). Left (a) and right (b) panels correspond to $\Omega=1$ and $\Omega=100$ respectively. The electron energy varies from $0.1\Omega$ to $0.9\Omega$. Apparently, the maximum electron polarization is reached at small electron energy. In the most likely case when particles have equal energy of $\varepsilon=0.5\Omega$ the polarization degree is $P_{e^-} = 0.7$. Electrons having energy approaching energy of the incident photon are practically unpolarized.

The total rate of the OPP process with fixed spin projections can be found by integrating the differential rate (2.185) over the electron energy $\varepsilon$ and its escape angle $\Psi$,

$$W^{\mu^-\mu^+} = \int_0^\Omega d\varepsilon \int_{-1}^1 d\Psi \frac{dW^{\mu^-\mu^+}}{d\varepsilon d\Psi}. \quad (2.195)$$

After summation of the final particles spin projections and averaging over the photon polarization this expression coincides with the known results [8].

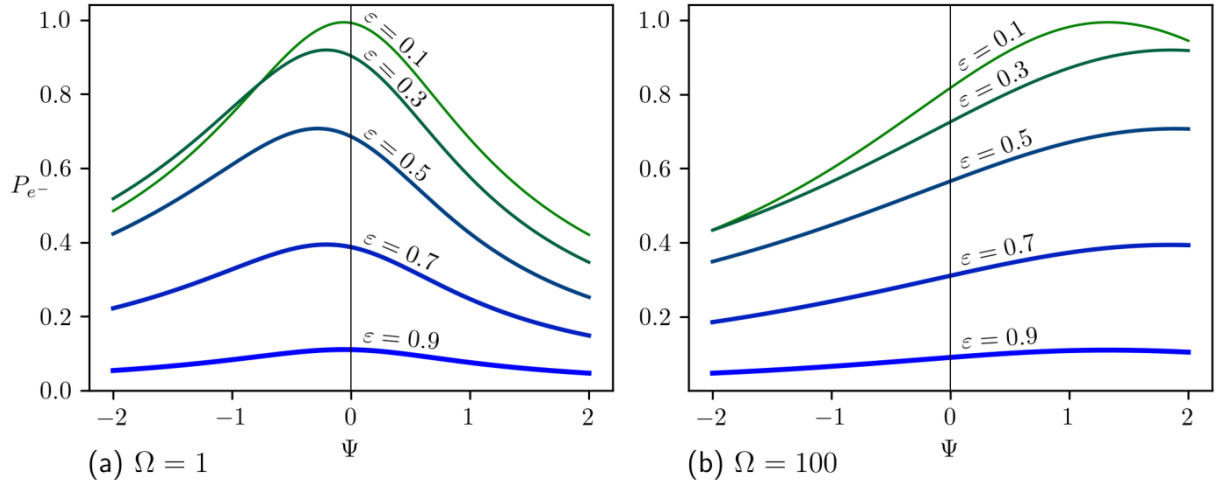

Fig.2.20. Dependence of the electron polarization degree on the electron escape angle, $\xi_2=1$. (*a*) $\Omega=1$; (*b*) $\Omega=100$.

The dependence of the total process rate on the inverse of the photon energy is shown in Fig. 2.21. Left and right panels show the cases $\xi_3=+1$ and $\xi_3=-1$ respectively. Here we denote $W_0=\alpha hm$, and $W_0$ is of order of $W_0 \sim 10^{18}\,\text{s}^{-1}$ for the field strength of $h=0.1$. The rate has maximum values when the dimensionless frequency is about $\Omega_{max} \approx 10$, that corresponds to photon energy about $\omega_{max} \approx 50 MeB$ for the field strength of $h=0.1$. The rate decreases when photon frequency exceeds $\omega_{max}$.

As shown in Fig. 2.21, the main channel is pair creation with opposite spin projections. Note that for anomalous photon polarization, the rate of the channel with inverse spin state of particles is by an order of magnitude higher than the rate of the channel process with ground spin state, $W^{+-} \gg W^{-+}$. At the same time, these channels have equal rates, $W^{+-} = W^{-+}$ in the case of normal photon polarization.

Thus, by changing the linear polarization of the initial photon (e.g. rotating the beam around its axis), it is possible to obtain electrons and positrons in a controlled spin state, ranging from fully unpolarized to inversely polarized state.

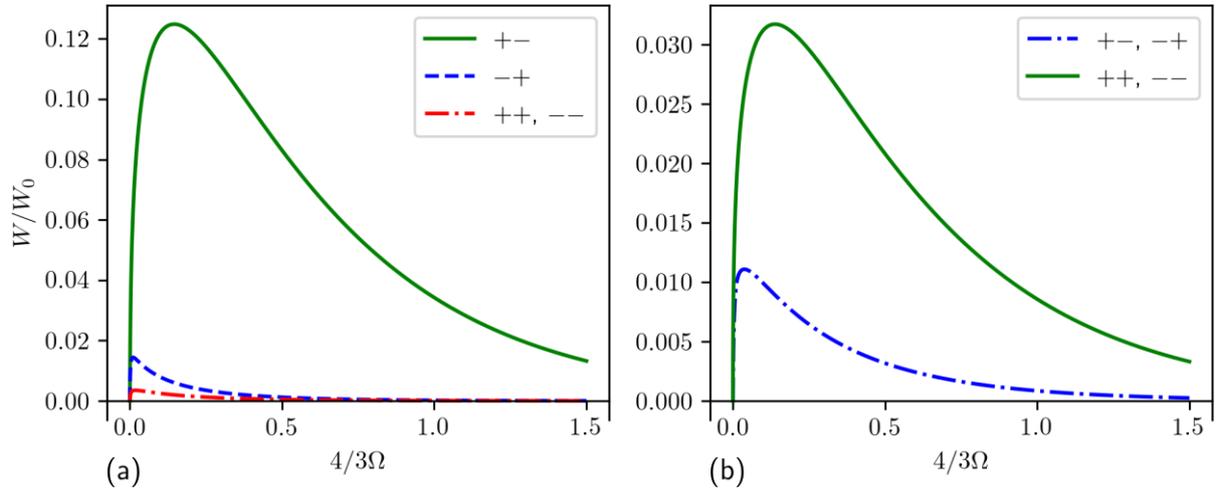

Fig.2.21. Dependence of the total rate on the parameter $4/3\Omega$. Left: $\xi_3=+1$, right: $\xi_3=-1$.

In laboratory conditions the maximum magnetic field is $10^6$ G. In this field, the photon energy of 100 GeV corresponds to the value $\Omega \sim 4.10^{-3}$, which is a small parameter. On the contrary, the $\Omega$ can reach very large values in a pulsar magnetosphere.

## 2.4. Conclusions to the Chapter 2

In this chapter we considered spin-polarization effects in SR and OPP process, i.e. the effects arising due to interaction between spin of particles polarization of photons. Let us now formulate some conclusions.

<u>1. SR process in the LLL approximation.</u>
i) The radiation polarization are the same as predicted by classical electrodynamics, if an electron does not change its spin orientation. The rates of the corresponding channels satisfy the condition $W^{--} > W^{++}$.

ii) The spin-flip radiative transition to the ground spin state ($\mu=-1$) changes the linear polarization of radiation from the normal ($\xi_3=-1$) to anomalous ($\xi_3=+1$) one. Its rate is less than the rate of the main channels by a factor of *h,* where *h* is a magnetic field strength in the units of the QED critical value, $h=H/H_0=e\hbar H/m^2c^3$. The other spin-flip channel with a transition to the inverted spin state is small and its fraction is of order of $h^3$. The account of the contribution of the spin-flip channels decreases by amount of ~*h*.

(iii) The emission of an electron with finite longitudinal momentum is fully linearly polarized when the frequency approach the maximum value.

(iv) An ultrarelativistic electron moving along the magnetic field and occupy a low Landau level emits photons of right circular polarization.

2. SR process in the ultrarelativistic approximation.

(i) The angular dependence of the radiation polarization qualitatively resembles the results of the LLL approximation, but "contracts" into a narrow radiation cone of the width of ~$m/\varepsilon$.

In the channels without spin inversion, the radiation in the orbit plane has pure normal linear polarization ($\sigma$ polarization). In the spin-flip processes, the linear polarization changes its sign ($\pi$ polarization).

(ii) The spin-flip process has small probability in moderate relativistic energy range, $z\ll1$ ($z=H\varepsilon/H_0m$). Its contribution is of the order of ~$z^2/8$ compared to the main channel near the spectral maximum. If $z>1$, the spin-flip channel is comparable with the channel without spin inversion.

(iii) In the case of ultrarelativistic electron energy $z>1$, the polarization of radiation of an initially polarized electron beam depends essentially on energy and its spin projection of electrons, as well as on the radiation frequency. If initially electron spins are oriented opposite to the field, the polarization degree of the radiation decreases monotonically with increase of *z*. For electrons in the inverted spin state, the polarization degree of radiation is a nonmonotonic function of *z*. Radiation in the orbit plane changes the polarization from normal to anomalous and back when the electron energy and the parameter *z* increase.

(iv) In the case of *z*>1, the radiation of an initially unpolarized electron beam is partially polarized with the polarization degree of 2.3/*z*.

3\. OPP process in the LLL approximation.

(i) The most probable channel is pair production to the ground spin states. The rate is maximum for production of electron and positron with equal energies.

(ii) In the case of anomalous linear photon polarization ($\xi_3=+1$), the created e$^+$e$^-$ pairs are fully polarized to the ground spin state. Except the small interval of ~$h$ near $\xi_3=-1$, the particle polarization degree deviates from unity by the small value of order of $h$. Within a small vicinity of photon polarization near the value of $\xi_3=-1$, the polarization of the produced particles depends on the difference of their energies. The particles are unpolarized if their energies are equal. The less energetic particles tend to be created in the ground spin state.

4\. OPP process in the ultrarelativistic approximation.

(i) If the photon propagates perpendicular to the field, the main channels are (1) creation of a pair with the same spin projections by a photon of normal polarization, and (2) creation of a pair with opposite spin projections by a photon of anomalous polarization.

(ii) The total process rates satisfy the conditions $W^{+-} \gg W^{-+}$ for a photon of anomalous polarization and $W^{+-} = W^{-+}$ for a photon of normal linear polarization. Thus, changing the linear polarization of the initial photon (e.g. rotating the beam around its axis), it is possible to obtain electron and positron beams a controlled spin state, ranging from unpolarized to inversely polarized state.

The main scientific results of this chapter are published in [280-283].

# CHAPTER 3
# RESONANT AND SPIN-POLARIZATION EFFECTS IN THE PROCESS OF PHOTON SCATTERING ON AN ELECTRON

## 3.1. Introduction

In the chapter in the process of photon scattering by an electron, ie in Compton scattering (CS) in a magnetic field, spin-polarization effects under resonant conditions are studied, ie, the effect of initial photon polarization on both radiation polarization and spin states of final electron for different spin states of initial electron. The process of emission of two photons by an electron, ie the process of double synchrotron radiation (DSR) in a magnetic field under resonant conditions with polarized photons and certain values of particle spin projections, has been studied.

Adding a single photon in the initial or final states to the SR process, which was discussed in Chapter 2, gives the CS and DSR processes, which are the object of study in this section. The problems of the chapter consider QED processes, where in addition to particles in the initial and final states, particles in intermediate states also participate. Such particles are described by a Green's function.

<u>Green's function of an electron in a magnetic field.</u> For the Green's function of an electron in an external magnetic field, we use the expression obtained in [284]:

$$G_H(\mathbf{x}_1,\mathbf{x}_2) = \frac{1}{(2\pi)^3}\int d^3\mathbf{g}\, e^{-i\mathbf{g}(\mathbf{x}_1-\mathbf{x}_2)} G_H(\tilde{g};\rho_1,\rho_2),\qquad (3.1)$$

where $\mathbf{x}=(t,x,y,z)$ is 4-dimensional coordinate, $\mathbf{g}=(g_0,0,g_y,g_z)$, $\tilde{g}=(g_0,0,0,g_z)$,

$$\rho = \sqrt{hm}(x+g_y/hm^2),\quad G_H(\tilde{g};\rho_1,\rho_2) = \sum_{n=0}^{\infty}\frac{G_H(\rho_1,\rho_2)}{(\tilde{g}^2-m_g^2)},$$

$$G_H(\rho_1,\rho_2) = \sqrt{hm}[U_n(\rho_1)U_n(\rho_2)(\gamma\tilde{g}+m)\alpha_{12} + U_{n-1}(\rho_1)U_{n-1}(\rho_2)(\gamma\tilde{g}+m)\alpha_{21} +$$

$$+i\sqrt{2nhm}(U_{n-1}(\rho_1)U_n(\rho_2)\gamma^1\alpha_{12} - U_n(\rho_1)U_{n-1}(\rho_2)\gamma^1\alpha_{21})],\qquad (3.2)$$

$m_g^2 = m^2(1+2nh)$, $\alpha_{12} = \frac{1}{2}(1-i\gamma^1\gamma^2)$, $\alpha_{21} = \frac{1}{2}(1-i\gamma^2\gamma^1)$, $\gamma^\mu$ are the Dirac gamma matrices, $U_n(\rho)$ is Hermite function (2.8).

Green's function of an electron in a Redmond configuration field.

The study of QED processes in an external magnetic field, along which a plane electromagnetic wave is directed (Redmond field), allows, leaving a fixed number of photons of the wave, to study the QED processes of higher order perturbation theory in a purely magnetic field. The expression for the Green's function of the electron in the field of this configuration was found in [285] and has the form:

$$G_R(\mathbf{x}_1,\mathbf{x}_2) = \frac{-1}{(2\pi)^3}\int dg_0 dg_y dg_z \sum_{n=0}^{\infty} B_g(\varphi_1)\frac{G_H(\tilde{\rho}_1,\tilde{\rho}_2)}{g_0^2-\varepsilon^2}\bar{B}_g(\varphi_2)e^{-i(\Phi(\mathbf{x}_1)-\Phi(\mathbf{x}_2))}, \quad (3.3)$$

where $B_g(\varphi) = 1 - \frac{1}{2\kappa}e(\gamma^0-\gamma^3)(\gamma\tilde{A})$, $\varphi = t-z$, $\kappa = g^0 - g_z$,

$e\tilde{A} = (0, eA_x^{ext} - hm^2 K_y, eA_y^{ext} + hm^2 K_x, 0)$, $\tilde{\rho} = \sqrt{h}m^2(x+\frac{g_y}{hm^2}-K_x)$,

$\Phi(\mathbf{x}) = \mathbf{gx} + hm^2 K_x K_y + \sqrt{h}mK_y\tilde{\rho} + \int\frac{J(\varphi)d\varphi}{2\kappa}$,

$J(\varphi) = h\kappa m^2(\dot{K}_x K_y - K_x \dot{K}_y) - e^2\tilde{A}^2$, $\varepsilon^2 = m_g^2 + g_z^2$. Functions $K_x(\varphi), K_y(\varphi)$ are defined by the following equations:

$$\kappa\dot{K}_x = e\tilde{A}_y, \quad \kappa\dot{K}_y = e\tilde{A}_x.$$

The vector potential $\vec{A}^{ext} = (A_x^{ext}(\varphi), A_y^{ext}(\varphi))$ of the external field of a plane wave is perpendicular to the magnetic field, which is directed along the z axis.

**3.2. Spin polarization effects in the process of photon scattering by an electron**

Amplitude of probability of the process of photon scattering by an electron. The expression for the amplitude of the process corresponds to Feynman diagrams shown in Fig.3.1, and has the form:

$$A_{if} = ie^2 \int d^4\mathbf{x}_1 d^4\mathbf{x}_2 \bar{\Psi}'(\mathbf{x}_1)[\gamma^i A_i(\mathbf{x}_1) G_{H1}(\mathbf{x}_1,\mathbf{x}_2)\gamma^j A'^{*}_j(\mathbf{x}_2) +$$
$$+\gamma^j A'^{*}_j(\mathbf{x}_1) G_{H2}(\mathbf{x}_1,\mathbf{x}_2)\gamma^i A_i(\mathbf{x}_2)]\Psi(\mathbf{x}_2), \qquad (3.4)$$

where $\Psi(\mathbf{x}_2) = S^{-1/2} e^{-i\mathbf{r}_2 \mathbf{p}} \psi_l(\varsigma_2)$, $\bar{\Psi}(\mathbf{x}_1) = S^{-1/2} e^{i\mathbf{r}_1 \mathbf{p}'} \bar{\psi}_{l'}(\varsigma_1)$ are wave functions of the initial and final electrons, the explicit form of which is given by expressions (2.6) and (2.27); $\mathbf{r}_{1,2}=(t_{1,2}, 0, y_{1,2}, z_{1,2})$; $A_i(\mathbf{x})$, $A'^{*}_j(\mathbf{x})$ are wave functions of the initial and final photons (2.12); $G_{H1}(\mathbf{x}_1,\mathbf{x}_2)$, $G_{H2}(\mathbf{x}_1,\mathbf{x}_2)$ are the Green's functions of the electron in the intermediate state, which correspond to the diagrams $g$ and $f$ (see Fig. 3.1), and are given by expression (3.1).

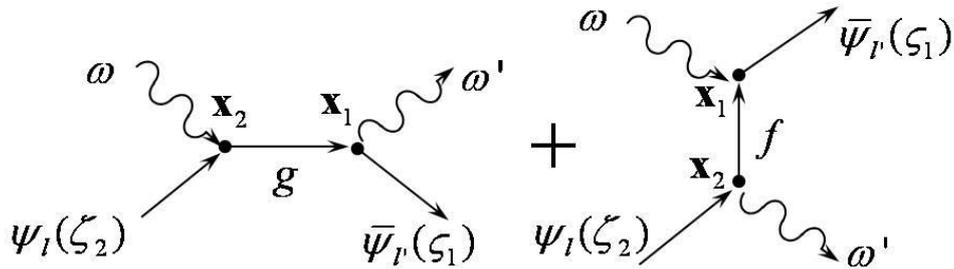

Fig.3.1. Feynman diagrams of the process of photon scattering by an electron

The external magnetic field H, which is directed along the z axis, is given similarly to the previous chapter with the field potential (2.11). In a constant magnetic field, the laws of conservation of energy and longitudinal component of momentum are fulfilled, which have the form:

$$\varepsilon + \omega = \varepsilon' + \omega', \quad p + \omega v = p' + \omega' u, \qquad (3.5)$$

where $\omega, v = \cos\theta$ and $\omega', u = \cos\theta'$ are frequencies and cosines of polar angles of the initial and final photons; $\varepsilon \equiv \varepsilon_l, p \equiv p_z$ і $\varepsilon' \equiv \varepsilon'_{l'}, p' \equiv p'_z$ are energies and longitudinal momenta of the initial and final electrons, respectively, which are related by the laws of dispersion as in the expression (2.6).

For the fixed Landau levels of particles $l, l'$ and the fixed angle of the final photon $u$, the conservation laws (3.5) specify the frequency of the final photon similar to the expression (2.30)

$$\omega' = \frac{1}{1-u^2}(\mathcal{E} - \mathcal{P}u - \sqrt{(\mathcal{E} - \mathcal{P}u)^2 - (\mathcal{E}^2 - \mathcal{P}^2 - m^2 - 2l'hm^2)(1-u^2)}), \quad (3.6)$$

where $\mathcal{E} = \varepsilon + \omega$, $\mathcal{P} = p + \omega v$. In the case of low frequencies $\omega \ll m$ or emission along the direction of the field $u \to 1$, the expression (3.6) takes the form:

$$\omega' = \frac{\mathcal{E}^2 - \mathcal{P}^2 - m^2 - 2l'hm^2}{2(\mathcal{E} - \mathcal{P}u)} \stackrel{h \ll 1}{=} \omega + (l - l')hm. \quad (3.7)$$

In the ultraquantum approximation, difference between the frequencies of the initial and final photons is equal to the distance between the Landau levels of the initial and final electrons. In particular, the Landau level of the electron does not change during Thomson scattering ($\omega = \omega'$).

In the ultrarelativistic case, it is convenient to analyze dependence of the process probability on the frequency and the polar angle of the final photon $\omega', u$. The Landau level of the final photon in this case has the form:

$$l' = \frac{1}{2hm^2}[(\varepsilon + \omega - \omega')^2 - (\omega v - \omega' u)^2]. \quad (3.8)$$

After taking the integrals over 4-dimensional coordinates in (3.4) and introducing special functions defined by expression (2.37), the amplitude of the CS process with polarized initial and final photons, as well as initial and final electrons with given spins in the general case can be reduced to:

$$A_{if} = (2\pi)^4 \frac{M_{if}}{SV} \delta^3(p + k - p' - k'), \quad (3.9)$$

$$M_{if} = \frac{-ie^2 e^{i\Phi}}{4\sqrt{\omega\omega'\varepsilon\varepsilon' m_l m_{l'}}} \left[ e^{i\Delta\Phi} \sum_{n_g=0}^{\infty} \frac{e^{-i\Phi_g} \sum_{a=1}^{10} Q_{ga}}{g_0^2 - \varepsilon_g^2} + \sum_{n_f=0}^{\infty} \frac{e^{i\Phi_f} \sum_{a=1}^{10} Q_{fa}}{f_0^2 - \varepsilon_f^2} \right]. \quad (3.10)$$

Two terms in expression (3.10) correspond to the two Feynman diagrams in Fig.3.1. Zero and longitudinal to field components of the 4-pulses **g** and **f** of the intermediate particle in these diagrams, respectively, are equal to:

$$g_0 = \varepsilon + \omega, \quad g = p + \omega v, \quad f_0 = \varepsilon - \omega', \quad f = p - \omega' u. \qquad (3.11)$$

The energies $\varepsilon_g$ and $\varepsilon_f$ of intermediate states at fixed Landau levels $n_g$ and $n_f$ are equal to:

$$\varepsilon_g = \sqrt{m^2 + 2n_g hm^2 + g^2}, \quad \varepsilon_f = \sqrt{m^2 + 2n_f hm^2 + f^2}. \qquad (3.12)$$

The general phase $\Phi$, phase difference $\Delta\Phi$, phases $\Phi_g$ and $\Phi_f$ have the form:

$$\Phi = \frac{1}{2hm^2}(k_x k_y - k'_x k'_y - 2p_y(k_x - k'_x)) + \frac{\pi}{2}(l'-l) + l\varphi' - l'\varphi, \qquad (3.13)$$

$$\Delta\Phi = \frac{k_y(k_x - k'_x)}{hm^2} + (l+l')(\varphi - \varphi'), \quad \Phi_g = n_g(\varphi - \varphi'), \quad \Phi_f = n_f(\varphi - \varphi'), \qquad (3.14)$$

where $k_x, k_y, k'_x, k'_y$ are transverse components of momenta of the initial and final photons, $\varphi, \varphi'$ are azimuthal angles of photons. The values $Q_{ga}$ in the first term (3.10) have the form:

$$Q_{g1} = J_{g1}^{++} J_{g2}^{++}[m\tilde{C} + g_0 C + gD], \quad Q_{g2} = J_{g1}^{+-} J_{g2}^{+-}[mC + g_0 \tilde{C} - gD],$$

$$Q_{g3} = J_{g1}^{-+} J_{g2}^{-+}[-mC + g_0 \tilde{C} - gD], \quad Q_{g4} = J_{g1}^{--} J_{g2}^{--}[-mC + g_0 \tilde{C} + gD],$$

$$Q_{g5} = J_{g1}^{-+} J_{g2}^{++}[-mD - g_0 \tilde{D} + gC], \quad Q_{g6} = J_{g1}^{++} J_{g2}^{-+}[-mD + g_0 \tilde{D} + gC],$$

$$Q_{g7} = J_{g1}^{--} J_{g2}^{+-}[mD + g_0 \tilde{D} + gC], \quad Q_{g8} = J_{g1}^{+-} J_{g2}^{--}[mD - g_0 \tilde{D} + gC],$$

$$Q_{g9} = [J_{g1}^{++} J_{g2}^{+-} + J_{g1}^{+-} J_{g2}^{++} + J_{g1}^{-+} J_{g2}^{--} + J_{g1}^{--} J_{g2}^{-+}] D_h,$$

$$Q_{g10} = [J_{g1}^{-+} J_{g2}^{+-} + J_{g1}^{+-} J_{g2}^{-+} - J_{g1}^{--} J_{g2}^{++} - J_{g1}^{++} J_{g2}^{--}] C_h, \qquad (3.15)$$

where the notations are entered

$$J_{g1}^{++} = J_1(l, n_g) M_m e_z, \quad J_{g1}^{--} = J_1(l-1, n_g-1) \mu M_p e_z,$$

$$J_{g1}^{+-} = J_1(l, n_g-1) M_m H_m, \quad J_{g1}^{-+} = J_1(l-1, n_g) \mu M_p H_p. \qquad (3.16)$$

The values $M_m$, $M_p$, $C$, $D$, $H_m$, $H_p$ are defined by expressions (2.40) - (2.43), and the values $\tilde{C}, \tilde{D}, C_h, D_h, e_z$ are equal, respectively:

$$\tilde{C} = E_m E_m' + \mathrm{sgn}(p_z)\mathrm{sgn}(p_z') E_p E_p', \quad \tilde{D} = \mathrm{sgn}(p_z') E_m E_p' - \mathrm{sgn}(p_z) E_p E_m', \qquad (3.17)$$

$$C_h = \sqrt{2n_g hm} C, \quad D_h = \sqrt{2n_g hm} D, \quad e_z = -\sin\theta\cos\alpha. \qquad (3.18)$$

The functions $J_{g2}^{++}, J_{g2}^{--}, J_{g2}^{+-}, J_{g2}^{-+}$ are defined by expressions (3.16), where you want to replace $1 \to 2$. The values $l$, $M_m$, $M_p$, $\mu$ related to the initial electron should be replaced by the corresponding hatched ones describing the final electron. The values $e_z$, $H_m$, $H_p$ describing the initial photon should be replaced by hatched complex conjugate analogues related to the final photon. $J_1(l,n_g)$, $J_2(l',n_g)$ are special functions (2.37), whose arguments $\eta$, $\eta'$ are:

$$\eta = \frac{\omega^2(1-v^2)}{2hm^2}, \quad \eta' = \frac{\omega'^2(1-u^2)}{2hm^2}. \qquad (3.19)$$

The values of $Q_{fa}$ in the second term of the amplitude (3.9) have the form (3.15), where you want to replace $g \to f$, as well as $J^{+-} \leftrightarrow J^{-+}$, with

$$J_{f1}^{++} = J_1(n_f, l')M'_m e_z, \quad J_{f1}^{--} = J_1(n_f-1, l'-1)\mu' M'_p e_z,$$

$$J_{f1}^{+-} = J_1(n_f, l'-1)\mu' M'_p H_m, \quad J_{f1}^{-+} = J_1(n_f-1, l')M'_m H_p. \qquad (3.20)$$

The expressions for $J_2$ are similar (3.20) and depend on parameters of the initial electron and the final photon.

<u>Cross section of the CS process.</u> The number of final states of the CS process coincides with that similar to the SR process (2.148). The product of the squared modulus of the amplitude (3.9) by the number of final states and divided by the time $T$ and the flux of the initial photons $j = 1/V$ gives the expression for the differential cross section CS

$$d\sigma \equiv d\sigma_{l'} = |M_{if}|^2 \delta(\varepsilon + \omega - \varepsilon' - \omega')\omega'^2 d\omega' du d\varphi', \qquad (3.21)$$

where $dud\varphi'$ is a solid angle element of the final photon. The index $l'$ in $d\sigma_{l'}$ shows that expression (3.21) is a partial scattering cross section of the CS process, as a result of which the electron passes to a fixed Landau level $l'$. In a general case, the total scattering cross section is equal to:

$$\sigma_{tot} = \int d\omega' du d\varphi' \sum_{l'=0}^{l'\max} \frac{d\sigma_{l'}}{d\omega' du d\varphi'}, \qquad (3.22)$$

where $l'_{max}$ is the maximum possible Landau level of the finite electron, which is determined by the law of conservation of energy with the condition $\omega'=0$. Only one term in the sum over the parameter $l'$ in the total cross section (3.22), which corresponds to the transition of an electron to the neighboring Landau level, plays the main role in ultraquantum approximation. In expression (3.21), which contains the Dirac delta function, integration over frequency $\omega'$ is easily carried out

$$d\sigma = \frac{|M_{if}|^2 \varepsilon' \omega'^2 \, du \, d\varphi'}{|\varepsilon' - u(p + \omega v - \omega' u)|}. \tag{3.23}$$

Assuming that only one term in the sum of the Landau levels $n_g$ in the expression (3.10) gives the main contribution to the cross section (resonance in the diagram g Fig.3.1), the amplitude (3.10) can be represented as

$$M_{if} = \frac{N_{if}}{g_0^2 - \varepsilon_g^2} = \frac{e^2}{4\sqrt{\omega \omega' \varepsilon \varepsilon' m_l m_{l'}}} \frac{Q_g}{g_0^2 - \varepsilon_g^2}, \tag{3.24}$$

where the insignificant phase is rejected. In the case of resonance

$$g_0^2 - \varepsilon_g^2 = 2\varepsilon_g (g_0 - \varepsilon_g).$$

Breit's phenomenological rule is used to eliminate divergence at the resonance point [113]

$$\varepsilon_g \to \varepsilon_g - i\Gamma/2, \tag{3.25}$$

where $\Gamma$ is a resonance width, which is defined as the total probability of decay of an intermediate state. Then the differential cross section (3.22) can be represented as:

$$\frac{d\sigma}{du} = \frac{\pi \omega'^2 |N_{if}|^2}{2m^2[(g_0 - \varepsilon_g)^2 + \Gamma^2/4]}. \tag{3.26}$$

If resonance takes place in the *f* diagram, the denominator of the cross section (3.26) must be replaced $g_0-\varepsilon_g \to f_0-\varepsilon_f$.

In ultrarelativistic approximation, the factor $dl'$ is added to the number of final states, and integration by this value removes the Dirac delta function, as a result, the differential cross section has the form:

$$d\sigma = \frac{1}{hm^2} |M_{if}|^2 \varepsilon' \omega'^2 \, d\omega' du \, d\varphi'. \tag{3.27}$$

The sum in the first term of the amplitude (3.10) over the Landau levels of the intermediate particle $n_g$ is replaced by an integral, which is easy to take because it contains a pole $(g_0-\varepsilon_g+i\Gamma/2)$:

$$\sum_{n_g}\frac{Q_g}{g_0^2-\varepsilon_g^2} = -\int\frac{dn_g Q_g}{2\varepsilon_g(\varepsilon_g-g_0-i\Gamma/2)} = \frac{-\pi i}{2hm^2}Q_g\big|_{n_{g0}}$$

and similarly in the second term of the amplitude (3.10). $Q_g$ is the sum of terms (3.15). Landau level numbers $n_g$, $n_f$ in the obtained amplitude should be replaced by the expressions:

$$n_{g0} = \frac{(\varepsilon+\omega)^2-\omega^2 v^2}{2hm^2}, \quad n_{f0} = \frac{(\varepsilon-\omega')^2-\omega'^2 u^2}{2hm^2}. \tag{3.28}$$

Neglecting the interference term of two amplitudes $g$ and $f$ (see Fig. 3.1), the differential scattering cross section in ultrarelativistic approximation can be reduced to the form:

$$\frac{d\sigma}{d\omega'du} = \frac{\pi^3 e^4\varepsilon'[|Q_g|^2+|Q_f|^2]}{32h^3m^6\omega\varepsilon\sqrt{\varepsilon'^2-p'^2}}, \tag{3.29}$$

where the energy and momentum of the final electron are determined by the conservation laws (3.5).

<u>Resonance conditions in the CS process.</u> Resonance in the process of photon scattering by an electron corresponds to transition of the intermediate state on the mass shell. The resonance in the $g$ diagram (see Fig. 3.1) corresponds to hitting the pole point in the first term of the amplitude (3.10), that is, fulfillment of the equality

$$\varepsilon_g-\varepsilon-\omega = 0. \tag{3.30}$$

In general, you can put the longitudinal momentum of the initial electron to zero:

$$p = 0. \tag{3.31}$$

Condition (3.30) specifies value of the frequency of the initial photon to enter the resonance at a fixed Landau level $n_g$:

$$\omega_g = \frac{1}{1-v^2}[\sqrt{\varepsilon^2+2(n_g-1)hm^2(1-v^2)}-\varepsilon]. \tag{3.32}$$

It can be seen that the resonant conditions do not depend on the parameters of the final particles and are determined only by the initial particles, while the polarization of the initial particles does not affect the resonant conditions.

In the ultraquantum approximation, the frequency of the initial photon in resonant conditions is equal to (up to $h^2$):

$$\omega_g = (n_g - l)hm - h^2 m(n_g - l)(n_g + l - (n_g - l)v^2)/2. \qquad (3.33)$$

Under such conditions, the final photon is emitted with frequency

$$\omega'_g = (n_g - l')hm - h^2 m(n_g - l')[n_g + l' + (n_g - l')u^2 - 2(n_g - l)vu]/2. \qquad (3.34)$$

Thus, up to the first degree of magnetic field $h$, the resonant frequencies of the initial and final photons are multiples of the cyclotron frequency $hm$. They are equal to the distances between the Landau levels of the initial and intermediate electrons, and the intermediate and final electrons, respectively.

The resonance in the *f* diagram (see Fig. 3.1) corresponds to hitting the pole point in the second term of the amplitude (3.10), that is, fulfillment of one of the two equalities

$$\varepsilon_f - \varepsilon + \omega' = 0, \quad \varepsilon_f + \varepsilon - \omega' = 0. \qquad (3.35)$$

The first condition (3.35) is satisfied if the initial electron emits a final photon with frequency

$$\omega'_f = \frac{1}{1-u^2}[\varepsilon - \sqrt{\varepsilon^2 - 2(l - n_f)hm^2(1-u^2)}]. \qquad (3.36)$$

Since the frequency of the final photon in any case must be equal to (3.6), equation (3.36) determines the frequency of the initial photon under resonant conditions, which in this case is equal to

$$\omega_f = \frac{1}{a}[-b + \sqrt{b^2 + ac}], \qquad (3.37)$$

where $a = (1-v^2)(1-u^2)$, $b = k(1-vu) + \varepsilon(vu - u^2)$, $c = 2(l' - n_f)hm^2(1-u^2)$, $k = (\varepsilon^2 - 2(l - n_f)hm^2(1-u^2))^{1/2}$.

In ultraquantum approximation, the resonant frequencies (3.36), (3.37) have the form:

$$\omega'_f = (l - n_f)hm - h^2 m(l - n_f)[l + n_f + (l - n_f)u^2]/2, \qquad (3.38)$$

$$\omega_f = (l' - n_f)hm - h^2 m(l' - n_f)[l' + n_f - (l' - n_f)v^2 + 2(l - n_f)vu]/2. \quad (3.39)$$

Note, up to $h^2$, the frequency of the initial photon resonant in the diagram $f$, in addition to the Landau level numbers of particles $l$, $l'$, $n_f$ and the polar angle of the initial photon $v$, also depends on the outlet angle of the final photon $u$, except the case v = 0. To realize this resonance, one can choose a fixed value of the frequency of the initial photon, a multiple of the cyclotron frequency with a given detuning $-\delta h^2 m(l' - n_f)/2$, and enter the resonance by choosing the angle of emission of the final photon $u_f$:

$$\omega_f = hm(l' - n_f)(1 - h\delta/2), \quad u_f = \frac{\delta + (l' - n_f)v^2 - l' - n_f}{2(l - n_f)v}. \qquad (3.40)$$

The second condition (3.35) for appearance of resonance in the $f$ diagram determines the frequency of the final photon, which has the form

$$\omega'_f = \frac{1}{1 - u^2}[\varepsilon + \sqrt{\varepsilon^2 - 2(l - n_f)hm^2(1 - u^2)}]. \qquad (3.41)$$

Upon entering such a resonance, the initial photon generates an electron positron pair, and the intermediate positron subsequently annihilates with the initial electron into the final photon. The frequency of the initial photon is determined by expression (3.37), where you want to replace the value b→-b that is equal to $b = k(1 - vu) - \varepsilon(vu - u^2)$.

In ultraquantum case, the frequencies of photons in the zero approximation (up to $h^0$) have the form:

$$\omega_f = \omega'_f \frac{1 + u^2 - 2vu}{(1 - v^2)}, \quad \omega'_f = \frac{2m}{1 - u^2}. \qquad (3.42)$$

In particular, for the case $v = u = 0$ the frequencies of photons up to $h$ are equal to:

$$\omega_f = 2m + (l' + n_f)hm, \quad \omega'_f = 2m + (l + n_f)hm, \qquad (3.43)$$

that is, the frequency of the photon, both initial and final, is equal to the sum of the energies of particles produced at fixed Landau levels with zero longitudinal pulses.

Interference of resonances in two diagrams can take place at simultaneous performance of condition (3.30) and the first condition (3.35) which lead to expression

$$\varepsilon_g + \varepsilon_f = \varepsilon + \varepsilon', \qquad (3.44)$$

which in ultraquantum approximation leads to the following conditions:

$$n_g + n_f = l + l', \quad vu = 1. \qquad (3.45)$$

that is, the initial and final photons move along the field.

<u>The CS process in ultraquantum approximation.</u> First, we analyze spin-polarization effects in the process of photon scattering by an electron under resonant conditions (3.33) (resonance in the *g* diagram of Fig. 3.1). Differential cross sections of the process with certain values of spins projections of the initial and final electrons $\sigma^{\mu\mu'}$ can be reduced to the form:

$$\frac{d\sigma_{\xi\xi'}^{--}}{du} = \frac{2\pi}{\omega^2} \frac{\frac{dW_\xi^{--}}{dv} \cdot \frac{dW_{\xi'}^{--}}{du}}{[(\omega-\omega_g)^2 + \Gamma^2/4]}, \quad \frac{d\sigma_{\xi\xi'}^{++}}{du} = \frac{2\pi}{\omega^2} \frac{\frac{dW_\xi^{++}}{dv} \cdot \frac{dW_{\xi'}^{++}}{du}}{[(\omega-\omega_g)^2 + \Gamma^2/4]}, \qquad (3.46)$$

$$\frac{d\sigma_{\xi\xi'}^{+-}}{du} = \frac{2\pi}{\omega^2} \frac{\frac{dW_\xi^{++}}{dv} \cdot \frac{dW_{\xi'}^{+-}}{du}}{[(\omega-\omega_g)^2 + \Gamma^2/4]}, \quad \frac{d\sigma_{\xi\xi'}^{-+}}{du} = \frac{2\pi}{\omega^2} \frac{\frac{dW_\xi^{+-}}{dv} \cdot \frac{dW_{\xi'}^{++}}{du}}{[(\omega-\omega_g)^2 + \Gamma^2/4]}, \qquad (3.47)$$

where differential probabilities $dW_\xi^{++}/dv$, $dW_\xi^{--}/dv$, $dW_\xi^{+-}/dv$ описуються виразами (2.46) - (2.48), in which replacements are made

$$u \to v, \ l \to n_g, \ l' \to l. \qquad (3.48)$$

Such probabilities can be interpreted as differential probabilities of radiation of the initial photon with polarization $\xi$ (ie with Stokes parameters $\xi_1$, $\xi_2$, $\xi_3$) by the intermediate electron at the certain Landau level $n_g$ and in certain spin state $\mu_g$ (the first sign in the probability *W*), with transition of the electron to initial state at the Landau level *l*. Differential probabilities $dW_{\xi'}^{++}/du$, $dW_{\xi'}^{--}/du$, $dW_{\xi'}^{+-}/du$ in the cross sections (3.46), (3.47) are described by the same expressions (2.46) - (2.48), in which replacements are made

$$l \to n_g, \quad \eta \to \eta', \qquad (3.49)$$

that is, it is the probability of emission of the final photon with polarization $\xi'$ by the intermediate electron, which is also at the Landau level $n_g$ and in the certain spin state, with transition of the electron to the final state at the Landau level $l'$.

From the obtained cross sections $\sigma^{+-}$, $\sigma^{-+}$, which correspond to spin-flip processes, it follows that the intermediate electron is in the inverse spin state ($\mu_n=+1$). This is due to the lack of probability $W^{-+}$, described by expression (2.49), which has a higher degree of small parameter $h$. All cross sections are factorized, which means that in resonance the polarization of the final photon does not depend on the polarization of the initial one.

Differential cross sections (3.46), (3.47), averaged over the polarizations of the initial photon and summed over the polarizations of the final photon (for this, all Stokes parameters must be zeroed in them and the result must be doubled), correspond to the process of scattering of an unpolarized photon by an electron and coincide with the results [125].

The analysis and the obtained results of influence of electron spin states on polarization of the final photon in processes with a fixed values of particle spins coincide with those in the process of photon emission by an electron. In the CS process without spin flip ($\sigma^{++}$, $\sigma^{--}$), the polarization of radiation is determined by the Stokes parameters (2.57). In the spin-flip process to the ground spin state ($\sigma^{+-}$), the polarization of the final photon is determined by expressions (2.60), in which the sign of the linear polarization is changed in comparison with (2.57). In the spin-flip process to the inverse spin state ($\sigma^{-+}$), the polarization of the final photon is the same as in the process without spin-flip (2.57). A significant difference between the spin-flip of the CS process from the SR process is that the cross sections (3.47) have the same degree $h$, that is, the probabilities of the spin-flip in the CS to the ground and inverse states are close in magnitude, while in the SR process, the flip to the inverse spin state is suppressed.

The degree of polarization of the spin-flip of the SR process by electrons, which are both in the ground ($\mu=-1$) and in the inverse ($\mu=+1$) spin states, has the form:

$$P_{\xi'} = \max \frac{\sigma_{\xi,\xi'}^{+-} + \sigma_{\xi,\xi'}^{-+} - (\sigma_{\xi,-\xi'}^{+-} + \sigma_{\xi,-\xi'}^{-+})}{\sigma_{\xi,\xi'}^{+-} + \sigma_{\xi,\xi'}^{-+} + \sigma_{\xi,-\xi'}^{+-} + \sigma_{\xi,-\xi'}^{-+}}, \quad (3.50)$$

whence the expressions for the Stokes parameters of the emitted photon follow:

$$\xi'_{1CS} = 0, \; \xi'_{2CS} = \frac{2u}{1+u^2}, \; \xi'_{3CS} = \frac{1-u^2}{1+u^2} \cdot \frac{l(n_g - l')^2 \Pi - l'(n_g - l)^2 \tilde{\Pi}}{l(n_g - l')^2 \Pi + l'(n_g - l)^2 \tilde{\Pi}}, \quad (3.51)$$

$$\Pi = 1 - \frac{1-v^2}{1+v^2}\xi_3 + \frac{2v}{1+v^2}\xi_2, \quad \tilde{\Pi} = 1 + \frac{1-v^2}{1+v^2}\xi_3 + \frac{2v}{1+v^2}\xi_2. \quad (3.52)$$

In such a process, the circular polarization of the final photon has the same form as in the process without spin flip. The linear polarization of the final photon substantially depends both on the Landau levels of the electron and on the polar angle and polarization of the initial photon. In particular, for the elastic channel (Thomson scattering $l=l'$) $\xi'_{3CS}$ has the form:

$$\xi'_{3CS} = -\frac{1-u^2}{1+u^2} \cdot \frac{1-v^2}{1+v^2+2v\xi_2}\xi_3, \quad (3.53)$$

that is, parameter $\xi'_{3CS}$ is proportional to the degree of linear polarization of the initial photon and is opposite in sign. Figure 3.2 shows the dependence of the degree of polarization on a) the angle of the final photon, b) the angle of the initial photon. Wherein $l=2$, $n_g=4$, $\xi_2=0.3$, $\xi_3=0.5$; $v=0.1$ in Figure 3.2.a; $u=0.1$ in Figure 3.2.b. The three curves in Fig. 3.2 correspond to the cases: 1. $l'=1$ is inverse Compton scattering $\omega < \omega'$, 2. $l'=2$ is Thomson scattering $\omega = \omega'$, 3. $l'=3$ is direct Compton scattering $\omega > \omega'$. From Fig. 3.2 it follows that the smallest degree of polarization of the final photon corresponds to inverse Compton scattering in the direction perpendicular to the magnetic field.

Suppose that in the initial state the spin of the electron has a certain orientation, and the final states are summed over the spins. Let us analyze the influence of the spin-flip process on the degree of radiation polarization. The cross sections of the CS process are proportional to the values:

$$\sigma^-_{\xi\xi'} \sim W^{--}_\xi \cdot W^{--}_{\xi'} + W^{+-}_\xi \cdot W^{++}_{\xi'}, \quad \sigma^+_{\xi\xi'} \sim W^{++}_\xi \cdot W^{++}_{\xi'} + W^{++}_\xi \cdot W^{+-}_{\xi'}, \quad (3.54)$$

where $\sigma^-$ corresponds to the process with the electron with spin initially oriented against the field, $\sigma^+$ - corresponds to the process with the electron with spin along the field. The degree of polarization of the final photon in these cases is equal to:

$$P^-_{\xi'} = \max \frac{\sigma^-_{\xi,\xi'} - \sigma^-_{\xi,-\xi'}}{\sigma^-_{\xi,\xi'} + \sigma^-_{\xi,-\xi'}} = 1, \quad (3.55)$$

$$P^+_{\xi'} = \max \frac{\sigma^+_{\xi,\xi'} - \sigma^+_{\xi,-\xi'}}{\sigma^+_{\xi,\xi'} + \sigma^+_{\xi,-\xi'}}, \quad P^+_{\xi'}\big|_{l'\neq 0} = 1 - \frac{2h(n_g - l')^2(1-u^2)^2}{l'(1+u^2)^2}, \quad P^+_{\xi'}\big|_{l'=0} = 1. \quad (3.56)$$

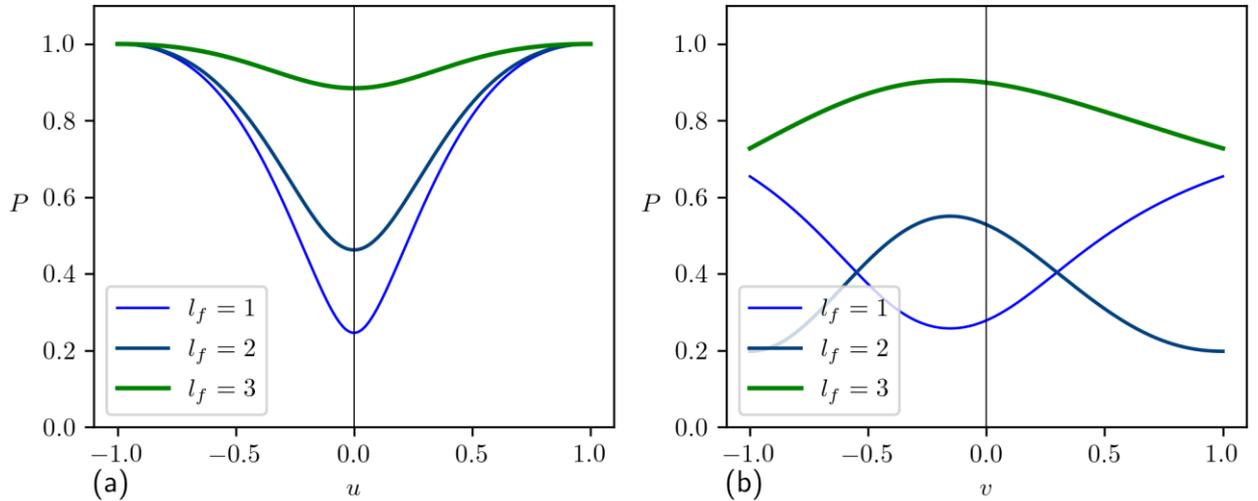

Fig.3.2. The degree of polarization of the radiation in the spin-flip process of CS from the ground and inverse spin states as a function of a) the angle of the final photon, b) the angle of the initial photon; 1. $l'=1$, 2. $l'=2$, 2. $l'=3$

Thus, in the CS process, electrons with spins initially oriented against the field emit fully polarized photons with radiation polarization as in the JI process (2.64). Photons scattered by electrons, initially oriented along the field, are partially polarized.

The linear polarization of the final photons is partially violated by ~ $h$ and depends on the Landau level of the final electron.

The polarization of the final photon in the process with electrons, unpolarized in both the initial and final states, is determined by analyzing the sum $\sigma^- + \sigma^+$ of expressions (3.54). In this case, the Stokes parameters of the final photon are

$$\xi'_{2CS} = \frac{2u}{1+u^2}, \quad \xi'_{3CS} = -\frac{1-u^2}{1+u^2}(1 - h\frac{(n_g - l')^2 l}{n^2 + ll'}). \tag{3.57}$$

The linear polarization of radiation in the CS process differs from the linear polarization of the SR process (2.64) by ~ $h$, which depends on the Landau levels of both the initial and final electrons. The polarization of the final photon does not depend on the polarization of the initial one.

Now let us consider the question of the orientation of the electron spins as a result of the CS process, or vice versa, the depolarization of the initially oriented spins and the influence of the polarization of the initial photon on the spin states of the final electron. Let the spin of the initial electron be directed against the field (ground spin state), ie, a beam of the initial electrons is completely polarized against the field P = -1. The degree of polarization of the electron beam due to the spin-flip process in CS differs from -1 and is determined by the expression:

$$P_{e^-}^- = \frac{d\sigma_{\xi\xi'}^{-+}/du - d\sigma_{\xi\xi'}^{--}/du}{d\sigma_{\xi\xi'}^{-+}/du + d\sigma_{\xi\xi'}^{--}/du}, \tag{3.58}$$

which, taking into account the explicit form of the cross sections (3.46), (3.47), is equal to

$$P_{e^-}^- = \frac{-\Pi + hl'(n_g - l)^2 \tilde{\Pi}/2n_g^2}{\Pi + hl'(n_g - l)^2 \tilde{\Pi}/2n_g^2}, \tag{3.59}$$

where the values $\Pi, \tilde{\Pi}$ are determined by expressions (3.52). If $\Pi \neq 1$, then to the first degree $h$:

$$P_{e^-}^- = -1 + hl'\frac{(n_g - l)^2 \tilde{\Pi}}{n_g^2 \Pi}. \tag{3.60}$$

In the case when the initial photon is directed perpendicular to the field ($v=0$), the values $\Pi, \tilde{\Pi}$ are equal to:

$$\Pi = 1 - \xi_3, \quad \tilde{\Pi} = 1 + \xi_3. \qquad (3.61)$$

The expression for the degree of polarization of the final electrons (3.60) is similar to the degree of polarization of an electron in the process of one-photon $e^+e^-$ pair production (2.169). Except for a narrow cone $\vartheta < \vartheta_c$, where $\xi_3 = 1 - 2\vartheta$ (see Fig.2.18), the degree of polarization is $P \approx -1$ and less than one by an amount of the order of $h$. The magnitude of nonpolarization (depolarization) of electrons is proportional to the Landau level number of the final electron. In a narrow cone $\vartheta < h$, the spins of the final electrons are oriented along the field.

If the initial photon is directed along the field ($v=1$), the values $\Pi, \tilde{\Pi}$ have the form:

$$\Pi = \tilde{\Pi} = 1 + \xi_2, \qquad (3.62)$$

that is, the polarization of the initial photon does not affect the orientation of the electron spins. Note that if the initial photon has left-hand circular polarization, then $\Pi = \tilde{\Pi} = 0$. the CS process does not take place.

Now let the spin of the initial electron be directed along the field (inverse spin state), ie $P = +1$. The degree of polarization of the electron beam due to the spin-flip process in CS is less than one and is equal to:

$$P_{e^-}^+ = \frac{d\sigma_{\xi\xi'}^{++}/du - d\sigma_{\xi\xi'}^{+-}/du}{d\sigma_{\xi\xi'}^{++}/du + d\sigma_{\xi\xi'}^{+-}/du} = \frac{\Pi' - h(n_g - l')^2 \tilde{\Pi}'/2l'}{\Pi' + h(n_g - l')^2 \tilde{\Pi}'/2l'}, \qquad (3.63)$$

$$\Pi' = 1 - \frac{1-u^2}{1+u^2}\xi'_3 + \frac{2u}{1+u^2}\xi'_2, \quad \tilde{\Pi}' = 1 + \frac{1-u^2}{1+u^2}\xi'_3 + \frac{2u}{1+u^2}\xi'_2. \qquad (3.64)$$

If we do not fix the polarization of the final photon, that is, we put the value $\xi'_i=0$, then $\Pi' = \tilde{\Pi}' = 1$. As a result, the degree of polarization of the final electron depends only on the Landau level numbers and up to $h$ is equal to:

$$P_{e^-}^+ = 1 - h\frac{(n_g - l')^2}{l'}. \qquad (3.65)$$

The spins of the electrons in the initially unpolarized beam due to the self-polarization effect acquire a predominant direction. The effect of self-polarization was first discovered by AA Sokolov, IM Ternov in the JI process [4], where the spins of electrons are oriented mainly in the ground spin state. This effect in the CS process leads to the orientation of the electron spins both along and against the field, which is also affected by the polarization of the initial photon. The degree of self-polarization of the electron is equal to:

$$P_{e^-} = \frac{d\sigma_{\xi\xi'}^{+-}/du - d\sigma_{\xi\xi'}^{-+}/du}{d\sigma_{\xi\xi'}^{+-}/du + d\sigma_{\xi\xi'}^{-+}/du} = \frac{\delta-1}{\delta+1}, \quad \delta = \frac{d\sigma_{\xi\xi'}^{+-}/du}{d\sigma_{\xi\xi'}^{-+}/du} = \frac{(n_g-l')^2 l \Pi \tilde{\Pi}'}{(n_g-l)^2 l' \tilde{\Pi} \Pi'}. \quad (3.66)$$

If the initial photon is unpolarized and the polarization of the final photon is not fixed, then the degree of self-polarization depends only on the Landau levels

$$\delta = \frac{(n_g-l')^2 l}{(n_g-l)^2 l'}. \quad (3.67)$$

In the "elastic" channel $l = l'$ (Thomson scattering) $\delta=1$ and $P = 0$, ie self-polarization is absent. In direct Compton scattering $l<l'$, $\delta<1$, $P<0$, ie the spins of the electrons are mainly oriented against the field. In the inverse Compton scattering $l>l'$, $\delta>1$, $P>0$, ie the spins of the electrons are mainly oriented along the field.

Let $l=l'$ (Thomson scattering) and the initial photon is polarized, then the degree of polarization of the final electron is equal to:

$$P_{e^-} = -\frac{(1-v^2)\xi_3}{1+v^2+2v\xi_2}. \quad (3.68)$$

If the initial photon is directed perpendicular to the field $v=0$, then $P^-=-\xi_3$. Thus, rotating the beam of initial photons about its axis, that is, changing the linear polarization of photons from normal to abnormal polarization, it is possible to orient the spins of electrons from the direction against the field in the direction along the field. A significant difference between this effect in the CS process and a similar effect in the OPP process is that in the CS process the change in the degree of polarization of electrons is proportional to the change in the degree of polarization of photons. This makes the CS process more promising for obtaining electron beams with variable polarization.

Now we analyze the spin-polarization effects in the CS process under resonant conditions (3.39) (resonance in the *f* diagram of Fig.3.1). Differential cross sections of the process with certain values of the spin projections of the initial and final electrons $\sigma^{\mu\mu'}$ have a form similar to the expressions (3.46), (3.47):

$$\frac{d\sigma^{--}_{\xi\xi'}}{du} = \frac{2\pi}{\omega^2} \frac{\dfrac{dW^{--}_{\xi}}{dv} \cdot \dfrac{dW^{--}_{\xi'}}{du}}{[(\omega-\omega_f)^2 + \Gamma^2/4]}, \quad \frac{d\sigma^{++}_{\xi\xi'}}{du} = \frac{2\pi}{\omega^2} \frac{\dfrac{dW^{++}_{\xi}}{dv} \cdot \dfrac{dW^{++}_{\xi'}}{du}}{[(\omega-\omega_f)^2 + \Gamma^2/4]}, \quad (3.69)$$

$$\frac{d\sigma^{+-}_{\xi\xi'}}{du} = \frac{2\pi}{\omega^2} \frac{\dfrac{dW^{--}_{\xi}}{dv} \cdot \dfrac{dW^{+-}_{\xi'}}{du}}{[(\omega-\omega_f)^2 + \Gamma^2/4]}, \quad \frac{d\sigma^{-+}_{\xi\xi'}}{du} = \frac{2\pi}{\omega^2} \frac{\dfrac{dW^{+-}_{\xi}}{dv} \cdot \dfrac{dW^{--}_{\xi'}}{du}}{[(\omega-\omega_f)^2 + \Gamma^2/4]}, \quad (3.70)$$

where the differential probabilities $dW^{++}_{\xi}/dv$, $dW^{--}_{\xi}/dv$, $dW^{+-}_{\xi}/dv$ are described by the expressions (2.46) - (2.48), in which substitutions are made:

$$u \to v, \; l \to l', \; l' \to n_f, \quad (3.71)$$

and to obtain differential probabilities $dW^{++}_{\xi'}/du$, $dW^{--}_{\xi'}/du$, $dW^{+-}_{\xi'}/du$ the following substitution is made in the expressions (2.46) - (2.48):

$$l' \to n_f, \; \eta \to \eta'. \quad (3.72)$$

Note that in spin-flip processes, the intermediate electron is in the ground spin state.

The analysis of spin-polarization effects is similar to the analysis performed for the CS process in resonance in the g diagram. The polarization of the final photon for the process without spin flip and in the spin-flip process to the ground state is determined by formulas (2.57). In the spin-flip process in the inverse spin state, it is determined by formula (2.60). The Stokes parameters of the final photons in the CS process, where electrons are located both in the ground ($\mu = 1$) and in the inverse ($\mu = + 1$) spin states, with a flip of their spins, have the form:

$$\xi'_{1CS} = 0, \; \xi'_{2CS} = \frac{2u}{1+u^2}, \; \xi'_{3CS} = \frac{1-u^2}{1+u^2} \cdot \frac{l'(l-n_f)^2 \Pi - l(l'-n_f)^2 \tilde{\Pi}}{l'(l-n_f)^2 \Pi + l(l'-n_f)^2 \tilde{\Pi}}. \quad (3.73)$$

From a comparison of this expression with a similar one (3.51) it follows that the polarizations of the final photons coincide in both channels (diagrams *g* and *f*) for elastic scattering *l=l'*.

In the case where the electrons have spins initially oriented against the field, the finale photons are completely polarized, as in the CS process (2.64). If the spins of the electrons are initially oriented along the field, the degree of polarization of the final photons is equal to:

$$P_{\xi'}^{+} = 1 - \frac{2h(l-n_f)^2(1-u^2)^2}{n_f(1+u^2)^2}, \qquad (3.74)$$

which is analogous to the expression (3.56).

The Stokes parameters describing the polarization of the radiation of final photons in the CS process with unpolarized electrons have a form similar to expressions (3.57):

$$\xi'_{1CS} = 0, \quad \xi'_{2CS} = \frac{2u}{1+u^2}, \quad \xi'_{3CS} = -\frac{1-u^2}{1+u^2}(1-h\frac{(l-n_f)^2 n_f}{n^2+ll'}). \qquad (3.75)$$

We now turn to the question of the influence of polarization of the initial photon on the spin states of the final electron in the CS process at resonance in the diagram f (see Fig. 3.1). The initially oriented spins against the field are depolarized due to the spin-flip process and the degree of polarization of the final electron is equal to:

$$P_{e^-}^{-} = \frac{d\sigma_{\xi\xi'}^{-+}/du - d\sigma_{\xi\xi'}^{--}/du}{d\sigma_{\xi\xi'}^{-+}/du + d\sigma_{\xi\xi'}^{--}/du} = -\frac{2l'\Pi - h(l'-n_f)^2\tilde{\Pi}}{2l'\Pi + h(l'-n_f)^2\tilde{\Pi}}. \qquad (3.76)$$

If the initial photon moves perpendicular to the field (*v* = 0) and *l'*Π≠0, the degree of polarization has a simple form similar to (3.60)

$$P_{e^-}^{-} = -1 + h\frac{(l'-n_f)^2(1+\xi_3)}{l'(1-\xi_3)}. \qquad (3.77)$$

If the electrons in the initial state have spins oriented along the field, then the degree of polarization of the final electrons is determined similarly to (3.63) and is equal to:

$$P_{e^-}^{+} = +1 - h\frac{l'(l-n_f)^2}{n_f^2}. \qquad (3.78)$$

The degree of self-polarization of the electron as a result of CS is determined by the first expression in (3.66) and in the resonance in the diagram f is equal to:

$$P_{e^-} = \frac{\delta - 1}{\delta + 1}, \quad \delta = \frac{l'(l - n_f)^2 \Pi}{l(l' - n_f)^2 \tilde{\Pi}}. \tag{3.79}$$

For an elastic channel ($l=l'$) the degree of self-polarization $P_{e^-}$ is determined by expression (3.68). Thus, the two diagrams $f$ and $g$ make the same contribution to the self-polarization process in the elastic channel, which is the most probable. In direct ($l<l'$) Compton scattering the electrons are oriented against the field, in reverse ($l>l'$) Compton scattering they are oriented towards the field.

Let us consider the RFE process when the resonance conditions are realized in both diagrams (see Fig. 3.1). Since the cross sections of the process in resonances in the diagrams $f$ and $g$ are proportional to product of the probabilities of the SR process, that is, they are proportional, respectively, to the factors

$$(1-v^2)^{n_g - l - 1}(1-u^2)^{n_g - l' - 1}, \quad (1-v^2)^{l' - n_f - 1}(1-u^2)^{l - n_f - 1}, \tag{3.80}$$

then, taking into account conditions (3.45), in order for the cross sections to be nonzero, it is necessary to zero the degrees in expressions (3.80). It means that interference of the resonances of the two diagrams takes place for the elastic channel ($l=l'$), if the electron in the process goes over to the neighboring Landau level, with the initial photon directed parallel to the magnetic field:

$$l' = l, \quad n_g = l + 1, \quad n_f = l - 1, \quad vu = 1. \tag{3.81}$$

The differential cross section of the process without spin-flip in these conditions is equal to:

$$\frac{d\sigma_{\text{int}}^{\mu\mu}}{du} = \frac{\pi}{2}\alpha^2 h^2 \Pi\Pi' \left| \frac{l + 1/2 - \mu/2}{\omega - \omega_g + i\Gamma_g/2} + \frac{l - 1/2 - \mu/2}{\omega_f - \omega + i\Gamma_f/2} \right|^2, \tag{3.82}$$

where $\omega_g$, $\omega_f$ are the resonant frequencies of the initial photon under resonant conditions (3.33) and (3.39), respectively; $\Gamma_g$, $\Gamma_f$ are resonance widths in diagrams g and f, respectively. Under condition (3.81) the frequencies $\omega_g$ and $\omega_f$ coincide and are equal to the cyclotron frequency accurate to h:

$$\omega_g = \omega_f \approx hm. \qquad (3.83)$$

The widths of the resonances is defined as the total probabilities of the SR process at the corresponding Landau levels, averaged over the polarizations of the initial particles and summed over the polarizations of the final particles. In resonance in the diagram g, the width is equal to the sum of the probabilities of emission of the initial and final photons by the intermediate electron. In resonance in the diagram f, the width is equal to the sum of the probabilities of emitting the final photon by the initial electron and emitting the initial photon by the final electron. As a result, the same width for both diagrams is equal to:

$$\Gamma_g = \Gamma_f = \frac{4}{3}\alpha h^2 m(2l+1). \qquad (3.84)$$

Differential cross sections of spin-flip processes in the ground and inverse spin states at interference of the resonances coincide and have the form:

$$\frac{d\sigma_{int}^{+-}}{du} = \frac{d\sigma_{int}^{-+}}{du} = \frac{\pi}{4}\alpha^2 h^3 \Pi\Pi' \left| \frac{1}{\omega - \omega_g + i\Gamma_g/2} + \frac{1}{\omega_f - \omega + i\Gamma_f/2} \right|^2. \qquad (3.85)$$

At the point of resonance, the differential cross sections are equal to:

$$\frac{d\sigma_{int}^{--}}{du} = \frac{9\pi\Pi\Pi'}{8h^2 m^2}, \quad \frac{d\sigma_{int}^{++}}{du} = \frac{9\pi\Pi\Pi'}{8h^2 m^2}\left(\frac{2l-1}{2l+1}\right)^2, \qquad (3.86)$$

$$\frac{d\sigma_{int}^{+-}}{du} = \frac{d\sigma_{int}^{-+}}{du} = \frac{9\pi\Pi\Pi'}{16hm^2}\frac{1}{(2l+1)^2}. \qquad (3.87)$$

Due to the equality of the cross sections of the spin-flip processes in the ground and inverse spin states, the effect of electron self-polarization is absent under the considered conditions.

All cross sections are proportional to the factor ПП', which implies that in the interference of two resonances, the polarization of the initial photons does not affect the orientation of the electron spins, ie, the spin-polarization effects are absent.

When the initial photon propagates along and against the field, the factors ПП' are equal, respectively:

$$\Pi\Pi' = \begin{cases} (1+\xi_2)(1+\xi'_2), & e.v = u = +1 \\ (1-\xi_2)(1-\xi'_2), & e.v = u = -1 \end{cases}. \qquad (3.88)$$

This means that if the initial photon is directed along the field, then the cross section is maximum when this photon is right-handed circularly polarized. The probability of the CS process with the initial photon of the left-handed circular polarization is zero. And vice versa, if the initial photon is directed against the field, then the cross section is maximum if this photon has left-handed circular polarization and is equal to zero at right-handed circular polarization of the photon. The final photon is circularly polarized, and the direction of polarization coincides with the initial one.

<u>Electron polarizer.</u> The ultraquantum approximation in the case of small magnetic fields, of the laboratory order, passes into the dipole approximation, and the Landau level numbers take large values of $l \gg 1$. Therefore, the results obtained on the orientation of electron spins in the CS process in the ultraquantum approximation can be used to create an electron beam polarizer under laboratory conditions. Scheme of the polarizer shown in Fig.3.3.

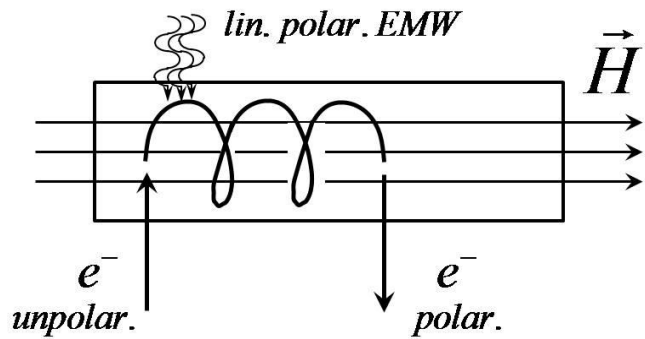

Fig.3.3. Scheme of electron beam polarization under action of an electromagnetic wave in a magnetic field

This setting polarizes the electron beam, and the degree of polarization changes smoothly, in proportion to change in the polarization of the electromagnetic wave (EMW) over the entire range. According to fig. 3.3 unpolarized electron beam is introduced into a region with a uniform magnetic field almost perpendicular to the field. A beam of polarized electromagnetic radiation, which irradiates a section of the path of electrons, is also perpendicular to the magnetic field. The design must be able to change the polarization of the EMW beam (rotation relative to its axis by an angle of $90^0$).

In the dipole approximation, that is, the approximation: $l>>1$, $hl<<1$, $n_g=l+1$, $n_f=l-1$, after integrating the differential cross sections (3.46), (3.47) and (3.69), (3.70) over the emission angle of the final photon at the resonance point with the width (3.84), the cross sections of the CS process have the form:

$$\sigma^{--} = \sigma^{++} = \sigma_0(1-\xi_3), \ \sigma^{+-} = \frac{h}{2l}\sigma_0(1-\xi_3), \ \sigma^{-+} = \frac{h}{2l}\sigma_0(1+\xi_3), \ \sigma_0 = \frac{3\pi}{2\omega^2}. \quad (3.89)$$

An initial photon flux is determined by power of the electromagnetic wave $W_\gamma$:

$$j = W_\gamma / \omega S, \quad (3.90)$$

where $S$ is EMW beam cross section. The polarization time of the electron beam is equal to inverse rate of the CS process:

$$\tau = 1/j\sigma^{+-} = \hbar\omega S / W_\gamma \sigma^{+-}. \quad (3.91)$$

The irradiation power $W_\gamma$ and the cross section $S$ must be large enough so that the polarization time is less than the flight time of the irradiation region.

As an example, we will give the following values of physical quantities in the polarizer: irradiation wavelength $\lambda=2.5mm$, it corresponds to the energy of the photon $\hbar\omega = 5 \cdot 10^{-4} \ eV$; to enter the cyclotron resonance, the magnetic field must be equal to $h=10^{-9}$ ($H=40kGs$); let the Landau level number be equal to $l=2 \cdot 10^6$ (the condition is fulfilled $lh<<1$), then the kinetic energy of the electron is equal to $\varepsilon=1keV$, the velocity of the electron is equal to $\upsilon=0.06 \cdot c$; the power of the electromagnetic wave generator is equal to $W_\gamma=10kW$; the cross section of the beam is equal to $S=0.04cm^2$. The cross section of the spin-flip process of CS according to expression (3.89) is equal to $\sigma^{+-}=3.8 \cdot 10^{-18}cm^2$. Then the polarization time of the electrons according to (3.91) $\tau=10^{-10}c$. During this time, the electron will fly along the direction of the magnetic field, the distance $\upsilon\tau=1.8mm$, which is comparable to the transverse dimensions of the EMW beam. Note that the effect of Sokolov-Ternov self-polarization [6] is very small for given parameters. Time of self-polarization $3/\alpha m h^3 \approx 5 \cdot 10^8 c$.

## 3.3. Spin polarization effects in the process of two photons emission by an electron

Probability of two-photon synchrotron radiation (Double Synchrotron Radiation, DSR). The process of two-photon radiation in a strong magnetic field in the LLL approximation was considered in [286]. Feynman diagrams of the DSR process are shown in Fig.3.4. Since the CS and DSR processes are cross-channel, the amplitude of the DSR process can be obtained from expression (3.9) by substituting

$$k(\omega,\vec{k}) \to -k_1(\omega_1,\vec{k}_1), \quad k'(\omega',\vec{k}') \to k_2(\omega_2,\vec{k}_2). \tag{3.92}$$

The resonant conditions of the DSR process in the LLL approximation with respect to replacement (3.92) coincide with the resonant conditions for the CS process. Thus, the resonance in the $g$ diagram is realized if the frequency of the first photon $\omega_{1g}$ is equal to expression (3.33) with the opposite sign, the frequency of the second photon $\omega_{2g}$ is determined by expression (3.34). The resonance in the $f$ diagram is realized if the frequency of the first photon $\omega_{1f}$ is equal to expression (3.39) with the opposite sign, and the frequency of the second photon $\omega_{2g}$ is equal to expression (3.38).

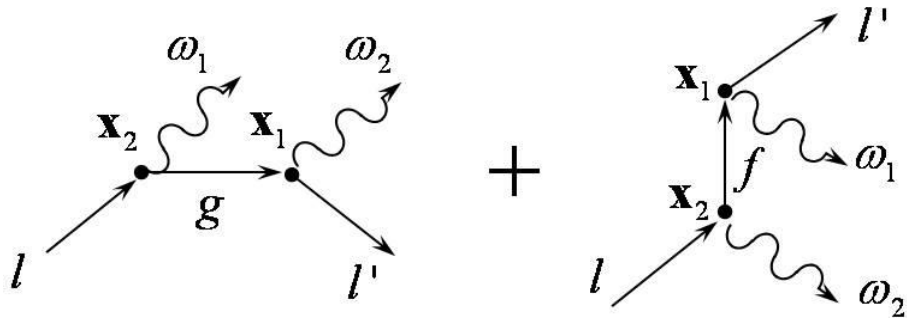

Fig.3.4. Feynman diagrams of the DSR process

Equating the frequencies $\omega_{1g} = \omega_{2g}$ and similarly $\omega_{1f} = \omega_{2f}$ leads to the conditions

$$n_g = n_f = l - 1 = l' + 1, \quad vu = 1. \tag{3.93}$$

This is fulfilled under conditions (3.45) and means the interference of resonances in both diagrams, while the photon frequencies are equal to the cyclotron frequency. Outside the interference conditions (3.93), the resonances in the $g$ and $f$ diagrams occur separately. On the one hand, photons $\omega_1$ and $\omega_2$ do not differ experimentally. It is impossible to specify which of the two experimental frequencies is $\omega_1$. On the other hand, the probability of the DSR process in the $f$ diagram can be obtained from the probability in the $g$ diagram by replacing $\omega_1 \leftrightarrow \omega_2$. This leads to the fact that the resulting probability of the process in one of the resonances must be doubled.

The differential probability of the DSR process $dW_D$ can be obtained from the differential cross section of the CS process $d\sigma_{CS}$, if we equate the amplitudes and take into account the number of final states in these processes, as a result

$$\frac{dW_D}{\omega_1^2 d\omega_1 dv} = 2 \cdot \frac{d\sigma_{CS}}{4\pi^2}. \tag{3.94}$$

The sum of the two differential probabilities, which correspond to the individual resonances in the $g$ and $f$ diagrams, for the DSR process without electron spin flip ($\mu'=\mu$) is equal to:

$$\frac{dW_D^{\mu\mu}}{d\omega_1 dv du} = \frac{1}{\pi} \frac{\dfrac{dW_{l\to n_g}^{\mu\mu}}{dv} \cdot \dfrac{dW_{n_g\to l'}^{\mu\mu}}{du}}{[(\omega_1 - \omega_{1g})^2 + \Gamma_g^2/4]} + \frac{1}{\pi} \frac{\dfrac{dW_{l\to n_f}^{\mu\mu}}{du} \cdot \dfrac{dW_{n_f\to l'}^{\mu\mu}}{dv}}{[(\omega_1 - \omega_{1f})^2 + \Gamma_f^2/4]}, \tag{3.95}$$

where the resonance widths $\Gamma_g$, $\Gamma_f$ are defined as the sum of the two total probabilities of the SR process of the initial and intermediate electrons, respectively, and have the form:

$$\Gamma_g = \frac{4}{3}\alpha h^2 m(l + n_g - 1), \quad \Gamma_f = \frac{4}{3}\alpha h^2 m(l + n_f - 1), \tag{3.96}$$

which become the same at $n_g = n_f$. The differential probabilities of SR of the first and second photons written in the numerator of expression (3.96) are defined by expressions (2.46), (2.47) with corresponding renotation. The integration of expression (3.95) over the frequency of the first photon and the polar angles of both photons, as well as the averaging over the initial spin states and the summation over the final polarizations of the particles gives the full probability of the process under resonant conditions. The most probable case in the DSR process is the case of an electron transition to neighboring Landau levels $l \to l-1 \to l-2$, for which the total probability has the form:

$$W_D = \frac{2 W_{l \to l-1} W_{l-1 \to l-2}}{\Gamma} = \alpha h^2 m \frac{2(2l-1)(2l-3)}{6(l-1)}. \tag{3.97}$$

For the lowest possible level of the initial electron $l = 2$, the probability is equal to $W_D = \alpha h^2 m$. The process of emission of one photon by an electron from the level $l = 2$ is equal to $W_{\text{opp}} = 2\alpha h^2 m$, ie in resonance the process of the second order becomes comparable to the process of the first order.

<u>Polarization of radiation in the DSR process.</u> Note that in the differential probability DSR (3.95) the dependence on polarization parameters of the photons is the same as in the SR process. Thus, in the process without an electron spin flip, the emission of two photons has the same polarization as the emission of one photon.

Consider the spin-flip process of DSR in resonance in the g diagram, when an electron passes into the ground spin state. According to (3.47), the differential probability of such a process should be proportional to the value:

$$\frac{dW_D^{+-}}{d\omega_1 dv du} \propto \frac{dW_{l \to n_g}^{++}}{dv} \cdot \frac{dW_{n_g \to l'}^{+-}}{du}.$$

However, there is a second channel for the implementation of such a process, the probability of which has the same degree of h:

$$\frac{dW_D^{+-}}{d\omega_1 dv du} \propto \frac{dW_{l \to n_g}^{+-}}{dv} \cdot \frac{dW_{n_g \to l'}^{--}}{du}.$$

Therefore, you need to take into account both channels and add not the probabilities, but the amplitudes of the probabilities. As a result, taking into account (2.46) - (2.48) and (2.43), the differential probability finally looks like:

$$\frac{dW_D^{+-}}{d\omega_1 dv du} = \frac{1}{\pi} \frac{A_g D_g}{(\omega_1 - \omega_{1g})^2 + \Gamma_g^2 / 4}, \tag{3.98}$$

$$A_g = \frac{1}{8l'} \alpha^2 h^3 \omega_1 \omega_2 R_{1g}^2 R_{2g}^2, \tag{3.99}$$

$$D_g = \frac{(1+v^2)(1+u^2)}{4}\left[a_g^2\tilde{\Pi}_1\Pi_2 + b_g^2\Pi_1\tilde{\Pi}_2\right] + a_g b_g H, \qquad (3.100)$$

$$H = 2vu + u(1+v^2)\xi_2 + v(1+u^2)\xi'_2 + (1-v^2)(1-u^2)\xi_1\xi'_1 + (1+v^2)(1+u^2)\xi_2\xi'_2, \qquad (3.101)$$

$$a_g = \sqrt{l - n_g}, \quad b_g = \sqrt{n_g - l'}, \qquad (3.102)$$

where the factors $R_{1g}^2$, $R_{2g}^2$ are determined by expression (2.50), while the first factor refers to the first photon and in (2.50) you need to replace $l' \to n_g$, and the second factor refers to the second photon and in (2.50) you need to replace $l \to n_g$; $\Pi_1, \tilde{\Pi}_1$ are polarization functions, which are determined by expression (3.52), and $\Pi_2, \tilde{\Pi}_2$ are determined by expression (3.64); $\xi_i$ are the Stokes parameters, they refer to the first photon, and the hatched parameters $\xi'_i$ refer to the second photon.

Note that the found probability (3.98) depends on the Stokes parameters $\xi_1$, $\xi'_1$. For a single photon process, the SR process, the presence of the parameter $\xi_1$ in the probability would mean that the probabilities for the emitted photon to have a plane of polarization oriented at angles $+45^0$ and $-45^0$ relative to the plane ($\vec{k}, \vec{H}$) are the same, which is impossible due to the symmetry of these situations. If two photons directed in different directions are measured at the same time, there is no symmetry with respect to these angles, which explains the appearance of Stokes parameters $\xi_1$, $\xi'_1$.

Let's analyze the polarization of one of the photons (let for certainty $\omega_1$). If you do not fix the second photon, you need to integrate the probability (3.98) on its polar angle $u$ and sum up the polarizations. As a result, the factor $D_g$ has the form (the most probable case $l'=n_g-1$ is taken):

$$D_g = \frac{4}{3}(1+v^2)(l-n_g+1)\left[1 + \frac{l-n_g-1}{l-n_g+1}\cdot\frac{1-v^2}{1+v^2}\xi_3 + \frac{2v}{1+v^2}\xi_2\right], \qquad (3.103)$$

whence, according to (2.3), the form of the Stokes parameters of the emitted photon follows:

$$\xi_{1\,DSR} = 0, \quad \xi_{2\,DSR} = \frac{2v}{1+v^2}, \quad \xi_{3\,DSR} = \frac{l-n_g-1}{l-n_g+1}\cdot\frac{1-v^2}{1+v^2}. \qquad (3.104)$$

From expressions (3.104) it follows that the Stokes parameter $\xi_{2DSR}$, responsible for the circular polarization of the emitted photon, has the same form as in the process of emission of one photon. The degree of linear polarization of the emitted photon in the spin-flip process depends on the value of $l-n_g$, ie on the multiplicity of the cyclotron frequency. For the most probable case of the transition of an electron to the adjacent Landau level $l-n_g=1$, the final photon has no linear polarization, in particular, the photon emitted perpendicular to the field is completely unpolarized.

The spin-flip process of DSR in resonance in the f diagram, when the electron goes into the ground spin state, is described by the differential probability (3.98), where the index $g$ must be replaced by $f$ with the corresponding replacement of Landau level numbers, the factors $a_f$, $b_f$ are equal to:

$$a_f = \sqrt{n_f - l'}, \quad b_f = \sqrt{l - n_f}. \qquad (3.105)$$

If you fix the parameters of only one of the photons, the value of $D_f$ has the form similar to (3.103):

$$D_f = \frac{4}{3}(1+v^2)(n_f - l'+1)\left[1 + \frac{n_f - l'-1}{n_f - l'+1} \cdot \frac{1-v^2}{1+v^2}\xi_3 + \frac{2v}{1+v^2}\xi_2\right], \qquad (3.106)$$

whence it follows that the polarization of the final photon is the same as for the process in resonance in the diagram g. The linear polarization of the photon in the spin-flip process is absent during the transition of the electron to the neighboring Landau level. The probability of the spin-flip process of DSR with the transition of the electron to the inverse spin state $W_D^{-+}$ is extremely small.

### 3.4. Conclusions to the Chapter 3

The process of photon scattering by an electron (the CS process) is considered under resonant conditions in the LLL approximation in the nonrelativistic case, taking into account the polarization of all particles. In the CS process, the influence of the polarization of the initial photons on the degree of orientation of the electron spins and the polarization of the radiation was studied. In the process of resonant double synchrotron radiation ( the DSR process) the influence of the spin-flip process on the polarization of radiation was studied. As a result, it was shown:

1. Resonant conditions in the CS process occur if the photon frequencies are multiples of the cyclotron frequency, the polarization of the photons does not affect these conditions. Interference of two resonances in two Feynman diagrams occurs if both photons have the same frequency and are directed along the magnetic field.

2. The differential cross section of the CS process in resonance outside the interference region is factorized and represented as a Breit-Wigner formula. The electron in the intermediate state has a certain value of the spin direction.

3. For the processes with fixed values of the spins projections of the initial and final electrons, the polarization of the final photon does not depend on the polarization of the initial one. In the process without spin flip (with cross section $\sigma^{--}$, $\sigma^{++}$) and spin-flip process in the inverse spin state (with cross section $\sigma^{-+}$) the polarization of radiation is the same as in the SR process, and in spin-flip process in the ground spin state (with cross section $\sigma^{+-}$) radiation has anomalous linear polarization.

4. For the spin-flip process in the ground and inverse states, the degree of linear polarization of the radiation is proportional to the degree of linear polarization of the initial photon and opposite in sign. The lowest degree of general polarization of the radiation corresponds to the inverse Compton scattering in the direction perpendicular to the field.

5. Electrons with spins initially oriented against the field (in the ground state) emit fully polarized photons with polarization as in the SR process. If the initial photons have a normal linear polarization $\xi_3=-1$, then the violation of the degree of electron polarization is proportional to h. If the initial photons are abnormally linearly polarized $\xi_3=+1$, the spins of the electrons are oriented along the field.

6. Electrons with spins oriented along the field (in the inverse state) emit partially polarized photons, the degree of depolarization of the radiation is proportional to h and depends on the Landau level of the final electron. The degree of depolarization of the electron beam is also proportional to h, depending on the values of the Landau levels of the electron in the initial, intermediate and final states.

7. The degree of self-polarization of the initially unpolarized electron beam depends on the Landau level numbers of the initial and final electrons, for the Thomson process ($\omega=\omega'$) it is equal to the degree of linear polarization of the initial photon with the opposite sign: $P^-=-\xi_3$. The proportionality of a change in the degree of polarization of electrons to a change in the degree of polarization of photons is a significant difference between this effect in the CS process and a similar effect in the OPP process. In the region of resonance interference in both Feynman diagrams, the polarization of the initial photon does not affect the orientation of the electron spins, and the self-polarization effect is absent.

8. A scheme of an electron beam polarizer is proposed, where the directions of the electron spins change during CS process in a magnetic field in proportion to the change in the polarization of the electromagnetic wave.

8. A linearly polarized electromagnetic wave of the millimeter range ($\lambda=2.5mm$) with a power of 10*kW* in a magnetic field of 40*kGs* completely polarizes the electron beam in a time $\tau=10^{-10}c$ in a 2*mm* area.

9. The probability of the DSR process without an electron spin flip is factored in resonance conditions. Each of the two emitted photons has the same polarization as the photon in the SR process.

10. The probability of a spin-flip of the DSW process with the transition of an electron to the ground spin state is not factorized under resonance conditions. The degree of linear polarization of radiation in such a process depends on the parameter, which is a multiple of the cyclotron frequency. For the most probable case of the transition of an electron to the neighboring Landau level in the spin-flip process, the final photons do not have linear polarization, in particular, the radiation perpendicular to the field is completely unpolarized.

      The main scientific results of this chapter are published in [125], [126], [284-286].

# CHAPTER 4
# e⁺e⁻ PAIR PRODUCTION BY TWO POLARIZED PHOTONS IN RESONANCE CONDITIONS

## 4.1. Introduction

In this chapter, two-photon e⁺e⁻ pair production (TPP) is studied, taking into account spin particles and photon polarizations in the resonance conditions, when an intermediate particle comes to the mass surface. Influence of polarization of the initial photons on the degree of polarization of the final particles is analyzed. Influence of the field of cyclotron photons on e⁺e⁻ pair creation in the conditions of magnetosphere of the X-ray pulsars is analyzed and the polarization of electrons and positrons in the generation of pair plasma radiation from pulsar magnetosphere is taken into account.

## 4.2. Cross section of the TPP process

Amplitude of two-photon e⁺e⁻ pair production. The expression for the process amplitude corresponds to the Feynman diagrams shown in Fig.4.1,

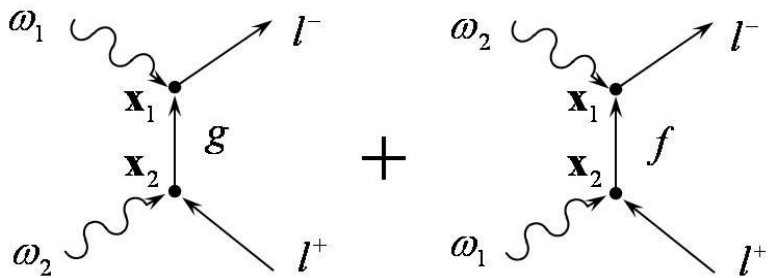

Fig.4.1. Feynman diagrams of two-photon e⁺e⁻ pair production.

and has the form

$$A_{if} = ie^2 \int d^4\mathbf{x}_1 d^4\mathbf{x}_2 \bar{\Psi}^-(\mathbf{x}_1)[\gamma^i A_{1i}(\mathbf{x}_1) G_{H1}(\mathbf{x}_1,\mathbf{x}_2)\gamma^j A_{2j}(\mathbf{x}_2) +$$
$$+ \gamma^j A_{2j}(\mathbf{x}_1) G_{H2}(\mathbf{x}_1,\mathbf{x}_2)\gamma^i A_{1i}(\mathbf{x}_2)]\Psi^+(\mathbf{x}_2), \qquad (4.1)$$

where $\bar{\Psi}^-(\mathbf{x}_1)$, $\Psi^+(\mathbf{x}_2)$ are the wave functions of electron and positron in the final states; $A_{1i}(\mathbf{x}), A_{2j}(\mathbf{x})$ are the wave functions of photons in the initial states; $G_{H1}(\mathbf{x}_1,\mathbf{x}_2)$, $G_{H2}(\mathbf{x}_1,\mathbf{x}_2)$ are the electron propagator in a magnetic field, according to the diagrams $g$ and $f$ (see Fig.4.1). The TPP process is a cross-channel to the CS process, which differs from the latter by replacing the final positron with the initial electron and one of the initial photons by the final one.

The laws of conservation of energy and the longitudinal component of momentum are similar to (3.5)

$$\omega_1 + \omega_2 = \varepsilon^- + \varepsilon^+, \quad \omega_1 v + \omega_2 u = p^- + p^+, \qquad (4.2)$$

where $v = \cos\theta_1$, $u = \cos\theta_2$ are the cosines of polar angles of the first and second photons, respectively. If we denote the total energy and total longitudinal momentum of photons as

$$\omega = \omega_1 + \omega_2, \quad k = \omega_1 v + \omega_2 u, \qquad (4.3)$$

then the conservation laws (4.2) coincide with the laws (2.120) of the OPP process.

The expressions for threshold values of the total frequency of photons, energies and longitudinal momentum of electron and positron have the form (2.123), where we need to replace

$$u \to U = \frac{\omega_1 v + \omega_2 u}{\omega_1 + \omega_2}. \qquad (4.4)$$

The threshold values of energies and momentum of particles with fixed Landau levels can be reduced to the form:

$$\varepsilon^\pm = \frac{m^\pm \omega}{m^- + m^+}, \quad p^\pm = \frac{m^\pm k}{m^- + m^+}, \qquad (4.5)$$

where $m^\pm = m(1 + 2l^\pm h)^{1/2}$, hence the threshold condition looks like

$$(\omega_1^{th} + \omega_2^{th})^2 - (k_1^{th} + k_2^{th})^2 = (m^+ + m^-)^2. \qquad (4.6)$$

As can be seen, the condition (4.6) does not hold, if both photons move parallel to the field in the same direction.

For the photon frequency above the threshold value $\omega > \omega_m$, taking into account (4.3) and (4.4), the expressions for energies and momentum of electron have the form (2.130). Without loss of generality, we can eliminate the longitudinal momentum of photons by appropriately choosing the reference frame so that

$$k = k_1 + k_2 = 0. \tag{4.7}$$

In this case, the energies and momentum of particles are determined by expressions (2.132), (2.133). Also for an analysis of the process in resonant conditions, it is convenient to choose the reference frame in which the hard photon ($\omega > m^+ + m^-$) is directed perpendicular to a magnetic field.

After integration of the expression (4.1), in the general case, the amplitude of the TPP process can be represented as

$$A_{if} = (2\pi)^4 \frac{M_{if}}{SV} \delta^3(k_1 + k_2 - p^- - p^+), \quad M_{if} = M_{if}^{(g)} + M_{if}^{(f)}, \tag{4.8}$$

where the two terms $M_{if}^{(g)}$ and $M_{if}^{(f)}$ correspond to $g$ and $f$ Feynman diagrams, respectively (see Fig. 4.1). The term $M_{if}^{(g)}$ has the form

$$M_{if}^{(g)} = \frac{-ie^2}{4\sqrt{\omega_1 \omega_2 \varepsilon^- \varepsilon^+ m^- m^+}} \sum_{n_g=0}^{\infty} \frac{e^{-i\Phi_g} Q^{(g)}}{g_0^2 - \varepsilon_g^2}, \quad Q^{(g)} = \sum_{a=1}^{10} Q_a. \tag{4.9}$$

The zero and longitudinal to the field components of the 4-momentum of $g$ and $f$ intermediate particle are equal to

$$g_0 = \varepsilon^- - \omega_1, \quad g = p^- - \omega_1 v, \quad f_0 = \varepsilon^- - \omega_2, \quad f = p^- - \omega_2 u. \tag{4.10}$$

The phase $\Phi_g$ is determined

$$\Phi_g = \frac{k_{1x}(2p_y^- - k_{1y})}{2hm^2} + \frac{k_{2x}(2g_y - k_{2y})}{2hm^2} + (l^- - n_g)\varphi_1 + (n_g - l^+)\varphi_2 + \frac{\pi}{2}(l^+ - l^-), \tag{4.11}$$

where $k_{1x}$, $k_{1y}$, $k_{2x}$, $k_{2y}$ are the transverse components of photon momentum, $\varphi_1$, $\varphi_2$ are azimuthal angles, $l^+$, $l^-$, $n_g$ are the Landau levels of the final positron, final electron, intermediate particle. The phase $\Phi_f$ in the amplitude of the Feynman diagram $f$ differs from expression (4.11) by a sign and after re-designation of parameters is equal to

$$\Phi_f = -\frac{k_{1x}(2p_y^+ - k_{1y})}{2hm^2} - \frac{k_{2x}(2p_y^- - k_{2y})}{2hm^2} + (n_f - l^+)\varphi_1 + (n_f - l^-)\varphi_2 + \frac{\pi}{2}(l^- - l^+). \quad (4.12)$$

The terms $Q_a$ in the amplitude $M_{if}^{(g)}$ have the form

$$Q_1 = J_1^{++}J_2^{++}[-m\tilde{C} - g_0 C - gD], \quad Q_2 = J_1^{++}J_2^{-+}[-mD + g_0\tilde{D} + gC],$$

$$Q_3 = J_1^{-+}J_2^{+-}[-mC - g_0\tilde{C} + gD], \quad Q_4 = J_1^{-+}J_2^{--}[mD - g_0\tilde{D} + gC],$$

$$Q_5 = J_1^{+-}J_2^{++}[mD + g_0\tilde{D} - gC], \quad Q_6 = J_1^{+-}J_2^{-+}[-mC + g_0\tilde{C} - gD],$$

$$Q_7 = J_1^{--}J_2^{+-}[-mD - g_0\tilde{D} - gC], \quad Q_8 = J_1^{--}J_2^{--}[-mC + g_0\tilde{C} + gD],$$

$$Q_9 = [-J_1^{-+}J_2^{++} - J_1^{++}J_2^{+-} + J_1^{--}J_2^{-+} + J_1^{+-}J_2^{--}]D\sqrt{2n_g hm},$$

$$Q_{10} = [J_1^{-+}J_2^{-+} - J_1^{++}J_2^{--} + J_1^{--}J_2^{++} - J_1^{+-}J_2^{+-}]C\sqrt{2n_g hm}, \quad (4.13)$$

where the functions $J_1$ have the form similar to (3.16)

$$J_1^{++} = J_1(n_g, l^-)M_m^- e_{1z}, \quad J_1^{--} = J_1(n_g - 1, l^- - 1)\mu^- M_p^- e_{1z},$$

$$J_1^{+-} = J_1(n_g, l - 1)\mu^- M_p^- H_{1m}, \quad J_1^{-+} = J_1(n_g - 1, l^-)M_m^- H_{1p}, \quad (4.14)$$

and similarly the expressions for $J_2$ can be written

$$J_2^{++} = J_2(l^+, n_g)M_p^+ e_{2z}, \quad J_2^{--} = J_2(l^+ - 1, n_g - 1)\mu^+ M_m^+ e_{2z},$$

$$J_2^{+-} = J_2(l^+, n_g - 1)M_p^+ H_{2m}, \quad J_2^{-+} = J_2(l^+ - 1, n_g)\mu^+ M_m^+ H_{2p}, \quad (4.15)$$

where $\mu^+$, $\mu^-$ are the signs of spin projections of positron and electron. The indices 1, 2 correspond to the first and second photons. The amplitude $M_{if}^{(f)}$ is obtained by replacing indices 1↔2 in the expression $M_{if}^{(g)}$, that corresponds to a replacement of the initial photons. Taking into account (4.10), the energies of intermediate states $\varepsilon_g$, $\varepsilon_f$ for the fixed Landau levels $n_g$, $n_f$ are equal to

$$\varepsilon_g = \sqrt{m^2 + 2n_g hm^2 + (p^- - \omega_1 v)^2}, \quad \varepsilon_f = \sqrt{m^2 + 2n_f hm^2 + (p^+ - \omega_1 v)^2}. \quad (4.16)$$

Conditions of resonances in the TPP process. For the occurrence of a resonance in the diagram $g$ (see. Fig.4.1) it is necessary to satisfy the conditions

$$g_0 = \varepsilon_g \text{ that is } \varepsilon^- - \omega_1 = \sqrt{m^2 + 2n_g h m^2 + (p^- - \omega_1 v)^2}. \quad (4.17)$$

First, consider the case of propagation of the initial photons perpendicular to a magnetic field: $v = 0$, $u = 0$. In this case, based on expressions (2.130) and taking into account (4.3), the energy and momentum of electron are equal to

$$\varepsilon^- = a^- / 2\omega, \quad p^- = b^- / 2\omega, \quad (4.18)$$

where $a^- = \omega^2 + (m^-)^2 - (m^+)^2$, $b^- = \sqrt{(a^-)^2 - 4(m^-)^2 \omega^2}$. Expanding in a series by the small parameter $h$ of expressions $a^-$, $b^-$ and leaving only the zero term (essentially, assuming $h = 0$) we obtain

$$a^- = \omega^2, \quad b^- = \omega\sqrt{\omega^2 - 4m^2}, \quad (4.19)$$

accordingly, the equation (4.17) can be rewritten as

$$\omega_2 - \omega_1 = \omega_2 + \omega_1. \quad (4.20)$$

This equality occurs if in the zero approximation on the field $h$ the frequency is $\omega_1 = 0$, taking into account (4.19) we can show that $\omega_2 \geq 2$. Taking into account the first degree by $h$, we have $\omega_1 \propto h$. Thus, in the resonance (4.17), two initial photons play a fundamentally different role. The hard photon $\omega_2$ create $e^+e^-$ pair and the intermediate state is electronic. The soft photon $\omega_1$ is absorbed by this intermediate electron, which can be represented by the Feynman diagram in Fig.4.2. Note that for the realization of resonance in addition to conditions (4.17) the following equality may also occur: $g_0 = -\varepsilon_g$. This condition coincides with (4.17) when we mutually replace $\omega_1$ by $\omega_2$.

Choose the frame reference in which the hard photon $\omega_2$ is directed perpendicular to the field, while the soft photon $\omega_1$ is directed arbitrarily

$$u = 0, \quad \forall v. \quad (4.21)$$

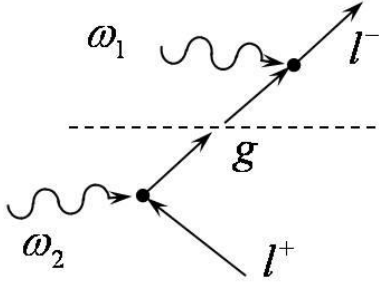

Fig.4.2. Feynman diagram $g$ of the resonant two-photon $e^+e^-$ pair production.

The frequency of hard photon $\omega_2$ is chosen arbitrary, but it is sufficient for the creation of $e^+e^-$ pair at the levels $l^+$, $l^-$. The frequency of soft photon is written in the form: $\omega_1 = \beta_1 h$. Expanding equation (4.17) in a series by the parameter $h$ with accuracy of the first degree, we have, the frequency of soft photon in resonance within a given accuracy:

$$\omega_{1g} = \frac{2(l^- - n_g)hm^2}{\omega_2 - vR}, \quad R = \sqrt{\omega_2^2 - 4m^2}. \qquad (4.22)$$

This frequency corresponds to the energies and momentum of the final particles, which have the form:

$$\varepsilon^- = \frac{\omega_2}{2} + \frac{hm^2}{\omega_2}[(l^- - n_g)\frac{\omega_2 + vR}{\omega_2 - vR} + l^- - l^+], \qquad (4.23)$$

$$p^- = \frac{R}{2} + \frac{hm^2}{R}[(l^- - n_g)\frac{\omega_2 + vR}{\omega_2 - vR} - l^- - l^+], \qquad (4.24)$$

$$\varepsilon^+ = \frac{\omega_2}{2} + \frac{hm^2}{\omega_2}(l^+ - n_g), \quad p^+ = -\frac{R}{2} + \frac{hm^2}{R}(l^+ + n_g). \qquad (4.25)$$

Note that according to the expression (4.22), the frequency of soft photon $\omega_1$ is proportional to the difference between the Landau level numbers of the intermediate and finale electrons $l^- - n_g$ and it does not depend on the Landau level number of positron.

It should be emphasized that the condition (4.22) determines the resonance only with accuracy of the first degree by $h$, and a detuning from resonance has the following order $\Delta \sim h^2 m$. The resonance width is determined mainly by the radiation width (total probability of synchrotron radiation), which is equal in magnitude $\Gamma \sim \alpha h^2 m$, where $\alpha$ is the fine-structure constant. A ratio of the TPP probability in the case (4.22) to the probability at the maximum point as a result is equal to $(\Gamma/\Delta)^2 \sim \alpha^2 \sim 10^{-4}$. On the other hand, the ratio of the TPP probability in the case (4.22) to the nonresonant probability has the order $(m/\Delta)^2 \sim h^{-4}$.

Expanding the equation (4.17) by parameter $h$ up to the second degree, the frequency of soft photon has the form

$$\omega_{1g} = \frac{2(l^- - n_g)hm^2}{\omega_2 - vR} + \beta_g h^2 m, \qquad (4.26)$$

where $\beta_g = \dfrac{4(l^- - n_g)m^3[2v\omega_2^2 l^+ + R\omega_2(l^- - l^+ - v^2(l^- + l^+)) + 4vm^2(n_g - l^+)]}{3v\omega_2^2(\omega_2^2 - 4m^2)(\omega_2 - vR) - 8m^2\omega_2 v^3(\omega_2^2 - 2m^2) + \omega_2^4(v^3\omega_2 - R)}$.

The special case of resonance is the case near a threshold. Then, the small parameter is $\sqrt{h}$. Let us represent the frequency of hard photon in the form

$$\omega_2 = 2m + (l^+ + n_g + \alpha_g)hm. \qquad (4.27)$$

In this case, the hard photon creates a positron and an intermediate electron at levels $l^+$ and $n_g$, respectively. The small additive $\alpha_g hm$ is transmitted to the longitudinal momentum of particles. The resonance condition (4.17) determines the frequency of soft photon that with accuracy of the second degree by $h$ has the form

$$\omega_{1g} = (l^- - n_g)hm\left\{1 + v\sqrt{\alpha_g h} + \frac{h}{2}\left[v^2(l^- - n_g) - l^- - n_g + \alpha_g(2v^2 - 1)\right]\right\}. \qquad (4.28)$$

The selected frequencies (4.27), (4.28) set the values of energies and momentum of the final particles in the form

$$\varepsilon^- = m + (l^- + \frac{\alpha_g}{2})hm + (l^- - n_g)v\alpha_g^{1/2}h^{3/2}m +$$

$$+ \frac{h^2 m}{4}\left[2v^2(l^- - n_g)^2 + l^{+2} + n_g^2 - 2l^{-2} + \alpha_g(l^+ + n_g - 2l^-) + 4\alpha_g v^2(l^- - n_g)\right], \qquad (4.29)$$

$$p^- = \alpha_g^{1/2}h^{1/2} + (l^- - n_g)vhm +$$

$$+ \frac{h^{3/2}m}{8}\left[8\alpha_g^{1/2}v^2(l^- - n_g) + 2\alpha_g^{1/2}(l^+ + n_g) + 2\alpha_g^{-1/2}(l^{+2} + n_g^2) + \alpha_g^{3/2}\right], \qquad (4.30)$$

$$\varepsilon^+ = m + (l^+ + \frac{\alpha_g}{2})hm - \frac{h^2 m}{2}\left[\alpha_g(l^+ - n_g) + l^{+2} - n_g^2\right], \qquad (4.31)$$

$$p^+ = -\alpha_g^{1/2}h^{1/2}m - \frac{h^{3/2}m}{8}\left[2\alpha_g^{1/2}(l^+ + n_g) + 2\alpha_g^{-1/2}(l^{+2} + n_g^2) + \alpha_g^{3/2}\right]. \qquad (4.32)$$

In conclusion of the question of resonant conditions in the $g$ diagram (see Fig.4.2), we illustrate the exact dependences of the resonant frequency of soft photon $\omega_1$ on its polar angle $\theta_1$ ($v=\cos\theta_1$) and on the detuning from the threshold value of the frequency of hard photon $\delta\omega=\alpha_g hm$, which are shown in Fig.4.3.a and Fig.4.3.b, respectively. At the same time we have $h = 0.1$ and the numbers of energy levels of particles are equal to $l^-=1$, $l^+=0$. As can be seen from Fig.4.3, the resonant frequency of soft photon near the process threshold ($\delta\omega=0$) does not depend on the polar angle and with increasing the detuning $\delta\omega$ it has a maximum value if the photon is directed along the field $\theta_1=0$.

Next, we analyze the conditions of resonance in the $f$ diagram (see Fig.4.1). In this case, the following condition is met

$$f_0 = \varepsilon_f \quad \text{that is} \quad \varepsilon^+ - \omega_1 = \sqrt{m^2 + 2n_f hm^2 + (p^+ - \omega_1 v)^2}. \qquad (4.33)$$

$f$ diagram of resonant process is shown in Fig.4.4.

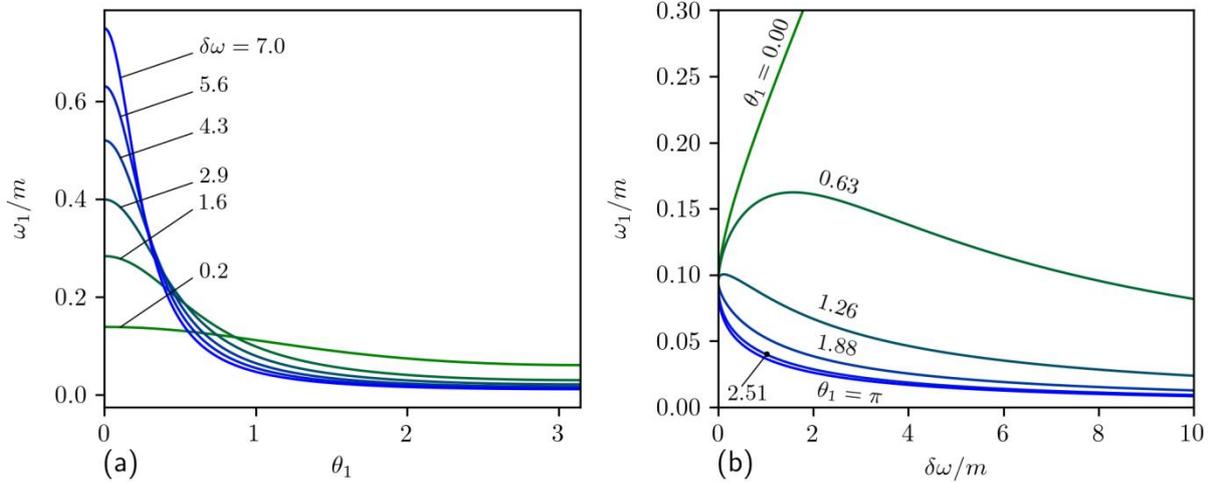

Fig.4.3. Dependence of the frequency of soft photon $\omega_1$ on its polar angle (*a.*) and on the detuning of frequency from the threshold value $\delta\omega$ (*b.*).

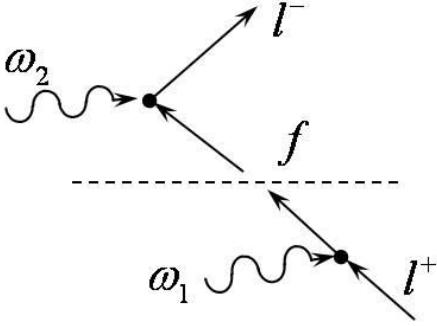

Fig.4.4. Feynman diagram *f* of the resonant two-photon e⁺e⁻ pair production.

In resonance, a hard photon $\omega_2$ create e⁺e⁻ pair with an electron in the final state, at the same time the intermediate state is a positron and a soft photon $\omega_1$ is absorbed by this intermediate positron.

Performing the expansion of equation (4.33) in a series by the parameter *h* with accuracy of the first degree, we have the expression for the frequency of soft photon in resonance

$$\omega_{1f} = \frac{2(l^+ - n_f)hm^2}{\omega_2 + vR}, \quad R = \sqrt{\omega_2^2 - 4m^2}, \qquad (4.34)$$

where $\omega_2$ is the frequency of hard photon that takes an arbitrary value greater than the threshold value. The energies and momentum of the final particles in resonance have the form

$$\varepsilon^- = \frac{\omega_2}{2} + \frac{hm^2}{\omega_2}(l^- - n_f), \quad p^- = \frac{R}{2} - \frac{hm^2}{R}(l^- + n_f), \qquad (4.35)$$

$$\varepsilon^+ = \frac{\omega_2}{2} + \frac{hm^2}{\omega_2}[(l^+ - n_f)\frac{\omega_2 - vR}{\omega_2 + vR} + l^+ - l^-], \qquad (4.36)$$

$$p^+ = -\frac{R}{2} + \frac{hm^2}{R}[(n_f - l^+)\frac{\omega_2 - vR}{\omega_2 + vR} + l^- + l^+]. \qquad (4.37)$$

As noted earlier, the resonance width $\Gamma \sim \alpha h^2 m$ requires that the resonance condition (4.17) be fulfilled to the second order of the parameter $h$ and the frequency of soft photon has the form similar to expression (4.26)

$$\omega_{1f} = \frac{2(l^+ - n_f)hm^2}{\omega_2 + vR} + \beta_f h^2 m, \qquad (4.38)$$

where $\beta_f = \dfrac{4(l^+ - n_f)m^3[2v\omega_2^2 l^- + R\omega_2(l^- - l^+ + v^2(l^- + l^+)) + 4vm^2(n_f - l^-)]}{3v\omega_2^2(\omega_2^2 - 4m^2)(\omega_2 + vR) - 8m^2\omega_2 v^3(\omega_2^2 - 2m^2) + \omega_2^4(v^3\omega_2 + R)}$.

Near the threshold, the frequency of hard photon is represented as

$$\omega_2 = 2m + (l^- + n_f + \alpha_f)hm, \qquad (4.39)$$

which determines the frequency of soft photon in resonance taking into account (4.17) in the form

$$\omega_{1f} = (l^+ - n_f)hm\left\{1 - v\sqrt{\alpha_f h} + \frac{h}{2}\left[v^2(l^+ - n_f) - l^+ - n_f + \alpha_f(2v^2 - 1)\right]\right\}. \qquad (4.40)$$

The frequencies (4.39), (4.40) correspond to the following energies and momentum of the final particles

$$\varepsilon^- = m + (l^- + \frac{\alpha_f}{2})hm - \frac{h^2}{4}\left[\alpha_f(l^- - n_f) + l^{-2} - n_f^2\right], \qquad (4.41)$$

$$p^- = \alpha_f^{1/2}h^{1/2}m + \frac{h^{3/2}m}{8}\left[2\alpha_f^{1/2}(l^- + n_f) + 2\alpha_f^{-1/2}(l^{-2} + n_f^2) + \alpha_f^{3/2}\right], \qquad (4.42)$$

$$\varepsilon^+ = m + (l^+ + \frac{\alpha_f}{2})hm - (l^+ - n_f)v\alpha_f^{1/2}h^{3/2}m +$$

$$+ \frac{h^2 m}{4}\left[2v^2(l^- - n_f)^2 + l^{-2} + n_f^2 - 2l^{+2} + \alpha_f(l^- + n_f - 2l^+) + 4\alpha_f v^2(l^+ - n_f)\right], \qquad (4.43)$$

$$p^+ = \alpha_f^{1/2}h^{1/2}m + (l^+ - n_f)vhm -$$

$$- \frac{h^{3/2}m}{8}\left[8\alpha_f^{1/2}v^2(l^+ - n_f) + 2\alpha_f^{1/2}(l^- + n_f) + 2\alpha_f^{-1/2}(l^{-2} + n_f^2) - \alpha_f^{3/2}\right]. \qquad (4.44)$$

Resonance interference in both diagrams $g$ and $f$ can occur if the conditions (4.17) and (4.33) are satisfied simultaneously. Far from the threshold, this corresponds to the equality of the frequencies of soft photon (4.26), (4.38), which sets the resonant value of the frequency of hard photon. Figure 4.5 shows the dependence of the resonant frequencies (4.26), (4.38) on $\omega_2$. At the same time selected $h=0.1$, $v=0.6$, $l^+=5$, $l^-=4$. Curves 1,2,3,6 correspond to expression (4.26) with the Landau level numbers $n_g=0$, 1, 2, 3, respectively, and curves 4, 5, 7, 8, 9 correspond to expression (4.38) with $n_f=0$, 1, 2, 3, 4, respectively. In Fig.4.5, the points A1-A8 of the intersection of the curves correspond to the interference of resonances.

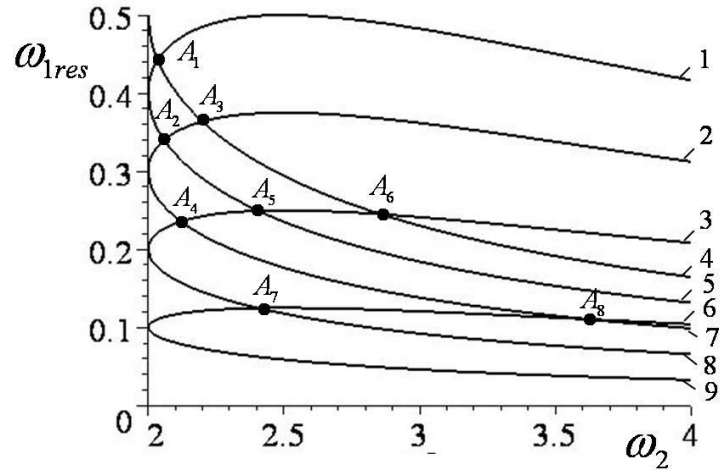

Fig.4.5. Dependence of the soft photon frequency in the resonant conditions on the hard photon frequency.

In the special case $v=0$, interference occurs when the Landau levels of final particles are the same and $n_g=n_f$:

$$v = 0, \quad l^+ = l^-, \quad n_g = n_f, \qquad (4.45)$$

in this regard, the frequency $\omega_2$ takes any value above the threshold. For $l^+ \neq l^-$, the resonances $g$ and $f$ can only approach each other to the following value

$$\omega_{1f} - \omega_{1g} = 2\beta h^2 m, \quad \beta = \beta_f = -\beta_g = \frac{4(l^- - n_g)(l^- - l^+)m^3}{\omega_2^3}, \qquad (4.46)$$

if the condition (4.47) is met

$$l^- - n_g = l^+ - n_f. \qquad (4.47)$$

To realize the interference of resonances of two diagrams near the threshold, it is necessary to fulfill the equality of expressions (4.27) and (4.39), as well as (4.28) and (4.40). There is no interference of diagrams for $v \neq 0$. Under condition (4.47) and taking into account an equality $\alpha_g = \alpha_f$, the difference of the resonant frequencies of soft photon is

$$\omega_{1f} - \omega_{1g} = -2(l^- - n_g)vh^{3/2}m. \qquad (4.48)$$

The interference still occurs if conditions (4.45) are met.

<u>Cross section of the TPP process in the LLL approximation.</u> A number of the final states in the TPP process is the same as in the OPP process (2.148), so in the general case, the process probability per unit time is determined similarly to expression (2.151). The delta function conforms to the law of conservation of energy and removes the last integral over the longitudinal component of momentum using rule (2.153), where $p_i^-$ are two values of longitudinal momentum of an electron, which are determined by expression (2.130) taking into account (4.3) and (4.4). As noted earlier, a role of two initial photons is different: one (hard) photon creates a pair of particles, and the other (soft) is absorbed by one of these particles. In doing so, in the general case, the frequencies of photons are arbitrary, i.e. there is a scattering of the flux of soft photons on the process of $e^+e^-$ pair production by the hard photons. The cross section of the TPP process is defined as the ratio of the process probability per unit time to the flux of the initial photons $j$ ($j=(1-\cos\chi)/V$, where $\chi$ is the angle between the directions of motion of the photons) and has the form

$$\sigma = \frac{W}{Tj} = \frac{2\pi h m^2}{(1-\cos\chi)} |M_{if}|^2 \sum_{i=1}^{2} \frac{\varepsilon_i^- \varepsilon_i^+}{\varepsilon_i^+ p_i^- - \varepsilon_i^- p_i^+}. \qquad (4.49)$$

Taking into account (4.9), the expression (4.49) in the LLL approximation is written in the form

$$\sigma = \frac{e^4 \pi h}{8m^2 p^- \omega_1 \omega_2 (1-\cos\chi)} \left| \sum_{n_g} \frac{e^{-i\Phi_g} Q^{(g)}}{g_0^2 - \varepsilon_g^2} + \sum_{n_f} \frac{e^{-i\Phi_f} Q^{(f)}}{f_0^2 - \varepsilon_f^2} \right|^2. \qquad (4.50)$$

In the resonance of the diagram $g$, in the sum of the Landau level numbers of the intermediate particle there is only one term, as a result we have

$$\sigma = \frac{e^4 \pi h |Q^{(g)}|^2}{32 m^4 p^- \omega_1 \omega_2 (1 - \cos \chi) \left[ (g_0 - \varepsilon_g)^2 + \Gamma_g^2 / 4 \right]}. \qquad (4.51)$$

The cross section in the resonance of the diagram $f$ with amplitude $Q^{(f)}$ looks similar.

Amplitudes $Q^{(g)}$ in the LLL approximation with fixed values of projections of spins of the final particles can be written as

$$Q^{-+} = 4\sqrt{2} \sqrt{\frac{l^- h}{n_g}} m^3 G_1 G_2 H_{1m} e_{2z}, \qquad (4.52)$$

$$Q^{--} = -4 \sqrt{\frac{l^- l^+}{n_g}} h m^3 G_1 G_2 H_{1m} H_{2p}, \qquad (4.53)$$

$$Q^{++} = \frac{4 h m^3}{\sqrt{n_g}} G_1 G_2 (A_g + B_g), \quad A_g = (n_g - l^-) \tilde{H}_{1m} e_{2z}, \quad B_g = n_g H_{1m} H_{2m}, \qquad (4.54)$$

$$Q^{+-} = 2\sqrt{2}(l^- - n_g) \sqrt{\frac{l^+}{n_g}} h^{3/2} m^3 G_1 G_2 \tilde{H}_{1m} H_{2p}, \qquad (4.55)$$

where $H_m$, $H_p$, $e_z$ are the quantities defined by expressions (2.43), (3.18), respectively, while the indices 1, 2 denote the first and second photons,

$$\tilde{H}_m = \cos \alpha - i \cos \theta \cdot \sin \alpha \cdot e^{i\beta}, \quad \tilde{H}_p = \cos \alpha + i \cos \theta \cdot \sin \alpha \cdot e^{i\beta}, \qquad (4.56)$$

$$G_1 = \frac{e^{-\eta_1/2} \eta_1^{\frac{l^- - n_g - 1}{2}}}{(l^- - n_g - 1)!} \sqrt{\frac{(l^- - 1)!}{(n_g - 1)!}}, \quad G_2 = \frac{(-1)^{l^+} e^{-\eta_2/2} \eta_2^{\frac{l^+ + n_g}{2}}}{\sqrt{l^+! n_g!}}, \qquad (4.57)$$

$$\eta_1 = \frac{\omega_1^2 (1 - v^2)}{2 h m^2}, \quad \eta_2 = \frac{\omega_2^2}{2 h m^2}. \qquad (4.58)$$

The signs + and − in the quantity $Q$ denote the sign of the projections of spins of the final particles, with the first sign corresponds to the spin of electron and the second to the spin of positron.

The amplitude (4.52) contains the smallest degree of the parameter $h$ and it corresponds to the most probable case of particles creation in the main spin states, when the electron spin is directed opposite to the field direction and the positron spin is co-directed to the field. Taking into account the amplitude (4.52), the cross section (4.51) has the form

$$\sigma^{-+} = \frac{\pi e^4 h^2 m^3 (l^-/n_g) G_1^2 G_2^2 |H_{1m}|^2 e_{2z}^2}{p^- \omega_1 \omega_2 (1-\cos\chi) \left[(g_0 - \varepsilon_g)^2 + \Gamma_g^2/4\right]}. \qquad (4.59)$$

In the cross section (4.59), the resonance width $\Gamma_g$ is equal to the total probability of SR of the final electron, in which the main contribution is made by the transition of electron to the neighboring Landau level

$$\Gamma_g = W_{CB}(l^- \to l^- - 1) = \frac{2}{3}(2l^- - 1)\alpha h^2 m, \qquad (4.60)$$

also similarly to (3.52), (3.61), we have

$$|H_{1m}|^2 = \frac{1}{2}(1+v^2)\Pi_1 = \frac{1}{2}(1+v^2)(1 - \frac{1-v^2}{1+v^2}\xi_3^{(1)} + \frac{2v}{1+v^2}\xi_2^{(1)}), \qquad (4.61)$$

$$e_{2z}^2 = \frac{1}{2}(1+\xi_3^{(2)}). \qquad (4.62)$$

The cross section (4.59) is obviously factorized, which means an independence of the process of $e^+e^-$ pair creation by hard photon and absorption of soft photon by electron in the resonant conditions. Taking into account the expressions for probabilities of SR $dW_{SR\,e^-}^{--}/dv$ (2.46) and the OPP process $W_{OPP}^{-+}$ (2.156), the expression (4.59) can be reduced to the Breit-Wigner form

$$\sigma^{-+} = \frac{2\pi}{\omega_1^2 (1-\cos\chi)} \frac{\frac{dW_{SR\,e^-}^{--}}{dv} \cdot W_{OPP}^{-+}}{(\omega_1 - \omega_{1g})^2 + \Gamma_g^2/4}. \qquad (4.63)$$

In the case of pair production at the lowest possible Landau levels ($l^+=0$, $l^-=1$) in the magnetic field $h=0.1$ by unpolarized photons directed towards each other perpendicular to the field and with frequencies of hard and soft photons equal to $\omega_2=2m+1,25hm$, $\omega_1=hm$, respectively, an estimation of the cross section of the TPP process gives

$$\sigma_{10}^{-+} = \frac{\pi e^{-2/h}}{h^3 p^- m} = \frac{2\pi e^{-2/h}}{h^4 m^2} \approx 2\cdot 10^{-25}\, cm^2. \qquad (4.64)$$

For the TPP process, in which the spins of the final particles are directed opposite to the field, the resonant cross section $\sigma^{--}$ has the form

$$\sigma^{--} = \frac{\pi e^4 h^3 m^3 (l^- l^+ / n_g) G_1^2 G_2^2 |H_{1m}|^2 |H_{2p}|^2}{2 p^- \omega_1 \omega_2 (1 - \cos\chi)\left[(\omega_1 - \omega_{1g})^2 + \Gamma_g / 4\right]}, \qquad (4.65)$$

where
$$|H_{2p}|^2 = \frac{1}{2}(1 + u^2)\Pi_2 = \frac{1}{2}(1 - \xi_3^{(2)}). \qquad (4.66)$$

Similar to expression (4.63) and taking into account the expressions for the probabilities of SR $dW_{SR}^{--}/dv$ (2.46) and OPP $W_{OPP}^{--}$ (2.157), the resonant cross section can be reduced to the Breit-Wigner form

$$\sigma^{--} = \frac{2\pi}{\omega_1^2 (1 - \cos\chi)} \frac{\dfrac{dW_{SR\, e^-}^{--}}{dv} \cdot W_{OPP}^{--}}{(\omega_1 - \omega_{1g})^2 + \Gamma_g^2 / 4}. \qquad (4.67)$$

The found cross section $\sigma^{--}$ is an order of magnitude $h$ smaller than the cross section $\sigma^{-+}$.

For the TPP process, in which the final particles are in the inverse spin states (the spin of electron is directed opposite to the field, and the spin of positron is so-directed to the field), the resonant cross section $\sigma^{+-}$ has the form

$$\sigma^{+-} = \frac{\pi e^4 h^4 m^3 (l^- - n_g)^2 (l^+ / n_g) G_1^2 G_2^2 |\tilde{H}_{1m}|^2 |H_{2p}|^2}{4 p^- \omega_1 \omega_2 (1 - \cos\chi)\left[(\omega_1 - \omega_{1g})^2 + \Gamma_g / 4\right]}, \qquad (4.68)$$

where
$$|\tilde{H}_{1m}|^2 = \frac{1}{2}(1 + v^2)\tilde{\Pi}_1 = \frac{1}{2}(1 + v^2)\left(1 + \frac{1 - v^2}{1 + v^2}\xi_3^{(1)} + \frac{2v}{1 + v^2}\xi_2^{(1)}\right), \qquad (4.69)$$

or in the Breit-Wigner form

$$\sigma^{+-} = \frac{2\pi}{\omega_1^2 (1 - \cos\chi)} \frac{\dfrac{dW_{SR\, e^-}^{+-}}{dv} \cdot W_{OPP}^{--}}{(\omega_1 - \omega_{1g})^2 + \Gamma_g^2 / 4}. \qquad (4.70)$$

The cross sections (4.63), (4.67), (4.70) can be written in a single way

$$\sigma^{\mu^-\mu^+} = \frac{2\pi}{\omega_1^2(1-\cos\chi)} \frac{\dfrac{dW_{SR\,e^-}^{\mu^-\mu_g}}{dv}\cdot W_{OPP}^{\mu_g\mu^+}}{(\omega_1-\omega_{1g})^2+\Gamma_g^2/4}, \tag{4.71}$$

where $\mu_g = -1$ is the spin of the intermediate electron directed opposite to the field.

Finally, for the TPP process, in which the spins of the final particles are co-directed to the field, the resonant cross section $\sigma^{++}$ has the form

$$\sigma^{++} = \frac{\pi e^4 h^3 m^3 (1/n_g) G_1^2 G_2^2 |A_g + B_g|^2}{2p^-\omega_1\omega_2(1-\cos\chi)\left[(\omega_1-\omega_{1g})^2+\Gamma_g^2/4\right]}. \tag{4.72}$$

The found expression is not factorized, because the components $A_g$ and $B_g$ have the same degree of the parameter $h$. If only the term $A_g$ is left in the expression (4.72), then the cross section is factorized and in the Breit-Wigner form looks like

$$\sigma_A^{++} = \frac{2\pi}{\omega_1^2(1-\cos\chi)} \frac{\dfrac{dW_{SR\,e^-}^{+-}}{dv}\cdot W_{OPP}^{-+}}{(\omega_1-\omega_{1g})^2+\Gamma_g^2/4}. \tag{4.73}$$

Similarly, if in the expression (4.72) the term $B_g$ is left, then the cross section is also factorized

$$\sigma_B^{++} = \frac{2\pi}{\omega_1^2(1-\cos\chi)} \frac{\dfrac{dW_{SR\,e^-}^{++}}{dv}\cdot W_{OPP}^{++}}{(\omega_1-\omega_{1g})^2+\Gamma_g^2/4}. \tag{4.74}$$

Thus, the intermediate electron in this process of particles production with spins in the direction of the field is in a mixed spin state. Two process channels with cross section $\sigma^{++}$ have the same order of magnitude. In channel $A$, a hard photon create $e^+e^-$ pair in the most favorable spin states $\mu^+=+1$, $\mu_g=-1$ and then the intermediate electron absorbs a soft photon with a change of spin to the opposite value, which adds an additional degree of the parameter $h$ compared to the absorption process without changing a spin. As a result, the spin of the final electron is co-directed to the field $\mu_g=-1 \to \mu^-=+1$. In channel $B$, a hard photon immediately create $e^+e^-$ pair with spins in the direction of the field $\mu^+=+1$, $\mu_g=+1$. Herewith, the probability contains an additional degree of the parameter *h* in comparison with the previous case, and then the intermediate electron absorbs a soft photon without changing the spin $\mu_g=+1 \rightarrow \mu^-=+1$.

The cross section (4.72) can be reduced to the form

$$\sigma^{++} = \sigma_A^{++} + \sigma_B^{++} + \sigma_{int\,AB}^{++},  \qquad (4.75)$$

$$\sigma_{int\,AB}^{++} = \frac{\pi e^4 h^3 m^3 (l^- - n_g) G_1^2 G_2^2 \Xi_g}{2p^- \omega_1 \omega_2 (1-\cos\chi)\left[(\omega_1 - \omega_{1g})^2 + \Gamma_g/4\right]}, \qquad (4.76)$$

$$\Xi_g = v\xi_2^{(2)} - \frac{1}{2}(1-v^2)\xi_1^{(1)}\xi_1^{(2)} + \frac{1}{2}(1+v^2)\xi_2^{(1)}\xi_2^{(2)}. \qquad (4.77)$$

In the case of the unpolarized initial photons, we have $\Xi=0$ and thus the interference term (4.76) is absent.

Figure 4.6 shows the dependence of the resonant cross section of two-photon $e^+e^-$ pair production on a magnetic field for different projections of particle spins. As the magnetic field increases, the ratio of cross section of the main process $\sigma^{-+}$ to the rest decreases.

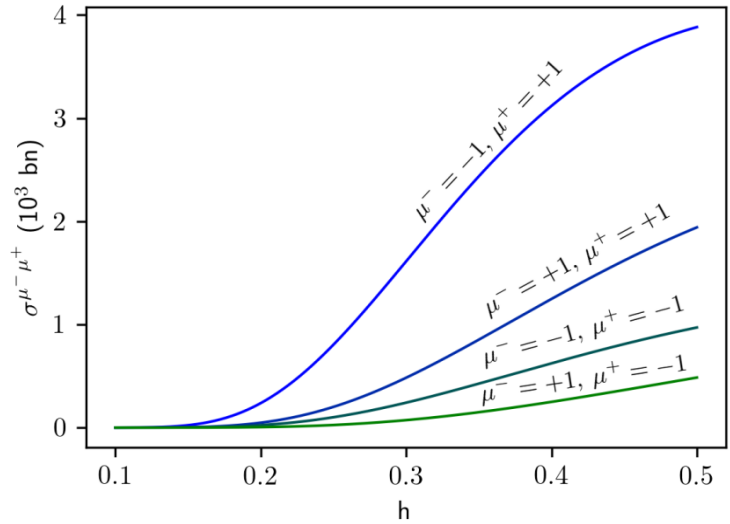

Fig.4.6. Dependence of the cross section of pair production by unpolarized photons on the field strength *h* in $l^-=2$, $l^+=1$, $\omega_1=\delta\omega=hm$.

We proceed to find the cross section of the TPP process in the resonant conditions in the diagram $f$. The cross section is defined by an expression similar to (4.51)

$$\sigma = \frac{e^4 \pi h |Q^{(f)}|^2}{32 m^4 p^- \omega_1 \omega_2 (1-\cos\chi)\left[(f_0 - \varepsilon_f)^2 + \Gamma_f / 4\right]}, \quad (4.78)$$

where the resonance width (the probability of the SR process of final positron) is given by the Landau level of positron

$$\Gamma_f = W_{SR}(l^+ \to l^+ - 1) = \frac{2}{3}(2l^+ - 1)\alpha h^2 m. \quad (4.79)$$

The amplitudes $Q^{(f)}$ in the LLL approximation with fixed values of the projections of spins of the final particles $Q^{\mu^- \mu^+}$ have the form

$$Q^{-+} = -4\sqrt{2}\sqrt{\frac{l^+ h}{n_f}} m^3 F_1 F_2 H_{1p} e_{2z}, \quad (4.80)$$

$$Q^{++} = -4\sqrt{\frac{l^- l^+}{n_f}} h m^3 F_1 F_2 H_{1p} H_{2m}, \quad (4.81)$$

$$Q^{--} = \frac{4hm^3}{\sqrt{n_f}} F_1 F_2 (A_f + B_f), \quad A_f = (n_f - l^+)\tilde{H}_{1p} e_{2z}, \quad B_f = n_f H_{1p} H_{2p}, \quad (4.82)$$

$$Q^{+-} = -2\sqrt{2}(l^+ - n_f)\sqrt{\frac{l^-}{n_f}} h^{3/2} m^3 F_1 F_2 \tilde{H}_{1p} H_{2m}, \quad (4.83)$$

where $\displaystyle F_1 = \frac{e^{-\eta_1/2} \eta_1^{\frac{l^+ - n_f - 1}{2}}}{(l^+ - n_f - 1)!} \sqrt{\frac{(l^+ - 1)!}{(n_f - 1)!}}, \quad F_2 = \frac{(-1)^{n_f} e^{-\eta_2/2} \eta_2^{\frac{l^- + n_f}{2}}}{\sqrt{l^-! n_f!}}, \quad (4.84)$

For the case of fixed spins of the final particles and taking into account the amplitudes (4.80), (4.81), (4.83), the cross section (4.78) can be reduced to the form similar to the expressions (4.63), (4.67), (4.70), respectively

$$\sigma^{-+} = \frac{2\pi}{\omega_1^2 (1-\cos\chi)} \frac{\dfrac{dW^{++}_{SR\,e^+}}{dv} \cdot W^{-+}_{OPP}}{(\omega_1 - \omega_{1f})^2 + \Gamma_f^2 / 4}, \quad (4.85)$$

$$\sigma^{++} = \frac{2\pi}{\omega_1^2(1-\cos\chi)} \frac{\dfrac{dW_{SR\,e^+}^{++}}{dv} \cdot W_{OPP}^{++}}{(\omega_1-\omega_{1f})^2 + \Gamma_f^2/4}, \tag{4.86}$$

$$\sigma^{+-} = \frac{2\pi}{\omega_1^2(1-\cos\chi)} \frac{\dfrac{dW_{SR\,e^+}^{-+}}{dv} \cdot W_{OPP}^{++}}{(\omega_1-\omega_{1f})^2 + \Gamma_f^2/4}. \tag{4.87}$$

The cross sections (4.85) - (4.87) are factorized, the spin states of the intermediate positron are pure, the spin is co-directed to the field. These cross sections are combined in one expression, which can be obtained from expression (4.71) if we introduce the replacements $W_{SR\,e^-}^{\mu^-\mu_g} \to W_{SR\,e^+}^{\mu^+\mu_f}$ and $W_{OPP}^{\mu_g\mu^+} \to W_{OPP}^{\mu_f\mu^-}$, with $\mu_f = +1$. Finally, we will write the cross section of the TPP process with spins directed opposite to the direction of the field $\sigma^{--}$ in resonance. This cross section is not factorized due to the fact that the intermediate positron is in the mixed spin state, similar to the cross section (4.75) it can be represented as

$$\sigma^{--} = \sigma_A^{--} + \sigma_B^{--} + \sigma_{\text{int}\,AB}^{--}, \tag{4.88}$$

$$\sigma_A^{--} = \frac{2\pi}{\omega_1^2(1-\cos\chi)} \frac{\dfrac{dW_{SR\,e^+}^{-+}}{dv} \cdot W_{OPP}^{-+}}{(\omega_1-\omega_{1f})^2 + \Gamma_f^2/4}, \quad \sigma_B^{--} = \frac{2\pi}{\omega_1^2(1-\cos\chi)} \frac{\dfrac{dW_{SR\,e^+}^{--}}{dv} \cdot W_{OPP}^{--}}{(\omega_1-\omega_{1f})^2 + \Gamma_f^2/4}, \tag{4.89}$$

$$\sigma_{\text{int}\,AB}^{--} = \frac{\pi e^4 h^3 m^3 (l^+ - n_f) F_1^2 F_2^2 \Xi_f}{2 p^- \omega_1 \omega_2 (1-\cos\chi)\left[(\omega_1-\omega_{1f})^2 + \Gamma_f/4\right]}, \tag{4.90}$$

$$\Xi_f = -v\xi_2^{(2)} - \frac{1}{2}(1-v^2)\xi_1^{(1)}\xi_1^{(2)} + \frac{1}{2}(1+v^2)\xi_2^{(1)}\xi_2^{(2)}. \tag{4.91}$$

The expression (4.91) differs from the similar one (4.77) only by the sign of the first term.

### 4.3. Spin-polarization effects in the TPP process

In this section we analyze the spin-polarization effects in the TPP process in the resonant conditions, using obtained expressions for the cross sections (4.63), (4.67), (4.70), (4.75), (4.85) - (4.88). The degree of polarization of the final electrons is determined by the expression

$$P_{e^-} = \frac{\sigma^{++} + \sigma^{+-} - \sigma^{-+} - \sigma^{--}}{\sigma^{++} + \sigma^{+-} + \sigma^{-+} + \sigma^{--}}. \qquad (4.92)$$

In the resonance conditions of the diagram $g$ and taking into account (4.63), (4.67), (4.70), (4.75), the degree of polarization can be reduced to the form

$$P_{e^-} = \frac{-2l^-\Pi_1\tilde{\Pi}_2 - l^-l^+ h\Pi_1\Pi_2 + hK_g}{2l^-\Pi_1\tilde{\Pi}_2 + l^-l^+ h\Pi_1\Pi_2 + hK_g}, \qquad (4.93)$$

$$K_g = (l^- - n_g)^2 \tilde{\Pi}_1\tilde{\Pi}_2 + n_g^2 \Pi_1\Pi_2 + \frac{4(l^- - n_g)n_g}{1+v^2}\Xi_g. \qquad (4.94)$$

For comparison, we write the degree of polarization in the OPP process $P_{OPPe^-}$ with the final particles in the same states (2.167)

$$P_{OPPe^-} = -\frac{2\tilde{\Pi}_2 + (l^+ - l^-)h\Pi_2}{2\tilde{\Pi}_2 + (l^+ + l^-)h\Pi_2}. \qquad (4.95)$$

Let us consider few individual cases. If a hard photon has the abnormal linear polarization $\xi_3^{(2)} = +1$, in this case we have

$$\Pi_2 = 0, \ \tilde{\Pi}_2 = 2, \ \xi_1^{(2)} = 0, \ \xi_2^{(2)} = 0, \ \Xi_g = 0,$$

then the degree of polarization of the final electrons in the TPP process is equal to

$$P_{e^-} = -\frac{2l^-\Pi_1 - h(l^- - n_g)\tilde{\Pi}_1}{2l^-\Pi_1 + h(l^- - n_g)\tilde{\Pi}_1}. \qquad (4.96)$$

The degree of polarization of electrons in the OPP process in such conditions is equal to $P_{OPPe^-} = -1$, in other words, the spin of all electrons is directed opposite to the field. From the expression (4.96) it follows that for $\Pi_1 \neq 0$ the degree of polarization $P_{e^-}$ close to the value -1 and only in the narrow interval $\Pi_1 < h$, the degree $P_{e^-}$ significantly depends both on the polarization of soft photon, its angle of incidence, and on the Landau level of the final electron. If the polarization of soft photon is such that $\tilde{\Pi}_1 = 0$, then the degree of polarization of electrons is equal to $P_{e^-} = -1$, as in the OPP process. For this purpose it is necessary, according to the expression (4.61), that Stokes parameters of polarization of soft photon are equal to

$$\xi_1^{(1)} = 0, \quad \xi_2^{(1)} = -\frac{2v}{1+v^2}, \quad \xi_3^{(1)} = -\frac{1-v^2}{1+v^2}. \qquad (4.97)$$

In particular, for the case of soft photon perpendicular to the field $\xi_3^{(1)} = -1$, i.e. its polarization must be completely normal. If we choose the opposite case $\Pi_1 = 0$, then $P_{e^-} = +1$, in other words, all electrons are in the inverse spin states. In this case, it is necessary to change the sign of linear polarization $\xi_3^{(1)}$

$$\xi_1^{(1)} = 0, \quad \xi_2^{(1)} = -\frac{2v}{1+v^2}, \quad \xi_3^{(1)} = +\frac{1-v^2}{1+v^2}. \qquad (4.98)$$

Thus, under changing the linear polarization of soft photon, which is perpendicular to the field, in the entire range from $\xi_3^{(1)} = -1$ to $\xi_3^{(1)} = +1$, the orientation of electron spins can be changed from fully polarized in the ground spin state to fully polarized in the inverse state.

Now let a hard photon have the normal linear polarization $\xi_3^{(2)} = -1$, in this case we have

$$\Pi_2 = 2, \quad \tilde{\Pi}_2 = 0, \quad \xi_1^{(2)} = 0, \quad \xi_2^{(2)} = 0, \quad \Xi_g = 0$$

and the degree of polarization of the final electrons is equal to

$$P_{e^-} = \frac{n_g^2 - l^- l^+}{n_g^2 + l^- l^+}. \qquad (4.99)$$

For the OPP process according to the expression (2.170), a similar quantity has the form:

$$P_{OPPe^-} = \frac{l^- - l^+}{l^- + l^+}. \qquad (4.100)$$

In the case of the particles production at the same Landau levels $l^-=l^+$, the degree of polarization is $P_{e^-}<0$, because $n_g<l^-$ is always satisfied, while in the OPP process the degree of polarization is equal to $P_{OPPe^-}=0$. The most probable is the TPP process, when a hard photon creates pairs at the same energy levels, i.e. $n_g=l^+$ and $l^->l^+$. In this case, we have

$$P_{e^-}=-P_{OPPe^-}<0,$$

i.e. in the TPP process with normally polarized photon, the electrons are created mainly in normal spin states.

Let us analyze the dependence of the degree of polarization $P_{e^-}$ on the circular polarization of photons. Consider the Stokes parameters of photons in the form

$$\xi_2^{(1,2)}=\pm 1,\ \xi_1^{(1,2)}=0,\ \xi_3^{(1,2)}=0,$$

that is, both photons are completely circularly polarized. In this case, the degree of electron polarization is equal to

$$P_{e^-}=-1+\frac{h}{l^-}(l^--n_g+n_g\xi_2^{(1)}\xi_2^{(2)})^2, \qquad (4.101)$$

in particular, under the transition to the neighboring Landau level $n_g=l^--1$

$$P_{e^-}=-1+\frac{h}{l^-}(1+n_g\xi_2^{(1)}\xi_2^{(2)})^2,$$

that is, preferably the electrons are in the ground spin state. The violation of the polarization of electrons is proportional to the parameter $h$ and it is greatest if photons have the same direction of circular polarization. It should be noted that the dependence of the degree of orientation of particle spins on the circular polarization of photons distinguishes the TPP process from the OPP process and it is the consequence of mixed intermediate state of the TPP process with the spins of the final particles, which are co-directed to the field.

Let us analyze the degree of polarization of the final positrons, that is determined by the expression

$$P_{e^+}=\frac{\sigma^{++}+\sigma^{-+}-\sigma^{+-}-\sigma^{--}}{\sigma^{++}+\sigma^{-+}+\sigma^{+-}+\sigma^{--}}. \qquad (4.102)$$

Taking into account (4.63), (4.67), (4.70), (4.75), the degree of polarization can be reduced to the form

$$P_{e^+} = \frac{2l^-\Pi_1\tilde{\Pi}_2 - l^-l^+h\Pi_1\Pi_2 + hK_g}{2l^-\Pi_1\tilde{\Pi}_2 + l^-l^+h\Pi_1\Pi_2 + hK_g}. \qquad (4.103)$$

If an initial hard photon is abnormally polarized $\xi_3^{(2)} = +1$ ($\Pi_2 = 0$), then all final positrons have spins in the direction of the field $P_{e^+} = +1$, regardless of the type of polarization of soft photon. Note that the polarization of soft photon significantly affects the degree of electron polarization, changing it to the opposite in the narrow interval near the polarization (4.98).

In the case of normal polarization of hard photon $\xi_3^{(2)} = -1$ ($\tilde{\Pi}_2 = 0$), the expression for the degree of polarization of positrons coincides with a similar expression for electrons (4.99) $P_{e^+} = P_{e^-}$. As mentioned earlier, this quantity in the TPP process with the most probable values of Landau levels of final particles is negative, i.e. positrons are mainly in the inverse spin state.

If both photons are completely circularly polarized $\xi_2^{(1)}\xi_2^{(2)} = \pm 1$, the degree of polarization of positrons takes the simple form

$$P_{e^+} = 1 - l^+h, \qquad (4.104)$$

that coincides with a similar degree of polarization (2.169) in the OPP process. Thus, the circular polarization of the initial photons does not affect the degree of orientation of spins of the final positrons.

Consider the spin-polarization effects in the TPP process in the resonance of diagram $f$ with resonant conditions (4.39), (4.40). In this case, taking into account the expressions for the cross sections (4.85) - (4.88), the degree of polarization of the final electrons (4.92) and positrons (4.102) can be reduced to the form

$$P_{e^-} = \frac{-2l^+\Pi_1^+\tilde{\Pi}_2 + l^-l^+h\Pi_1^+\Pi_2 - hK_f}{2l^+\Pi_1^+\tilde{\Pi}_2 + l^-l^+h\Pi_1^+\Pi_2 + hK_f}, \qquad (4.105)$$

$$P_{e^+} = \frac{2l^+\Pi_1^+\tilde{\Pi}_2 + l^-l^+h\Pi_1^+\Pi_2 - hK_f}{2l^+\Pi_1^+\tilde{\Pi}_2 + l^-l^+h\Pi_1^+\Pi_2 + hK_f}, \qquad (4.106)$$

$$K_f = (l^+ - n_f)^2 \tilde{\Pi}_1^+ \tilde{\Pi}_2 + n_g^2 \Pi_1^+ \Pi_2 + \frac{4(l^+ - n_f)n_f}{1+v^2} \Xi_f, \qquad (4.107)$$

where the polarization functions of soft photon $\Pi_1^+, \tilde{\Pi}_1^+$ differ from similar ones $\Pi_1, \tilde{\Pi}_1$ (4.61), (4.69) by changing the sign in the term with $\xi_2^{(1)}$, due to the fact that changing the sign of the charge changes the sign of the circular polarization of the SR process.

If the first terms of the numerators in (4.105), (4.106) are not equal to zero (more precisely, much larger than the parameter $h$), the degree of polarization in the linear approximation on the parameter $h$ take a simpler form

$$P_{e^-} = -1 + l^- h \frac{\Pi_2}{\tilde{\Pi}_2}, \qquad P_{e^+} = 1 - h \frac{K_f}{l^- \Pi_1^+ \tilde{\Pi}_2}. \qquad (4.108)$$

Note that the degree of electron polarization in the TPP process coincides with a similar expression in the OPP process (2.169). This is understandable, because in the resonance of the diagram $f$, a hard photon immediately creates a final electron and an intermediate positron, and the latter absorbs an additional soft photon. In particular, if $\Pi_2 = 0$ (anomalous linear polarization of the hard photon)

$$P_{e^-} = -1, \qquad P_{e^+} = 1 - \frac{(l^+ - n_f)^2 h}{l^+} \cdot \frac{\tilde{\Pi}_1^+}{\Pi_1^+}, \qquad (4.109)$$

the spins of electrons are completely oriented opposite to the field, and the degree of polarization of positron is slightly disturbed (by the value $\sim h$). The spins of positrons are completely oriented to the field $P_{e^+} = +1$, if $\tilde{\Pi}_1^+ = 0$ that is realized in the conditions

$$\xi_1^{(1)} = 0, \quad \xi_2^{(1)} = \frac{2v}{1+v^2}, \quad \xi_3^{(1)} = -\frac{1-v^2}{1+v^2}. \qquad (4.110)$$

When choosing the polarization of soft photon, when $\Pi_1^+ = 0$ that is realized if

$$\xi_1^{(1)} = 0, \quad \xi_2^{(1)} = \frac{2v}{1+v^2}, \quad \xi_3^{(1)} = \frac{1-v^2}{1+v^2}, \qquad (4.111)$$

the final particles are completely polarized

$$P_{e^-} = -1, \ P_{e^+} = -1,$$

moreover, all electrons are in the ground spin state and positrons are in the inverse state.

In the case, if $\tilde{\Pi}_2 = 0$ (normal linear polarization of the hard photon), the degrees of polarization of the particles are the same

$$P_{e^-} = P_{e^+} = \frac{l^- l^+ - n_f^2}{l^- l^+ + n_f^2}. \tag{4.112}$$

Because the most probable Landau levels of the final particles are $n_f = l^-$ and $l^- < l^+$, then $P_{e^-} = P_{e^+} > 0$, i.e. in the TPP process with the normally polarized hard photon, the electrons are mainly create in the inverse spin states, and the positrons are in the normal ones.

Finally, when both photons are completely circularly polarized $\xi_2^{(1)}, \xi_2^{(2)} = \pm 1$, the degree of polarization of final particles is equal to

$$P_{e^-} = -1 + l^- h, \quad P_{e^+} = 1 + \frac{h}{l^+}(l^+ - n_f + n_f \xi_2^{(1)} \xi_2^{(2)})^2. \tag{4.113}$$

The first expression in (4.113) coincides with a similar expression in the OPP process (2.169) when the hard photon is not polarized. The degree of polarization of positron has a form similar to expression (4.101) for the degree of polarization of electron in the resonance TPP process of the diagram $g$.

### 4.4. Production of e$^+$e$^-$ pair by photon in a magnetic field and SR field in a pulsar magnetosphere

It is known that the production of e$^+$e$^-$ pairs is an important element of the model of X-ray pulsars, because the existence of an electron-positron plasma in magnetized magnetosphere with the magnetic field $\sim 10^{12} Гс$ is necessary for the generation of pulsar radiation. To date, it is considered that the main source of pairs is the process of e$^+$e$^-$ pair production by one photon [134-137]. e$^+$e$^-$ pairs are also create as a result of the interaction of two photons with the total energy $>2m$, which appear due to the inverse Compton scattering of thermal X-rays from the surface of neutron star. Since the attenuation length of $\gamma$ photons beam for two-photon process is greater than for one-photon process, traditionally the process of production of a pair by two photons is considered insignificant in comparison with one-photon process. The only exceptions are magnetars, where in the supercritical magnetic fields ~$10^{15}$Gs the one-photon generation of a pair is suppressed by the process of photon splitting.

It should be emphasized that this opinion is not correct, because it does not take into account the influence of a strong magnetic field on the process of two-photon pair production. In the TPP process, as shown above, there are resonances in which the cross section is several orders of magnitude larger than the cross section of the process without the field, which significantly increases the contribution to plasma generation.

In the first stage of e$^+$e$^-$ plasma formation, when its density is not high, the one-photon process is indeed the most significant. However, over time, one hard photon will create e$^+$e$^-$ pair not in an empty vacuum, but in the field of photons with cyclotron frequency and multiple of it, as electrons and positrons appear, which move in a strong magnetic field accompanied by cyclotron radiation. Thus, e$^+$e$^-$ pair can be create by one hard photon with the capture of soft cyclotron photon, i.e. we get a resonance production of a pair by two photons. As follows from the previous section, the resonance conditions of this process near the threshold (4.27), (4.28) are automatically fulfilled. The hard photon must have the above-threshold frequency (4.27) to creates a pair at given Landau levels. This condition is necessary for both second-order and first-order processes. According to (4.28) the frequency of soft photon should be multiple of the cyclotron frequency that takes place.

Let us make an estimated comparison of these processes, when the particles are created at the lowest possible energy levels in the main spin states, provided that the two-photon process takes place in the resonant conditions of the $g$ diagram. Herewith, the Landau levels, spin values of particles and Stokes parameters of photon polarizations are equal to

$$l^- = 1, \ l^+ = n_g = 0, \ \mu^- = -1, \ \mu^+ = 1, \ \xi^{(1)}_{1,2,3} = \xi^{(2)}_{1,2,3} = 0. \qquad (4.114)$$

Taking into account (4.114), the cross section of the two-photon pair production (4.59) is equal to

$$\sigma = \frac{\pi e^4 h e^{-2/h}}{2(1-\cos\chi)\Gamma^2}\sqrt{\frac{m}{\delta\omega}}, \qquad (4.115)$$

where the resonance width is taken as the radiation width $\Gamma = 2e^2 mh^2/3$.

The probability of the process per unit time is related to the cross section by the expression

$$W_{2\gamma} = n_\gamma \sigma(1-\cos\chi), \qquad (4.116)$$

where $n_\gamma$ is the density of cyclotron photons in the pulsar magnetosphere. The rate of the one-photon pair production in the ground energy states (2.156) with momentum $p = \sqrt{\delta\omega m}$ is equal to

$$W_{1\gamma} = \frac{e^2 h m e^{-2/h}}{4\sqrt{\delta\omega/m}}. \qquad (4.117)$$

Equality of probabilities (4.116) and (4.117) gives the expression for the critical density

$$n_{\gamma c} = 2e^2 h^4 m^3 / 9\pi, \qquad (4.118)$$

as we see, it is determined only by the magnetic field. If the density of cyclotron photons exceeds the critical value (4.118), the two-photon process dominates over the one-photon process. When the magnetic field strength is $h=0.1$ then the critical density is $n_{\gamma c} \sim 10^{24}$ см$^{-3}$, which is an order of magnitude higher than the estimated characteristic density of photons in the magnetosphere of pulsars [134]. Thus, if in the first stages the electron-positron plasma is generated due to the one-photon pair production, then at the final stage of its formation the resonant two-photon process dominates.

**4.5. Influence of particle polarization on the intensity of pulsar synchrotron radiation**

In the previous section, we considered the formation of e⁺e⁻ plasma in the magnetosphere of X-ray pulsar. When e⁺e⁻ pair is formed by a hard photon or by picking up a soft one, the electrons (positrons) at excited energy levels in a magnetic field of the pulsar generate synchrotron radiation, which is the object of observation. In the modern pulsar model, the particles are considered unpolarized. Consider the influence of particle polarization on the intensity of synchrotron radiation. According to expression (2.151), the degree of polarization of particles strongly depends on the polarization of the initial hard photon and as follows from (2.44), itself significantly affects the probability of emission of the final photons.

We introduce the ratio of intensities $R$:

$$R = <I>_{pol} / <I>_{therm}, \qquad (4.119)$$

where $<I>_{pol}$ is the total intensity of radiation by electron, which is averaged with the weight fractions of electrons in the inverse and ground spin states $x_+$, $x_-$, respectively. $<I>_{therm}$ is the total intensity of radiation, averaged over the spins of the initial electron and summed over the spins of the final one. Thus, under using the magnitude of $<I>_{therm}$, it is considered a common view of the thermodynamic preparation of plasma particles, that are not polarized and under using expression $<I>_{pol}$, the polarization of particles formed in the OPP process or resonant TPP is taken into account. These averaged intensities of the SR process are determined as follows

$$<I>_{pol} = I^{-} x_{-} + I^{+} x_{+}, \quad <I>_{therm} = (I^{-} + I^{+})/2, \qquad (4.120)$$

where $I^{-}$, $I^{+}$ are the total intensities of the SR process of polarized initial electron with spin opposite and in the direction of the field, respectively, which are taken from (2.51). The weight fractions $x_+$, $x_-$ are defined by expressions

$$x_{-} = (W^{-+} + W^{--})/\sum W^{\mu^{-}\mu^{+}}, \quad x_{+} = (W^{++} + W^{+-})/\sum W^{\mu^{-}\mu^{+}}, \qquad (4.121)$$

where $W^{\mu^{-}\mu^{+}}$ is the total probability of e⁺e⁻ pair production, which is obtained in the general case from expression (2.151) after integration on the longitudinal component of the momentum of the final electron, while the obvious ratio is fulfilled

$$x_{-} + x_{+} = 1. \qquad (4.122)$$

The LLL approximation. In this approximation, the relation (4.119) with the fixed value of the parameter $x_+$ is obtained after averaging the expressions (2.46) - (2.49) and it has the form

$$R = 2(l - x_+(l - l'))/(l + l'). \qquad (4.123)$$

In the case of unpolarized particles $x_+=x_-=1/2$, the intensity ratio is obviously equal to $R=1$. If the initial electrons are completely polarized with the spin directed opposite to the field $x_+=0$, the ratio is equal to $R=2l/(l+l')$, in particular, under the transition of electrons to the ground energy state we have $l'=0$: $R=2$ and the intensity of polarized electrons is twice the intensity of unpolarized ones.

The probability of the OPP process with the fixed values of the projections of the spins of particles created at given Landau levels in the LLL approximation is determined by the expressions (2.156) - (2.159). Then part of the electrons with spins co-directed to the field $x_+$ in the case when $e^+e^-$ pairs are created at the same Landau level $l^+ = l^- = l$ has the form

$$x_+ = lh(1-\xi_3)/2(1+\xi_3 + lh(1-\xi_3)), \qquad (4.124)$$

where $\xi_3$ is the Stokes parameter of the linear polarization of the initial hard photon.

As follows from the obtained expression (4.124), the ratio of intensities $R$ (4.123) significantly depends on the linear polarization of the initial photon. In the case of the unpolarized initial photon $\xi_3=0$, the fraction $x_+=lh/2$ is small, the spins of electrons are preferably directed opposite to the field. The ratio $R$ does not depend on the frequency of the initial photon. Figure 4.7 (*a*) shows the dependence of the ratio of the intensities $R$ on the linear polarization of the initial photon and the magnitude of magnetic field in the case of the transition of electron from $l=5$ to $l'=0$. Fig.4.7 (b) shows the dependence of the magnitude $R$ on the polarization of the photon and the Landau level number of the final electron, $h=0.1$. As can be seen from Fig.4.7, $R > 1$ is always satisfied if $\xi_3 \neq -1$ and $R = 1$ when $\xi_3 = -1$. Thus, taking into account the polarization of the particles of a pulsar magnetosphere leads to higher values of the SR intensity in the LLL approximation.

<u>Ultrarelativistic approximation.</u> In this approximation, the intensity ratio $R$ can be obtained by integrating expressions (2.90), (2.185). Fig.4.8 (a) shows the dependence of the ratio R on the linear polarization of the initial photon $\xi_3$ and the parameter $\Omega = h\omega/m$ (product of the field on the frequency of the initial photon), with $\omega = 100m$, and the magnetic field takes values from $h=0.001$ to $h=0.1$. Fig.4.8 (b) shows the dependence of $R$ on the Stokes parameter $\xi_3$ and the frequency of the finite photon $y=\omega/\omega_c$, where $h=0.1$.

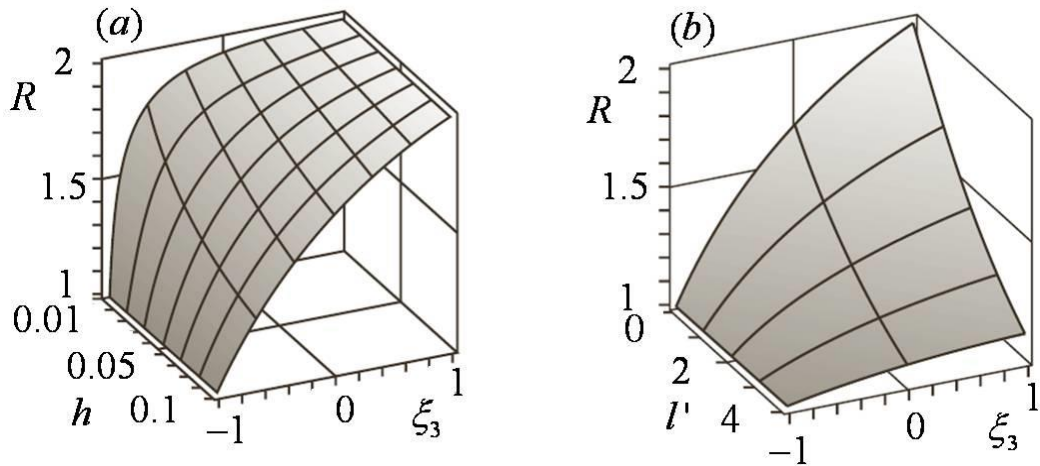

Fig.4.7. Dependence of the ratio of intensities of synchrotron radiation of polarized and unpolarized electrons: (*a*) on the Stokes parameter $\xi_3$ of the initial photon and the magnetic field $h$, (*b*) on the Landau level number of the final electron and the Stokes parameter $\xi_3$

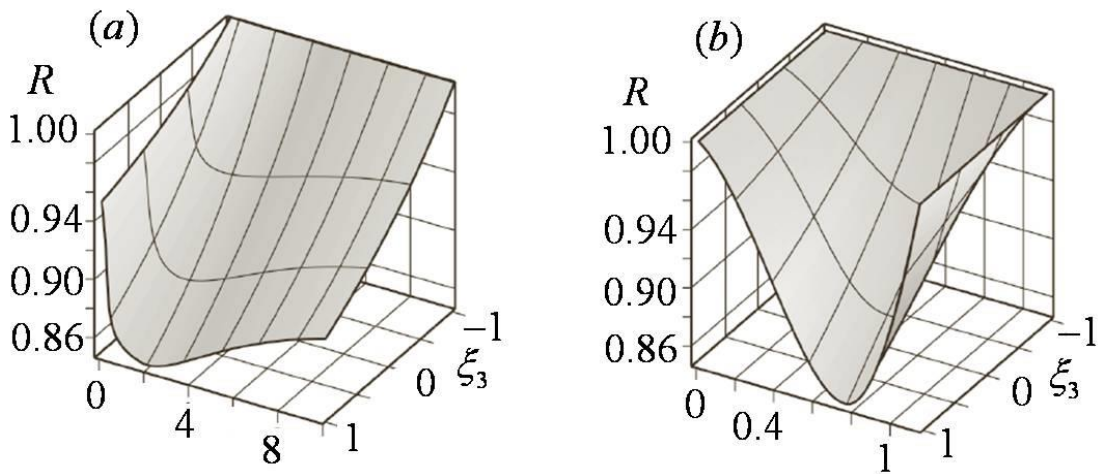

Fig.4.8. Dependence of the ratio of the SR intensities of polarized and unpolarized electrons: (a) on the Stokes parameter $\xi_3$ of the initial photon and on the parameter $\Omega=\hbar\omega/m$, (b) on the Stokes parameter $\xi_3$ and the frequency of the final photon $y=\omega/\omega_c$

As can be seen from Fig.4.8 in the ultrarelativistic case R<1. The minimum value of $R=0.86$ corresponds to the anomalous linear polarization of the initial photon and monotonically goes to $R=1$ when $\xi_3=-1$.

Thus, taking into account the spin population of $e^+e^-$ plasma leads to a change in the SR spectrum, increases the low-frequency part of the spectrum and decreases the high-frequency part.

### 4.6. Conclusions to the Chapter 4

The process of an electron-positron pair production by two polarized photons taking into account the spin of particles in a strong external magnetic field was first studied in resonant conditions if one photon (hard) creates a pair and the other (soft) causes resonance. As a result, it was shown:

1. In the analyzed process of two-photon $e^+e^-$ pair production (TPP), the resonance is possible near the threshold if the frequency of hard photon exceeds the sum of the energies of pair, and the frequency of soft photon is multiple of the cyclotron frequency. Far from the threshold, the frequency of soft photon depends on its polar angle and it is maximum in the case of movement of this photon along the field. Interference of two resonances near the process threshold occurs if both photons are directed perpendicular to the field and the relation for Landau levels of the final and intermediate particles $l^--n_g=l^+-n_f$ is fulfilled.

2. The largest cross section of the process corresponds to the particles production in the main spin states $\sigma^{-+}$. It has the maximum value if a soft photon is normally polarized and a hard photon is abnormally polarized. In the case when the magnetic field strength is $H=10^{12} Гc,$ the cross section of the process has the order of the Thomson cross section, and the width of resonance is $30eB$. Cross sections of processes in which the particles have the same spin direction $\sigma^{--}$, $\sigma^{++}$ have an additional degree of small parameter $h$. The smallest is the cross section with particles in the inverse spin states $\sigma^{+-}$. In resonance, the cross sections with the electronic intermediate state $\sigma^{-+}$, $\sigma^{--}$, $\sigma^{+-}$ are factorized and they can be represented in the Breit-Wigner form. The cross section $\sigma^{++}$ is not factorized because the intermediate electron is in the mixed state. For resonance with the positron intermediate state, the cross section $\sigma^{--}$ is not factorized.

3. Spin-polarization effects in the resonant TPP process with an intermediate electron are expressed as follows:

a. In the process with abnormally linear polarized of hard photons ($\xi_3=1$), the change in the linear polarization of soft photons in the whole range changes the orientation of the electron spins from fully oriented opposite to the field to fully oriented to the field without changing the spin direction of positron.

b. In the process with normally linearly polarized of hard photon ($\xi_3=-1$), the degree of orientation of the electron spins does not depend on the polarization of soft photon and it is determined only by the Landau levels of the intermediate and final particles. For the process with the lowest possible energy levels, the electrons are completely unpolarized and the first excited levels correspond mainly to the normal spin population.

c. If both photons of the process are circularly polarized, the particles are preferably in the ground spin state. Violation of the polarization of particles $\sim h$ and the maximum if the photons have the same direction of circular polarization.

4. Taking into account the field of cyclotron photons on the process of $e^+e^-$ plasma formation in the magnetosphere of the X-ray pulsar showed the dominant role of resonances in the field $H=10^{12} Gs$ under the characteristic photon concentration, that refutes the generally accepted view of the dominant role of the OPP process in magnetosphere formation.

5. Taking into account the spin population of electrons and positrons in the process of $e^+e^-$ plasma generating in a pulsar magnetosphere leads to a change in the SR spectrum, increases the low-frequency part of the spectrum and decreases the high-frequency.

The main scientific results of this chapter are published in [283], [287-289].

# CHAPTER 5
# ONE-PHOTON PRODUCTION OF AN ELECTRON-POSITRON PAIR WITH EMISSION OF A PHOTON

## 5.1. Introduction

The process of production of an e⁺e⁻ pair by a photon with subsequent emission of a final photon (OPPE) in a strong magnetic field is studied as a single second-order process. The kinematics of the process and the conditions for the occurrence of resonances are analyzed. The probabilities of the process are calculated in the resonant and nonresonant cases, taking into account the polarization of the particles in the nonrelativistic LLL approximation. Spin-polarization effects are studied. The possibility of the existence of mixed spin states of an intermediate electron (positron) under resonance conditions is analyzed.

## 5.2. Kinematics of the OPPE process

A feature of the process under study, in contrast to those previously considered, is the presence of three particles in the final state. As a result of the increase in the phase space of the final particles, the kinematics of the process becomes more complicated and requires separate consideration. The CS, TPP and OPPE processes are cross-channels of one generalized reaction. Therefore, the laws of conservation of energy and longitudinal momentum component are constructed from the same parameters for these processes. Similarly to expressions (3.5), (4.2) they can be written in the form:

$$\omega = \varepsilon^- + \varepsilon^+ + \omega', \quad \omega v = p^- + p^+ + \omega' u, \qquad (5.1)$$

where $\omega, v = \cos\theta$ and $\omega', u = \cos\theta'$ are frequencies and cosines of the polar angles of the initial and final photons; $\varepsilon^-$, $p^-$ and $\varepsilon^+$, $p^+$ are energies and longitudinal momenta of the electron and the positron, respectively. As before, the energies and momenta of the particles are related by relations:

$$\varepsilon^{\pm} = ((m^{\pm})^2 + (p^{\pm})^2)^{1/2} = (m^2 + 2l^{\pm}hm^2 + (p^{\pm})^2)^{1/2}. \qquad (5.2)$$

For fixed Landau levels of final particles $l^+$, $l^-$, longitudinal momentum of the electron $p^-$, and the angle of emission of the final photon $u$, conservation laws (5.1) specify the frequency of the final photon as follows:

$$\omega' = \frac{1}{1-u^2}(\omega_\varepsilon - k_p u - \sqrt{(\omega_\varepsilon - k_p u)^2 - (\varepsilon_\varepsilon^2 - k_p^2 - (m^+)^2)(1-u^2)}), \qquad (5.3)$$

where $\omega_\varepsilon = \omega - \varepsilon^-$, $k_p = \omega v - p^-$. The frequency of the final photon is set in the range:

$$0 \leq \omega' \leq \omega - (m^- + m^+). \qquad (5.3)$$

The upper limit is the threshold value, it corresponds the next condition $p^- = p^+ = 0$.

To determine the limiting values of the energy and momentum of the electron, it is convenient, similarly to (2.121), to introduce function $f(p)$, which tends to zero for a real process:

$$f(p) = \omega - \omega' - \sqrt{(m^-)^2 + p^2} - \sqrt{(m^+)^2 + (p + \omega'u - \omega v)^2}. \qquad (5.4)$$

The dependence of this function on the longitudinal momentum of the electron is presented in Fig.5.1.

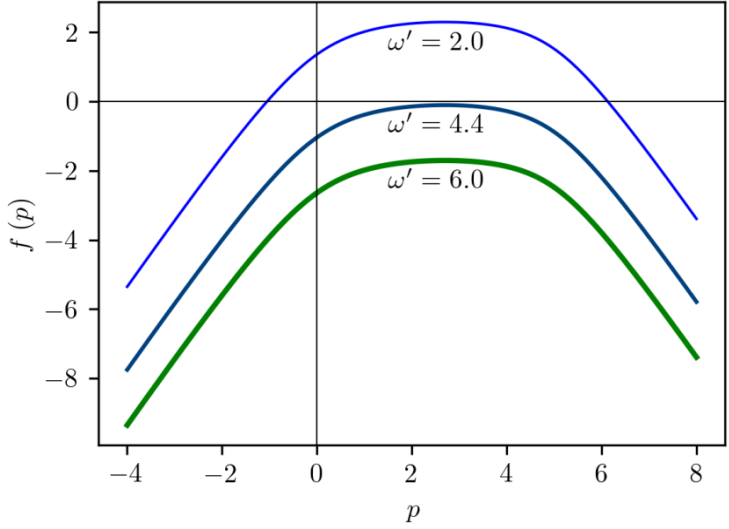

Figure 5.1. Graph of the function $f(p)$ for various values of the frequency of the final photon, $h=0.3$, $l^-=2$, $l^+=1$, $\omega=10$, $v=0.5$, $u=0$

As follows from Fig.5.1. the limit values $\varepsilon^-$ and $p$ correspond to the zero of the maximum of the function $f(p)$, are determined by the system of equations $\partial f/\partial p=0$, $f=0$ and are equal to:

$$\varepsilon_m^\pm = m^\pm W/(m^+ + m^-), \quad p_m^\pm = \beta\varepsilon_m^\pm, \qquad (5.5)$$

where
$$W = \omega - \omega', \quad \beta = (\omega v - \omega' u)/(\omega - \omega'). \qquad (5.6)$$

Expressions (5.5) obviously follow from (2.123) by replacing:

$$\omega \to W, \quad u \to \beta. \qquad (5.7)$$

Note that taking into account the laws of conservation (5.1)

$$\beta = (p^- + p^+)/(\varepsilon^- + \varepsilon^+), \quad |\beta| < 1. \qquad (5.8)$$

In the general case, the values of the energies and momenta of the final electron and positron follow from expressions (2.130) by replacing (5.7).

Let us determine the limiting values of the emission angle of the final photon by analyzing the dependence of the electron momentum $p^-=p(u)$ on the cosine of the emission angle $u$ at fixed values of the Landau levels of particles and the frequency of the final photon:

$$p(u) = p_{1,2}^- = (a^-(u)\cdot\beta \pm b^-(u))/2(\omega - \omega')(1-\beta^2), \qquad (5.9)$$

where $a^-(u) = W^2(1-\beta^2) + (m^-)^2 - (m^+)^2$, $b^-(u)^2 = a^-(u)^2 - 4(m^-)^2 W^2(1-\beta^2)$. The sought interval of angles is determined by the condition $p_1 = p_2$, which is equivalent to the equation $b(u)=0$. The dependence $b^2(u)$ is shown in Fig.5.2.

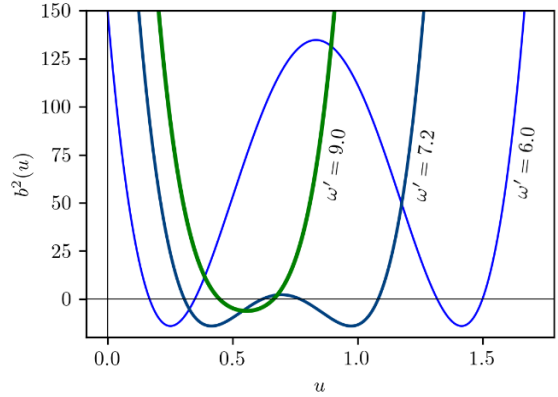

Fig.5.2. Graph of the function $b^2(u)$ for different values of the final photon frequency, $h$=0.3, $l^-$=2, $l^+$=1, $\omega$=10$m$, $v$=0.5

The following parameters are selected: $h$=0.3, $l^-$=2, $l^+$=1, $\omega$=10$m$, $v$=0.5. The three curves correspond to the frequencies of the final photon $\omega'/m$: 9.0, 7.25, 6.0. In the general case, the $b^2(u)$ curves have four roots (points of intersection of the curve with the abscissa), and the interval of the photon emission angles is determined by two internal roots. For example, the curve $\omega'/m$= 9.0 has only two intersections with the abscissa, and radiation is absent in this case. The curve $\omega'/m$= 7.25 corresponds to the threshold of the process, the final photon is emitted in a narrow cone with $u = 0.69$. Finally, for the curve $\omega'/m$=6.0, the inner roots are equal: $u_{min}$=0.35, $u_{max}$=1.32, hence 0.35 <$u$ <1. In the general case, the boundaries of the interval of polar angles of the final photon are determined by expressions:

$$u_{min} = \frac{1}{\omega'}(\omega v - \sqrt{W^2 - m_\Sigma^2}), \; u_{max} = \frac{1}{\omega'}(\omega v + \sqrt{W^2 - m_\Sigma^2}), \; m_\Sigma = m^- + m^+, \quad (5.10)$$

further, if necessary, you need to redefine

$$u_{min} = \max\{u_{min}, -1\}, \; u_{max} = \min\{u_{max}, 1\}. \quad (5.11)$$

Figure 5.3 shows the dependence of the longitudinal momentum of the electron on the cosine of the emission angle of the final photon.

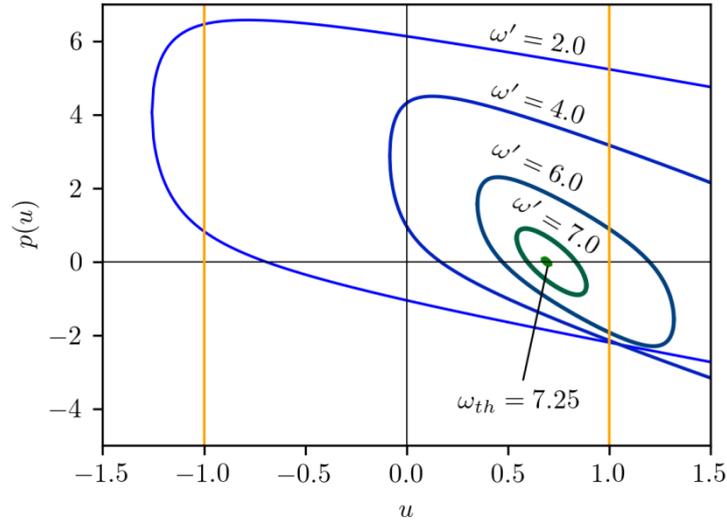

Figure 5.3. Dependence of the electron momentum on the emission angle of the final photon, $h=0.3$, $l^-=2$, $l^+=1$, $\omega=10m$, $v=0.5$

As can be seen from Fig.5.3, in the case of emission of a low-frequency photon (curve $\omega'/m=2$ for the chosen parameters $h=0.3$, $l^-=2$, $l^+=1$, $\omega=10m$, $v=0.5$), the final photon can be emitted in any direction, while the electron momentum weakly depends on u. At the threshold of the process, when $\omega'_{th}/m= 7.25$, $u_{th}=0.69$, the longitudinal momentum of the electron (and the positron) is zero. That is, particles at the threshold are produced motionless at fixed Landau levels. The principal difference between the threshold in the OPPE process and the similar threshold in the OPP process is that it is possible for any frequency and polar angle of the initial photon (sufficient for the production of a pair) and corresponds to the threshold maximum possible frequency of the final photon. The threshold frequency and cosine of the emission angle of the final photon are determined by the expressions:

$$\omega'_{th} = \omega - m_\Sigma, \quad u_{th} = \omega v / \omega'_{th}. \qquad (5.12)$$

The cosine of the emission angle of the final photon $u$ is proportional to the cosine of the angle $v$ of the initial photon. The condition $u \leq 1$ defines constraint $v \leq 1-m_\Sigma/\omega$. Inverse to relation (5.9), the dependence $u(p)$ has the form:

$$u(p) = (\omega v - p \pm \sqrt{(W - \sqrt{(m^-)^2 + p^2})^2 - (m^+)^2})/\omega'. \qquad (5.13)$$

Let us analyze the properties of the characteristic limiting values of the electron momentum and the corresponding emission angles of the final photon. In the general case, there are eight such quantities, they correspond to the points $P_1$-$P_8$, indicated in Fig. 5.4., where characteristic dependence $p(u)$ is shown.

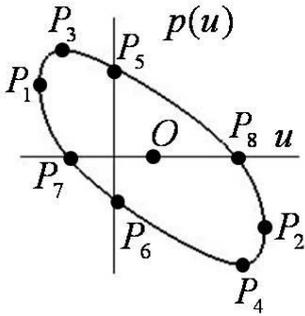

Figure 5.4. Characteristic limiting values of electron momentum on the $p(u)$ curve

A similar closed curve also takes place for the dependence of positron momentum on $u$. Points $P_1$, $P_2$ correspond to the limiting value of the cosine of the final photon angle, which are determined by expressions (5.10): $u_1=u_{\min}$, $u_2=u_{\max}$. The longitudinal momentum of the electron and positron at these points is equal to:

$$p_1^- = -p_2^- = m^-\sqrt{W^2 - m_\Sigma^2}/m_\Sigma, \quad p_1^+ = -p_2^+ = m^+\sqrt{W^2 - m_\Sigma^2}/m_\Sigma. \qquad (5.14)$$

Points $P_3, P_4$ correspond to the maximum and minimum values of electron momentum, respectively:

$$p_3^- = -p_4^- = \sqrt{(W - m^+)^2 - (m^-)^2}, \quad p_{3,4}^+ = 0, \qquad (5.15)$$

in this case, the longitudinal momentum of the positron is zero. Thus, the excess of the initial photon energy over the limiting value is completely converted into the electron momentum.

$$u_{3,4} = (\omega v \mp \sqrt{(W - m^+)^2 - (m^-)^2})/\omega'. \qquad (5.16)$$

Points $P_5, P_6$ correspond to the case when the final photon emits perpendicular to the direction of the magnetic field, that is $u_5=u_6=0$. The longitudinal momentum of the electron is determined from (5.9), where $\beta = \omega v/W$. Finally, the points $P_7, P_8$ correspond to the zero value of the longitudinal momentum of electron. They are similar to points $P_3, P_4$, only now the electron and positron are swapped. The positron momentum at points $P_7, P_8$ is maximum and is expressed by relation (5.15), where it is necessary to replace $m^+ \leftrightarrow m^-$. The angles of emission of the photon are determined by the relation (5.16) with the replacement $m^+ \leftrightarrow m^-$. As seen in Figure 5.4. the curve $p(u)$ has central symmetry about the point $O$ with coordinates $u_0 = \omega v/\omega'$, $p_0 = 0$.

Let us analyze the range of emission angles of the final photon near the reaction threshold. Let, for definiteness, the initial photon be perpendicular to the field $v = 0$. Let certain meanings have $l^-$, $l^+$, $\omega'$. At the threshold $W = W_m = m_\Sigma$, longitudinal momenta of particles are absent $p^- = p^+ = 0$, the final photon, as well as the initial one, is perpendicular to the magnetic field $u = 0$. We choose the value of the frequency of the initial photon close to the threshold value, so that

$$W = W_m + \delta W = m_\Sigma + \delta W, \quad \delta W \ll W. \tag{5.17}$$

Under these conditions, the limiting values of the angles $u_{1,2}$ and the corresponding momenta of the electron and positron have the form:

$$u_{1,2} = \frac{\mp 1}{\omega'}\sqrt{2m_\Sigma \delta W}, \quad p^-_{1,2} = \pm m^- \sqrt{\frac{2\delta W}{m_\Sigma}}, \quad p^+_{1,2} = \pm m^+ \sqrt{\frac{2\delta W}{m_\Sigma}}. \tag{5.18}$$

The range of angles is $\Delta u \equiv u_2 - u_1 = 2\sqrt{2m_\Sigma \delta W}/\omega'$. For the radiation frequency, which is equal to the cyclotron frequency $\omega' = hm$ and for the lowest Landau levels $m^- \approx m^+ \approx m$, the interval of radiation angles is $\Delta u = 4\sqrt{\delta W/m}/h$. In the case of a power-law dependence of the above-threshold addition $\delta W$ on a small parameter $h$, that is, when $\delta W \sim h^k m$ ($k$ is a natural number), the angle interval is of the order of magnitude: $\Delta u \sim h^{k/2-1}$. For $k > 2$, we have $\Delta u \ll 1$, that is, radiation occurs in a narrow cone along the direction perpendicular to the field. If $k = 2$, then $\Delta u \sim 1$, that is, radiation occurs in a wide range of directions. Finally, if $k < 2$, then the estimates give $\Delta u \gg 1$, but the physical interval is $\Delta u = 2$, that is, radiation occurs in all directions, while the longitudinal momenta of the particles weakly depend on the angle of emission of the photon.

## 5.3. Probability amplitude and resonance conditions of the OPPE process

<u>The probability amplitude of OPPE.</u> The expression for amplitude of the process corresponds to the Feynman diagrams shown in Figure 5.5.,

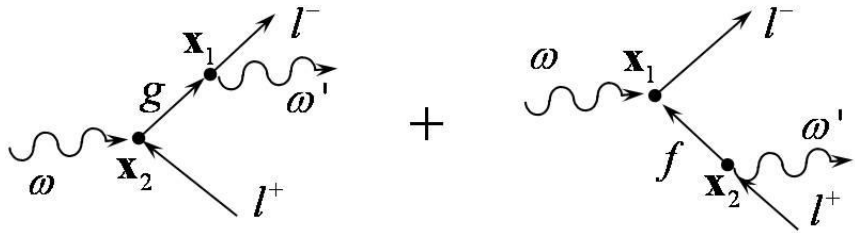

Figure 5.5. Feynman diagrams of the process of one-photon production of an $e^+e^-$ pair with emission of a photon

and has the form:

$$A_{if} = ie^2 \int d^4\mathbf{x}_1 d^4\mathbf{x}_2 \bar{\Psi}^-(\mathbf{x}_1)[\gamma^i A'_i(\mathbf{x}_1)G_{H1}(\mathbf{x}_1,\mathbf{x}_2)\gamma^j A_j(\mathbf{x}_2) +$$
$$+ \gamma^j A_j(\mathbf{x}_1)G_{H2}(\mathbf{x}_1,\mathbf{x}_2)\gamma^i A'_i(\mathbf{x}_2)]\Psi^+(\mathbf{x}_2), \quad (5.19)$$

where $\bar{\Psi}^-(\mathbf{x}_1)$, $\Psi^+(\mathbf{x}_2)$ are wave functions of the final electron and positron; $A_j(\mathbf{x})$, $A'_i(\mathbf{x})$ are wave functions of the initial and final photons; $G_{H1}(\mathbf{x}_1,\mathbf{x}_2)$, $G_{H2}(\mathbf{x}_1,\mathbf{x}_2)$ are Green's functions of the electron in an intermediate state, corresponding to the *g* and *f* diagrams (see Fig. 5.5). Expression (5.19) is obtained from amplitude (4.1) by replacing $A_{2j}(\mathbf{x}) \to A_j(\mathbf{x})$, $A_{1i}(\mathbf{x}) \to A'_i(\mathbf{x})$.

Similarly to the TPP process, after taking the integrals in (5.19), the amplitude of the OPPE probability in the general case can be represented as:

$$A_{if} = (2\pi)^4 \frac{M_{if}}{SV} \delta^3(k-k'-p^- -p^+), \quad M_{if} = M_{if}^{(g)} + M_{if}^{(f)}. \quad (5.20)$$

The value $M_{if}^{(g)}$ corresponds to the first diagram in Figure 5.5 and has the form:

$$M_{if}^{(g)} = \frac{-ie^2}{4\sqrt{\omega_1\omega_2\varepsilon^-\varepsilon^+m^-m^+}} \sum_{n_g=0}^{\infty} \frac{e^{i\Phi_g}Q^g}{g_0^2 - \varepsilon_g^2}, \quad Q^g = \sum_{a=1}^{10} Q_a^g, \quad (5.21)$$

where

$$Q_1^g = -J_1^{++}J_2^{++}[m\tilde{C} + g_0 C + gD], \quad Q_2^g = J_1^{-+}J_2^{++}[-mD + g_0\tilde{D} + gC],$$

$$Q_3^g = J_1^{+-}J_2^{+-}[-mC - g_0\tilde{C} + gD], \quad Q_4^g = J_1^{--}J_2^{+-}[mD - g_0\tilde{D} + gC],$$

$$Q_5^g = J_1^{++}J_2^{-+}[mD + g_0\tilde{D} - gC], \quad Q_6^g = -J_1^{-+}J_2^{-+}[mC - g_0\tilde{C} + gD],$$

$$Q_7^g = -J_1^{+-}J_2^{--}[mD + g_0\tilde{D} + gC], \quad Q_8^g = J_1^{--}J_2^{--}[-mC + g_0\tilde{C} + gD],$$

$$Q_9^g = [-J_1^{++}J_2^{+-} - J_1^{+-}J_2^{++} + J_1^{-+}J_2^{--} + J_1^{--}J_2^{-+}]D\sqrt{2n_g hm},$$

$$Q_{10}^g = [J_1^{-+}J_2^{+-} - J_1^{--}J_2^{++} + J_1^{++}J_2^{--} - J_1^{+-}J_2^{-+}]C\sqrt{2n_g hm}, \quad (5.22)$$

functions $J_1$, $J_2$ have a form similar to (4.14), (4.15):

$$J_1^{++} = J_1(l^+, n_g)M_p^+ e_z, \qquad J_2^{++} = J_2(l^-, n_g)M_m^- e'_z,$$

$$J_1^{-+} = J_1(l^+ - 1, n_g)\mu^+ M_m^+ H_p, \qquad J_2^{-+} = J_2(l^- - 1, n_g)\mu^- M_p^- H_p^{'*},$$

$$J_1^{+-} = J_1(l^+, n_g - 1)M_p^+ H_m, \qquad J_2^{+-} = J_2(l^-, n_g - 1)M_m^- H_m^{'*},$$

$$J_1^{--} = J_1(l^+ - 1, n_g - 1)\mu^+ M_m^+ e_z, \quad J_2^{--} = J_2(l^- - 1, n_g - 1)\mu^- M_p^- e'_z. \quad (5.23)$$

The definition of the notation in (5.22), (5.23) is given in the previous chapters. The parameters $\eta$, $\eta'$, which are defined by expression (3.19), are the arguments of the functions $J_1(l^+, n_g)$, $J_2(l^-, n_g)$. The zero $g_0$ and longitudinal $g$ components of the 4-momentum of the intermediate state in the $g$ diagram, as well as the energy of the intermediate state $\varepsilon_g$, which is at a fixed Landau level $n_g$, respectively, are equal to:

$$g_0 = \omega - \varepsilon^+, \quad g = \omega v - p^+, \quad \varepsilon_g = \sqrt{m^2 + 2n_g hm^2 + g^2}. \quad (5.24)$$

The phase $\Phi_g$ is equal to:

$$\Phi_g = \frac{k_x k_y - k'_x k'_y}{2hm^2} + \frac{g_y(k'_x - k_x)}{hm^2} + \frac{\pi}{2}(l^- - l^+) - (n_g - l^+)\varphi + (n_g - l^-)\varphi'. \quad (5.25)$$

The amplitude $M_{if}^{(f)}$ corresponding to the second diagram in Figure 5.5 has the form (5.21) with the replacement $g \to f$, where

$$Q_1^f = -J_1^{++} J_2^{++} [m\tilde{C} - f_0 C - fD], \quad Q_2^f = -J_1^{++} J_2^{+-} [mD + f_0 \tilde{D} + fC],$$

$$Q_3^f = -J_1^{-+} J_2^{-+} [mC - f_0 \tilde{C} + fD], \quad Q_4^f = J_1^{-+} J_2^{--} [mD + f_0 \tilde{D} - fC],$$

$$Q_5^f = J_1^{+-} J_2^{++} [mD - f_0 \tilde{D} + fC], \quad Q_6^f = -J_1^{+-} J_2^{+-} [mC + f_0 \tilde{C} - fD],$$

$$Q_7^f = -J_1^{--} J_2^{-+} [mD - f_0 \tilde{D} - fC], \quad Q_8^f = -J_1^{--} J_2^{--} [mC + f_0 \tilde{C} + fD],$$

$$Q_9^f = [-J_1^{-+} J_2^{++} - J_1^{++} J_2^{-+} + J_1^{--} J_2^{+-} + J_1^{+-} J_2^{--}] D\sqrt{2n_f hm},$$

$$Q_{10}^f = [J_1^{-+} J_2^{+-} - J_1^{++} J_2^{--} + J_1^{--} J_2^{++} - J_1^{+-} J_2^{-+}] C\sqrt{2n_f hm}, \quad (5.26)$$

functions $J_1$, $J_2$ have the form:

$$J_1^{++} = J_1(n_f, l^-) M_m^- e_z, \qquad J_2^{++} = J_2(n_f, l^+) M_p^+ e'_z,$$

$$J_1^{+-} = J_1(n_f, l^- - 1)\mu^- M_p^- H_m, \qquad J_2^{+-} = J_2(n_f, l^+ - 1)\mu^+ M_m^+ H_m^{'*},$$

$$J_1^{-+} = J_1(n_f - 1, l^-) M_m^- H_p, \qquad J_2^{-+} = J_2(n_f - 1, l^+) M_p^+ H_p^{'*},$$

$$J_1^{--} = J_1(n_f - 1, l^- - 1)\mu^- M_p^- e_z, \quad J_2^{--} = J_2(n_f - 1, l^+ - 1)\mu^+ M_m^- e'_z. \quad (5.27)$$

The zero $f_0$ and longitudinal $f$ components of the 4-momentum of the intermediate state in the $f$ diagram, as well as the energy of the intermediate state $\varepsilon_f$, which is at a fixed Landau level $n_f$, respectively, are equal to:

$$f_0 = \omega - \varepsilon^-, \quad f = \omega v - p^-, \quad \varepsilon_f = \sqrt{m^2 + 2n_f hm^2 + f^2}. \quad (5.28)$$

The phase $\Phi_f$ is equal to:

$$\Phi_f = \frac{k_x k_y - k'_x k'_y}{2hm^2} - \frac{p_y^- k_x + p_y^+ k'_x}{hm^2} + \frac{\pi}{2}(l^- - l^+) - (l^- - n_f)\varphi + (l^+ - n_f)\varphi'. \quad (5.29)$$

<u>Resonant conditions of the OPPE process.</u> Let us consider the question of realization of resonance conditions in the ultraquantum approximation. With an accuracy to the first power of $h$, neglecting the longitudinal momenta of the particles, the energies of the particles can be written in the form:

$$\omega = 2m + \kappa h m, \quad \omega' = \kappa' h m, \quad \varepsilon^{\pm} = m + l^{\pm} h m, \quad \varepsilon_g = m + n_g h m. \qquad (5.30)$$

Energy conservation law (5.1) for expressions (5.30) gives:

$$\kappa - \kappa' = l^+ + l^-. \qquad (5.31)$$

The condition for a resonant process with resonance in the first Feynman diagram *g* (see Figure 5.5) is similar to expression (4.17):

$$g_0 = \varepsilon_g \qquad (5.32)$$

and defines an expression for $\kappa, \kappa'$:

$$\kappa = n_g + l^+, \quad \kappa' = n_g - l^-. \qquad (5.33)$$

Thus, resonance in the first diagram at the threshold of the process (without longitudinal momenta of particles) occurs if the frequency of the initial photon is equal to the sum of the energies of the intermediate electron at the Landau level $n_g$ and the final positron at the Landau level $l^+$, and the frequency of the final photon is equal to the distance between the Landau levels intermediate and final electron.

Similarly, the resonance in the second diagram of Figure 5.5 at the threshold of the process occurs if
the next condition is fulfilled

$$f_0 = \varepsilon_f. \qquad (5.34)$$

In this case, the frequency of the initial photon is equal to the sum of the energies of the intermediate positron at the Landau level $n_f$ and the final electron at the Landau level $l^-$, and the frequency of the final photon is equal to the distance between the Landau levels of the intermediate and final positron

$$\kappa = n_f + l^-, \quad \kappa' = n_f - l^+. \qquad (5.35)$$

The interference of two resonances is determined by the simultaneous fulfillment of conditions (5.33) and (5.35), which leads to such a relation between the Landau levels

$$n_g - l^- = n_f - l^+.  \qquad (5.36)$$

This relation is illustrated in Figure 5.6. Thus, resonance is possible for any above-threshold frequency of the initial photon and, up to $h$, does not depend on the emission angle of the final photon. In resonance, the frequency of the final photon is a multiple of the cyclotron frequency. Resonance conditions with an accuracy of $h$ in both diagrams are fulfilled simultaneously, that is, at resonance there is always "interference" of two diagrams.

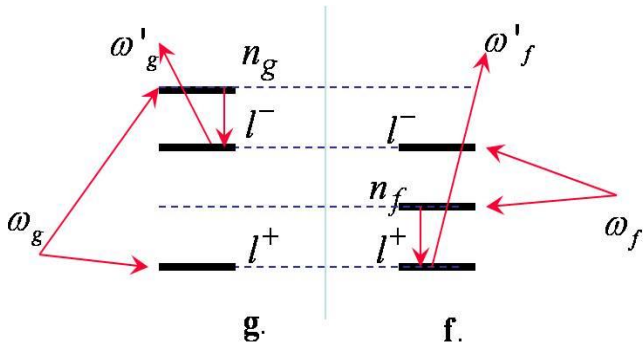

Figure 5.6. Arrangement of Landau levels of particles with interference of resonances of the first and second Feynman diagrams

It should be noted that resonance conditions (5.33) and (5.35) were obtained up to the first power of the small parameter $h$. However, the resonance width is of the order of magnitude $\Gamma \sim e^2 h^2 m$. Therefore, the analysis of conditions (5.32) and (5.34) should be carried out more correctly up to the value of $h^2$ inclusive. Since the momenta of particles in the ultraquantum approximation are multiples of $h^{1/2}$, then $h^{1/2}$ will actually be a small parameter.

Let's start by analyzing the resonance condition in the first diagram (5.32) up to $h^2$ inclusive. Let's represent the frequencies of the initial and final photons in the form:

$$\omega = 2m + (n_g + l^+)hm + \alpha_g hm, \qquad (5.37)$$

$$\omega' = (n_g - l^-)hm + \beta_{1g} h^{3/2} m + \beta_{2g} h^2 m. \qquad (5.38)$$

The case $\alpha_g = 0$, $\beta_{1g} = 0$, $\beta_{2g} = 0$ corresponds to the resonance at the threshold (5.30), (5.33). Note that the two values of the momentum of the particles (5.9) correspond to the fixed values of the photon frequencies and the Landau level numbers of the final particles. Resonance conditions taking into account $h^2$ differ for these two values of the particle momenta. For definiteness, let us denote by $\alpha_{ga}, \beta_{1ga}, \beta_{2ga}$ the coefficients in (5.37), (5.58) corresponding to the upper sign in (5.9), and by $\alpha_{gb}, \beta_{1gb}, \beta_{2gb}$ corresponding to the lower sign. After substituting the photon frequencies of the form (5.37), (5.58) into the expression for the longitudinal momentum of the final electron (5.9) and similarly for the positron, it is easy to find expressions for $g_0$ and $\varepsilon_g$ from (5.24) in the form of an expansion in $h^{1/2}$ up to $h^2$ inclusive. The requirement that the factors are equal for each degree $h$ in the equation $g_0 = \varepsilon_g$ determines the desired coefficients. For the case a) they are equal:

$$\beta_{1ga} = \sqrt{\alpha_{ga}}(n_g - l^-)u, \tag{5.39}$$

$$\beta_{2ga} = -(n_g - l^-)[u^2(n_g - l^- - 2\alpha_{ga}) + n_g + l^- + \alpha_{ga}]/2. \tag{5.40}$$

Thus, the above-threshold value of the frequency of the initial photon $\alpha_{ga}hm$ does not affect the appearance of resonance, but only changes the parameters (5.39), (5.40). These parameters correspond to the momenta of the electron and positron in the form of expansion:

$$p_a^- = \sqrt{\alpha_{ga}}hm - (n_g - l^-)uh +$$

$$+h^{3/2}[-8\alpha_{ga}u^2(n_g - l^-) + 2\alpha_{ga}(n_g + l^+) + 2(n_g^2 + l^{+2}) + \alpha_{ga}^2]/8\sqrt{\alpha_{ga}}, \tag{5.41}$$

$$p_a^+ = -\sqrt{\alpha_{ga}}hm - h^{3/2}[\alpha_{ga}(n_g + l^+) + (n_g^2 + l^{+2}) + \alpha_{ga}^2/2]/4\sqrt{\alpha_{ga}}, \tag{5.42}$$

as well as the energies of these particles

$$\varepsilon_a^- = m + (l^- + \alpha_{ga}/2)hm - \sqrt{\alpha_{ga}}(n_g - l^-)uh^{3/2}m +$$

$$+\frac{h^2 m}{4}[(n_g^2 + l^{+2} - 2l^{-2}) + \alpha_{ga}(n_g + l^+ - 2l^-) + u^2(4\alpha_{ga}(l^- - n_g) + 2(l^- - n_g)^2)], \tag{5.43}$$

$$\varepsilon_a^+ = m + (l^+ + \alpha_{ga}/2)hm + h^2 m(n_g - l^+)(n_g + l^+ + \alpha_{ga})/4. \tag{5.44}$$

For the case b), ie the lower sign in (5.9) the parameters $\beta_{1gb}, \beta_{2gb}$ are equal to the previous value:

$$\beta_{1gb} = \beta_{1ga}, \ \beta_{2gb} = \beta_{2ga}, \qquad (5.45)$$

where you need to replace

$$\alpha_{ga} \to \alpha_{gb}, \ u \to -u. \qquad (5.46)$$

To find expressions for the energies and momenta of final particles, one can use relations (5.41) - (5.44), namely

$$p_b^- = -p_a^-, \ p_b^+ = -p_a^+, \ \varepsilon_b^- = \varepsilon_a^-, \ \varepsilon_b^+ = \varepsilon_a^+ \qquad (5.47)$$

with replacement (5.46).

When choosing photon frequencies (5.37), (5.38) with coefficients (5.39), (5.40) and selecting electrons and positrons with energies and momenta (5.41) - (5.44) (case a)), the denominator of the Green's function within a given accuracy is equal to zero, and when selecting particles with energies and momenta (5.47), it is proportional to $h^{3/2}$:

$$(g_0 - \varepsilon_g)\big|_a = 0(h^2), \ (g_0 - \varepsilon_g)\big|_b = 2\sqrt{\alpha_g}(n_g - l^-)uh^{3/2}. \qquad (5.48)$$

A similar analysis of the resonance conditions takes place for the second Feynman diagram (see Fig.5.5) using equation (5.34). The frequencies of the initial and final photons are given in the form:

$$\omega = 2m + (n_f + l^-)hm + \alpha_f hm, \qquad (5.49)$$

$$\omega' = (n_f - l^+)hm + \beta_{1f}h^{3/2}m + \beta_{2f}h^2 m. \qquad (5.50)$$

For the case a), that is, the upper sign in (5.9), equation (5.34) is realized up to $h^2$, inclusive, if the coefficients $\beta_{1fa}, \beta_{2fa}$ are:

$$\beta_{1fa} = -\sqrt{\alpha_{fa}}(n_f - l^+)u, \qquad (5.51)$$

$$\beta_{2fa} = -(n_f - l^+)[u^2(n_f - l^+ - 2\alpha_{fa}) + n_f + l^+ + \alpha_{fa}]/2. \qquad (5.52)$$

Note that formulas (5.49) - (5.52) are obtained from similar expressions from the previous analysis of the first Feynman diagram with the following replacement of parameters:

$$\alpha_{ga} \to \alpha_{fa}, \ n_g \to n_f, \ l^+ \leftrightarrow l^-, \ u \to -u. \qquad (5.53)$$

Longitudinal momenta and energies of the electron and positron are obtained from expressions (5.41) - (5.44) according to the rule

$$p_a^- \to -p_a^+, \ p_a^+ \to -p_a^-, \ \varepsilon_a^- \to \varepsilon_a^+, \ \varepsilon_a^+ \to \varepsilon_a^- \qquad (5.54)$$

with subsequent replacement (5.53).

When resonance is realized in the case of the lower sign in (5.9) (case b)), the coefficients $\beta_{1fb}, \beta_{2fb}$ differ from expressions (5.51), (5.52) by the sign of the cosine of the angle $u$, that is, equal

$$\beta_{1fb} = \beta_{1fa}, \ \beta_{2fb} = \beta_{2fa}, \qquad (5.55)$$

with replacement $\alpha_{ga} \to \alpha_{gb}$, $u \to -u$. The energies and momentum of the particles can be found by rule (5.47) with the same substitution.

The interference of resonances in the first and second Feynman diagrams takes place if the energies and momenta of all the corresponding particles coincide. For the equality of the photon frequencies, which are determined by expressions (5.37), (5.38) and (5.49), (5.50), the equality of the corresponding coefficients is necessary:

$$\alpha_g = \alpha_f, \ \beta_{1ga} = \beta_{1fa}, \ \beta_{2ga} = \beta_{2fa}. \qquad (5.56)$$

Here, for definiteness, case a) is written, that is, the upper sign in expression (5.9). From equations (5.56) follow the conditions of interference of resonances:

$$l^- = l^+, \ n_g = n_f, \ u = 0, \qquad (5.57)$$

which are much more stringent than conditions (5.36), which are necessary for the interference of resonances only up to $h$.

Note that the production of a final electron and positron at the same energy levels is not the most probable, since the process in resonance is two-stage. At the first stage, particles are born at equal levels, and at the second stage, one of the particles, emitting a final photon, goes to a lower Landau level.

The case of equality $\omega'_{ga} = \omega'_{fb}$ takes place if the first two equations in (5.57) hold for arbitrary u. In this case, the longitudinal momenta of the electron (positron) in the resonance of the first and second diagrams differ in sign. That is, this is not the interference of two resonances, although the resonances themselves on the graph of the dependence of the process probability on the radiation frequency of the final photon will coincide.

As an illustration, Figure 5.7. depicts the dependence of the denominators of Green's functions on the frequency of radiation (Figure 5.7.a.) and schematic arrangement of resonances (Figure 5.7. b.).

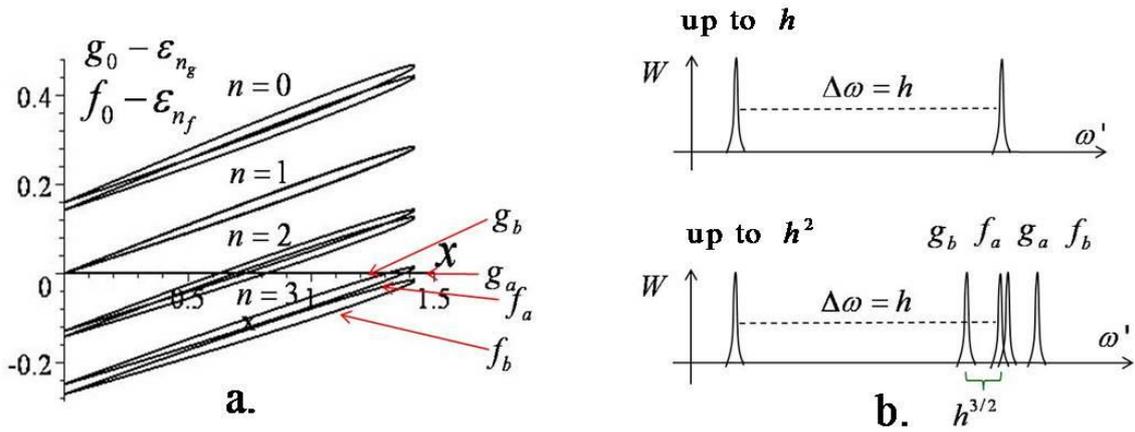

Figure 5.7. a. Dependence of the quantities $g_0$-$\varepsilon_g$ and $f_0$-$\varepsilon_f$ on the radiation frequency ($x=\omega'/hm$) b. Location of resonances on the graph of the dependence of the probability of the OPPE process on the radiation frequency

To construct the dependences of $g_0$-$\varepsilon_g$ and $f_0$-$\varepsilon_f$ on the radiation frequency ($x=\omega'/hm$) (see Fig. 5.7.a.) it was chosen: $l^-=2$, $l^+=1$, $n_f=n$, $n_g=n+1$ ($n=0..4$), $\omega=2m+(4+\alpha)$, $\alpha=1/20$, $u=1/20$, $h=0.2$. In Figure.5.7.a. 4 groups of curves are presented (cases $n = 0,1,2,3$). In each group there are four monotonously growing branches ($g_a$, $g_b$, $f_a$, $f_b$). In the approximation $h$ in the first power the curves of one group merge into one line. The intersection of zero corresponds to resonance. Therefore, up to $h^1$ there is only one resonance. With an accuracy of $h^2$, there are already four resonances (two diagrams for cases a and b). These zeros correspond to the resonant peaks schematically shown in Figure.5.7.b. Two pairs of resonances correspond to two different kinematic regions (cases a and b). One pair is the outer peaks (case b), for which the distance between the peaks is proportional $\sim h^{3/2}$. The second pair are internal peaks (case a), for which the distance between them can be $\ll h^{3/2}$.

Thus, taking into account the higher degree of the small parameter h in the analysis of resonance conditions leads to the appearance of pair resonances, the distance between which is much less than the distance between neighboring Landau levels. Pair resonances are clear evidence of the existence of two Feynman diagrams describing the OPPE process. The whole range of possible values of the radiation frequency can be divided into three sections (Fig.5.7.b.): 1) a narrow interval of a single resonance, its width $\Gamma \sim e^2 h^2 m$, 2) a region of pair resonances with the width $\sim h^{3/2} m$, 3) a nonresonant region with a width $\sim hm$.

### 5.4. Probability of the OPPE process

The amplitudes $Q^{(g)}$ (5.21) of the OPPE process in the ultraquantum approximation with fixed values of the projections of the electron and positron spins and photon frequencies close to the values (5.37), (5.38) are written in the form:

$$Q_{\text{OPPE}}^{-+(g)} = 4\sqrt{2}\sqrt{\frac{n_g h}{l^-}} m^3 G_p G_s e_z H_m^{'*} \operatorname{sgn}(p^-), \tag{5.58}$$

$$Q_{\text{OPPE}}^{--(g)} = -4\sqrt{\frac{n_g l^+}{l^-}} hm^3 G_p G_s H_p H_m^{'*}, \tag{5.59}$$

$$Q_{\text{OPPE}}^{++(g)} = 4\sqrt{n_g}\, hm^3 G_p G_s H_m H_m^{'*}, \tag{5.60}$$

$$Q_{\text{OPPE}}^{+-(g)} \sim h^2, \tag{5.61}$$

where the notation is similar to the expressions (4.52) - (4.55). Functions $G_p$, $G_s$ are similar (4.57):

$$G_p = \frac{(-1)^{l^+} \eta^{\frac{n_g + l^+}{2}} e^{-\eta/2}}{\sqrt{(l^+)!\,(n_g)!}}, \quad G_s = e^{-\eta'/2}\eta'^{\frac{n_g - l^- - 1}{2}}\sqrt{\frac{(n_g-1)!}{(l^- - 1)!}}\frac{1}{(n_g - l^- - 1)!} \tag{5.62}$$

and depend on the variables $\eta$, $\eta'$, respectively

$$\eta = \frac{\omega^2}{2hm^2}, \quad \eta' = \frac{\omega'^2(1-u^2)}{2hm^2}. \tag{5.63}$$

To establish the relationship between the amplitudes of the processes of OPPE and OPP, SR we write the latter in the ultraquantum approximation. For the SR process with the transition of the electron from the level of $n_g$ to $l^-$, the values of $Q_{SR}$ in the amplitude (2.39) with the same fixed electron spins are equal to:

$$Q_{SR}^{--} = -2\sqrt{2}\sqrt{\frac{n_g h}{l^-}} m^2 G_s H_m^{'*}, \quad Q_{SR}^{++} = -2\sqrt{2}\sqrt{h} m^2 G_s H_m^{'*}. \tag{5.64}$$

For the OPP process with the electron at the $n_g$ level and the positron at the $l^+$ level with fixed spins of particles, the value of $Q_{OPP}$ in the amplitude (2.142) is equal to:

$$Q_{OPP}^{-+} = 4m^2 e_z G_p \operatorname{sgn}(p^-), Q_{OPP}^{--} = -2\sqrt{2} m^2 \sqrt{l^+ h} G_p H_p, Q_{OPP}^{++} = +2\sqrt{2} m^2 \sqrt{l^- h} G_p H_m. \tag{5.65}$$

The amplitudes (5.58) - (5.60) are expressed in terms of (5.64), (5.65) by simple relations:

$$Q_{OPPE}^{-+(g)} = \frac{-1}{2m} Q_{OPP}^{-+} Q_{SR}^{--}, \quad Q_{OPPE}^{--(g)} = \frac{-1}{2m} Q_{OPP}^{--} Q_{SR}^{--}, \quad Q_{OPPE}^{++(g)} = \frac{-1}{2m} Q_{OPP}^{++} Q_{SR}^{++}. \tag{5.66}$$

The phase (5.25) of the OPPE process is obviously equal to the sum of phases (2.36) and (2.146) of the first-order processes

$$\Phi_{OPPE} = \Phi_{OPP} + \Phi_{SR}. \tag{5.67}$$

As a result, the sought-for relationship between the amplitudes for the first $g$ diagram has the form:

$$A_{OPPE}^{\mu^-\mu^+(g)} = \frac{-i2mS}{(2\pi)^3} \int d^3 g \sum_{n_g} \frac{A_{OPP}^{\mu_{n_g}\mu^+} A_{SR}^{\mu_{n_g}\mu^-}}{g_0^2 - \varepsilon_g^2}, \tag{5.68}$$

in this case, the spin of the intermediate electron is

$$\mu_{n_g} = \mu^-. \tag{5.69}$$

Note that in the OPPE process in the ultraquantum approximation, the intermediate electron emits a final photon without spin flip (spin flip occurred in the DPP process).

Similar to expression (5.68) there is a relationship of amplitudes for the second $f$ diagram

$$A_{\text{OPPE}}^{\mu^-\mu^+(f)} = \frac{-i2mS}{(2\pi)^3} \int d^3 f \sum_{n_f} \frac{A_{\text{OPP}}^{\mu^-\mu_{n_f}} A_{\text{SR}}^{\mu_{n_f}\mu^+}}{f_0^2 - \varepsilon_f^2}, \qquad (5.70)$$

where $\mu_{n_f} = \mu^+$; $A_{\text{OPP}}^{\mu^-\mu_{n_f}}$ is the amplitude of the process of one-photon e$^+$e$^-$ pair production with the final electron and the intermediate positron at the Landau levels $l^-$ and $n_f$, respectively. $Q_{\text{OPP}}$ values included in the amplitude have the form:

$$Q_{\text{OPP}}^{-+} = 4m^2 e_z F_p \, \text{sgn}(p^-), \; Q_{\text{OPP}}^{--} = -2\sqrt{2}m^2\sqrt{n_f h} F_p H_p, \; Q_{\text{OPP}}^{++} = +2\sqrt{2}m^2\sqrt{l^- h} F_p H_m. \quad (5.71)$$

$A_{\text{SR}}^{\mu_{n_f}\mu^+}$ is the amplitude of the process of synchrotron radiation of the positron with $Q_{\text{SR}}$ values that have the form

$$Q_{\text{SR}}^{++} = 2\sqrt{2}\sqrt{\frac{n_f h}{l^+}} m^2 F_s H_p^{'*}, \quad Q_{\text{SR}}^{--} = 2\sqrt{2}\sqrt{h} m^2 F_s H_p^{'*}. \qquad (5.72)$$

$F_p$, $F_s$ Functions have the form:

$$F_p = \frac{(-1)^{n_f} \eta^{\frac{n_f + l^-}{2}} e^{-\eta/2}}{\sqrt{(l^-)! \cdot (n_f)!}}, \quad F_s = e^{-\eta'/2} \eta'^{\frac{n_f - l^+ - 1}{2}} \sqrt{\frac{(n_f - 1)!}{(l^+ - 1)!}} \frac{1}{(n_f - l^+ - 1)!}. \quad (5.73)$$

The phases of the amplitudes are related as follows

$$\Phi_{\text{OPPE}} = \Phi_{\text{OPP}} + \Phi_{\text{SR}} + \pi(n_f - l^+). \qquad (5.74)$$

Values $Q^{(f)}$ similar to (5.66) can be written as:

$$Q_{\text{OPPE}}^{-+(f)} = \frac{-1}{2m} Q_{\text{OPP}}^{-+} Q_{\text{SR}}^{++}, \quad Q_{\text{OPPE}}^{--(f)} = \frac{-1}{2m} Q_{\text{OPP}}^{--} Q_{\text{SR}}^{--}, \quad Q_{\text{OPPE}}^{++(f)} = \frac{-1}{2m} Q_{\text{OPP}}^{++} Q_{\text{SR}}^{++}. \quad (5.75)$$

Thus, the general expression for the amplitude of the OPPE process in the ultraquantum approximation is the sum of two quantities (5.68) and (5.70).

The amplitude of the OPPE process near the resonance (5.36) with particles in the main spin states ($\mu^-$=-1, $\mu^+$=+1) without insignificant total phase has the form

$$A_{\text{OPPE}}^{-+} = N_1 e_z \Upsilon', \quad N_1 = \frac{(2\pi)^4 e^2 h^{1/2} R \cdot \delta^3(k - k' - p^- - p^+)}{2VS\sqrt{m\omega'}}, \qquad (5.76)$$

$$\Upsilon' = \left[ \frac{H_m^{'*} e^{i\kappa'\chi}}{(\omega' - \varepsilon_g + \varepsilon^-) + i\frac{n_g \Gamma}{2}} - \frac{H_p^{'*} e^{-i\kappa'\chi}}{(\omega' - \varepsilon_f + \varepsilon^+) + i\frac{n_f \Gamma}{2}} \right], \qquad (5.77)$$

where width $\Gamma = 4e^2 h^2 m/3$, $R = (n_g/l^-)^{1/2} |G_p| G_s$, the $\chi$ value has the form:

$$\chi = (\varphi' - \varphi) - \sin\theta' \cdot \sin(\varphi' - \varphi). \qquad (5.78)$$

The expression for the amplitude is simplified in the case of particle production at the same Landau levels ($l^+ = l^- = l, n_g = n_f = n$):

$$\Upsilon' = \frac{2i(u\cos\alpha' \cdot \sin\kappa'\chi + \sin\alpha' \cdot \cos\kappa'\chi \cdot e^{-i\beta'})}{(\omega' - \omega_{res}) + i\frac{n\Gamma}{2}}, \qquad (5.79)$$

The amplitudes of the process with particles in other spin states have a similar simple form:

$$A_{\text{OPPE}}^{--} = N_1 \sqrt{\frac{l^+ h}{2}} H_p \Upsilon', \quad A_{\text{OPPE}}^{++} = N_1 \sqrt{\frac{l^- h}{2}} H_m \Upsilon'. \qquad (5.80)$$

<u>Resonant probability of the OPPE process.</u> Let us write down an expression for the probability of the OPPE process at a point near the resonance of the first *g* diagram with an accuracy of $h^2$ (a narrow region of the individual resonance), when conditions (5.39), (5.40) are satisfied, and particles are produced in the ground spin states $W^{-+}$. The probabilities with other spin states $W^{--}$, $W^{++}$ are not difficult to obtain by replacing the expression for the amplitude (5.76) by (5.80). The second term in (5.77) can be neglected. The differential probability of the OPPE process is equal to the product of square of modulus of the amplitude (5.76) by the number of final states *dN*,

$$dN = \frac{VS^2}{(2\pi)^7} d^3k' \cdot d^2p^+ \cdot d^2p, \qquad (5.81)$$

and can be reduced to the form:

$$dW_{\text{OPPE}}^{-+(g)} = \frac{e^4}{64(2\pi)^2 \omega \omega' m^6} \left| \frac{Q_{\text{OPPE}}^{-+(g)}}{g_0 - \varepsilon_g + i\frac{n_g \Gamma}{2}} \right|^2 \delta(\omega - \omega' - \varepsilon^+ - \varepsilon^-) d^3k' \cdot dp_z \frac{p_y T}{L_x}. \qquad (5.82)$$

The $p_yT/L_y$ multiplier is traditionally interpreted as $hm^2$. The Dirac delta function of the particle energies removes the integral over the longitudinal momentum of the electron, $p \equiv |p_z| = (\alpha_g h)^{1/2} m$. In this case, the probability does not depend on the azimuthal angles of photons and is equal to:

$$\frac{dW_{OPPE}^{-+(g)}}{d\omega' du} = \frac{e^4 h\omega' |Q_{OPPE}^{-+(g)}|^2}{2^8 (2\pi) m^4 p} \left[ \frac{1}{(g_0 - \varepsilon_g)_a^2 + \frac{n_g^2 \Gamma^2}{4}} + \frac{1}{(g_0 - \varepsilon_g)_b^2 + \frac{n_g^2 \Gamma^2}{4}} \right]. \quad (5.83)$$

The subscripts $a$ and $b$ at the brackets in the denominator of the terms in square brackets of expression (5.83) correspond to the kinematics of the process described by expression (5.9) with upper and lower signs, respectively. Note that only one of the two terms remains essential in the square brackets of expression (5.83) at resonance up to $h^2$. These terms can be considered equal with an accuracy of the order of $h^1$. Taking into account the relationship between the amplitudes (5.66), as well as expressions for the probabilities SR of an electron $dW_{SR\,e^-}^{--}/du$ (2.46) and OPP $W_{OPP}^{-+}$ (2.156), the expression for the OPPE probability can be reduced to the Breit-Wigner form (for definiteness, the first term is left (case a)):

$$\left.\frac{dW_{OPPE}^{-+(g)}}{d\omega' du}\right|_a = \frac{1}{4\pi} \frac{W_{OPP}^{-+} \cdot \dfrac{dW_{SR\,e^-}^{--}}{du}}{(\omega' - (\varepsilon_g - \varepsilon^-))_a^2 + \dfrac{n_g^2 \Gamma^2}{4}}. \quad (5.84)$$

Factorization of expression (5.84) means the independence of the processes of the electron - positron pair production by the initial photon and emission of the final photon by the intermediate electron. Similar relations are valid for other spin states of particles:

$$\left.\frac{dW_{OPPE}^{\mu^-\mu^+(g)}}{d\omega' du}\right|_a = \frac{1}{4\pi} \frac{W_{OPP}^{\mu_{n_g}\mu^+} \cdot \dfrac{dW_{SR\,e^-}^{\mu_{n_g}\mu^-}}{du}}{(\omega' - (\varepsilon_g - \varepsilon^-))_a^2 + \dfrac{n_g^2 \Gamma^2}{4}}, \quad (5.85)$$

where $\mu_{n_g}$ is determined from (5.69).

Let us estimate the integral probability of the OPPE process in the region of resonances a) and b), where conditions (5.39), (5.40) are satisfied with accuracy $h^2$. Taking into account

$$\int_{\Gamma} \frac{A dx}{(x-x_0)^2+\Gamma^2/4} \approx A \int_{-\infty}^{+\infty} \frac{dx}{(x-x_0)^2+\Gamma^2/4} = \frac{2\pi A}{\Gamma} \quad (5.86)$$

the integration of expression (5.83) over the frequency $\omega'$ within the width $n_g\Gamma$ and over the polar angle of photon emission (over $u$ from -1 to 1) gives

$$\Delta W_{OPPE}^{-+(g)} = \frac{W_{OPP}^{-+} \cdot W_{SR\,e^-}^{--}}{n_g \Gamma}, \quad (5.87)$$

where the total probability of the SR process of an electron in the ground spin state $W_{SR\,e^-}^{--}$ is determined by the integral of expression (2.46) over the photon emission angle with subsequent summation over the polarizations of the final photon. The largest integral probability OPPE (5.87) corresponds to the transition of an electron to the neighboring Landau level $W_{CB\,e^-}^{--} = 4 n_g e^2 h^2 m/3 = n_g \Gamma$, then

$$\Delta W_{OPPE}^{-+(g)} = W_{OPP}^{-+}, \quad (5.88)$$

that is, we found that the second-order OPPE process in the resonance region is equiprobable with the first-order OPP process. In the case of the $e^+e^-$ pair production in the ground energy states $l^-=l^+=0$ ($n_g=1$) in a magnetic field of value $h=0.1$ ($H=4.4 \cdot 10^{12} Гс$), the probabilities of OPP and SR are equal in order of magnitude, respectively:

$$W_{OPP}^{-+} \sim 10^{10}[1/c], \quad W_{SR\,e^-}^{--} \sim 10^{17}[1/c]. \quad (5.89)$$

Near the resonance of the second Feynman diagram (diagram $f$ Fig.5.5), when the conditions (5.39), (5.40) are satisfied, the probability of ONPV is equal to expression (5.83) with replacement $g \to f$, $g_0 - \varepsilon_g \to f_0 - \varepsilon_f = \omega' - (\varepsilon_f - \varepsilon^+)$. The analysis performed for the first diagram is also valid in this case. The probability of the process in the resonance is factorized, the integral probability of OPPE is equal to the probability of OPP (5.89), ie taking into account both diagrams gives a double result (5.89). The difference is that the intermediate particle for the first diagram is an electron, and for the second diagram is a positron.

Synchrotron radiation of a positron. Since in the second Feynman diagram the intermediate state is positron, it makes sense to give some expressions for the probability of the SR of positron. The amplitude of the probability of the SR of positron is equal to (2.39), where

$$Q_1 = J(l',l)M_p^+ M_p\,'De\,'_z, \quad Q_2 = J(l'-1,l-1)\mu^+ M_m^+ \mu^+\,'M_m\,'De\,'_z,$$

$$Q_3 = J(l',l-1)\mu^+ M_m^+ M_p^+\,'CH\,'^*_p, \quad Q_4 = J(l'-1,l)M_p^+ \mu^+\,'M_m^+\,'CH\,'^*_m,$$

phase $\Phi^+ = -\dfrac{k_x\,'(2p_y^+ + k_y\,')}{2hm^2} - (l-l')(\varphi' + \dfrac{\pi}{2})$. The plus sign at the top left indicates the positron, the dash means that the particle is in the final state.

The differential probability of the process per unit time (rate of the process) is determined by expression (2.44), which in the ultraquantum approximation is equal to

$$\frac{dW}{du} = \frac{\alpha\omega'}{16m^3}\left|\sum Q\right|^2,$$

analogs of expressions (2.46) - (2.48) are

$$\frac{dW^{++}_{SR\,e^+}}{du} = e^2\omega'h\frac{l}{2l'}R_{ll'}^2\,|H\,'_p|^2, \quad Q^{++} = -2\sqrt{\frac{2lh}{l'}}m^2 R_{ll'}H\,'^*_p, \qquad (5.90)$$

$$\frac{dW^{--}_{SR\,e^+}}{du} = e^2\omega'h\frac{1}{2}R_{ll'}^2\,|H\,'_p|^2, \quad Q^{--} = 2\sqrt{2hm^2 R_{ll'}H\,'^*_p}, \qquad (5.91)$$

$$\frac{dW^{-+}_{SR\,e^+}}{du} = e^2\omega'h^2\frac{(l-l')^2}{4l'}R_{ll'}^2\,|\tilde{H}\,'_p|^2, \quad Q^{-+} = \frac{2(l-l')h}{\sqrt{l'}}m^2 R_{ll'}\tilde{H}\,'^*_p. \qquad (5.92)$$

Note, taking into account the form of the polarization functions (2.43), (4.56), that the SR of a positron differs from the SR of an electron only by the opposite sign of circular polarization.

Probability of the OPPE process in the region of pair resonances. As shown above, for any above-threshold frequency of the initial photon, the probability of emission of a photon with a frequency that is a multiple of the cyclotron frequency has a resonant character. In this case, conditions (5.36) determine the energy levels of intermediate particles $n_g$, $n_f$. Under these conditions, hitting the resonance has an accuracy $h$, that corresponds to the region of pair resonances (see Figure 5.7.b), excluding the narrow peaks themselves. There is interference of two Feynman diagrams, that is, it is necessary to take into account the contribution from both diagrams. For definiteness, we will choose the same energy levels of particles ($l^+ = l^- = l, n_g = n_f = n$), that simplifies analytical expressions, but does not detract from the generality of consideration. The square of the module $\Upsilon'$ (5.79) is equal to:

$$|\Upsilon'|^2 = \frac{K(\xi',\Omega')}{(\omega - \omega_{res})^2 + n^2\Gamma^2/4}, \qquad (5.93)$$

$$K(\xi',\Omega') = 2[u^2 \sin^2 \kappa'\chi + \cos^2 \kappa'\chi + \xi_3'(u^2 \sin^2 \kappa'\chi - \cos^2 \kappa'\chi) + \xi_1' u \sin 2\kappa'\chi]. \quad (5.94)$$

The second term with width $\Gamma$ in the denominator (5.93), in principle, should be discarded, because in this area it is very small. The differential probability of OPPE has the form:

$$\left.\frac{dW_{\text{ОНПВ}}^{-+}}{d\omega' du}\right|_a = \frac{B \cdot K(\xi',\Omega')}{(\omega' - \omega_{res})_a^2 + \frac{n^2\Gamma^2}{4}}, \quad B = \frac{(e^2 Rhme_z)\omega'}{2(4\pi)^2 p}, \qquad (5.95)$$

where the lower index $a$ corresponds to the longitudinal momentum of the electron directed along the field (upper sign (5.9)).

The angular dependence of the probability of OPPE is characterized by the function $K(\xi',\Omega')$ shown in Fig.5.8. The function $K(\xi',\Omega')$ with parameters $l=l'$, $\xi_1'=\xi_3'=2^{-1/2}$, $\kappa'=1$ has maxima at $u=1$ and $\varphi'-\varphi = n\pi/2$. The significant dependence of the process probability on the difference between the azimuthal angles of the initial and final photons in the region of an individual resonance is the difference between the region of pair resonances and the narrow region of an individual resonance.

The differential probability of the OPPE process (5.95), summed over the polarization of the final photon, is proportional to the quantity

$$\sum_{\xi'} K(\xi',\Omega') = 4[u^2 \sin^2 \kappa'\chi + \cos^2 \kappa'\chi]. \qquad (5.96)$$

The value (5.96) has a maximum value of $\Sigma K = 4$, when the radiation is directed along the magnetic field, it does not depend on the azimuthal angles of the photons. If the radiation is perpendicular to the field, the value (5.96) is maximum, when the difference in azimuthal angles $\Delta\varphi = \{0, \pi\}$ and is equal to zero, when $\Delta\varphi = 3\pi/4$.

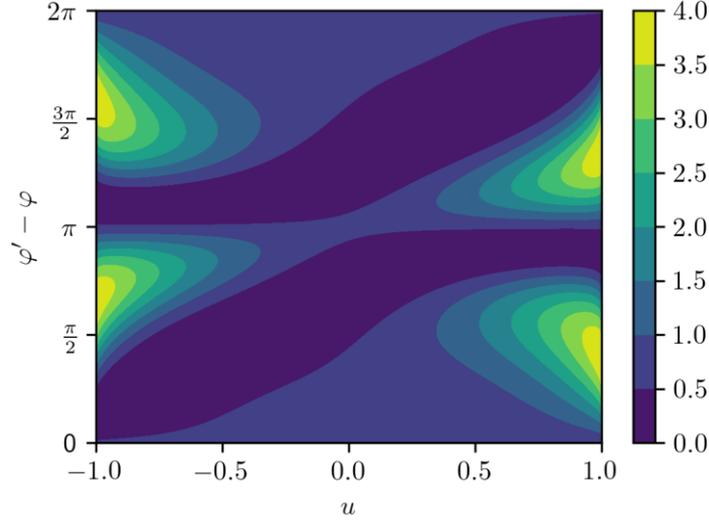

Figure 5.8. Angular dependence of the probability of OPPE in the region of pair resonances, $l=l'$, $\xi_1' = \xi_3' = 2^{-1/2}$, $\kappa' = 1$

Thus, when photons propagate $\perp \vec{H}$, the radiation in the plane perpendicular to the field in the region of pair resonances is maximum in the direction (parallel and antiparallel) of the initial photons and is absent at an angle $\Delta\varphi = 3\pi / 4$. In the region of individual resonance, the radiation does not depend on azimuthal angles.

The probability of the OPPE process in the nonresonant region. Consider the process near the threshold, when particles are produced in the ground energy states:

$$\omega = 2m + amh^2, \quad l^+ = l^- = 0, \quad \mu^+ = -\mu^- = 1, \qquad (5.97)$$

where $a \sim 1$. In this case, the frequency of the final photon and the momentum of the final electron can be written as:

$$\omega' = \kappa' m h^2, \quad |p^-| = \sqrt{a - \kappa'} hm, \quad 0 < \kappa' < a. \qquad (5.98)$$

Only two terms $n_g = 0, 1$ ($n_f = 0, 1$) remain in the amplitude (5.20), (5.21) in the sums over the energy levels of the intermediate particle. The amplitude can be reduced to the form

$$A_{if} = \frac{-ie^2(2\pi)^4 \delta^3(k-k'-p^+ -p^-)}{4VS\sqrt{\omega\omega' m^+ m^- \varepsilon^+ \varepsilon^-}} \Sigma, \quad \Sigma = 4m\frac{1}{h}e^{-1/h}e_z e^{-i\Lambda/2}Y, \quad (5.99)$$

$$Y = \left[\Delta e'_z \frac{\sqrt{a-\kappa'}}{\kappa'} - H_m^{'*} e^{i\Delta\varphi} + H_p^{'*} e^{i\Lambda - i\Delta\varphi}\right], \quad (5.100)$$

$$\Delta\varphi = \varphi' - \varphi, \quad \Lambda = 2\kappa' h \cdot \sin\theta' \cdot \sin\Delta\varphi, \quad \Delta = \mathrm{sgn}(p^-). \quad (5.101)$$

The differential probability of the process per unit time is equal to the square of the module (5.99) multiplied by the number of final states:

$$dW = \frac{\pi\alpha^2 h^2 e^{-2/h}}{\sqrt{a-\kappa'}} \kappa' |e_z|^2 \cdot |Y|^2 \, d\omega' d\Omega', \quad (5.102)$$

where

$$|Y|^2 = \frac{a-\kappa'}{\kappa'^2} \cdot K + \frac{\Delta\sqrt{a-\kappa'}}{\kappa'} \cdot L + M, \quad (5.103)$$

$$K = (1+\xi'_3)(1-u^2)/2, \quad (5.104)$$

$$L = (1+\xi'_3)\sin 2\theta' (\cos\Delta\varphi - \cos(\Delta\varphi - \Lambda))/2 +$$
$$+ \xi'_2 \sin\theta'(\cos\Delta\varphi - \cos(\Delta\varphi - \Lambda)) - \xi'_1 \sin\theta'(\sin\Delta\varphi + \sin(\Delta\varphi - \Lambda)), \quad (5.105)$$

$$M = (1+u^2) - \xi'_3(1-u^2) + ((1-u^2) - \xi'_3(1+u^2))\cos(2\Delta\varphi - \Lambda) + 2\xi'_1 u\sin(2\Delta\varphi - \Lambda). \quad (5.106)$$

The angular dependence of the differential probability of the OPPE process (the dependence of the value $|Y|^2$ on the angles $\theta'$ and $\varphi'$) is shown in Fig.5.9. To get the graphs Fig.5.9. the following parameters are selected $h=0.1$, $a=1$, $\kappa'=2/3$, $\Delta=1$.

After integrating expression (5.102) over the polar and azimuthal angles of the final photon, the differential probability takes the form:

$$\frac{dW}{d\omega'} = \frac{2\pi^2}{3} e^4 h^2 e^{-2/h}(1+\xi_3)Z, \quad Z = \left\{\frac{\sqrt{a-\kappa'}}{\kappa'}(1+\xi'_3) + \frac{2\kappa'}{\sqrt{a-\kappa'}}(2-\xi'_3)\right\}. \quad (5.107)$$

Figure 5.10 shows the spectral dependence of the probability of the OPPE process at different values of the Stokes parameter $\xi'_3$. with $h=0.1$, $a=1$. After summation over the polarizations of the final photon and averaging over the polarizations of the initial photon, the probability of the OPPE process per unit time has the form:

$$\frac{dW}{d\kappa'} = \frac{4\pi^2}{3} e^4 m h^4 e^{-2/h} \left\{ \frac{\sqrt{a-\kappa'}}{\kappa'} + \frac{4\kappa'}{\sqrt{a-\kappa'}} \right\}. \qquad (5.108)$$

Total probability

$$W = \int_0^a \frac{dW}{d\kappa'} d\kappa'$$

logarithmically diverges at the lower bound of integration. The reason for this so-called infrared divergence is that the emission of soft photons is not taken into account in the framework of the perturbation theory of the quantum theory of scattering [190].

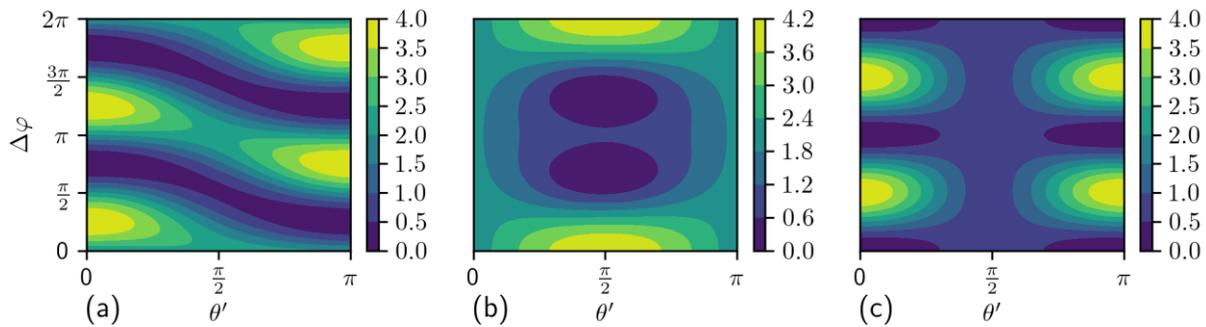

Figure 5.9. Angular dependence of the probability of the OPPE process in nonresonant region *a*) $\xi'_1 = 1$, *b*) $\xi'_2 = 1$, *c*) $\xi'_3 = 1$

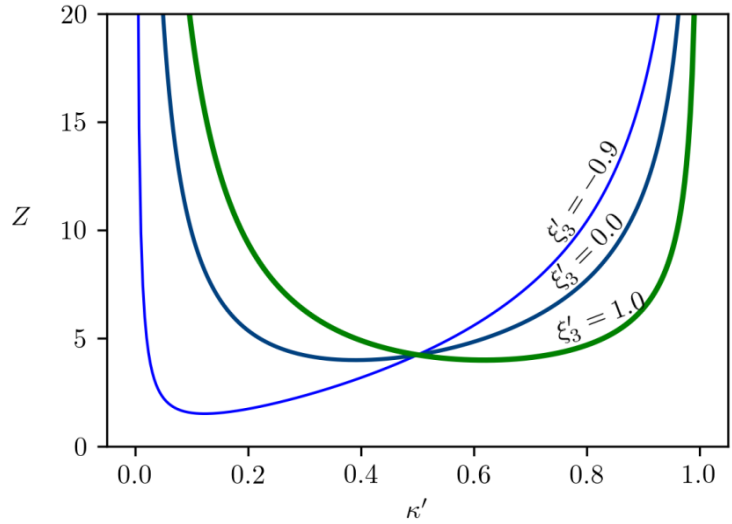

Figure 5.10. Spectral distribution of the probability of the OPPE at various values of the Stokes parameter $\xi'_3$

The elimination of this divergence is carried out by replacing at the lower boundary $0 \to \kappa_{min} = \omega_{min}/h^2 m$, that gives an expression for the total probability per unit time of the OPPE process:

$$W = \frac{4\pi^2}{3} e^4 m h^4 e^{-2/h} \sqrt{a} \left\{ \ln \frac{a}{\kappa_{min}} + \frac{16a}{3} \right\}. \qquad (5.109)$$

Estimation of the probability $W$ in the case when $h = 0.1$, $a = 1$, $\ln = 10$, gives

$$W \approx 10^6 s^{-1}. \qquad (5.110)$$

The integral probability of the process within the isolated resonance is equal to the total probability of the ONP (5.88) and for the parameters as in the estimation (5.110) is equal to $10^{10} s^{-1}$ (5.89), that is, 4 orders of magnitude greater than the obtained estimation for the nonresonant process (5.109).

### 5.5. Spin-polarization effects of the OPPE process

To analyze the spin-polarization properties of the particles involved in the OPPE process, it is convenient to write the probabilities of the process with explicitly selected polarization functions. Expressions for the differential probabilities of the OPPE process in the region of individual resonance of the first Feynman diagram $g$ have the form:

$$\frac{dW_{OPPE}^{-+(g)}}{d\omega' du} = C_g \frac{n_g}{l^-} \tilde{\Pi}\Pi', \quad \frac{dW_{OPPE}^{--(g)}}{d\omega' du} = C_g \frac{hl^+ n_g}{2l^-} \Pi\Pi', \quad \frac{dW_{OPPE}^{++(g)}}{d\omega' du} = C_g \frac{hn_g}{2} \Pi\Pi', \quad (5.111)$$

where the polarization functions $\Pi, \tilde{\Pi}, \Pi'$ are given by expressions (3.52), (3.64), with $u = 0$. The factor $C_g$ is a common factor for three expressions (5.111), which does not contain polarization parameters and is determined by the expression for probability (5.85). Similar expressions for the probabilities in the region of individual resonance of the second Feynman diagram $f$ have the form:

$$\frac{dW_{OPPE}^{-+(f)}}{d\omega' du} = C_f \frac{n_f}{l^+} \tilde{\Pi}\hat{\Pi}', \quad \frac{dW_{OPPE}^{++(f)}}{d\omega' du} = C_f \frac{hl^- n_f}{2l^+} \Pi\hat{\Pi}', \quad \frac{dW_{OPPE}^{--(f)}}{d\omega' du} = C_f \frac{hn_f}{2} \Pi\hat{\Pi}', \quad (5.112)$$

where $\hat{\Pi}'$ is defined as the function $\Pi'$ in which the replacement $\xi'_2 \to -\xi'_2$ is performed. Finally, the expressions for the probabilities of the OPPE process in the interference region and the region of pair resonances in the case $l^- = l^+$ can be reduced to the form:

$$\frac{dW_{OPPE}^{-+(pair)}}{d\omega' du} = C_p \tilde{\Pi} K', \quad \frac{dW_{OPPE}^{--(pair)}}{d\omega' du} = C_p \frac{hl^+}{2} \hat{\Pi} K', \quad \frac{dW_{OPPE}^{++(pair)}}{d\omega' du} = C_p \frac{hl^-}{2} \Pi K', \quad (5.113)$$

where $K'$ is defined by expression (5.94) and is more compactly written as:

$$K' = 2(u^2 S_\chi^2 + C_\chi^2)\left[1 + \frac{u^2 S_\chi^2 - C_\chi^2}{u^2 S_\chi^2 + C_\chi^2}\xi'_3 + \frac{2u S_\chi C_\chi}{u^2 S_\chi^2 + C_\chi^2}\xi'_1\right], \quad (5.114)$$

where $S_\chi = \sin \kappa'\chi$, $C_\chi = \cos \kappa'\chi$.

It should be noted that in all three cases (5.111) - (5.113) the dependence of the probabilities on the Stokes parameters of the final photon is the same for any spin states of the electron and positron. That is, the polarization of the radiation does not depend on the spin states of the particles.

The probabilities summed over the spins of the particles in the three indicated regions, respectively, are equal:

$$\sum_{\mu^-\mu^+}\frac{dW_{\text{OPPE}}^{\mu^-\mu^+(g)}}{d\omega' du}=C_g\frac{n_g}{l^-}\left[\tilde{\Pi}+\frac{h}{2}(l^++l^-)\Pi\right]\Pi', \qquad (5.115)$$

$$\sum_{\mu^-\mu^+}\frac{dW_{\text{OPPE}}^{\mu^-\mu^+(f)}}{d\omega' du}=C_f\frac{n_f}{l^+}\left[\tilde{\Pi}+\frac{h}{2}(l^++l^-)\Pi\right]\hat{\Pi}', \qquad (5.116)$$

$$\sum_{\mu^-\mu^+}\frac{dW_{\text{OPPE}}^{\mu^-\mu^+(pair)}}{d\omega' du}=C_p\left[\tilde{\Pi}+\frac{h}{2}(l^++l^-)\Pi\right]K'. \qquad (5.117)$$

From the written expressions (5.115) - (5.117) it follows that the polarization of the final photon is determined only by the functions $\Pi', \hat{\Pi}', K'$, respectively, and does not depend on the polarization of the initial photon. In the region of individual resonance of the first Feynman diagram, the polarization of the final photon coincides with the polarization of the classical synchrotron radiation of an electron. And in the region of individual resonance of the second Feynman diagram, the polarization of the final photon coincides with the polarization of the classical synchrotron radiation of a positron. In the region of pair resonances, in the case of $l^-=l^+$, the Stokes parameters of the emitted photon follow from expression (5.114) and are equal to:

$$\xi_1'=\frac{2uS_\chi C_\chi}{u^2 S_\chi^2+C_\chi^2}, \quad \xi_2'=0, \quad \xi_3'=\frac{u^2 S_\chi^2-C_\chi^2}{u^2 S_\chi^2+C_\chi^2}. \qquad (5.118)$$

The angular dependence of the radiation polarization (dependence $\xi_1', \xi_3'$ on $\varphi', \theta'$) is shown in Fig.5.11. As follows from Fig. 5.11, the polarization of radiation in the region of pair resonances substantially depends not only on the polar angle of radiation $\theta'$, but also on the difference between the azimuthal angles of the initial and final photons.

The degree of polarization of the radiation, as follows from (5.118), is equal to one

$$P_\xi'^2=\xi_1'^2+\xi_2'^2+\xi_3'^2=1,$$

that is, the radiation is completely polarized for all directions. In particular, in the interference region ($u = 0$) the radiation is normally linearly polarized: $\xi_1'=0, \ \xi_2'=0, \ \xi_3'=-1$. Note that in the region of pair resonances in all directions there is no circular polarization of radiation. This is due to the fact that, in the indicated region, two Feynman diagrams make the same contribution, and the circular polarization of radiation differs from these diagrams in sign.

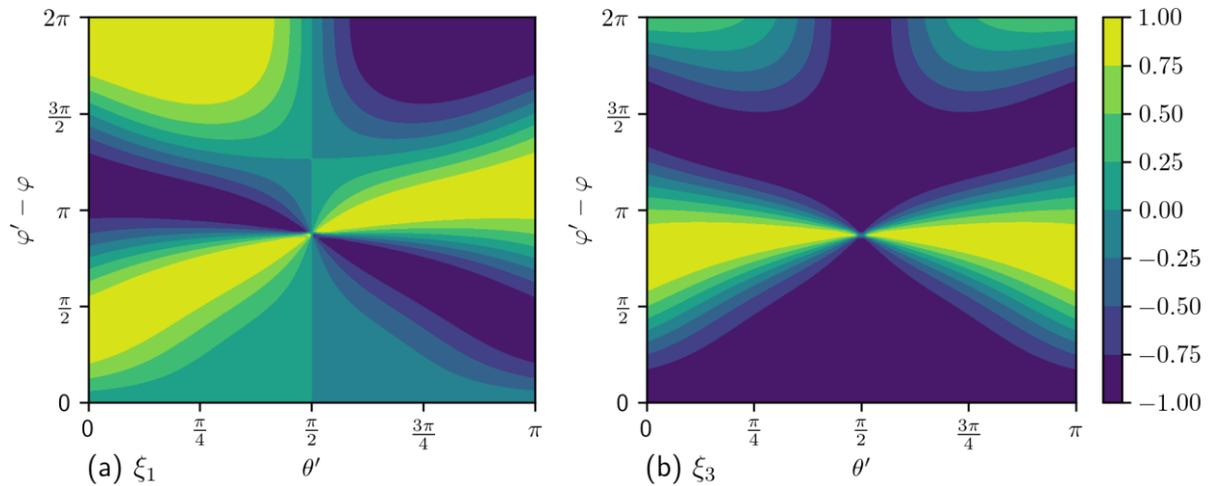

Figure 5.11. Angular dependence of the Stokes parameters of radiation

The degree of orientation of the spins of the final particles is determined by relations (2.167), (2.168). For all three cases (the region of individual resonances of the first and second diagrams, the region of pair resonances), for which the probabilities are given by relations (5.111) - (5.113), the degrees of orientation of the spins of electrons (positrons) are given by the same expressions, which have the form:

$$P_{e^-} = -\frac{2\tilde{\Pi} + h(l^+ - l^-)\Pi}{2\tilde{\Pi} + h(l^+ + l^-)\Pi}, \quad P_{e^+} = \frac{2\tilde{\Pi} + h(l^- - l^+)\Pi}{2\tilde{\Pi} + h(l^+ + l^-)\Pi}, \qquad (5.119)$$

in this case, as applied to the region of pair resonances in (5.119), one must set $l^+ = l^-$.

Expressions (5.119) coincide with those corresponding to the OPP process (4.95). Thus, the addition of a final photon in the OPP process (that is, the OPPE process) in the resonant kinematics does not affect the degree of orientation of the spins of the final particles and its dependence on the polarization of the initial photon, while the addition of an initial photon in the OPP process (the DPP process) significantly changes spin orientation. This is due to the fact that the spin states of the intermediate particles are pure in the OPPE process under resonance conditions, while for the DPP process they are mixed.

## 5.6. Mixed spin states of the intermediate electron (positron) under resonant conditions

Chapter 3-5 consider second-order QED processes, which are cross-channels of one generalized process with the presence of an intermediate electron (positron). Mixed spin states of intermediate particles were found in DSR and DPP processes under resonance conditions with certain values of the spin states of the initial and final particles. In such cases, the resonant probabilities are not factorized.

Let us analyze all possible situations of appearance of mixed spin states of intermediate particles under resonance conditions of the process. In total, there are ten second-order QED processes with one intermediate lepton state (intermediate electron or positron). Feynman diagrams of these processes are shown in Figure 5.12. These diagrams correspond to the following processes: Fig. 5.12.a is double synchrotron radiation by an electron (DSR), Fig. 5.12.b is absorption of two photons by an electron or double synchrotron absorption (DSA), Fig. 5.12.c is scattering of a photon by an electron or compton scattering (CS), Fig. 5.12.d is two-photon production of an $e^+e^-$ pair (TPP), Fig. 5.12.e is annihilation of an $e^+e^-$ pair into two photons or two photons annihilation of an $e^+e^-$ pair (TPA), Fig. 5.12.f is one-photon production of an $e^+e^-$ pair with emission of a photon (OPPE), Fig. 5.12.g is annihilation of an $e^+e^-$ pair into one photon with absorption of a photon or one photon annihilation with photon absorption (OPAA). Also note that there are three more QED processes with a positron, which are described by Feynman diagrams in Fig.5.12.a, Fig.5.12.b, Fig.5.12.c with a change in the direction of the arrows of solid lines.

Each QED process, which is depicted in Fig. 5.12, under resonant conditions can be represented as a cascade of two first-order processes, each of which is one of six: emission of a photon by an electron, emission of a photon by a positron, absorption of a photon by an electron, absorption of a photon by a positron, production of $e^+e^-$ pair by a photon, annihilation of an $e^+e^-$ pair into one photon, with amplitudes $A_{rad\ e^-}, A_{rad\ e^+}, A_{abs\ e^-}, A_{abs\ e^+}, A_{pr}, A_{ann}$, respectively. Cases without a change in direction of a spin of the particles are most probable in the processes of radiation, absorption of a photon. The particles are in the ground spin states ($\mu^+=+1$, $\mu^-=-1$) with a greater probability in the processes of production and annihilation of an $e^+e^-$ pair.

Let us write down the degrees of the small parameter $h$ of the amplitudes of the process of photon emission by an electron with respect to the most probable case ($\mu^-=\mu'^-$) for all spin states of the particles:

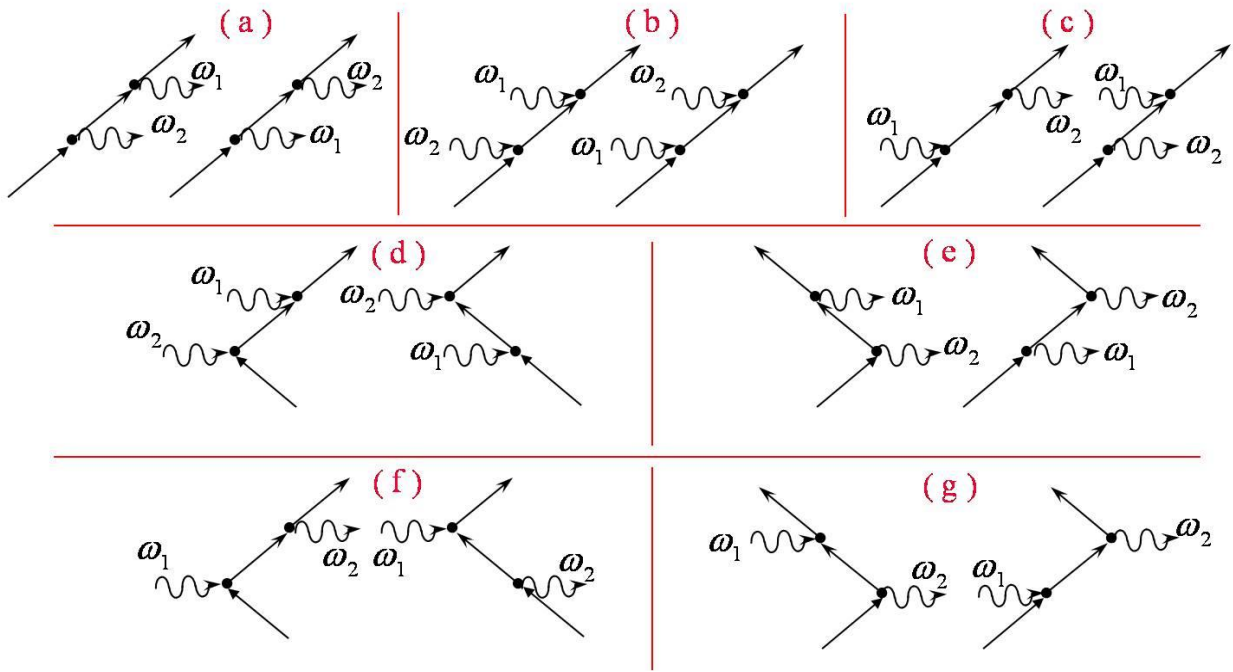

Figure 5.12. Feynman diagrams of QED processes of the second order with one intermediate lepton state

$$A_{rad\ e^-}^{--} \sim A_{rad\ e^-}^{++} \sim 1, \quad A_{rad\ e^-}^{+-} \sim \sqrt{h}, \quad A_{rad\ e^-}^{-+} \sim h\sqrt{h}, \qquad (5.120)$$

where the first and second superscript signs (+, -) correspond to the projections of the spins of the initial and final electron, respectively. The powers of parameter $h$ of amplitudes of the process of photon emission by a positron are determined by expressions (5.120) with the replacement of spin states $+ \leftrightarrow -$. The degrees $h$ of amplitudes of the processes of production and annihilation of an e⁺e⁻ pair with respect to the most probable case ($\mu^-=-1$, $\mu^+=+1$) are equal to:

$$A_{pr}^{-+} \sim 1, \quad A_{pr}^{--} \sim A_{pr}^{++} \sim \sqrt{h}, \quad A_{pr}^{+-} \sim h\sqrt{h}, \quad A_{ann}^{\mu^-\mu^+} \sim A_{pr}^{\mu^-\mu^+}. \qquad (5.121)$$

The powers of the parameter *h* of the amplitudes of absorption of a photon by an electron (positron) are determined from expression (5.120) by mutual replacement of the initial particles with the final ones

$$A_{abs\,e^-}^{\mu^-\mu^{'-}} \sim A_{rad\,e^-}^{\mu^{'-}\mu^-}, \quad A_{abs\,e^+}^{\mu^+\mu^{'+}} \sim A_{rad\,e^+}^{\mu^{'+}\mu^+}. \qquad (5.122)$$

Note that the intermediate state in the second-order processes under consideration is described by the Green's function of the electron, in which the summation over the spin states of the intermediate particle is carried out. Thus, in the general case, the spin states of the intermediate electron (positron) are mixed states. The amplitude of the QED process of the second order under resonance conditions is proportional to the sum (of two terms) of the product of two amplitudes determined by expressions (5.120) - (5.122). The latter, in turn, have different degrees of small parameter *h*. As a result, cases are possible when only one of the two terms has the smallest degree *h*. This corresponds to a pure intermediate state. The probability of the process in such situations is factorized and has the form of the Breit–Wigner formula. If both terms of the above sum have the same degree h, a mixed spin state takes place.

Let's move on to analyzing the amplitudes of the individual processes shown in Figure 5.12. The amplitudes of the DSR process of an electron $A_{DSR\,e^-}^{\mu\mu'}$, relative to the most probable cases, taking into account (5.120), have such degrees of the parameter h for the first diagram in Fig. 5.12.a

$$A_{DSR\,e^-}^{--} \sim A_{rad1\,e^-}^{--} A_{rad2\,e^-}^{--} + \cancel{A_{rad1\,e^-}^{+-} A_{rad2\,e^-}^{-+}} \sim 1, (-), \qquad (5.123)$$

$$A_{DSR\,e^-}^{++} \sim \cancel{A_{rad1\,e^-}^{-+} A_{rad2\,e^-}^{+-}} + A_{rad1\,e^-}^{++} A_{rad2\,e^-}^{++} \sim 1, (+), \qquad (5.124)$$

$$A_{DSR\,e^-}^{+-} \sim A_{rad1\,e^-}^{+-} A_{rad2\,e^-}^{++} + A_{rad1\,e^-}^{--} A_{rad2\,e^-}^{+-} \sim \sqrt{h}, (\pm), \qquad (5.125)$$

$$A_{DSR\,e^-}^{-+} \sim A_{rad1\,e^-}^{++} A_{rad2\,e^-}^{-+} + A_{rad1\,e^-}^{-+} A_{rad2\,e^-}^{--} \sim h\sqrt{h}, (\pm), \qquad (5.126)$$

where the subscript numbers 1,2 denote the first and second photons, respectively, the terms with a greater degree of *h* are crossed out, the sign of projection of spin of the intermediate electron is indicated at the end of the expressions in parentheses. Thus, in DSW processes without electron spin flip (5.123), (5.124), the spin intermediate state is a pure state, and in spin-flip processes it is a mixed state. This result coincides with the direct calculations of the amplitude of the DSR process performed in Chapter 3. The second Feynman diagram of this process in Fig.5.12.a is obtained from the first as a result of replacing the photons in places, which does not change the spin states of the particles, that is, the previous result will be true for it as well.

The amplitudes of the process of absorption of two photons by an electron $A_{DSA\,e^-}^{\mu\mu'}$, relative to the most probable cases, have degrees of the parameter h for the first and second diagrams in Fig. 5.12.b, which are determined by expressions (5.123) - (5.126) taking into account rule (5.122). As a result, the pure spin states take place for a process without spin flip, and mixed states for a spin-flip process.

The process of scattering of a photon by an electron described by the first Feynman diagram in Fig. 5.12.c (a straight diagram in which the initial particles meet at one point) has relative amplitudes $A_{CS}^{\mu\mu'}$ (referred to the magnitude of the amplitude in the most probable spin configuration), which are proportional to such powers of *h*:

$$A_{CS}^{--} \sim A_{rad1\,e^-}^{--} A_{rad2\,e^-}^{--} + \cancel{A_{rad1\,e^-}^{+-} A_{rad2\,e^-}^{-+}} \sim 1, (-), \qquad (5.127)$$

$$A_{CS}^{++} \sim \cancel{A_{rad1\,e^-}^{-+} A_{rad2\,e^-}^{+-}} + A_{rad1\,e^-}^{++} A_{rad2\,e^-}^{++} \sim 1, (+), \qquad (5.128)$$

$$A_{CS}^{+-} \sim A_{rad1\,e^-}^{++} A_{rad2\,e^-}^{+-} + \cancel{A_{rad1\,e^-}^{-+} A_{rad2\,e^-}^{--}} \sim \sqrt{h}, (+), \qquad (5.129)$$

$$A_{CS}^{-+} \sim A_{rad1\,e^-}^{+-} A_{rad2\,e^-}^{++} + \cancel{A_{rad1\,e^-}^{--} A_{rad2\,e^-}^{-+}} \sim \sqrt{h}, (+). \qquad (5.130)$$

For the second diagram in Fig. 5.12.c (exchange diagram), the amplitudes of the process without changing the electron spin are similar to expressions (5.127), (5.128). The relative amplitudes for the spin-flip process are of the following degrees *h*:

$$A_{CS}^{+-} \sim A_{rad1\,e^-}^{+-} A_{rad2\,e^-}^{--} + \cancel{A_{rad1\,e^-}^{++} A_{rad2\,e^-}^{-+}} \sim \sqrt{h}, (-), \qquad (5.131)$$

$$A_{CS}^{-+} \sim A_{rad1e^-}^{--} A_{rad2e^-}^{+-} + A_{rad1e^-}^{-+} A_{rad2e^-}^{++} \sim \sqrt{h}, (-). \qquad (5.132)$$

Thus, for all spin configurations, intermediate spin states in the CS process are pure, which coincides with the direct calculations in Chapter 3. For a spin-flip process, according to the direct Feynman diagram, the spin of the intermediate electron is directed along the field (inverse state), and according to the exchange diagram, the spin of the intermediate electron is directed against the field (ground spin state). It is the presence in the general case of two terms in the amplitudes that explains the same degree of amplitudes for spin-flip processes with spin flip along and against the field $A_{CS}^{-+} \sim A_{CS}^{+-}$, which differs significantly from the spin-flip process of the SR.

An analysis of the amplitudes of the processes of emission of two photons by a positron, absorption of two photons by a positron, and scattering of a photon by a positron gives the same corresponding expressions (5.127) - (5.132) with mutual replacement of spin states $+ \leftrightarrow -$.

The relative amplitudes of the DPP process, corresponding to the first diagram in Fig. 5.12.d, contain the following powers of $h$:

$$A_{DPP}^{-+} \sim A_{pr2}^{-+} A_{rad1e^-}^{--} + A_{pr2}^{++} A_{rad1e^-}^{-+} \sim 1, (-), \qquad (5.133)$$

$$A_{DPP}^{+-} \sim A_{pr2}^{--} A_{rad1e^-}^{+-} + A_{pr2}^{+-} A_{rad1e^-}^{++} \sim h, (-), \qquad (5.134)$$

$$A_{DPP}^{--} \sim A_{pr2}^{--} A_{rad1e^-}^{--} + A_{pr2}^{+-} A_{rad1e^-}^{-+} \sim \sqrt{h}, (-), \qquad (5.135)$$

$$A_{DPP}^{++} \sim A_{pr2}^{-+} A_{rad1e^-}^{+-} + A_{pr2}^{++} A_{rad1e^-}^{++} \sim \sqrt{h}, (\pm) \qquad (5.136)$$

and similarly for the second diagram of Fig.5.12.d

$$A_{DPP}^{-+} \sim A_{pr2}^{-+} A_{rad1e^+}^{++} + A_{pr2}^{--} A_{rad1e^+}^{+-} \sim 1, (+), \qquad (5.137)$$

$$A_{DPP}^{+-} \sim A_{pr2}^{++} A_{rad1e^+}^{-+} + A_{pr2}^{+-} A_{rad1e^+}^{--} \sim h, (+), \qquad (5.138)$$

$$A_{DPP}^{++} \sim A_{pr2}^{++} A_{rad1e^+}^{++} + A_{pr2}^{+-} A_{rad1e^+}^{+-} \sim \sqrt{h}, (+), \qquad (5.139)$$

$$A_{\text{DPP}}^{--} \sim A_{pr2}^{-+}A_{rad1e^+}^{-+} + A_{pr2}^{--}A_{rad1e^+}^{--} \sim \sqrt{h}, (\pm). \qquad (5.140)$$

Here the second photon products a pair, and the first photon is emitted. As a result, in both diagrams, three out of four variants contain pure spin states of intermediate particles, and one variant contains a mixed state. A mixed spin state occurs when particles are produced with identically directed spins, the spin of an intermediate electron (positron) is directed along the field (against the field), that is, it is in an inverse state. Pure states are ground spin states. The result coincides with the direct calculations of Chapter 4.

For the process of annihilation of an e⁺e⁻ pair into two photons (Fig. 5.12.e), the same expressions (5.133) - (5.140), taking into account the property (5.121), will be suitable. That is, the previous conclusions are valid in this case as well.

Finally, we will give estimates of the degrees of $h$ for the relative amplitudes of the OPPE process, which proceeds in accordance with the first diagram in Fig. 5.12.f. In this case, the first photon products a pair, and the second photon is emitted

$$A_{\text{OPPE}}^{-+} \sim A_{pr1}^{-+}A_{rad2e^-}^{--} + \cancel{A_{pr1}^{++}A_{rad2e^-}^{+-}} \sim 1, (-), \qquad (5.141)$$

$$A_{\text{OPPE}}^{+-} \sim A_{pr1}^{+-}A_{rad2e^-}^{++} + \cancel{A_{pr1}^{--}A_{rad2e^-}^{-+}} \sim h\sqrt{h}, (+), \qquad (5.142)$$

$$A_{\text{OPPE}}^{--} \sim A_{pr1}^{--}A_{rad2e^-}^{--} + \cancel{A_{pr1}^{+-}A_{rad2e^-}^{+-}} \sim \sqrt{h}, (-), \qquad (5.143)$$

$$A_{\text{OPPE}}^{++} \sim A_{pr1}^{++}A_{rad2e^-}^{++} + \cancel{A_{pr1}^{-+}A_{rad2e^-}^{-+}} \sim \sqrt{h}, (+) \qquad (5.144)$$

and for the second Feynman diagram of Fig.5.12.f

$$A_{\text{OPPE}}^{-+} \sim A_{pr1}^{-+}A_{rad2e^+}^{++} + \cancel{A_{pr1}^{--}A_{rad2e^+}^{-+}} \sim 1, (+), \qquad (5.145)$$

$$A_{\text{OPPE}}^{+-} \sim A_{pr1}^{+-}A_{rad2e^+}^{--} + \cancel{A_{pr1}^{++}A_{rad2e^+}^{+-}} \sim h\sqrt{h}, (-), \qquad (5.146)$$

$$A_{\text{OPPE}}^{--} \sim A_{pr1}^{--}A_{rad2e^+}^{--} + \cancel{A_{pr1}^{-+}A_{rad2e^+}^{+-}} \sim \sqrt{h}, (-), \qquad (5.147)$$

$$A_{\text{OPPE}}^{++} \sim A_{pr1}^{++}A_{rad2e^+}^{++} + \cancel{A_{pr1}^{+-}A_{rad2e^+}^{-+}} \sim \sqrt{h}, (+). \qquad (5.148)$$

Thus, in all variants, the spin states of the intermediate particles are pure. The spin-flip process when the second photon is emitted is suppressed; as a result, the direction of the spin of the intermediate particle coincides with the direction of the electron spin for the first diagram and the positron for the second one. The result is the same as the direct calculation done in Section 5.4.

For the process of annihilation of an $e^+e^-$ pair into one photon with absorption of a photon (Fig. 5.12.g), the same conclusions are valid as for the OPPE process.

Thus, the analysis of the spin states of an intermediate particle in ten QED processes under resonance conditions showed that mixed spin states take place in processes where two photons are involved in the initial or final state, that is, in the processes: emission of two photons by an electron (positron), absorption of two photons by an electron (positron), production of an $e^+e^-$ pair by two photons, annihilation of an $e^+e^-$ pair into two photons.

### 5.7. Conclusions to the Chapter 5

A theory of the process of one-photon production of an electron-positron pair with photon emission (OPPE) is constructed taking into account the spin of particles in a strong external magnetic field under resonant and nonresonant conditions. As a result, it was shown:

1. The principal difference between the threshold in the OPPE process and the analogous one in the OPP process is that it is possible for any frequency and polar angle of the initial photon (sufficient for the production of a pair) in the first case and corresponds to the threshold maximum possible frequency of the final photon.

2. In the OPPE process, there are pair resonances (a consequence of the presence of two Feynman diagrams), the distance between which is much less than the distance between neighboring Landau levels. Resonance is realized for any above-threshold frequency of the initial photon. In the region of pair resonances, the frequency of the final photon is equal to the distance between the Landau levels of the intermediate and final electron (positron), and there is an "interference" of the two diagrams. To get into the region of individual resonance, it is necessary to select the frequency of the final photon with an accuracy of $h^2$ inclusive. Pair resonances merge into one when particles are produced at the same energy levels and when a photon is emitted perpendicular to the field.

3. The differential probability of the OPPE process per unit time with a fixed value of the frequency and angle of photon emission at resonance is factorized and reduced to the Breit-Wigner form for any values of the projections of the particle spin. The integral probability of the OPPE process (second-order process) in the region of resonances coincides with the probability of the $e^+e^-$ pair production by one photon (first-order process). In the case of production of an $e^+e^-$ pair in the ground energy states $l^-=l^+=0$ in a magnetic field of $h=0.1$ ($H=4.4 \cdot 10^{12} Gs$), the OPP and SR probabilities are, respectively, in order of magnitude

$$W_{OPP}^{-+} \sim 10^{10}[1/c], \ W_{SR\,e^-}^{--} \sim 10^{17}[1/c].$$

4. The difference between the region of pair resonances and the narrow region of an individual resonance is significant dependence of probability of the process in the first case on the difference between the azimuthal angles $\Delta\varphi$ of the initial and final photons. When the initial photon propagates $\perp \vec{H}$ in the region of pair resonances, the radiation is maximum in the plane, that is perpendicular to the field, in the direction (parallel and antiparallel) of the initial photons and is absent at an angle $\Delta\varphi=3\pi/4$. Estimation of the probability $W$ in the nonresonant region near the threshold, when particles are produced in the ground energy states, for $h = 0.1$ gives a value $W \approx 10^6 s^{-1}$ that is four orders of magnitude less than the probability of the process in a region of the individual resonance.

5. In the region of individual resonance of the first Feynman diagram (see Fig. 5.5), the polarization of the final photon coincides with the polarization of the classical synchrotron radiation of an electron. In the region of individual resonance of the second Feynman diagram, the polarization of the final photon coincides with the polarization of the classical synchrotron radiation of a positron. In the region of pair resonances, if the e⁺e⁻ pair is born at the same Landau levels ($l^- = l^+$), the polarization of the radiation depends on both the polar angle and the azimuthal angle and is purely linear in any direction. The polarization of the initial photon does not affect the polarization of the final one. The degree of orientation of the electron and positron spins is the same as in the OPP process, that is, the addition of a final photon to the OPP process in resonant kinematics does not affect the spins of the final particles.

6. An analysis of the appearance of pure and mixed spin states of an intermediate particle under resonance conditions in all second-order QED processes with one intermediate lepton state (intermediate electron or positron) showed the following:

- mixed spin states take place in processes where two photons participate in the initial or final states, that is, in the processes: emission of two photons by an electron (positron), absorption of two photons by an electron (positron), production of an e⁺e⁻ pair by two photons, annihilation e⁺e⁻ pairs into two photons;

- in the DSR process without an electron spin flip, the spin intermediate state is pure, and in the spin-flip process it is mixed;

- a mixed spin state in the DPP process occurs when particles are produced with identically directed spins, while the spin of an intermediate electron (positron) is directed along the field (against the field), that is, it is in an inverse spin state. Pure states are ground spin states.

The main scientific results of this chapter are published in [290-292].

# CHAPTER 6
# CASCADE OF E⁺E⁻ PAIR PRODUCTION BY A PHOTON WITH SUBSEQUENT ANNIHILATION TO A SINGLE PHOTON

**6.1. Introduction**

In this chapter, the cascade of processes of e⁺e⁻ pair production and subsequent the pair annihilation (CPPPA) in a strong magnetic field is considered. The polarization operator of a photon in a magnetic field is found by the direct method of scattering theory. The resonant conditions of the CPPPA process are analyzed. A comparison of the amplitude of the CPPPA process with the probability of the OPP process (optical theorem) is performed, as well as an estimate of the ratio of the CPPPA probabilities in non-resonant and resonant cases in the LLL approximation in a subcritical field is found. Influence of the polarization of the initial photon on the polarization of the final one is studied. The effect of vacuum birefringence in a strong magnetic field is analyzed.

**6.2. Probability amplitude and resonant conditions of the CPPPA process**

<u>Probability amplitude of the CPPPA process.</u> The expression for the amplitude of the process corresponds to the Feynman diagram shown in Fig.6.1

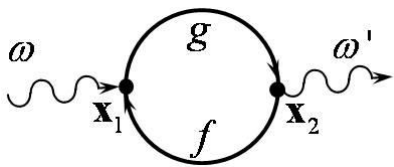

Fig.6.1. Feynman diagram of the cascade of one-photon e⁺e⁻ pair production and subsequent the pair annihilation to a single photon

and it has the form

$$A_{if} = -e^2 Tr \int d^4\mathbf{x}_1 d^4\mathbf{x}_2 (A'(\mathbf{x}_2)^* \gamma) G_{H\mathbf{g}}(\mathbf{x}_2, \mathbf{x}_1)(A(\mathbf{x}_1)\gamma) G_{H\mathbf{f}}(\mathbf{x}_1, \mathbf{x}_2), \quad (6.1)$$

where $A(\mathbf{x}_1)$, $A'(\mathbf{x}_2)^*$ are the wave functions of the initial and final photons, $G_{H\mathbf{g}}(\mathbf{x}_2, \mathbf{x}_1)$, $G_{H\mathbf{f}}(\mathbf{x}_1, \mathbf{x}_2)$ are the Green's function of electron in an external magnetic field, $\gamma$ are the Dirac gamma matrices, the symbol $Tr$ means a trace of spinor indices. In the general case, the amplitude of the CPPPA process after taking the integrals in (6.1) can be represented as

$$A_{if} = \frac{-8\pi^2 e^2 h m^2}{\omega V} \delta^4(k-k') \sum_{n_g=0}^{\infty} \sum_{n_f=0}^{\infty} \int dg_0 dg_z \frac{\sum_{j=1}^{5} Q_j}{(g_0^2 - \varepsilon_g^2)(f_0^2 - \varepsilon_f^2)}, \quad (6.2)$$

where $f_0 = \omega - g_0$, $f_z = k_z - g_z$. The energies of the intermediate states $\varepsilon_g$, $\varepsilon_f$, at fixed Landau levels $n_g$, $n_f$ are equal to

$$\varepsilon_g = \sqrt{m^2 + 2n_g h m^2 + g_z^2}, \quad \varepsilon_f = \sqrt{m^2 + 2n_f h m^2 + f_z^2}. \quad (6.3)$$

Note that the amplitude (6.2) contains the factor $\delta^4(k-k')$, which means that in this process, both the law of conservation of energy and the law of conservation of momentum are fulfilled, despite the presence of an external magnetic field. For all the processes discussed earlier in this paper, the amplitude contained only three Dirac delta functions ($\delta(k_x - k_x')$ is absent in the process amplitudes for the external field potential selected in the Landau calibration). However, given that the laws of dispersion $k^2 = 0$, $k'^2 = 0$ are satisfied for the initial and final photons, and taking into account that the amplitude always contains $\delta^3(k-k')$ for the QED processes in the external field, it is easy to obtain $k_x' = \pm k_x$. The case $k_x' = -k_x$ must be rejected as non-physical, because it corresponds to the case of reflection of a photon, regardless of the value of magnetic field strength and the frequency of photon.

The quantities $Q_j$ in the amplitude (6.2) have the form

$$Q_1 = (m^2 + gf)[J^{++}J'^{++} + J^{--}J'^{--}], \quad Q_2 = (m^2 + g\tilde{f})[J^{-+}J'^{-+} + J^{+-}J'^{+-}],$$

$$Q_3 = 2\sqrt{n_g n_f} h m^2 [J^{++}J'^{--} + J^{--}J'^{++} - J^{-+}J'^{+-} - J^{+-}J'^{-+}],$$

$$Q_4 = -2\sqrt{2n_f h} m g_z [J^{-+}J'^{++} + J^{++}J'^{-+} + J^{+-}J'^{--} + J^{--}J'^{+-}],$$

$$Q_5 = 2\sqrt{2n_g h} m \tilde{f}_z [J^{+-}J'^{++} + J^{++}J'^{+-} + J^{-+}J'^{--} + J^{--}J'^{-+}], \quad (6.4)$$

where $gf = g_0 f_0 - g_z f_z$, $\tilde{gf} = g_0 f_0 + g_z f_z$. The functions $J$ contain the parameters of the initial photon and they are defined as follows

$$J^{++} = J(n_f, n_g) e_z, \quad J^{--} = J(n_f - 1, n_g - 1) e_z,$$

$$J^{-+} = J(n_f - 1, n_g) H_p, \quad J^{+-} = J(n_f, n_g - 1) H_m. \quad (6.5)$$

Special functions $J(n_f, n_g)$ are defined by expression (2.37) and polarization functions $e_z$, $H_p$, $H_m$ by expressions (2.43). The functions $J'$ contain the parameters of the final photon and they can be obtained with the expression (6.5) by appropriate substitution and complex conjugation $H_p$, $H_m$. In this case, given the presence in the amplitude of the multiplier $\delta^4(k-k')$, we have

$$J'(n_f, n_g) = J(n_f, n_g).$$

Note that the integral in (6.2) diverges logarithmically on the upper bound of the variables $g_0, g_z$

$$\int dg_0 dg_z \frac{\sum_{j=1}^{5} Q_j}{(g_0^2 - \varepsilon_g^2)(f_0^2 - \varepsilon_f^2)} \xrightarrow{g \to \infty} \int \frac{dg}{g},$$

therefore, it is necessary to carry out the amplitude regularization procedure. We will use Bogolyubov's regularization method [293], according to which such a replacement of the denominator of the Green's function of the intermediate lepton should be performed

$$(g_0^2 - \varepsilon_g^2)^{-1} \to (g_0^2 - \varepsilon_g^2 + im\Gamma)^{-1} - (g_0^2 - \varepsilon_g^2 + m^2 - M^2 + im\Gamma)^{-1}, \quad (6.6)$$

where an additional mass $M$ is introduced, while the magnetic field $H$ should be taken equal to zero. When $M \to \infty$ the expression (6.6) goes to the initial one. In the denominator (6.6) a small imaginary additive $im\Gamma$ according to the Breit-Wigner rule is also introduced for the correct bypass of the poles, while the value $\Gamma$ makes sense of the width of intermediate state. A similar substitution is applied to the denominator $(f_0^2 - \varepsilon_f^2)$. Next, the found denominators are converted by the Schwinger's proper-time method according to the rule

$$\frac{1}{g_0^2 - \varepsilon_g^2 + im\Gamma} = \frac{1}{i}\int_0^\infty d\tau_1 e^{i\tau_1(g_0^2 - \varepsilon_g^2 + im\Gamma)}, \quad \frac{1}{f_0^2 - \varepsilon_f^2 + im\Gamma} = \frac{1}{i}\int_0^\infty d\tau_2 e^{i\tau_2(f_0^2 - \varepsilon_f^2 + im\Gamma)} \quad (6.7)$$

After the substitutions, the integrals over the variables $g_0, g_z$ in the amplitude (6.2) can be easily calculated using the equality

$$\int_{-\infty}^\infty dt e^{\pm i\alpha t^2 + 2i\beta t} = \sqrt{\frac{\pi}{\alpha}} e^{\pm\frac{\pi i}{4} \mp \frac{\beta^2 i}{\alpha}}; \alpha, \beta > 0.$$

Then, according to Bogolyubov's method, the variables are replaced

$$\tau_1, \tau_2 \to \zeta, \lambda: \quad \zeta = \frac{\tau_1}{\tau_1 + \tau_2}, \lambda = \tau_1 + \tau_2.$$

In what follows, we consider the propagation of a photon perpendicular to a magnetic field ($k_z=0$), which does not affect the generality of the problem.

Elementary integration over $\lambda$ leads to an amplitude with one integral over $\zeta$. Under $M \to \infty$ and discarding the terms containing the multiplier $\ln(M)$ in the obtained amplitude, the regular part of the amplitude can be reduced to the form

$$A_{if} = \frac{i8\pi^3 e^2 h m^2}{\omega V} \delta^4(k-k') \sum_{n_g=0}^\infty \sum_{n_f=0}^\infty \int_0^1 d\zeta \times$$

$$\times \left( a - \frac{b}{2}\ln\left|\frac{m^2\zeta(1-\zeta)}{\omega^2(\zeta-\zeta_1)(\zeta-\zeta_2)}\right| - \frac{m^2}{\omega^2}\frac{c(\zeta)}{(\zeta-\zeta_1)(\zeta-\zeta_2) - im\Gamma/\omega^2} \right), \quad (6.8)$$

where

$$a = J^{++}J'^{++} + J^{--}J'^{--} + J^{-+}J'^{-+} + J^{+-}J'^{+-}, \quad b = 2J^{-+}J'^{-+} + 2J^{+-}J'^{+-},$$

$$c(\zeta) = 2a(1+n_f h) + 4\sqrt{n_f n_g} h J^{++}J'^{--} + 2(n_f - n_g)ha\zeta.$$

The poles $\zeta_1, \zeta_2$ have the form

$$\zeta_{1,2} = \frac{1}{2} - \frac{m^2}{\omega^2} hN_- \pm \sqrt{\frac{\omega^2}{m^2}\left(\frac{\omega^2}{m^2} - 4(1+N_+h)\right) + 4h^2 N_-^2}, \quad (6.9)$$

where $N_\pm = n_g \pm n_f$.

The expression for the amplitude (6.8) is obtained in the general relativistic case. In what follows, it will be analyzed in the LLL approximation. Note that in the LLL approximation near the process threshold $\omega \approx 2m$, both poles $\zeta_{1,2} \approx 1/2$ are within the integration interval.

<u>Resonant conditions.</u> The integrand expression of the amplitude (6.8) generally contains two simple poles, in the case $\zeta_1 \neq \zeta_2$ the width can be neglected. In the case

$$\zeta_1 = \zeta_2, \qquad (6.10)$$

two simple poles merge into one second-order pole. The amplitude has a divergence (excluding the width process $\Gamma$). This is the resonance of the CPPPA process. The condition (6.10) determines the resonant frequency of initial photon

$$\omega_{res} = m\sqrt{1 + 2n_g h} + m\sqrt{1 + 2n_f h} \,. \qquad (6.11)$$

Note that the found expression for $\omega_{res}$ is accurate. In the previous sections, the resonant frequencies could be found only in the form of a series on the parameter $h$ with accuracy up to the second degree by $h$. The physical meaning of the resonance condition (6.11) is obvious: in resonance, the frequency of the initial photon is equal to the sum of the energies of the intermediate electron and positron, which are at fixed Landau levels with zero longitudinal momentum

$$\omega_{res} = (\varepsilon^- + \varepsilon^+)\big|_{p^-=p^+=0}. \qquad (6.12)$$

With a continuous increase of the photon frequency, the resonances will appear in intervals equal to the cyclotron frequency. If the Landau level number of electron is increased by the fixed value $l$, and the positron is reduced by the same value, the resonant frequency (6.11) changes slightly. With accuracy up to $h^2$, this change is equal to

$$\Delta\omega = \omega_{res}(n_g, n_f) - \omega_{res}(n_g + l, n_f - l) = l(n_g + l - n_f)h^2. \qquad (6.13)$$

That is, at distances multiple of the cyclotron frequency, a series of resonances are located for which the distance between adjacent peaks is the order of $h^2$. In the special case $n_g \neq n_f$, when choosing $l = n_f - n_g$, we can find $\Delta\omega = 0$, in other words, in the same resonant conditions are two different resonances and the amplitude of the process must contain both terms corresponding to them. This result is understandable for reasons of symmetry.

### 6.3. Probability of the CPPPA process in the resonant and non-resonant conditions

<u>Resonant amplitude and probability of the CPPPA process.</u> Let us take the frequency of the initial photon near the resonance in the form

$$\omega = \omega_{res} + \delta, \quad \delta \ll \Gamma. \tag{6.14}$$

When the resonant conditions (6.10) are satisfied, the last term in parentheses is the main one in the expression for the amplitude (6.8). After integration over $\zeta$ with the subsequent expanding into a power series in small parameter $\delta$ and keeping terms linear in $h$, the amplitude has the form

$$A_{if} = -i \frac{16\pi^4 e^2 h m^5}{V\omega^3 \sqrt{\Gamma m}} \delta^4(k-k')c(\zeta_{res})(1+i\frac{2^5 \delta}{\Gamma}). \tag{6.15}$$

The expression in parentheses can be rewritten as the denominator of Breit-Wigner

$$(1+i\frac{2^5 \delta}{\Gamma}) = \frac{i\Gamma}{2^5} \cdot \frac{1}{\omega - \omega_{res} - i\Gamma/2^5}$$

the probability amplitude of the CPPPA process in the LLL approximation can finally be reduced to the form

$$A_{if} = \frac{\pi^4 e^2 h m^5 \sqrt{\Gamma/m}}{2V\omega^3(\omega - \omega_{res} + \frac{i\Gamma}{2^5})} \delta^4(k-k') \times$$

$$\times J^2 \left[ e_z e'_z (2 + h(N_+ - 2n_g n_f)) + hN_+ H_m H_m^{'*} \right], \tag{6.16}$$

where $J^2 = e^{-\eta} \eta^{N_+}/(n_g! n_f!)$.

Recalling the probability of one-photon pair production $W_{OHП}$ (2.156), we can see that it is expressed through the imaginary part of the amplitude (6.16), thus

$$W_{OPP} = -\sqrt{\frac{4\Gamma}{\delta\omega}} \operatorname{Im} A_{if}(0), \qquad (6.17)$$

where $\delta\omega$ is the detuning of the frequency of the initial photon from the threshold of the one-photon pair creation. We note that it is at the threshold of the OPP process the frequency of the initial photon satisfies the resonant conditions (6.12). $A_{if}(0)$ is the amplitude of the CPPPA process without changing the parameters of the initial photon (with zero changes), in particular, this means

$$e'_z = e_z, \quad \delta^4(k-k') = VT/(2\pi)^4.$$

The expression (6.17) is a mathematical notation of the optical theorem. Thus, the optical theorem is valid in the resonant conditions.

After integration, the square of the modulus of amplitude (6.16) multiplied by the number of the final states $Vd^3k/(2\pi)^3$ and it determines the total probability of the CPPPA process in resonant conditions, which (ignoring the small correction of the order of $h$) can lead to the Breit-Wigner formula

$$W_{res} = \frac{\Gamma\delta\omega}{2^{12}} \cdot \frac{W_{OPP}W_{AP}}{[(\omega-\omega_{res})+\frac{\Gamma^2}{2^{10}}]}, \qquad (6.18)$$

where $W_{AP}$ is the probability of one-photon $e^+e^-$ pair annihilation, which differs from $W_{OPP}$ only by replacing the initial photon with the final one and the presence of the Dirac-delta function $\delta(\omega-\omega')$ instead of the multiplier $T/2\pi$. The obtained expression is factorized, which means the decay of the CPPPA process in the resonant conditions into two independent successive processes of the first order: the OPP process and the subsequent one-photon $e^+e^-$ pair annihilation.

In the resonance, expression (6.18) has a simple form

$$W_{res} = \frac{\delta\omega}{4\Gamma} \cdot W_{OPP}W_{AP}. \qquad (6.19)$$

We also write an explicit form of the resonant probability

$$W_{res} = \frac{\pi h^2 e^4}{2^5 \Gamma} J^4 m^3 T \delta(\omega - \omega') \times$$

$$\times \left[(1+\xi_3)(1+\xi'_3)(1+(N_+ - 2n_g n_f)h) + N_+ h(\xi_1 \xi'_1 + \xi_2 \xi'_2)\right]. \quad (6.20)$$

In Fig.6.2. the dependence of the probability of the CPPPA process (in relative units) on the value of magnetic field at different values of the polarization of the initial photon is presented ($n_g$=2, $n_f$=1). As follows from Fig. 6.2, the probability of the CPPPA process increases exponentially with increasing magnetic field strength, with the greatest probability of the cascade occurs for abnormally linear polarization of photons ($\xi_3$=1).

If the initial photon is set to be normally linearly polarized ($\xi_3$=-1), then the expression of probability (6.20) with accuracy up to the first degree by the parameter $h$ gives zero, i.e. the process is suppressed. This is a repetition of the situation like in the OPP process (see chapter 2). Taking into account the higher degree of the parameter $h$ gives a nonzero probability value in the case under consideration

$$W_{res}^{\xi_3=-1} = \frac{\pi h^4 e^4}{2^6 \Gamma} J^4 m^3 T \delta(\omega - \omega') N_+^2 (1 - \xi'_3). \quad (6.21)$$

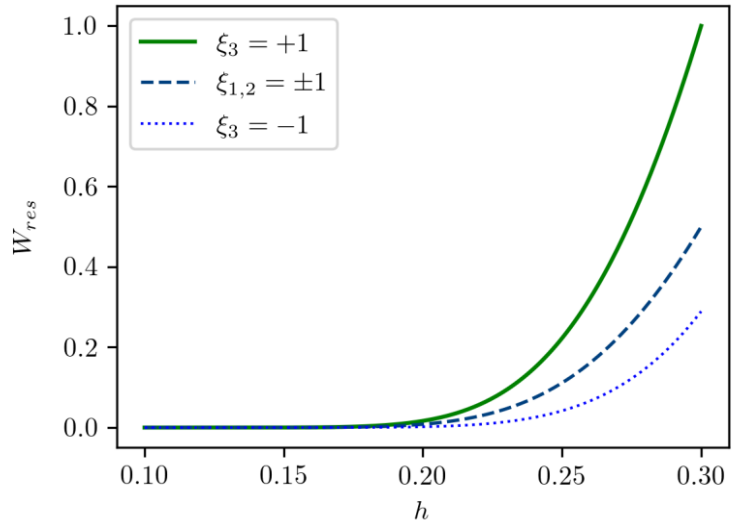

Fig.6.2. Dependence of probability of the CPPPA process on the magnetic field strength

In the second chapter it was shown that the probability of one-photon $e^+e^-$ pair production contains a divergence

$$W_{\text{ОНП}} \sim \sqrt{\frac{1}{\delta\omega}}$$

in cases when the pair is created with zero longitudinal momentum at fixed Landau levels, i.e. in the resonant conditions. However, since the Landau levels have the finite width $\Gamma$, the frequency of the initial photon can only be specified to the nearest width $\Gamma$. In other words, in resonance, the detuning of the frequency is approximately equal to the width $\delta\omega \approx \Gamma$.

It should be emphasized that the parameters $\delta\omega$, $\Gamma$ in the problem of the resonant CPPPA process are phenomenological and the exact proof of the relation between them is beyond the frame of considered approaches. Let us take the detuning of the photon frequency in the form

$$\delta\omega = 4\Gamma. \qquad (6.22)$$

This can be argued by the presence of two particles in the intermediate state at fixed Landau levels, remembering that each level is twice degenerate due to spin. Then the probability of the CPPPA process in the resonance (6.19), as well as the optical theorem (6.17) take the form

$$W_{res} = W_{OPP}W_{AP}, \quad W_{OPP} = -\operatorname{Im} A_{if}(0). \qquad (6.23)$$

Note that the probability of the CPPPA process according to expression (6.23) is proportional to the square of the distance traveled by the beam $W_{res} \sim L^2$, which was first noted in [13] for the process of $e^+e^-$ pair production by an electron (analysis of this process will be performed in the next chapter).

<u>Non-resonant CPPPA process.</u> In the range between the resonances, the frequency of the initial photon can be represented as

$$\omega = \omega_{res} + \delta\omega, \quad \delta\omega = \kappa hm, \qquad (6.24)$$

i.e. it consists of the resonant frequency (6.11) and the additional term $\delta\omega$ and which is part of the cyclotron frequency, in this case $0<\kappa<1$. In this case and in the LLL approximation, the probability amplitude (6.8) can be reduced to the form

$$A_{if} = \frac{-i\pi^4 e^2 hm^2}{V\omega\sqrt{\kappa h}} \delta^4(k-k')J^2(s_1 e_z e'_z + s_2 H_m H_m^{'*}), \qquad (6.25)$$

where $s_1 = 1 + \frac{24i}{\pi}\sqrt{\kappa h} - h(\frac{13}{8}\kappa + \frac{5}{4}N_+ + n_g n_f)$, $s_2 = \frac{N_+ h}{2}$.

For the amplitude (6.25), the optical theorem obviously holds $W_{OPP} = -\text{Im}\, A_{if}(0)$.

The probability of the CPPPA process in non-resonant conditions corresponding to the amplitude (6.25) has the form

$$W_{nonres} = \frac{\pi e^4 h}{2^{11}\kappa} J^4 m^2 T \delta(\omega-\omega')(1+\xi_3)(1+\xi'_3), \qquad (6.26)$$

where $s_1=1$, $s_2=0$.

It should be noted that the found probability (6.26) with accuracy up to a constant factor coincides with the resonant probability (6.20), where small components of the order of the parameter $h$ are discarded. The ratio of the non-resonant probability to the resonant one is equal to

$$\frac{W_{nonres}}{W_{res}} = \frac{\Gamma}{2^6 \kappa h m}. \qquad (6.27)$$

For the case $h=0.1$, $\kappa=0.5$, this ratio is equal to $W_{nonres}/W_{res}=3\cdot 10^{-5}$.

### 6.4. Polarization effects

<u>Change of photon polarization in the CPPPA process.</u> Let us consider how the polarization of the initial photon changes after the production and annihilation of a pair in a magnetic field. Note that in the expressions for the probabilities of the CPPPA process in resonant (6.20) and non-resonant conditions (6.26) in the largest terms (terms with the least degree of the parameter $h$), there is the same dependence on the Stokes parameters of the initial and final photons. Thus, the resonance does not actually change the photon polarization.

Let us analyze the photon polarization in the resonant conditions. The polarization degree of the final photon, defined as (2.4), using expressions (6.20) and in the case $\xi_3 \neq -1$ we can be obtained

$$P = \xi'_3 + N_+ h \frac{\xi_1 \xi'_1 + \xi_2 \xi'_2}{1 + \xi_3}, \tag{6.28}$$

whence the expressions for the Stokes parameters of the final photon have the following form

$$\xi'_3 = 1, \quad \xi'_1 = N_+ h \frac{\xi_1}{1 + \xi_3}, \quad \xi'_2 = N_+ h \frac{\xi_2}{1 + \xi_3}. \tag{6.29}$$

Thus, the polarization of final photon (in the linear approximation by $h$) is abnormally linear $\xi'_3 = 1$ and it is practically independent of the polarization of the initial photon, and the radiation of the final photon is completely polarized $P = 1$.

In the case when the initial photon is normally linearly polarized $\xi_3 = -1$, the process is suppressed (the probability contains an additional second-order factor of the parameter $h$). As follows from expression (6.21), the final photon is also normally polarized $\xi'_3 = -1$, i.e. the normal linear polarization of the photon does not change in the CPPPA process.

Note that if a photon is polarized as its own polarization mode ($\xi_3 = -1$ is normal mode, $\xi_3 = 1$ is anomalous mode), its polarization does not change when propagated in a magnetic field, which is a known result [171].

Vacuum birefringence (VB). As follows from an analysis of the above, the CPPPA process in a magnetic field is essentially "anomalous beam" in the process of birefringence when the beam propagates through an active medium. The active medium is an area of strong magnetic field. "Normal beam" should be considered the process of photon propagation without production and annihilation of $e^+e^-$ pair, which is described by the diagonal elements of the scattering matrix. The VB process in a magnetic field is described by Feynman diagrams shown in Fig.6.3. As noted above, the anomalous beam (Fig. 6.3.b) almost always has anomalous linear polarization $\xi'_3 = 1$. The normal beam has a polarization that coincides with the polarization of the initial photon $\vec{\xi}' = \vec{\xi}$, because this beam corresponds to the diagonal elements of the scattering matrix $S$.

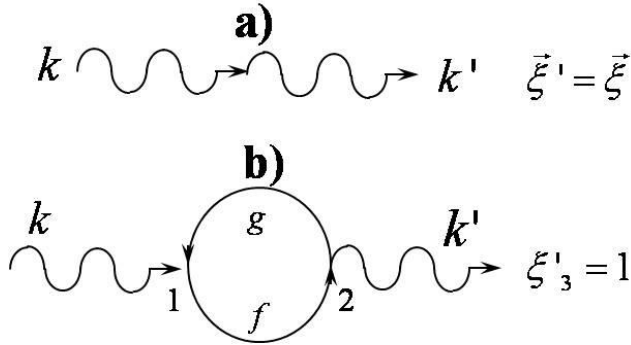

Fig.6.3. Feynman diagrams of normal a) and anomalous b) rays for the process of vacuum birefringence in a magnetic field

The amplitude of the VB process corresponding to the Feynman diagrams in Fig.6.3 consists of two terms

$$A_{if} = A_{if}^{(0)} + A_{if}^{(2)}, \qquad (6.30)$$

where $A_{if}^{(2)}$ is the amplitude of the CPPPA process in the general case has the form (6.8), $A_{if}^{(0)}$ is diagonal elements of the scattering matrix for one photon with fixed 4-momentum and polarization, which have the form

$$A_{if}^{(0)} = \frac{(2\pi)^4}{VT} \delta^4(k - k')(\vec{e}\vec{e}'), \qquad (6.31)$$

here $(\vec{e}\vec{e}')$ is the convolution of two polarization vectors of the initial and final photons. Let us write the expression for the square of modulus of this convolution

$$|\vec{e}\vec{e}'|^2 = \frac{1}{2}(1 + \vec{\xi}\vec{\xi}'). \qquad (6.32)$$

From expression (6.32) follow the obvious properties: 1) the polarization of the final photon is equal to the polarization of the initial one, 2) for the vectors $\vec{e}$ and $\vec{e}'$ such equality is fulfilled

$$\frac{1}{2}\sum_{polar}|\vec{e}\vec{e}\,'|^2 = 1, \tag{6.33}$$

where $1/2\sum_{polar}$ means averaging over initial and summation over final polarizations. The probability corresponding to the amplitude (6.31) has the form

$$W^{(0)} = \frac{\pi}{T}(1+\vec{\xi}\vec{\xi}\,')\delta(\omega-\omega') \tag{6.34}$$

and it is equal to

$$W^{(0)} = 1, \quad \omega' = \omega, \quad \vec{\xi}\,' = \vec{\xi}. \tag{6.35}$$

The probability of the VB process in a magnetic field corresponding to the amplitude (6.30) in the LLL approximation in resonant conditions (6.12) can be reduced to the form

$$W = \frac{2\pi}{T}\delta(\omega-\omega')M_{\xi\xi'} \tag{6.36}$$

where $M_{\xi\xi'} = \frac{1}{2}(1+\frac{B^2}{2}(1+\xi_3))\left[1+\vec{b}\vec{\xi}\,'\right]$, $B = \frac{e^2 h e^{-2/h}}{4\sqrt{\Gamma/m}}mT$. $\vec{b}$ determine the polarization of the final photon and they have the form

$$b_1 = \frac{\xi_1 - B\xi_2}{1+\frac{B^2}{2}(1+\xi_3)}, \quad b_2 = \frac{\xi_2 + B\xi_1}{1+\frac{B^2}{2}(1+\xi_3)}, \quad b_3 = \frac{2\xi_3 + B^2(1+\xi_3)}{2+B^2(1+\xi_3)}. \tag{6.37}$$

It should be noted if we choose for the initial photon $\xi_3 = \pm 1$, $\xi_1 = \xi_2 = 0$ (i.e. the photon is initially abnormally or normally linearly polarized), then from expression (6.37) it follows that $b_3 = \pm 1$, $b_1 = b_2 = 0$. The polarization does not change, which means the well-known above-mentioned result that normal and anomalous linear polarizations are eigenmodes of polarization.

The square of the degree of polarization of the finite photon is equal to

$$P'^2 = 1 - \frac{(1+B^2)(1-P^2)}{(1+B^2(1+\xi_3)/2)^2}, \tag{6.38}$$

where $P^2 = \xi_1^2 + \xi_2^2 + \xi_3^2$ is the square of the degree of polarization of the initial photon. If the initial photon is completely polarized $P=1$, then the final photon is also completely polarized regardless of the magnetic field strength $h$ and the propagation length of the photon in the magnetic field $L=cT$.

As can be seen from (6.37) the polarization of the finite photon is significantly influenced by $L$ and $h$, but only due to one parameter $B$. Let us consider a few limiting cases.

Let $B<<1$ (weak magnetic field and small area size), then the Stokes parameters of the final photon $\vec{b}$ differ slightly from the Stokes parameters of the initial photon $\vec{\xi}$. The change of polarization can be characterized by a change of these parameters $\Delta\vec{\xi} = \vec{b} - \vec{\xi}$, which is equal to

$$b_1 - \xi_1 = -B\xi_2, \quad b_2 - \xi_2 = B\xi_1, \quad b_3 - \xi_3 = 0. \tag{6.39}$$

If $\xi_1 = \pm 1$ (linear polarization at an angle of $\pm 45^0$), then

$$b_1 - \xi_1 = 0, \quad b_2 - \xi_2 = \pm B, \quad b_3 - \xi_3 = 0,$$

i.e. in this case, the ellipticity is maximum and proportional to the quantity $B$. The plane of linear polarization does not change orientation. If $\xi_2 = \pm 1$ (circular polarization), then

$$b_1 - \xi_1 = \mp B, \quad b_2 - \xi_2 = 0, \quad b_3 - \xi_3 = 0,$$

i.e. with accuracy up to the first degree by $B$, the circular polarization does not change, but the linear polarization at an angle of $\mp 45^0$ to the magnetic field appears. If $\xi_3 = \pm 1$ (abnormal, normal polarization), then

$$b_1 - \xi_1 = 0, \quad b_2 - \xi_2 = 0, \quad b_3 - \xi_3 = 0,$$

i.e. as mentioned earlier, the polarization does not change. The obtained results are in good agreement with the previously known results of the Cotton-Mouton effect in the framework of classical electrodynamics [294].

Let us consider the case when $B>>1$ (strong magnetic field and large area). Then for $\xi_3 \neq -1$ we have

$$b_1 = -\frac{2\xi_2}{(1+\xi_3)B}, \quad b_2 = \frac{2\xi_1}{(1+\xi_3)B}, \quad b_3 = 1. \quad (6.40)$$

Thus, for the case of large distances and a strong magnetic field, the final photon has predominantly anomalous linear polarization, which weakly depends on the polarization of the initial photon.

If the initial photon is unpolarized $\vec{\xi} = 0$, then after passing the area with magnetic field, the photon acquires the following Stokes parameters

$$b_1 = 0, \quad b_2 = 0, \quad b_3 = \frac{B^2}{2+B^2} > 0, \quad (6.41)$$

i.e. the final photon is abnormally linearly polarized with the degree of polarization $P' = b_3$, which can be reduced to the form

$$P' = \frac{e^4 h^2 e^{-4/h}}{32\Gamma} m^3 L^2, \quad (6.42)$$

where the width $\Gamma = 4e^2 h^2 m / 3$. In the case, when the linear size $L$ and magnetic field strength $H$ have values characteristic of laboratory conditions ($L\sim 1m$, $H < 10^6 Gs$), the degree $P'$ is negligible, i.e. initially unpolarized ray after passing through such area remains unpolarized, which is a well-known fact. The characteristic value of magnetic field strength for isolated X-ray pulsars is $H=10^{13}Gs$ or $h=0.2$. For such field values, the degree of polarization is equal to

$$P' \approx 4 \cdot 10^{-13} (L/R_c)^2,$$

where $R_c$ is Compton wavelength of an electron. The photon ray becomes completely polarized $P'=1$ at a distance $L=1\mu m$. Also, it is easy to see from expression (6.42) that for complete polarization of radiation after passing the area ~10km (characteristic size of neutron stars), the required magnetic field strength is $H=3 \cdot 10^{11}Gs$

### 6.5. Conclusions to the Chapter 6

Theory of the cascade of processes of one-photon an electron-positron pair production with subsequent annihilation to a photon taking into account the polarization of photons in a strong external magnetic field in resonant and non-resonant conditions is constructed. As a result, it was shown:

1. In the LLL approximation, the resonances in the CPPPA process occur if the frequency of the initial photon is equal to the sum of the energies of the intermediate electron and positron, which are at fixed Landau levels with zero longitudinal momenta. At distances multiple of the cyclotron frequency, a series of resonances are located for which the distance between adjacent peaks is the order of $h^2$.

2. The optical theorem for the probability amplitude of the CPPPA process (the imaginary part of the amplitude without changing the parameters of the initial photon is equal to the total probability of the OPP process with the opposite sign) both in resonant and non-resonant conditions is performed.

3. The expression for the probability of the CPPPA process near the resonance is reduced to the Breit-Wigner form. In the resonance case, the probability of the CPPPA process is equal to the product of the probabilities of the OPP process and one-photon $e^+e^-$ pair annihilation. The greatest probability of the cascade occurs for abnormally linear polarization of photons ($\xi_3=1$). The probability of the non-resonant CPPPA process (between resonances) for the field *h=0.1* is five orders of magnitude less than for the resonant process.

4. If the initial photon is not normally linearly polarized $\xi_3 \neq -1$, then in the CPPPA process the polarization of the final photon is abnormally linear $\xi'_3 = 1$ and practically it does not depend on the polarization of the initial photon. In the case when the initial photon is normally linearly polarized $\xi_3 = -1$, the process is suppressed (the probability contains an additional factor $h^2$) and the final photon is also normally polarized $\xi'_3 = -1$. In both cases, the radiation of the final photons is completely polarized $P = 1$.

5. Photons passing through an area with a magnetic field, both without interaction with the vacuum and through the CPPPA process, form two rays of vacuum birefringence.

Normally and abnormally linearly polarized photons do not change the polarization in the VB effect. After the photons pass an area of small size (when the change of the polarization of photons is weak), the VB effect coincides with the known quasiclassical result for birefringence in an anisotropic medium in the presence of a magnetic field. If the initial photons ray is unpolarized, then after passing the photons of the area with a magnetic field due to the VB effect, it becomes partially abnormally linearly polarized $\xi'_3 \neq 0$. The degree of polarization depends exponentially on the value of magnetic field strength. In the resonant conditions and the magnetic field $H=10^{13} Gs$, the photons are completely polarized after passing through the area of size $L=1 \mu m$.

The main scientific results of the section are published in [295-298].

# CHAPTER 7
# ELECTRON POSITRON PAIR PRODUCTION
# BY AN ELECTRON

## 7.1. Introduction

In this chapter, we study the process of $e^+e^-$ pair production by an electron (trident process), near the threshold in resonance conditions. We aim to calculate the resonant process rate and to analyze how the initial electron spin projection influences the process. As an application, we consider the SLAC experiment on the positron production in a collision of a 46.6 GeV electron beam with a terawatt laser pulse [236]. We use the Nikishov-Ritus theorem about the equivalence of the rate of QED processes with ultrarelativistic particles in arbitrary field configuration, and calculate the number of events in the experiment.

## 7.2. The trident process rate

<u>The probability amplitude of the trident process.</u> The probability amplitude of the process corresponds to Feynman diagrams shown in Fig. 7.1, and can be written as

$$A_{if} = A_1 - A_2, \qquad (7.1)$$

$$A_1 = ie^2 \iint d^4x \, d^4x' (\overline{\Psi}_1 \gamma^\mu \Psi) D_{\mu\nu} (\overline{\Psi}'_2 \gamma^\nu \Psi'_+), \qquad (7.2)$$

$$A_2 = ie^2 \iint d^4x \, d^4x' (\overline{\Psi}_2 \gamma^\mu \Psi) D_{\mu\nu} (\overline{\Psi}'_1 \gamma^\nu \Psi'_+), \qquad (7.3)$$

where $A_1, A_2$ are the amplitudes corresponding the first and the second diagrams respectively, $\Psi, \overline{\Psi}_1, \overline{\Psi}_2, \Psi_+$ are the wave functions of the initial and final elections and the final positron. Their explicit expressions are given in Chapter 2. The primed functions depend on the primed 4-vector $x'$; $D_{\mu\nu}$ is the Green function of the intermediate photon of the form [191]

$$D_{\mu\nu} = \frac{g_{\mu\nu}}{(2\pi)^4} \int d^4k \, e^{-ik(x-x')} D(k), \quad D(k) = \frac{2\pi}{k^2}. \qquad (7.4)$$

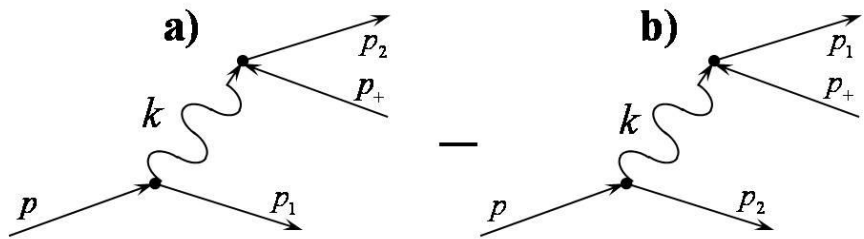

Fig 7.1. Feynman diagrams of the process of $e^+e^-$ pair production by an electron.

Note that when process kinematics allows the denominator of the Green function to vanish (the intermediate photon is on-shell), $k^2 = 0$, (the intermediate photon is on-shell), the process becomes resonant. To eliminate the resonant divergence we use the common Breit-Wigner prescription and introduce the small imaginary term in the photon frequency,

$$\omega \to \omega - i\Delta/2, \qquad (7.5)$$

where $\Delta$ is a spread in photon frequency (energy width) connected with finite widths of energy levels of the initial and final particles.

Let the magnetic field be directed along the $z$ axis. The expressions under the integrals in Eqs. (7.2) and (7.3) depend on time and ($y$, $z$) coordinates similarly to the considered processes. The integration over these variables yields $\delta$ functions expressing the conservation laws of energy and momentum,

$$\varepsilon_l = \varepsilon_1 + \varepsilon_2 + \varepsilon_+, \quad p = p_1 + p_2 + p_+, \qquad (7.6)$$

where $\varepsilon_l, p$ are the energy and the longitudinal momentum of the initial electron occupying a Landau level $l$, $\varepsilon_1, \varepsilon_2, \varepsilon_+$ are the energies of the final particles and $p_1, p_2, p_+$ are their longitudinal momenta.

Since magnetic field does not change after Lorentz transformation to a reference frame moving along $z$ axis, without loss of generality we set the longitudinal momentum of the initial electron to zero,

$$p = 0, \quad \varepsilon_l = m_l = m\sqrt{1 + 2lh}. \tag{7.7}$$

We consider the TRIDENT process near the threshold when the final particles occupy the ground energy level,

$$l_1 = l_2 = l_+ = 0. \tag{7.8}$$

After integration in Eq. (7.2) using the special function definition (2.33), the amplitude of the first Feynman diagram (Fig.7.1.a) can be transformed to

$$A_1 = \frac{ie^2\pi^2 M_m}{S^2 m_l \sqrt{2\varepsilon_1 \varepsilon_2 \varepsilon_+}} B_1^\mu \int dk_x D(k) I^{'*}(l_1, l) I(l_2, l_+) \cdot \delta^3(p - p_1 - p_2 - p_+), \tag{7.9}$$

where the factors $B_1^\mu$ are

$$B_1^\mu = \frac{E_m^2 p_1 + E_{1m}^2 p}{E_m E_{1m}} \frac{E_{2m}^2 p_+ + E_{+m}^2 p_2}{E_{2m} E_{+m}} - \frac{E_m^2 E_{1m}^2 + p p_1}{E_m E_{1m}} \frac{E_{2m}^2 E_{+m}^2 + p_2 p_+}{E_{2m} E_{+m}}, \tag{7.10}$$

$E_m$ and $M_m$ are the quantities defined by Eqs. (2.40) and (2.42); $S$ is the normalizing area in the $y, z$ plane; three $\delta$ functions express the conservation laws. The arguments of the special functions $I(l_2, l_+)$ and $I'(l_1, l)$ are, respectively,

$$\eta = \frac{\kappa_y^2 + k_y^2}{2hm^2}, \quad \eta' = \frac{\kappa_y'^2 + k_y^2}{2hm^2}, \tag{7.11}$$

where

$$\kappa_y = p_{2y} + p_{+y}, \quad \kappa'_y = p_y - p_{+y}. \tag{7.12}$$

Considering the $\delta$ functions in Eq. (7.9) we get $\eta = \eta'$, $\kappa_y = \kappa'_y$. Further, due to conditions (7.8), the quantities $B_1^\mu$ take the simple form

$$B_1^+ = 4m\sqrt{mm_l}\,\text{sgn}(p_{+z}), \quad B_1^- = 4p_{1z}\sqrt{mm_l}\,\text{sgn}(p_{+z}). \tag{7.13}$$

After the change of variables according to

$$q = \frac{k_x}{m\sqrt{2h}}, \quad s = \frac{p_y - p_{1y}}{m\sqrt{2h}}, \quad u = \frac{p_y - p_{2y}}{m\sqrt{2h}} \quad (7.14)$$

the probability amplitude (7.8) near the process threshold (7.7) finaly can be transformed to

$$A_1 = \frac{i 2\pi^3 e^2 M_m}{S^2 \sqrt{m_l \varepsilon_1 \varepsilon_2 \varepsilon_+} m\sqrt{l!h}} e^{-a^2} B_1^\mu X_1 \delta^3 (p - p_1 - p_2 - p_+), \quad (7.15)$$

Here, the function $X_1$ is

$$X_1 = \int_{-\infty}^{\infty} dq \frac{(s+iq)^l}{2/h - s^2 - q^2} e^{-q^2 - 2iuq}. \quad (7.16)$$

Amplitude $A_2$ that corresponds to the second Feynman diagram in Fig. 7.1. can be obtained from the expression (7.15) after the change of variables according to

$$e^{-s^2} B_1^\mu X_1 \to e^{-u^2} B_2^\mu X_2,$$

where $B_2^\mu$ and $X_2$ are given by Eqs. (7.13) and (7.15) with index replacements $1 \leftrightarrow 2$ and $s \leftrightarrow u$.

The process kinematics. The kinematics of the trident process are determined by the conservation laws (7.6). The process is possible when $\varepsilon_l \geq 3m$. At first glance it can be expected that at the process threshold the final particles occupy fixed Landau levels and have zero longitudinal momenta. With account of (7.6) it results in

$$p_{1z} = p_{2z} = p_{+z} = 0, \quad m_l = m_{l_1} + m_{l_2} + m_{l_+}. \quad (7.17)$$

However, in the general case these conditions are not fulfilled because the effective masses depend on level numbers, hence they are discrete quantities. Thus, at the process threshold the final particles can have nonzero longitudinal momenta, in contrast to the SPP process considered in Chapter 2.

Expanding the second equation in (7.17) into a series in small momenta results in an equation of a triaxial ellipsoid in the space of longitudinal momenta of the final particles,

$$\frac{p_{1z}^2}{b_1^2} + \frac{p_{2z}^2}{b_2^2} + \frac{p_{+z}^2}{b_+^2} = 1, \quad (7.18)$$

where $b_1^2 = 2m_{l_1}\delta\varepsilon$, $b_2^2 = 2m_{l_2}\delta\varepsilon$, $b_+^2 = 2m_{l_+}\delta\varepsilon$ and $\delta\varepsilon = m_l - m_{l_1} - m_{l_2} - m_{l_+}$.

The possible values of particle momenta correspond to the points $\lambda$ of an ellipsis defined by intersection of the ellipsoid (7.18) and a plane defined by the momentum conservation law (7.17) (see Fig. 7.2). The points $a$ and $b$ are the intersections of the ellipsis $\lambda$ with the plane ($p_{+z}$, $p_{1z}$), the points $c$ and $d$ are the intersections of the ellipsis $\lambda$ with the plane ($p_{2z}$, $p_{+z}$).

At the process threshold the final particles occupy the ground state (see (7.8)) and the following conditions are true,

$$\delta\varepsilon \leq hm, \quad p_{1z} \sim p_{2z} \sim p_{+z} \leq \sqrt{hm}. \quad (7.19)$$

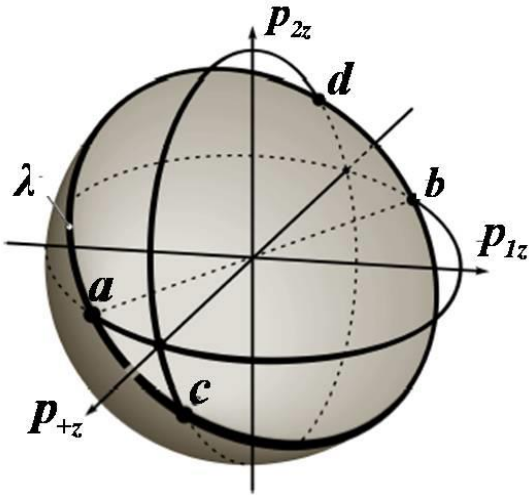

Fig. 7.2. Representation of the threshold momenta of the final particles (ellipsis $\lambda$) in the momentum space $p_{1z}, p_{2z}, p_{+z}$

The threshold value of the Landau level number of the initial electron is

$$l_{th} = [4/h] + 1, \quad (7.20)$$

where the square brackets denote the integer part of the number inside.

<u>The trident process rate.</u> The process rate equals to the product of the square of the absolute value of the process amplitude and the number of the final states,

$$dN = \frac{S^2 d^2 p_1}{(2\pi)^2} \frac{S^2 d^2 p_2}{(2\pi)^2} \frac{S^2 d^2 p_+}{(2\pi)^2}. \quad (7.21)$$

The differential rate of the trident process can be transformed to

$$dW^\mu = C_0 | e^{-s^2} X_1 B_1^\mu - e^{-u^2} X_2 B_2^\mu |^2 \, \delta^3(p - p_1 - p_2 - p_+) d^2 p_1 d^2 p_2 d^2 p_+, \tag{7.22}$$

where $\mu = \pm 1$ is spin direction of the initial electron and $C_0$ is a factor of the form

$$C_0 = \frac{e^4 M_m}{2^7 \pi^3 m^2 \varepsilon_1 \varepsilon_2 \varepsilon_+ m_l^2 l! h}. \tag{7.23}$$

The integration in $d^2 p_+$ can be carried out using the properties of two Dirac $\delta$ functions. After the change of variables $dp_{1y} dp_{2y} = 2m^2 l ds du$, the rate (7.22) takes on the form

$$dW^\mu = 2m^2 l C_0 \left[ B_1^{\mu 2} Y + B_2^{\mu 2} Y - 2 B_1^\mu B_2^\mu Y' \right] \delta(m_l - \varepsilon_1 - \varepsilon_2 - \varepsilon_+) dp_1 dp_2, \tag{7.24}$$

where

$$Y = \iint ds du \, e^{-2s^2} |X_1|^2 = \iint ds du \, e^{-2u^2} |X_2|^2, \tag{7.25}$$

$$Y' = \iint ds du \, e^{-s^2 - u^2} \operatorname{Re}(X_1 X_2^*). \tag{7.26}$$

Note that the quantity $Y'$ determines the interference term in the process rate.

Consider the case when the final particles occupy the ground level and have nonrelativistic energies, that is

$$p_1, p_2, p_+ \ll m,$$

Then, near the process threshold the total rate of the trident process can be expressed as

$$W^+ = \frac{e^4 (m_l - m)}{2\pi^3 m_l l!} (Y - Y') \iint \delta(m_l - \varepsilon_1 - \varepsilon_2 - \varepsilon_+) dp_1 dp_2, \tag{7.27}$$

$$W^- = \frac{e^4 (m_l + m)}{4\pi^3 m^2 m_l l!} (2Y + Y') \iint \delta(m_l - \varepsilon_1 - \varepsilon_2 - \varepsilon_+) dp_1 p_2^2 dp_2, \tag{7.28}$$

Integrals in Eqs. (7.27) and (7.28) can be calculated using the $\delta$ function that depends on particles energy. To do so, we transform it according to

$$\delta(f(p_1)) = \frac{\delta(p_1 - g_+)}{\left|\dfrac{df}{dp_1}\right|_{g_+}} + \frac{\delta(p_1 - g_-)}{\left|\dfrac{df}{dp_1}\right|_{g_-}}, \qquad (7.29)$$

where the function $f(p_1)$ is $f(p_1) = m_l - \varepsilon_1(p_1) - \varepsilon_2 - \varepsilon_+(p_1 + p_2)$ and $g_+$, $g_-$ are its the roots. Near the process threshold, this function and its derivative are

$$f(p_1) = \delta\varepsilon - \frac{p_1^2 + (p_1 + p_2)^2}{2m}, \quad \frac{df}{dp_1} = -\frac{2p_1 + p_2}{m}, \quad \delta\varepsilon = m_l - 3m. \qquad (7.30)$$

The roots of the function $f(p_1)$ look like

$$g_\pm = \frac{1}{2}(-p_2 \pm \sqrt{4m\delta\varepsilon - 3p_2^2}). \qquad (7.31)$$

It follows from (7.31) that the limit values of the momentum $p_2$ are

$$-p_{max} \le p_2 \le p_{max}, \quad p_{max} = \sqrt{4m\delta\varepsilon/3}. \qquad (7.32)$$

With this results, the last integration over $dp_2$ in (7.27) - (7.28) can be easily carried out, and double integrals equal to

$$\iint \delta(m_l - \varepsilon_1 - \varepsilon_2 - \varepsilon_+) dp_1 dp_2 = 2m\pi/\sqrt{3}, \qquad (7.33)$$

$$\iint \delta(m_l - \varepsilon_1 - \varepsilon_2 - \varepsilon_+) dp_1 p_2^2 dp_2 = 4\pi m^2 \delta\varepsilon/3\sqrt{3}. \qquad (7.34)$$

Finally, the total rate of the TRIDENT process is

$$W^+ = \frac{e^4 m}{3\sqrt{3}\pi^2 l!}(Y - Y'), \quad W^- = \frac{2e^4 \delta\varepsilon}{9\sqrt{3}\pi^2 l!}(2Y + Y'). \qquad (7.35)$$

Note, that the above expressions have been multiplied by the factor of 1/2! to account for particle identity.

Let us now calculate the quantities $X_1$ and $X_2$, and consequently $Y$ and $Y'$. The integral $X_1$ (7.16) can be easily transformed to

$$X_1 = \sum_{k=0}^{l} C_k^l s^{(l-k)} i^k D_k, \quad C_k^l = \frac{l!}{k!(l-k)!}, \qquad (7.36)$$

where $C_k^l$ are the binomial coefficients and $D_k$ are the integrals defined as

$$D_k = \int_{-\infty}^{\infty} \frac{q^k e^{-q^2-2iuq}}{r^2 - q^2} dq, \quad r^2 = \frac{2}{h} - s^2. \tag{7.37}$$

Note that the expression under the integral in (7.37) diverges if the condition $r^2 > 0$ is true. Accordingly, the value of $X_1$ is small when $r^2 < 0$ and can be neglected. Consequently, we consider the case when the $s$ variable satisfy the condition

$$-\sqrt{2/h} \leq s \leq \sqrt{2/h}. \tag{7.38}$$

The divergence is eliminated by introduction the width $\Delta$ of the intermediate state according to the prescription of Breit and Wigner (7.5),

$$r^2 \to \rho^2 = r^2 + ig, \quad g = \Delta/mh. \tag{7.39}$$

Let us first to calculate the integral $D_0$ defined by Eq. (7.37) with $k=0$. It is convenient to use the relation

$$\frac{e^{\rho^2-q^2}}{\rho^2 - q^2} = \int_0^1 dt\, e^{t(\rho^2-q^2)} + \frac{1}{\rho^2 - q^2},$$

Inserting this into Eq. (7.37) we obtain

$$D_0 = e^{-\rho^2}(I_1 + I_2), \quad I_1 = \int_{-\infty}^{\infty} \frac{e^{-2iuq} dq}{\rho^2 - q^2}, \quad I_2 = \int_0^1 \sqrt{\frac{\pi}{t}} e^{t\rho^2 - u^2/t} dt. \tag{7.40}$$

Taking into account the prescription (7.39), the first integral $I_1$ in (7.40) can be written as

$$I_1 = \int_{-\infty}^{\infty} \frac{e^{-2iuq} dq}{\rho^2 - q^2} = \frac{-\pi i}{\rho} e^{2i|u|q}. \tag{7.41}$$

To find the integral $I_2$, take into account the relation

$$\int^t \frac{e^{z-\frac{1}{z}} dz}{\sqrt{z}} = -i\int^t (\frac{i}{2\sqrt{z}} + \frac{1}{2z^{3/2}}) e^{z-\frac{1}{z}} dz - i\int^t (\frac{i}{2\sqrt{z}} - \frac{1}{2z^{3/2}}) e^{z-\frac{1}{z}} dz.$$

After a change of variables in the first and the second terms according to

$$y = \sqrt{z}i - \frac{1}{\sqrt{z}}, \quad y = \sqrt{z}i + \frac{1}{\sqrt{z}}$$

we get

$$\int_0^t \frac{e^{z-\frac{1}{z}}dz}{\sqrt{z}} = -i\frac{\sqrt{\pi}}{2}[e^{-2i}erf(\sqrt{t}i - \frac{1}{\sqrt{t}}) + e^{2i}erf(\sqrt{t}i + \frac{1}{\sqrt{t}})],$$

where $erf(x) = \frac{2}{\sqrt{\pi}}\int_0^x dy e^{-y^2}$ is the error function. The definite integral on an interval [0,1] is

$$\int_0^1 \frac{e^{z-\frac{1}{z}}dz}{\sqrt{z}} = \frac{\sqrt{\pi}}{2}[-2\sin 2 + ie^{-2i}erf(1-i) - ie^{2i}erf(1+i)], \quad (7.42)$$

which give us the sought values of the integral $I_2$,

$$I_2 = \frac{-\pi i}{2\rho}[e^{-2i\rho|u|}erfc(|u|-i\rho) - e^{2i\rho|u|}erfc(|u|+i\rho)], \quad (7.43)$$

Here, $erfc(x) = 1 - erf(x)$. Finally, inserting (7.41) and (7.43) in (7.40), the explicit expression of $D_0$ can be transformed to

$$D_0 = \frac{-i\pi e^{-\rho^2}}{2\rho}[e^{-2i\rho u}erfc(u-i\rho) + e^{2i\rho u}erfc(-u-i\rho)], \quad (7.44)$$

This expression is valid in both cases of $u > 0$ and $u < 0$.

To find the integral $D_k$ (7.37), we express it as a derivative of $D_0$ in respect to the parameter $u$:

$$D_k = \int_{-\infty}^{\infty} \frac{q^k e^{-q^2-2iuq}}{\rho^2 - q^2} dq = \frac{1}{(-2i)^k}\frac{\partial^k}{\partial u^k}D_0, \quad (7.45)$$

Finally, with account of the definition of the Hermite polynomial $H_n(x)$,

$$H_n(x) = (-1)^n e^{x^2}\frac{d^n}{dx^n}e^{-x^2}, \quad (7.46)$$

the expression of $D_k$ can be written in the form

$$D_k = \frac{-i\pi e^{-\rho^2}}{2\rho^{1-k}}[e^{-2i\rho u}erfc(u-i\rho) + (-1)^k e^{2i\rho u}erfc(-u-i\rho)] +$$

$$+ \frac{\sqrt{\pi}e^{-u^2}}{i(2i)^k \rho}\sum_{m=1}^{k} C_m^k (2i\rho)^{k-m}[H_{m-1}(u-i\rho) + (-1)^k H_{m-1}(-u-i\rho)]. \quad (7.47)$$

It is clear from Eq. (7.44) that $D_0$ diverges at the point $\rho = 0$ or $s = \sqrt{2/h}$. At the same time, in the case $k \geq 1$, the expression for $D_k$ (7.47) is finite at the point $\rho=0$ because it contains factors $\rho^{k-1}$ and $\rho^{k-m-1}$. It does not diverge in the both cases of $m < k$ (apparently), and $m = k$, $\rho=0$, because of the property of Hermite polynomials,

$$[H_{k-1}(u) + (-1)^k H_{k-1}(-u)] = 0.$$

Consequently, the main contribution in $X_1$ (7.36) comes from the first term with $k=0$, i.e.

$$X_1 = s^l D_0. \tag{7.48}$$

Consider now the quantity $Y$ (7.25). The quantity $X_1$, is an even function of $u$, hence the integral over $u$ can be written as

$$\int_{-\infty}^{\infty} du \, |e^{-s^2} X_1|^2 = \frac{\pi^2 s^{2l} e^{-4/h}}{2|\rho|^2}(J_1 + J_2), \tag{7.49}$$

$$J_1 = \int_{-\infty}^{\infty} du \, |e^{-2iu\rho} \mathrm{erfc}(u - i\rho)|^2, \quad J_2 = \int_{-\infty}^{\infty} du \, e^{-4iur} \mathrm{erfc}(u - ir) \cdot \mathrm{erfc}(-u + ir). \tag{7.50}$$

The integral $J_1$ transforms to

$$J_1 = \frac{1}{\sqrt{\pi} \, \mathrm{Im}(\rho)} \mathrm{Re}(e^{-2ig} j(\rho)), \quad j(\rho) = \int_{-\infty}^{\infty} du \, e^{-(u+i\rho)^2} \mathrm{erfc}(u - i\rho). \tag{7.51}$$

After the change $t=u+i\rho$, the derivative of $j(\rho)$ with respect to $\rho$ reduces to the Poisson integral,

$$dj(\rho)/d\rho = 2\sqrt{2} i e^{2\rho^2}. \tag{7.52}$$

With the initial condition $j(\rho) \xrightarrow{\rho \to +0} \sqrt{\pi}$, the solution of the differential equation (7.52) is

$$j(\rho) = \sqrt{\pi} + \sqrt{\pi} \mathrm{erf}(i\sqrt{2}\rho). \tag{7.53}$$

Since $\mathrm{Re}(e^{-2ig}) \approx 1$, the quantity $J_1$ takes the form

$$J_1 = \frac{1}{\mathrm{Im}(\rho)}[1 + \mathrm{Re}(e^{-2ig} \mathrm{erf}(i\sqrt{2}\rho))]. \tag{7.54}$$

Similarly, the integral $J_2$ (7.50) is

$$J_2 = -i \cdot erf(i\sqrt{2}\rho)/r. \tag{7.55}$$

After inserting the obtained expressions (7.54) and (7.55) in (7.49) and (7.25), we can write the sought quantity $Y$ in the form

$$Y = \frac{\pi^2 e^{\frac{-4}{h}}}{2} \int_{-\sqrt{2/h}}^{\sqrt{2/h}} ds \frac{s^{2l}}{|\rho|^2} \left( \frac{1}{\text{Im}\,\rho} + \frac{\text{Re}(e^{-2ig} erf(i\sqrt{2}\rho))}{\text{Im}\,\rho} + \frac{erf(i\sqrt{2}\rho)}{ir} \right). \tag{7.56}$$

In the integral over the $s$ variable (7.56), the main contribution comes from small vicinities of the points $s = \pm\sqrt{2/h}$ because of the factor $s^{2l}$. At these points, the first term in the brackets in (7.56) equals to $\sqrt{2}/\rho$, while the second and the third terms are equal to $\pm\sqrt{8/\pi}$ and can be neglected. The imaginary part of $\rho$ can be written in the form

$$\text{Im}\,\rho = \sqrt{\sqrt{\rho^4 + g^2} - \rho^2}/\sqrt{2} \tag{7.57}$$

Introducing a new variable $x = s/\sqrt{2/h}$, the quantity $Y$ can be transformed to

$$Y = \frac{\sqrt{2}\pi^2 2^l e^{\frac{-4}{h}}}{gh^l} \int_0^1 dx \cdot x^{2l} \sqrt{\frac{\sqrt{(1-x^2)^2 + \delta^2} + (1-x^2)}{(1-x^2)^2 + \delta^2}}, \tag{7.58}$$

where $\delta = g^2/\sqrt{2/h}$. If $\delta$ goes to zero, the integral in $x$ in Eq. (7.58) goes to the value of

$$\frac{\Gamma(1/2)\Gamma(l+1/2)}{\sqrt{2}\Gamma(l+1)}.$$

Finally, the quantity $Y$ takes the form

$$Y = \frac{\pi^2 \sqrt{\pi} 2^l e^{\frac{-4}{h}} \Gamma(l+1/2)}{gh^l l!}. \tag{7.59}$$

The integral $Y'$ (7.26) can be calculated in the same manner. Near the process threshold the condition $2/h \gg 1$ is true, and $Y'$ is much less than $Y$:

$$Y' \ll Y. \tag{7.60}$$

Finally, near the threshold, the rate of the trident process with the final particles occupying the ground state can be written as

$$W^+ = \frac{\sqrt{\pi}}{6\sqrt{3}} \cdot \frac{e^4 mh(2/h)^l e^{-4/h}\Gamma(l+\frac{1}{2})}{l!^2 \Delta/m}, \quad W^- = \frac{2\sqrt{\pi}}{9\sqrt{3}} \cdot \frac{e^4 \delta\varepsilon h(2/h)^l e^{-4/h}\Gamma(l+\frac{1}{2})}{l!^2 \Delta/m}.$$

(7.61)

The ratio between these rates equals to

$$W^- / W^+ = 4\delta\varepsilon / 3m, \qquad (7.62)$$

Thus, with account of Eq. (7.19) we conclude that near the process threshold the main channel is trident with the initial electron in the inverted spin state, ($\mu=+1$). In a particular case when magnetic field strength equals to $h=4/l$ and $\delta\varepsilon=0$, the trident reaction channel with the initial electron in the ground spin state is forbidden, $W^-=0$.

The obtained Eqs. (7.61) describe the resonant channel of the trident process, when the intermediate photon is on the mass shell. In this case, after averaging over the spin projection of the initial electron, the total process rate decomposes to the product of the rates of the one-vortex processes, namely SR and OPP. It follows from the threshold condition $\varepsilon_l = 3m$ that $hl=4$ and $l>>1$, so that the trident rate can be written as

$$W = \frac{\sqrt{\delta\varepsilon/m}}{3\sqrt{6}\Delta} \cdot W_{e\to\gamma e} \cdot W_{\gamma\to ee^+}, \qquad (7.63)$$

where $W_{e\to\gamma e}$ is the rate of the SR process for the case of radiative transition from a highly excited level $l>>1$ to the ground level $l'=0$, and $W_{\gamma\to ee^+}$ is the rate of OPP to the ground levels. With account of Eqs. (2.44), (2.156) and the threshold condition, the rates $W_{e\to\gamma e}$, $W_{\gamma\to ee^+}$ are

$$W_{e\to\gamma e} = \sqrt{\pi}e^2 m \frac{(2/h)^l e^{-2/h}}{\Gamma(l+1/2)l}, \quad W_{\gamma\to ee^+} = \frac{e^2 mh e^{-2/h}}{2\sqrt{2\delta\varepsilon/m}}. \qquad (7.64)$$

To obtain Eq. (7.63), we used the definition of the gamma function of a half-integer argument and the Stirling's formula,

$$\Gamma(l+1/2) = \sqrt{\pi}(2l)!/4^l l!, \quad l! = \sqrt{2\pi l}(l/e)^l.$$

Note that the expression for $W_{\gamma \to ee^+}$ in (7.64) does not account for the electron energy width and diverges if $\delta\varepsilon = \varepsilon_l - 3m$ goes to zero. Account of the corresponding width results in an estimation of the minimum value of $\delta\varepsilon$ of order of $\Delta \sim e^2 h^2 m$. On the other hand, to ensure pair production to the ground level we require the condition $\delta\varepsilon < hm$ to be met. Let be $\delta\varepsilon = khm$ with the constant coefficient $k$ and let magnetic field be $h = h_0 + \delta h$, where $h_0$ is determined by the threshold condition $kh_0 = \sqrt{1+2lh_0} - 3$. Then, $\delta h = 3kh_0/l$ which is apparently much less than $h_0$.

The estimation of the trident rate. As were said before, the quantity $\Delta$ entering (7.61) has the meaning of the intermediate state width. The main term in this width is the radiative width, i.e. the full SR rate of the initial electron.

As an example, let the initial parameters be:
$$l = 40, \ h_0 = 0.1 \ \to \ \delta\varepsilon \approx 0.05m, \ \delta h = 0.00375. \quad (7.65)$$

The width and trident rate estimations are
$$\Delta = 4 \cdot 10^{17} c^{-1}, \ W^+ = 1.2 \cdot 10^4 c^{-1}, \ W^- = 0. \quad (7.66)$$

According to Eq. (7.64), the corresponding rates of the single-vertex processes are
$$W_{e \to \gamma e} = 2.1 \cdot 10^{13} c^{-1}, \ W_{\gamma \to ee^+} = 7.9 \cdot 10^9 c^{-1}. \quad (7.67)$$

Note that in Ref. [13] the resonant $e^+e^-$ pair production by an electron is considered as a two-step cascade of the SR and OPP processes, i.e. the rate was defined as $W = W_{e \to \gamma e} \cdot W_{\gamma \to ee^+}$. The authors use the ultrarelativistic approximation for both the initial electron and the final particles. However, this approximation is not valid near the process threshold when the final particles occupy the ground level. Moreover, Ref. [13] does not take into account the radiative width, and the decay time of the intermediate photon is set to be equal to half the observation time. This results in overestimation of the process rate near the threshold. In particular, for the parameter values given by (7.56), the estimation is $4.7 \cdot 10^7 \ s^{-1}$, which is three order of magnitude greater than the corresponding value (7.66).

Figure 7.3 shows the dependence of the trident rate on the Landau level number of the initial electron (a) and magnetic field strength (b) for the parameter values given by (7.56). As can be seen in Fig. 7.3 (a), the rate $W^+$ is maximum at the threshold and decreases monotonically with increasing of the level number $l$. At the same time, the rate $W^-$ reaches its maximum value when the $l$ number exceeds the threshold value by a few levels and then decreases. Note that at sufficiently large values of $l$ the rate $W^-$ exceeds $W^+$. Figure shows moderate decrease of the rate at fixed $l$ while it significantly increases for lower values of the Landau level number of the initial electron.

Let us now compare the rate of trident process with the rates of other QED processes in magnetic field considered in the previous chapters. Table 7.1 shows rates of the SR, OPP, DSR, OPPE, CPPPA and trident processes in the LLL approximation for magnetic field strength of $h=0.1$. The first row lists the names of the processes, the second row shows the corresponding Feynman diagrams, the third row lists the initial parameter values and, finally, the last row lists the rate estimation.

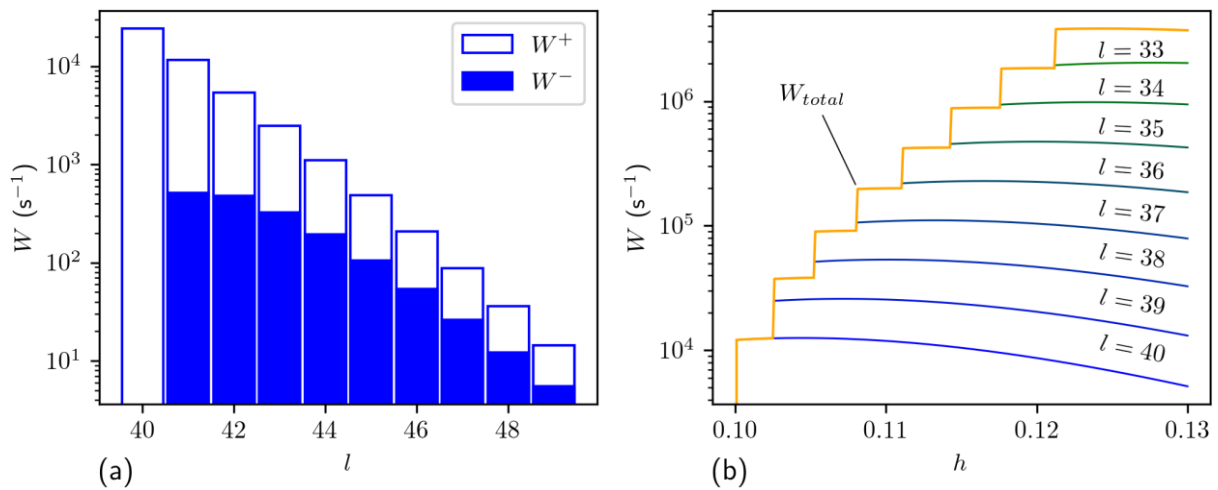

Fig. 7.3. Dependence of the total trident rate on the Landau level number of the initial electron (a) and magnetic field strength (b) for the parameter values given by (7.56).

To estimate the SR rate we use Eqs. (2.46) – (2.49), with integration over the emission angle, summation over the final polarizations and averaging over the initial electron spin. The initial electron energy is set to $3m$, so that the SR rate equals to $W_{e \to \gamma e}^{total} = 2.8 \cdot 10^{17}\ s^{-1}$.

Table 7.1. The rates ($s^{-1}$) of some QED processes in the magnetic field of $h=0.1$

| 1 | SR | OPP | DSR | OPPEB | CPPPA | trident |
|---|---|---|---|---|---|---|
| 2 | 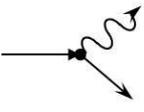 | 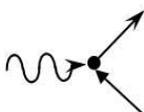 | 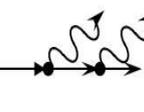 | 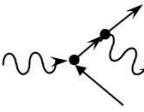 | 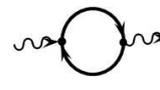 | 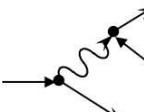 |
| 3 | $l=40$, $l'=0$ | $\omega=2m$, $l_+=l_-=0$ | the lowest levels | the lowest levels | the lowest levels | $l=40$, $\varepsilon_l = 2m$, $l_+=l_2=l_1=0$ |
| 4 | $W_{e \to \gamma e}^{total} \sim 10^{17}$ $W_{e \to \gamma e}^{\omega > 2m} \sim 10^{13}$ $W_{e \to \gamma e}^{1 \to 0} \sim 10^{16}$ | $W_{\gamma \to e^-e^+} \sim 10^{10}$ | $W_{e \to e\gamma\gamma}^{res} \sim W_{e \to e\gamma}$ | $W_{\gamma \to \gamma ee^+}^{nonres} \sim 10^{6}$ $W_{\gamma \to \gamma ee^+}^{res} \sim W_{\gamma \to ee^+}$ | $W_{\gamma \to \gamma}^{res} = W_{\gamma \to ee^+} W_{ee^+ \to \gamma}$ | $W_{e \to eee^+} \sim 10^{4}$ |

Excluding the photon radiation with energy insufficient to produce a pair, it is possible to estimate the rate (7.67) as $W_{e \to \gamma e}^{\omega > 2m} = 2.1 \cdot 10^{13}\ s^{-1}$. To complete the picture, we also present the estimation of the SR rate for an electron transition from the first excited level to the ground state, $W_{e \to \gamma e}^{1 \to 0} = 3 \cdot 10^{16}\ s^{-1}$. To estimate the OPP rate we use Eq. (7.64). It were shown in Chapter 3 and 5 that in the resonance conditions, the rates of DSR and OPPE are of the same order of magnitude as the rates of SR and OPP respectively. In other words, the addition of another final photon does not change the efficiency of SR and OPP processes at resonance conditions. In resonance conditions, the process of $e^+e^-$ pair production by a photon with subsequent annihilation are purely cascade one, and its rate equals to the product of the rates of the corresponding first-order processes. Note that CPPPA process is described by the probability instead of the rate, contrary to the other processes.

### 7.3. $e^+e^-$ pair production in the SLAC experiment

It is not yet possible to observe the QED processes in an external magnetic field of the strength approaching the critical value. On the other hand, as mentioned in Chapter 1, QED processes have already been observed in experiments with intense laser pulses at SLAC facility [235-237]. In Ref. [236], the process of $e^+e^-$ pair production by an electron in a pulsed laser field was studied. The schematic layout of the experiment is shown in Fig.1.6. About 100 positrons have been observed in 21962 collisions of a 46.6 GeV electron beam with green ($\lambda$=527 nm) terawatt laser pulses. The measure of the laser intensity is work of the field at the wavelength, which equals to the classical invariant parameter $\eta = e\sqrt{A^\mu A_\mu}/mc^2$, where $A_\mu$ is the electromagnetic 4-potential. In the SLAC experiment conditions, the invariant intensity parameter is $\eta$=0.36.

In Ref. [236], the positron appearance is interpreted as a result of two consequence processes. First, high-energy photons are generated by Compton backscattering in a head-on collision of GeV electrons and a laser pulse. Second, the hard photons create electron-positron pairs in the multiphoton Breit and Wheeler reaction,

$$e^- + n\omega_0 \to e^- + \omega, \qquad (7.68)$$

$$\omega + n'\omega_0 \to e^- + e^+, \qquad (7.69)$$

where $\omega_0$ denotes a laser photon. Note that in Ref. [236] the two-step process described by (7.68) and (7.69) is distinguished from less probable 'trident' process, or PPE process in a laser field,

$$e^- + n''\omega_0 \to e^- + e^-e^+. \qquad (7.70)$$

Let us consider in detail how differs description of positron production in both cases, namely two-step production in Compton backscattering and consequent Breit and Wheeler process, and the 'trident' production of Bethe-Heitler type. The first case is described by two Feynman diagrams (see Fig. 7.4). The solid lines correspond to wave functions of free electrons (positrons). Note that for the sake of clarity, Fig. 7.4 does not show the exchange diagrams needed for correct calculation of the process rate.

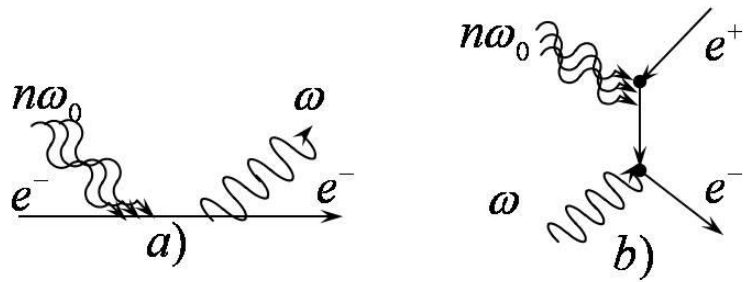

Fig.7.4. Feynman diagrams: a) Compton scattering of $n$ laser photons by an electron, b) multiphoton Breit-Wheeler process, i.e. pair production of an $e^+e^-$ pair by a hard photon and $n$ laser photons.

It should be noted that the description of generation of a hard photon with energy $> 2m_e c^2$ by Compton backscattering according to reaction (7.68) and Feynman diagram 7.4 is not accurate. A more accurate description is the representation of this process by a first-order diagram, namely the photon emission by an electron in an external laser wave. In this case, the Volkov functions are used as wave functions of electrons, which account the external laser field to all orders of perturbation theory (Fig. 7.5). In Fig. 7.5, the solid lines correspond to the Volkov functions of the electron. As was shown by A.I. Nikishov and V.I. Ritus [202, 203], if the intensity is small and only the absorption of a single photon has nonzero contribution to the expression of spontaneous photon radiation, then the emission rate coincides with the cross-section of Compton scattering. Thus, reaction (7.68) is entirely contained in the spontaneous emission process shown in Fig.7.5.

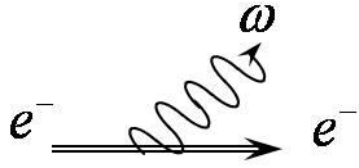

Fig.7.5. Feynman diagram of photon emission by an electron in a laser wave

Similarly, the Breit and Wheeler reaction (7.69) should be replaced by single photon $e^+e^-$ pair production in a laser field, Fig. 7.6.

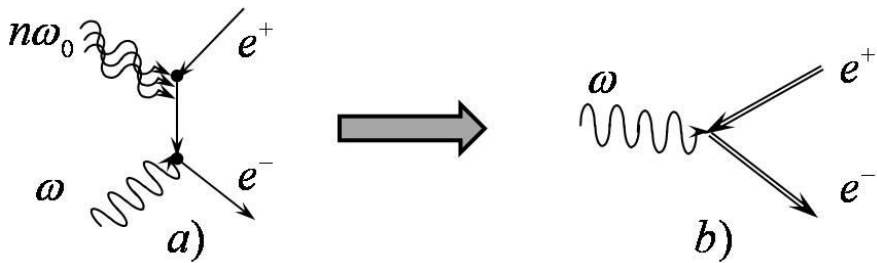

Fig.7.6. Feynman diagrams: a) the Breit-Wheeler process and b) the process single photon $e^+e^-$ pair production in a laser field.

Summarizing the above considerations, the $e^+e^-$ pair production in the SLAC experiment is described by a twostep process: an electron in a laser field emits a hard photon, which propagates in the same laser wave and converts to an $e^+e^-$ pair.

It is worth noting that in this picture the intermediate hard photon $\omega$ cannot be observed experimentally. This photon is described by a propagator in the probability amplitude in contrast to the case of a free photon. Therefore the intermediate photon can be either real (i.e., on-shell) or virtual. In particular, the emission and absorption moments are not defined for a virtual photon, hence the situation is possible when $e^+e^-$ pair production happens before emission of the virtual intermediate photon by the initial electron. Thus, the pair production in the SLAC experiment can be fully described by a Feynman diagram similar to one shown in Fig. 7.1. This lead us to the second scenario with the trident process indicated in Ref. [236]. In the resonant conditions the rate of the trident process decomposes to the product of the rates of the considered one-vertex processes with an additional factor, which can only be obtained within the consistent theoretical treatment.

It is possible to apply the theory of the trident process in a magnetic field (trident process in a magnetic field) to the description of the SLAC experiment [236] by using the Nikishov and Ritus theorem. In Refs. [202,203], Nikishov and Ritus considered QED processes with relativistic particle energies and proved analytically that the process rates expressed in terms of gauge invariants are the same for any configuration of external electromagnetic field. In particular, they obtained the rate of spontaneous emission by an electron in a laser field. If the laser field varies slowly, in the limit of vanishing laser frequency, these rates can be applied to the corresponding processes in the crossed electric and magnetic fields when $\vec{E} \perp \vec{H}$ and $|\vec{E}|=|\vec{H}|$. The electromagnetic field enters the total probability of these processes in the form of a single invariant parameter $e^2(F_{\mu\nu}p^{\nu})^2/m^6$, where $F_{\mu\nu}$ is the electromagnetic field tensor and $p^{\nu}$ is 4-momentum. This allow us to extend the consideration to the case of arbitrary external electromagnetic field. Note that in the general case the process rates depend on the additional invariants $e^2(F_{\mu\nu})^2/m^4$ and $ie^2\varepsilon_{\mu\nu\rho\sigma}F^{\mu\nu}F^{\rho\sigma}$. In the considered case of crossed fields, $\vec{E} \perp \vec{H}$ and $|\vec{E}|=|\vec{H}|$, the additional parameters equal to zero. In the general case, they have non-zero values, however, they are much less than unity for feasible field strength. Moreover, the additional invariants can be safely neglected in comparison with $e^2(F_{\mu\nu}p^{\nu})^2/m^6$ if particle energy is large. In particular, considering $F^{\mu\nu}$ as a magnetic field, Nikishov and Ritus reproduced the results of Klepikov [8] concerning the radiation intensity and pair production efficiency in a magnetic field. The Nikishov and Ritus theorem has the simple physical meaning: after Lorentz transformation to the rest frame of an ultrarelativistic particle, any field transforms to almost equal and almost perpendicular electric and magnetic fields.

Let us find the magnetic field $H_{eq}$, that would be equivalent to a laser field of the strength $E_L$ in the sense described above. If a high-energy electron propagates opposite to an electromagnetic wave of the strength $E_L$, then in its rest frame it experiences the field of the strength of $E_0 = 2\gamma E_L$, where $\gamma$ is the gamma factor. On the other hand, if an electron moves perpendicular to a magnetic field $H_{eq}$, then the electric field strength in the rest frame is $E_{0eq} = \gamma H_{eq}$. Comparing $E_0$ and $E_{0eq}$ we conclude that the equivalent magnetic field in the laboratory reference frame is

$$H_{eq} = 2E_L. \qquad (7.71)$$

The factor 2 emerges in (7.71), apparently, because the equivalent magnetic field should 'replace' both the electric and magnetic field components of a laser wave.

To account the time dependent character of an electromagnetic wave, we need to average the obtained expressions of the PPE rate in a magnetic field (7.61) $W(H)$ over the wave period. This defines the rate of the equivalent PPE process in laser wave [205],

$$W_{eq}(E_L) = \frac{2}{\pi} \int_0^{\pi/2} W(H_{eq} \sin\varphi) d\varphi. \qquad (7.72)$$

The above equation allows to compare the QED process rates in a magnetic field and in a field of an intense laser wave.

Note that Eqs. (7.61) have been obtained near the process threshold, when the condition $\varepsilon_l \approx 3m$ is met. Therefore, we need to pass to a moving 'threshold' frame where the electron energy satisfies the same condition $\varepsilon \approx 3m$. Note that this 'threshold' reference frame should not be confused with the rest frame, where $\varepsilon = m$. After some calculation we find that in the experiment conditions the equivalent magnetic field in the threshold frame is

$$H_{eq} = 6.1 \cdot 10^{12}\, G, \text{ or } h = 0.14. \qquad (7.73)$$

Note that in the SLAC experiment, $e^+e^-$ pair production has been observed near the reaction threshold too [236]. Despite the high energy of the electron beam of about 46.6 GeV, the large part of this energy is the energy of the rectilinear motion of the center of mass.

Let us estimate the interaction time $\Delta t_L$ between the electron beam and a laser pulse, and the electron number $N_{int}$ in the interaction region. According to [236], electron beam size is about $\sim 25 \times 40 \ \mu m^2$, the electron number in a bunch is $\sim 7 \cdot 10^9$, the focal area of the laser beam is $\sim 30 \ \mu m^2$, and the intersection angle between electron and laser beams is $\sim 17^0$. As a result, we find

$$\Delta t_L = 50 \text{ fs}, \quad N_{int} = 2.8 \cdot 10^8. \tag{7.74}$$

Note that both radiative width $\Delta_{rad}$ and the width associated with finite interaction time $1/\Delta t_T$ contribute to the total width of the intermediate photon $\Delta$ entering (7.61),

$$\Delta = \Delta_{rad} + 1/\Delta t_T = \Delta_{rad} + \gamma/\Delta t_L, \tag{7.75}$$

where $\Delta t_T$ is the time of beam-laser interaction in the threshold reference frame.

The number of created $e^+e^-$ pairs is given by

$$N_{e^+e^-} = k \cdot N_{int}(1 - e^{-W_{eq}\Delta t_T}), \tag{7.76}$$

where $k=21962$ is the number of collisions between laser pulses and electron bunches [236]. Apparently, the number of pairs (7.76) is Lorentz invariant and stays the same in the laboratory frame, though it is defined in the threshold reference frame. With the parameter values given above, numerical estimation of the event number given by Eq. (7.76) is

$$N_{e^+e^-} = 80. \tag{7.77}$$

The obtained value is in reasonable agreement with the experimental result of 106±14 positron events and the numerical estimation given in Ref. [239]. Note that it is indicated in Ref. [236], that the experimental result possibly includes residual background of $\sim 2 \cdot 10^{-3}$ positrons per laser pulse due to the interaction of backscattered photons with residual gas. If the experimental data restricted to events with $\eta > 0.216$, this reduces positron number to 69±9, which is in an even better agreement with theoretical estimation (7.77).

### 7.4. Conclusions to the Chapter 7

In this chapter, we considered the process of pair production by an electron (trident process) in a strong magnetic field within consistent relativistic theory. The following conclusions can be formulated.

1. In the LLL approximation near the process threshold the final particles are created at fixed Landau levels and their longitudinal momenta are of order of $p_{1z} \sim p_{2z} \sim p_{+z} \leq \sqrt{h}m$.

2. The main contribution to the PPE rate near the threshold comes from the resonant channel with on-shell intermediate photon. In this case, the trident rate factorizes to the product of the rates of the SR and OPP processes.

3. The maximum rate has the channel with the spin projection of the initial electron oriented along the field (inverted spin state). If the initial electron is in the ground spin state (i.e. spin is directed opposite to the field), then the process rate is less by the order of magnitude of the small parameter $h$.

4. The trident rate is in inverse proportion to the energy width of the intermediate photon. In particular, if at the process threshold the field is $h = 0.1$ and Landau level number of the initial electron is $l = 40$, then the estimation of the trident rate and the corresponding width are $W = 1.2 \cdot 10^4 \, \text{s}^{-1}$ and $\Delta = 4 \cdot 10^{17} \, \text{s}^{-1}$ respectively.

5. The Nikishov and Ritus theorem allows us to estimate the number of events in the SLAC experiment on observation of positron production in collisions of relativistic electron beam and laser pulse. Corresponding value of 80 events is in reasonable agreement with the experimental result of 106±14 events.

The main scientific results of this chapter are published in [299-301].

# LIST OF SYMBOLS AND ABBREVIATIONS

QED        - Quantum Electrodynamics

QFT         - Quantum Field Theory

SR           - Syncrotron Radiation, emission of a photon by an electron

OPP        - $e^+e^-$ Pair Production by One photon

CS           - Compton Scattering, scattering of a photon by an electron

DSR        - Double Syncrotron Radiation, emission of two photons by an electron

TPP         - $e^+e^-$ Pair Production by Two photons

OPPE - $e^+e^-$ Pair Production by One photon with photon Emission

VB           - Vacuum Birefringence

CPPPA     - Cascade of processes of $e^+e^-$ Pair Production and subsequent the Pair Annihilation

PPE        - Trident process, $e^+e^-$ Pair Production by an Electron

LLL        - Lowest Landau Level

FAIR - Facility for Antiproton and Ion Research

SPARC     - Stored Particles Atomic Physics Research Collaboration

HESR      - High Energy Storage Ring

SLAC      - Stanford Linear Accelerator Center

**Unit of measurement**

The paper uses a relativistic system of units, in which the Planck constant and the speed of light are equal to one. The following units are used:

for energy – *eV* (electron volts),

for magnetic induction – *Gs* (Gauss).

# АНОТАЦІЯ

*Холодов Р.І., Новак О.П., Дяченко М.М.* Резонансні і поляризаційні ефекти в процесах квантової електродинаміки в сильному магнітному полі – монографія – Інститут прикладної фізики Національної академії наук України, Суми, 2020.

 Робота присвячена теоретичному дослідженню елементарних процесів квантової електродинаміки (КЕД): (синхротронне випромінювання (СВ), однофотонне народження $e^+e^-$ (позитрон-електронної) пари (ОНП), розсіяння фотона на електроні (РФЕ), двофотонне синхротронне випромінювання (ДСВ), двофотонне народження $e^+e^-$ пари (ДНП), однофотонне народження $e^+e^-$ пари з випромінюванням фотона (ОНПВ), народження $e^+e^-$ пари електроном (НПЕ), каскадне народження $e^+e^-$ пари фотоном з подальшою анігіляцією в один фотон (КНПАП)) в сильному магнітному полі з поляризованими частинками і фотонами. Розроблено методи аналізу спін-поляризаційних ефектів (ефектів впливу поляризації початкових фотонів на напрямок спінів кінцевих частинок і навпаки вплив спінів початкових частинок на поляризацію кінцевих фотонів), резонансних ефектів (ефектів пов'язаних з виходом частинок у проміжних станах на массову оболонку) в процесах КЕД в сильному магнітному полі та створено теорію руху зарядженої частинки в замагніченому електронному газі з анізотропною температурою з врахуванням впливу знака заряду.

 В першому розділі проведено огляд робіт і проаналізовано сучасний стан досліджень процесів КЕД в сильному магнітному полі.

 В другому розділі розроблено методику вивчення спін-поляризаційних ефектів. Вивчено спін-поляризаційні ефекти у процесах випромінювання фотона електроном (СВ) і народження електрон-позитронної пари фотоном (ОНП). Показано, що в ультраквантовому наближенні (з частинками на низьких рівнях Ландау) в процесі СВ електрона поляризація випромінювання збігається з поляризацією, одержаною в рамках класичної електродинаміки, якщо електрон не змінює напрямку спіна і знаходиться або в основному, або в інверсному спіновому стані, при цьому імовірність в першому випадку більша. Спін-фліп процес в основний спіновий стан (з проекцією спіна проти поля $\mu$=-1) змінює лінійну поляризацію випромінювання з нормальної (з параметром Стокса $\xi_3$=-1) на аномальну ($\xi_3$=+1) і в $h$ разів ($h=H/H_0=e\hbar H/m^2c^3$, $H$-магнітне поле) менший за основний процес. Спін-фліп процес в інверсний спіновий стан дуже малий і становить частку ~$h^3$ від основного процесу. Врахування спін-фліп процесу зменшує ступінь поляризації випромінювання на величину ~$h$. Показано, що в ультрарелятивістському наближенні в процесі СВ у випадку $z$>1 ($z=H\varepsilon/H_0m$, $\varepsilon$ - енергія електрона, $z$~1 відповідає енергії $\varepsilon$ ~10$TeB$ в магнітному полі ~$10^6 Гс$) поляризація випромінювання від спочатку поляризованого пучка електронів суттєво залежить від енергії електрона, проекції його спіна і частоти випромінювання. Для початкових електронів зі спінами, що направлені проти поля, ступінь поляризації випромінювання монотонно падає з ростом $z$. Для електронів в інверсному спіновом стані ступінь поляризації випромінювання як функція $z$ має суттєво немонотонний характер. Випромінювання в площині орбіти змінює поляризацію від нормальної до аномальної і, навпаки, при поступовому збільшенні енергії електрона (з ростом $z$). Показано, що в процесі ОНП для аномально поляризованого фотона ($\xi_3$=+1) народжені $e^+e^-$ пари повністю орієнтовані в основні спінові стани. За винятком малого інтервалу ~$h$ поблизу $\xi_3$=-1 ступінь орієнтації злегка порушений на величину $h$. У вузькому інтервалі значень поблизу нормальної поляризації фотона $\xi_3$=-1 спінова орієнтація народжених частинок залежить від різниці їх енергій. Частинки не поляризовані, якщо їх енергії однакові. З меншою енергією частинки народжуються переважно в основний спіновий стан.

В третьому розділі в процесі розсіяння фотона на електроні (РФЕ) в магнітному полі вивчено спін-поляризаційні ефекти в резонансних умовах, тобто вивчено вплив поляризації початкових фотонів як на поляризацію випромінювання, так і на спінові стани кінцевих електронів для різних спінових станів початкового електрона. Вивчено процес випромінювання двох фотонів електроном в магнітному полі (ДСВ) в резонансних умовах з поляризованими фотонами і певними значеннями проекцій спіна електронів. Аналіз процесів РФЕ і ДСВ проведено в ультраквантовому наближенні. Показано, що з точністю до $h$ резонансна частота початкового (кінцевого) фотона кратна циклотронній і дорівнює відстані між рівнями Ландау початкового (кінцевого) і проміжного електронів. З точністю до $h^2$ частота початкового фотона в резонансі крім номерів рівнів Ландау частинок, також залежить від полярних кутів початкового і кінцевого фотонів. Врахування поляризації фотонів не впливає на умови виникнення резонансів. Проміжний електрон в резонансних умовах має певне значення проекції спіна. Диференціальні імовірности процесу РФЕ і процесу ДСВ (без перевороту спіна) поблизу резонансних умов факторизуються і приведені до форми Брейта-Вігнера. Відмінністю "спін-флипа" РФЕ від СВ є те, що імовірності перевороту спіна в РФЕ в основний і інверсний стани близькі за величиною, в той час як в процесі СВ переворот в інверсний спіновий стан пригнічений. Електрони зі спінами орієнтованими спочатку проти поля (в основному стані) випромінюють повністю поляризовані фотони з поляризацією як в процесі СВ. Якщо початкові фотони мають нормальну лінійну поляризацію $\xi_3 = -1$, то порушення ступеня поляризації електронів пропорційно $h$. Якщо початкові фотони аномально лінійно поляризовані $\xi_3 = +1$, спіни електронів орієнтуються по полю. Електрони зі спінами орієнтованими по полю випромінюють частково поляризовані фотони, ступінь деполяризації випромінювання пропорційний $h$ і залежить від рівня Ландау кінцевого електрона.

Запропоновано схему поляризатора пучка електронів, де напрямки спінів електронів змінюються в процесі РФЕ в магнітному полі пропорційно зміні поляризації електромагнітної хвилі. Лінійно поляризована електромагнітна хвиля міліметрового діапазону ($\lambda$=2.5*мм*) потужністю 10*кВт* в магнітному полі 40*кГс* повністю поляризує пучок електронів за час $\tau$=10$^{-10}$*с* на ділянці розміром 2*мм*.

В четвертому розділі вивчено процес двофотонного народження $e^+e^-$ пари (ДНП) з урахуванням спінів частинок і поляризацій фотонів в області резонансу. Аналізується вплив поляризації початкових фотонів на ступінь поляризації пучків кінцевих частинок. Проведено порівняння процесів ОНП і ДНП в магнітному полі $H \sim 10^{12} Гс$, характерному для магнітосфери рентгенівських пульсарів і враховано поляризацію електронів і позитронів при генерації випромінювання $e^+e^-$ плазми магнітосфери пульсара. Показано, що в процесі ДНП резонанс можливий поблизу порога, якщо частота жорсткого фотона перевищує суму енергій пари, а частота м'якого фотона кратна циклотронній. Найбільший переріз процесу відповідає народженню частинок в основні спінові стани. Він має максимальне значення, якщо м'який фотон нормально поляризований, а жорсткий фотон поляризований аномально. У випадку поля $H=10^{12} Гс$ переріз процесу порядку томсонівського, ширина резонансу 30 $eB$. В процесі з аномально лінійними поляризованими жорсткими фотонами ($\xi_3=1$) зміна лінійної поляризації м'яких фотонів у всьому діапазоні змінює орієнтування спінів електронів від повністю орієнтованих проти поля до повністю орієнтованих по полю, не змінюючи напрямку спіна позитрона. В процесі з нормально лінійно поляризованими жорсткими фотонами ($\xi_3=-1$) ступінь орієнтування спінів електронів не залежить від поляризації м'якого фотона і визначається тільки рівнями Ландау проміжної і кінцевих частинок. Для процесу з найнижчими можливими енергетичними рівнями електрони повністю неполяризовані, а першим збудженим рівням відповідає переважно нормальна спінова заселеність.

Врахування поля циклотронних фотонів на процес формування $e^+e^-$ плазми в магнітосфері рентгенівського пульсара показало домінуючу роль резонансів в полі $H=10^{12} Гс$ при характерній концентрації фотонів, що спростовує загальноприйняту точку зору про домінуючу роль процесу ОРП у формуванні магнітосфери. Врахування поляризації електронів і позитронів при генерації $e^+e^-$ плазми магнітосфери пульсара змінює спектр СВ, збільшує низькочастотну частину спектру і зменшує високочастотну.

В п'ятому розділі побудована теорія процесу однофотонного народження електрон-позитронної пари з випромінюванням фотона (ОНПВ) з урахуванням спіна частинок в сильному зовнішньому магнітному полі в резонансних і нерезонансних умовах. Показано, що в процесі мають місце парні резонанси (наслідок наявності двох фейнманівських діаграм), відстань між якими багато менша відстані між сусідніми рівнями Ландау. Резонанс реалізується для будь-якої надпорогової частоти початкового фотона. В області парних резонансів частота кінцевого фотона дорівнює відстані між рівнями Ландау проміжного і кінцевого електрона (позитрона), при цьому має місце "інтерференція" двох фейнманівських діаграм. Повна імовірність процесу ОНПВ (процесу другого порядку) в області резонансу збігається з імовірністю однофотонного народження пари (процесу першого порядку). Оцінка імовірності в одиницю часу $W$ в нерезонансній області поблизу порогу, коли частинки народжуються в основні енергетичні стани, для $h$=0.1 дає величину $W \approx 10^6 s^{-1}$, що на 4-и порядки менше імовірності процесу в межах окремого резонансу. Додавання кінцевого фотона в процесі ОНП в резонансній кінематиці не впливає на спіни кінцевих частинок.

В шостому розділі вивчено процес розповсюдження фотона в сильному магнітному полі, коли відбувається каскадне народження e⁺e⁻ пари з подальшою анігіляцією (КНПАП). Проаналізовано ефект вакуумного подвійного променезаломлення (ВПП) в сильному магнітному полі. Показано, що в процесі КНПАП мають місце резонанси, якщо частота початкового фотона дорівнює сумі енергій проміжних електрона і позитрона, які знаходяться на фіксованих рівнях Ландау з нульовими поздовжніми імпульсами, при цьому на інтервалах рівних циклотронній частоті розташовані непоодинокі резонанси, а серія резонансів, для яких відстань між сусідніми піками ~$h^2$. В точці резонансу імовірність КНПАП дорівнює добутку імовірностей ОНП і анігіляції e⁺e⁻ пари в фотон. Найбільша імовірність каскаду відбувається з фотонами аномально лінійної поляризації ($\xi_3$=1). Імовірність нерезонансного процесу КНПАП (між резонансами) для поля $h$ = 0.1 на 5 порядків менше резонансного процесу.

Фотони, що проходять область з магнітним полем, як без взаємодії з вакуумом, так і за участю в процесі КНПАП, складають два променя вакуумного подвійного променезаломлення. Після проходження області невеликих розмірів (коли зміна поляризації фотонів слабка), ефект ВПП збігається з класичною границею. Якщо первинний промінь фотонів неполяризований, тоді після проходження області з сильним магнітним полем через ВПП він придбає часткову аномальну лінійну поляризацію $\xi'_3 \neq 0$. В резонансних умовах в магнітному полі $H=10^{13} Гс$ фотони повністю поляризуються після проходження області розміром $L=1 мкм$.

В сьомому розділі вивчено процес народження $e^+e^-$ пари електроном (НПЕ), тридент процес, поблизу порога в резонансних умовах. Знайдено імовірності процесу в резонансному випадку і проаналізовано вплив напрямку спіна початкового електрона на процес. Показано, що в імовірність НПЕ в одиницю часу поблизу порога головний внесок дає резонансна мода, коли проміжний фотон виходить на масову поверхню. У цьому випадку повна імовірність НПЕ факторизується і виражена через добуток імовірностей процесів першого порядку: СВ і ОНП. Найбільшу імовірність має випадок, якщо спін початкового електрона спрямований уздовж поля. Імовірність процесу НПЕ зі спіном початкового електрона проти поля має меншу величину на порядок малого параметра $h$. Імовірність НПЕ обернено пропорційна енергетичній ширині проміжного фотона. Зокрема, на порозі реакції, коли магнітне поле дорівнює $h = 0.1$, номер рівня Ландау початкового електрона $l = 40$, для імовірності НПЕ і ширини знайдено оцінку $W = 1.2 \cdot 10^4 c^{-1}$, $\Delta = 4 \cdot 10^{17} c^{-1}$. Побудовану теорія народження $e^+e^-$ пари електроном в магнітному полі використано для пояснення SLAC експерименту, в якому було зафіксовано близько 100 позитронів в 21962 подіях при зіткненнях пучка електронів з енергією 46.6 *ГеВ* з променем тераватного імпульсного лазера, з використанням теореми Нікішова-Рітуса, згідно з якою вигляд формул для імовірностей процесів КЕД в зовнішньому електромагнітному полі, виражених через калібрувальні інваріанти, для випадку ультрарелятивістських початкових частинок однаковий для будь-якої конфігурації зовнішнього електромагнітного поля. Здобуте значення 80 $e^+e^-$ пар задовільно узгоджується з експериментальними результатами (106 ± 14 подій).

**Ключові слова:** Квантова електродинаміка, сильне магнітне поле, поляризація, спін-фліп, циклотронний резонанс, найнижчі рівні Ландау.

# ABSTRACT

*Kholodov R.I., Novak O.P., Dyachenko M.M.* Resonance and polarization effects in quantum electrodynamics processes in a strong magnetic field. – Manuscript. – Institute of Applied Physics, National Academy of Sciences of Ukraine, Sumy, 2020.

The papier is devoted to the theoretical research of the elementary processes of quantum electrodynamics (synchrotron radiation (SR), one-photon production of a $e^+e^-$ pair (OPP), photon scattering by an electron (PhSE), double synchrotron radiation (DSR), two-photon production of a $e^+e^-$ pair (TPP), one-photon production of the $e^+e^-$ pair with the emission of a photon (OPPE), a production of a $e^+e^-$ pair by an electron (PPE), a cascade production of a $e^+e^-$ pair by a photon followed by annihilation into one photon (CPPA)) in a strong magnetic field with polarized particles and photons. The methods of analyzing spin-polarization effects (influence of a polarization of the initial photons on a spin direction of the final particles and vice versa), resonance effects (effects, when the particle in an intermediate state goes to the mass shell) are developed in QED processes in a strong magnetic field. The theory of motion of a charged particle in a magnetized electron gas with an anisotropic temperature taking into account the influence of a charge sign is created.

In the first section, an overview of the works is carried out and the current state is analyzed for researches of processes of quantum electrodynamics in a strong magnetic field.

In the second section, the technique of studying spin-polarization effects is developed. Spin-polarization effects in a photon emission by an electron (SR) and electron-positron pair production by one photon (OPP) are studied. It is shown that in the ultraquantum approximation (with particles at low Landau levels) in the process of SR by an electron, the radiation polarization coincides with the polarization obtained in the classical electrodynamics, if the electron does not change the direction of spin and is located either in the main or inverse spin state, with the greater probability in the first case. The spin-flip process to the main spin state (with spin projection against the field $\mu$=-1) changes the linear polarization of radiation from normal (with the Stokes parameter $\xi_3$=-1) to abnormal one ($\xi_3$=+1) and in $h$ times ($h=H/H_0=e\hbar H/m^2c^3$, $H$- magnetic field) less than the main process. The spin-flip process to the inverse spin state is very small and is a fraction about $h^3$ of the main process. Consideration of the spin-flip process reduces the degree of radiation polarization by an amount ~ $h$. It is shown that in SR process in ultrarelativistic approximation in the case $z$>1 ($z=H\varepsilon/H_0m$, $\varepsilon$ - electron energy, $z$~1 corresponds to energy $\varepsilon$ ~ 10$TeV$ in a magnetic field ~$10^6 Tc$) the polarization of radiation from the initially polarized electron beam essentially depends on the electron energy, the projection of its spin, and the frequency of radiation. For the initial electrons with spins against the field the degree of polarization of radiation decreases monotonically with increasing $z$. For electrons in an inverse spin state, the degree of polarization of radiation as a function of $z$ has a substantially non-monotonous character. Radiation in the orbital plane changes polarization from normal to abnormal and vice versa with a gradual increase in the energy of the electrons (with increasing $z$). It is shown that in the OPP process for abnormally polarized photons ($\xi_3$=+1) the e$^+$e$^-$ pairs are produced fully oriented to the main spin states. With exception of a small interval ~$h$ near $\xi_3$=-1, the degree of orientation is slightly disturbed by magnitude $h$. In the narrow range of values near the normal polarization of photons $\xi_3$=-1, the spin orientation of the produced particles depends on difference in their energies. The particles are unpolarized if their energies are the same. With less energy, the particles are produced predominantly in the main spin state.

In the third section, the spin-polarization effects in resonant conditions are studied in the process of scattering a photon on an electron (PhSE) in a magnetic field. That is, the effect of the polarization of the initial photons on both the polarization of radiation and on the spin states of the final electrons is studied for various spin states of the initial electron. The process of radiation of two photons by an electron in a magnetic field (DSR) is studied in resonant conditions with polarized photons and certain values of projections of electron spin. The analysis of the processes of PhSE and DSR is carried out in the Lower Landau Level approximation. It is shown that with accuracy up to magnitude $h$ the resonance frequency of the initial (final) photon is a multiple cyclotron frequency and is equal to the distance between the Landau levels of the initial (final) and the intermediate electrons. With accuracy up to magnitude $h^2$ the frequency of the initial photon in resonance, in addition to the numbers of the Landau levels of the particles, also depends on the polar angles of the initial and final photons. Taking into account the polarization of photons does not affect the conditions for the occurrence of resonances. Intermediate electron in resonant conditions has a certain value of the spin projection. Differential probabilities of the PhSE process and the DSR process (without a spin flip) near the resonant conditions are factored into Breit-Wigner form. A distinctive feature of particle spin flip in the PhSE process from the SR process is that the probability of a spin flips in the RFE to the main and inverse states are close in magnitude, while in the SR process the spin flip to the inverse spin state is suppressed. Electrons in the initial state with spins oriented against field emit completely polarized photons with polarization as in the SR process. If the initial photons have normal linear polarization $\xi_3 = -1$, then the violation of degree of electron polarization is proportional to $h$. If the initial photons have abnormal linear polarization $\xi_3 = 1$, then the electron spins are oriented along the field. Electrons with spins oriented along the field emit partially polarized photons, degree of radiation depolarization is proportional to $h$ and depends on the Landau level of the finite electron. The scheme of a polarizer of an electron beam is proposed, where directions of electron spins are changed in the PhSE process in magnetic field in proportion to the change in polarization of the electromagnetic wave. The linearly polarized electromagnetic wave of the millimeter range ($\lambda = 2.5 mm$) with a power of 10 $kW$ in a magnetic field of 40 $kGs$ completely polarizes the beam of electrons during the time $\tau = 10^{-10} s$ in an area with the size $2mm$.

In the fourth section, the process of two-photon production of $e^+e^-$ pairs (TPP) is studied, taking into account spin particles and photon polarizations in the resonance region. Influence of polarization of the initial photons on the degree of polarization of the final particles is analyzed. Comparison of the OPP and TPP processes is carried out in the magnetic field $H \sim 10^{12} Gs$, which is characteristic for a magnetosphere of X-ray pulsars, and the polarization of electrons and positrons is taken into account in the process of generation of radiation of $e^+e^-$ plasma of magnetosphere of pulsar. It is shown that in the TPP process a resonance near threshold is possible if the frequency of a rigid photon exceeds the sum of pair energies, and the frequency of a soft photon is a multiple cyclotron. The largest cross section of the process corresponds to particle production in the main spin states. It has the maximum value if the soft photon is normally polarized and the rigid photon is polarized abnormally. In the case of field $H=10^{12}$ *Gs*, the cross section of the process is Thomson order, the resonance width is 30 *eV*. In the process with abnomal linear polarized rigid photons ($\xi_3$=1), the change in linear polarization of soft photons in entire range changes the orientation of electron spins from fully oriented against the field to fully oriented along the field without changing the direction of positron spin. In the process with nomal linear polarized rigid photons ($\xi_3$=-1), the degree of orientation of the electron spins does not depend on polarization of the soft photon and is determined only by Landau levels of the intermediate and final particles. For a process with the lowest possible energy levels, the electrons are completely unpolarized, and the first excited levels correspond mostly to normal spin populations. Taking into account of the field of cyclotron photons on the formation of the electron-positron plasma in the magnetosphere of the X-ray pulsar showed the dominant role of the resonances in the field $H = 10^{12}$Gs with the characteristic concentration of cyclotron photons, that refutes the generally accepted view about dominant role of the OPP process in the magnetosphere formation. Taking into account the polarization of electrons and positrons in the $e^+e^-$ plasma generation of pulsar magnetosphere changes the synchrotron radiation spectrum, it increases the low-frequency part of the spectrum and reduces the high-frequency.

In the fifth section, the theory of a process of one-photon production of an electron-positron pair with photon emission (OPPE) is constructed taking into account spin particles in a strong external magnetic field under resonance and nonresonance conditions. It is shown that in the process there are pair resonances (due to the presence of two Feynman diagrams), the distance between which is much smaller than the distance between adjacent Landau levels. The resonance is realized for any over-threshold frequency of the initial photon. In the region of pair resonance the frequency of the final photon is equal to the distance between Landau levels of the intermediate and the final electron (positron), and the "interference" of the two Feynman diagrams takes place. The total probability of the OPPE process (second order process) in the resonance region coincides with the probability of a one-photon pair production (the first-order process). Estimation of the probability in the unit of time $W$ in nonresonance region near the threshold when the particles are produced in the main energy states for $h = 0.1$ gives a value $W \approx 10^6 s^{-1}$, that is 4 times lower than the probability of the process within the boundary of the individual resonance. The addition of a final photon in the OPP process in resonance kinematics does not affect on spins of final particles.

In the sixth section the process of propagation of a photon in a strong magnetic field is studied when cascading production of $e^+e^-$ pair occurs with subsequent annihilation of pair (CPPA). The effect of vacuum birefringence (VB) in a strong magnetic field has been analyzed. It is shown that in the CPPA process the resonances occur if the frequency of the initial photon is equal to the sum of energies of the intermediate electrons and positrons, which are at fixed Landau levels with zero longitudinal momenta. On the interval, that is equal to the cyclotron frequency, the resonance is not single, but is a series of resonances for which the distance between adjacent peaks $\sim h^2$. At the resonance, the probability of CPPA process is equal to product of probabilities of processes of the OPP and the annihilation of $e^+e^-$ pair to a photon. The most probable cascade occurs with photons of abnormally linear polarization ($\xi_3=1$). The probability of a nonresonance CPPA process (between resonances) for a field $h = 0.1$ is 5 orders of magnitude smaller than the resonant process. Photons passing through an area with a magnetic field, both without interaction with the vacuum, and with participation in the CPPA process, make up two beams of a vacuum birefringence. The effect of a vacuum birefringence coincides with the classic limits if the area is small in size (when the change in the photon polarization is weak). If the primary photon beam is not polarized, then after passing through an area with a strong magnetic field due to the VB effect it will acquire a partial abnormal linear polarization $\xi'_3 \neq 0$. In resonant conditions, the photons are completely polarized in a magnetic field $H=10^{13}Gs$ after passing through an area of size $L = 1$ $\mu m$.

In the seventh section, the process of a production of a e⁺e⁻ pair by electron (PPE, trident process) is studied near the threshold in resonant conditions. Probability of the process in the resonant case is found and the influence of direction of spin of the initial electron on the process is analyzed. It is shown that the main contribution gives the resonance mode in the PPE probability per unit of time near the threshold, when the intermediate photon reaches a mass surface. In this case, the total probability of PPE process is factorized and expressed through the product of probabilities of first-order processes: SR and OPP. The most probable case is if the initial electron spin is directed along the field. The probability of the PPE process with the initial electron spin against the field is smaller by an order of the small parameter $h$. The PPE probability is inversely proportional to the energy width of an intermediate photon. In particular, at the threshold of reaction when the magnetic field is $h = 0.1$, the Landau level number of the initial electron is $l = 40$, such an estimate is found for PPE probability and width: $W = 1.2 \cdot 10^4 c^{-1}$, $\Delta = 4 \cdot 10^{17} c^{-1}$. The constructed theory of the PPE process in a magnetic field is used to explain the SLAC experiment, in which about 100 positrons in 21962 events were detected in collisions of an electron beam with an energy of 46.6 *GeV* with a beam of terawatt pulse laser. Nikishov-Ritos's theorem is used, according to which the formulas for the probabilities of QED processes in an external electromagnetic field expressed by the gauge invariants, for the case of ultrarelativistic initial particles, are the same for any configuration of the external electromagnetic field. The obtained value of 80 e⁺e⁻ pair is satisfactorily consistent with the experimental results (106 ± 14 events).

**Keywords:** Quantum electrodynamics, strong magnetic field, polarization, spin-flip, cyclotron resonance, lowest Landau levels.